\documentclass[twocolumn,traditabstract]{aa}  

\usepackage{amsmath}
\usepackage{fixltx2e}
\usepackage[english]{babel}
\usepackage{graphicx}
\usepackage{epstopdf}
\usepackage{epsf,color}
\usepackage[mathscr]{eucal}
\usepackage{amsmath}
\usepackage{amssymb,amsfonts}
\usepackage{natbib}
\usepackage{graphicx}
\usepackage{txfonts}
\usepackage{dsfont}
\definecolor{Mygreen}{rgb}{0.75, 0.0, 0.0}
\definecolor{Mypink}{rgb}{1.0, 0.0, 0.5}
\definecolor{Myred}{rgb}{0.7, 0.0, 0.0}
\usepackage[breaklinks, citecolor=blue, linkcolor=Myred, urlcolor=Myred, colorlinks=true, debug, baseurl=' ']{hyperref}
\usepackage{float} 
\usepackage{color}
\usepackage{ulem}

\bibpunct{(}{)}{;}{a}{}{,}
\newfont{\gwpfont}{cmssq8 scaled 1000}
\newcommand{\rexcess}{{\gwpfont REXCESS}}

\begin{document}
\title{Sub-structure and merger detection in resolved NIKA Sunyaev-Zel'dovich images of distant clusters \thanks{Based on observations carried out under project number 237-13, 110-14, and 222-14, with the NIKA camera at the IRAM 30 m Telescope. IRAM is supported by INSU/CNRS (France), MPG (Germany) and IGN (Spain). Data products, including the FITS file of the published maps are available at the NIKA2 SZ Large Program web page via \url{http://lpsc.in2p3.fr/NIKA2LPSZ/nika2sz.release.php}}}

\author{R.~Adam \inst{\ref{OCA},\ref{LPSC},\ref{CEFCA}}\thanks{Corresponding author: R\'emi Adam, \url{remi.adam@oca.eu}}
\and O.~Hahn\inst{\ref{OCA}}						
\and  F.~Ruppin \inst{\ref{LPSC}}
\and  P.~Ade \inst{\ref{Cardiff}}
\and  P.~Andr\'e \inst{\ref{CEA}}
\and M.~Arnaud\inst{\ref{CEA}}					
\and I.~Bartalucci\inst{\ref{CEA}}					
\and  A.~Beelen \inst{\ref{IAS}}
\and  A.~Beno\^it \inst{\ref{Neel}}
\and  A.~Bideaud \inst{\ref{Neel}}
\and  N.~Billot \inst{\ref{IRAME}}
\and  O.~Bourrion \inst{\ref{LPSC}}
\and  M.~Calvo \inst{\ref{Neel}}
\and  A.~Catalano \inst{\ref{LPSC}}
\and  G.~Coiffard \inst{\ref{IRAMF}}
\and  B.~Comis \inst{\ref{LPSC}}
\and  A.~D'Addabbo \inst{\ref{Neel},\ref{Roma}}
\and  F.-X.~D\'esert \inst{\ref{IPAG}}
\and  S.~Doyle \inst{\ref{Cardiff}}
\and C.~Ferrari\inst{\ref{OCA}}						
\and  J.~Goupy \inst{\ref{Neel}}
\and  C.~Kramer \inst{\ref{IRAME}}
\and  G.~Lagache \inst{\ref{LAM}}
\and  S.~Leclercq \inst{\ref{IRAMF}}
\and  J.-F.~Lestrade \inst{\ref{LERMA}}
\and  J.F.~Mac\'ias-P\'erez \inst{\ref{LPSC}}
\and G.~Martinez Aviles\inst{\ref{OCA}}				
\and D.~Martizzi\inst{\ref{Berkeley}}					
\and S.~Maurogordato\inst{\ref{OCA}}				
\and  P.~Mauskopf \inst{\ref{Cardiff},\ref{Arizona}}
\and  F.~Mayet \inst{\ref{LPSC}}
\and  A.~Monfardini \inst{\ref{Neel}}
\and  F.~Pajot \inst{\ref{IAS}}
\and  E.~Pascale \inst{\ref{Cardiff}}
\and  L.~Perotto \inst{\ref{LPSC}}
\and  G.~Pisano \inst{\ref{Cardiff}}
\and E.~Pointecouteau\inst{\ref{IRAP}, \ref{UniToulouse}}
\and  N.~Ponthieu \inst{\ref{IPAG}}
\and G.W.~Pratt\inst{\ref{CEA}}					
\and  V.~Rev\'eret \inst{\ref{CEA}}
\and M.~Ricci\inst{\ref{OCA}}						
\and  A.~Ritacco \inst{\ref{IRAME}}
\and  L.~Rodriguez \inst{\ref{CEA}}
\and  C.~Romero \inst{\ref{IRAMF}}
\and  H.~Roussel \inst{\ref{IAP}}
\and  K.~Schuster \inst{\ref{IRAMF}}
\and  A.~Sievers \inst{\ref{IRAME}}
\and  S.~Triqueneaux \inst{\ref{Neel}}
\and  C.~Tucker \inst{\ref{Cardiff}}
\and H.-Y.~Wu\inst{\ref{CalTech}}					
\and  R.~Zylka \inst{\ref{IRAMF}}}

\institute{
  Laboratoire Lagrange, Universit\'e C\^ote d'Azur, Observatoire de la C\^ote d'Azur, CNRS, Blvd de l'Observatoire, CS 34229, 06304 Nice cedex 4, France
  \label{OCA}
  \and
  Laboratoire de Physique Subatomique et de Cosmologie, Universit\'e Grenoble Alpes, CNRS/IN2P3, 53, avenue des Martyrs, Grenoble, France
  \label{LPSC}
    \and
  Centro de Estudios de F\'isica del Cosmos de Arag\'on (CEFCA), Plaza San Juan, 1, planta 2, E-44001, Teruel, Spain
  \label{CEFCA}
  \and
Institut de RadioAstronomie Millim\'etrique (IRAM), Grenoble, France
  \label{IRAMF}
\and
Laboratoire AIM, CEA/IRFU, CNRS/INSU, Universit\'e Paris Diderot, CEA-Saclay, 91191 Gif-Sur-Yvette, France 
  \label{CEA}
\and
Astronomy Instrumentation Group, University of Cardiff, UK
  \label{Cardiff}
\and
Institut d'Astrophysique Spatiale (IAS), CNRS and Universit\'e Paris Sud, Orsay, France
  \label{IAS}
\and
Institut N\'eel, CNRS and Universit\'e Grenoble Alpes, France
  \label{Neel}
\and
Institut de RadioAstronomie Millim\'etrique (IRAM), Granada, Spain
  \label{IRAME}
\and
Dipartimento di Fisica, Sapienza Universit\`a di Roma, Piazzale Aldo Moro 5, I-00185 Roma, Italy
  \label{Roma}
\and
Univ. Grenoble Alpes, CNRS, IPAG, F-38000 Grenoble, France 
  \label{IPAG}
    \and
Aix Marseille Universit\'e, CNRS, LAM (Laboratoire d'Astrophysique de Marseille) UMR 7326, 13388, Marseille, France
  \label{LAM}
\and
School of Earth and Space Exploration and Department of Physics, Arizona State University, Tempe, AZ 85287
  \label{Arizona}
\and
Universit\'e de Toulouse, UPS-OMP, Institut de Recherche en Astrophysique et Plan\'etologie (IRAP), Toulouse, France
  \label{IRAP}
\and
CNRS, IRAP, 9 Av. colonel Roche, BP 44346, F-31028 Toulouse cedex 4, France 
  \label{IRAP2}
\and
University College London, Department of Physics and Astronomy, Gower Street, London WC1E 6BT, UK
  \label{UCL}
\and 
Institut d'Astrophysique de Paris, Sorbonne Universit\'es, UPMC Univ. Paris 06, CNRS UMR 7095, 75014 Paris, France 
  \label{IAP}
\and 
LERMA, CNRS, Observatoire de Paris, 61 avenue de l'Observatoire, Paris, France
  \label{LERMA}
  \and  
Department of Astronomy, University of California, Berkeley, CA 94720-3411, USA
  \label{Berkeley}
    \and
Universit\'e de Toulouse, UPS-OMP, Institut de Recherche en Astrophysique et Plan\'etologie (IRAP), Toulouse, France
  \label{IRAP}
\and
CNRS, IRAP, 9 Av. colonel Roche, BP 44346, F-31028 Toulouse cedex 4, France 
  \label{UniToulouse}
  \and
California Institute of Technology, MC 367-17, Pasadena, CA 91125, USA.
  \label{CalTech}  
}

\date{Received \today \ / Accepted --}
\abstract {Sub-structures in the hot gas atmosphere of galaxy clusters are related to their formation history and to the astrophysical processes at play in the intracluster medium (ICM). The thermal Sunyaev-Zel'dovich (tSZ) effect is directly sensitive to the line-of-sight integrated ICM pressure, and is thus particularly adapted to study ICM sub-structures. In this paper, we apply structure-enhancement filtering algorithms to high resolution tSZ observations (e.g. NIKA) of distant clusters, in order to search for pressure discontinuities, compressions, as well as secondary peaks in the ICM. The same filters are applied to toy-model images and to synthetic tSZ images extracted from RHAPSODY-G cosmological hydrodynamic simulations, in order to better interpret the extracted features. We also study the noise propagation trough the filters and quantify the impact of systematic effects, such as data processing induced artifacts and point source residuals, the latter being identified as the dominant potential contaminant. In three of our six NIKA-observed clusters we identify features at high signal-to-noise that show clear evidence for merger events. In \mbox{MACS~J0717.5+3745} ($z=0.55$), three strong pressure gradients are observed on the east, southeast and west sectors, and two main peaks in the pressure distribution are identified. We observe a lack of tSZ compact structure in the cool-core cluster \mbox{PSZ1~G045.85+57.71} ($z=0.61$), and a tSZ gradient ridge dominates in the southeast. In the highest redshift cluster, \mbox{CL~J1226.9+3332} ($z=0.89$), we detect a $\sim 45$ arcsec (360 kpc) long ridge pressure gradient associated with a secondary pressure peak in the west region. Our results show that current tSZ facilities have now reached the angular resolution and sensitivity to allow an exploration of the details of pressure sub-structures in clusters, even at high redshift. This opens the possibility to quantify the impact of the dynamical state on the relation between the tSZ signal and the mass of clusters, which is important when using tSZ clusters to test cosmological models. This work also marks the first NIKA cluster sample data release.}
\titlerunning{Detection of sub-structures in high resolution tSZ maps}
\authorrunning{R. Adam, O. Hahn, F. Ruppin et al.}
\maketitle

\section{Introduction}
The internal structure of the hot ionized gas in galaxy clusters reflects their formation through the hierarchical merging of smaller structures and groups, and the accretion of surrounding material \citep[e.g.][and references therein]{Kravtsov2012}. It is tightly connected to turbulence, shocks and sloshing \citep[e.g.][]{Markevitch2007} in the intracluster medium (ICM), as well as various non-gravitational physical processes such as feedback from compact sources \citep[e.g.][]{Fabian2012}. The study of the structure of the ICM is therefore a unique way to understand how clusters form, and to assess connections with the astrophysics at play. As the intracluster gas is commonly used to trace the overall mass distribution of clusters, an investigation of cluster astrophysics is in turn essential to handle scatter and biases that arise in the mass--observable relations, which are fundamental when using clusters as cosmological probes \citep[see, e.g.][for a review]{Allen2011}.

The thermal Sunyaev-Zel'dovich \citep[tSZ,][]{Sunyaev1972} effect provides a direct probe of the integrated electron pressure along the line-of-sight in clusters. It is thus an excellent diagnostic of the thermodynamical properties of the ICM that are induced by gravitational dynamics and complements well X-ray imaging, which is sensitive to the squared electron density with weak temperature dependance. X-ray surface brightness provides a high density contrast, but cannot distinguish between cold front and shocks. The latter are accessible via the tSZ surface brightness through the pressure, but discontinuities that are in quasi-pressure equilibrium (e.g. cavities, cold fronts) will remain invisible in the tSZ signal, while they will show up in the X-ray. In addition, the tSZ effect is well suited to study distant clusters since, unlike other probes, its surface brightness is insensitive to distance.

The study of sub-structures in the ICM is now routinely applied to X-ray imaging. Dedicated filtering techniques have been developed in the literature, such as unsharp masking, or gradient filtering \citep[see, for example, recent results by][]{Sanders2016a}, highlighting ongoing physical processes in clusters, which would be missed otherwise (e.g. cold/shock front, jet cavity, sound waves, etc). However, only few applications of such procedures have been performed using tSZ data \citep[see e.g.][using simulated nearby Planck clusters]{Bourdin2015}. Indeed, these methods require both high angular resolution and high sensitivity observations, in order to obtain significant detections, and the corresponding data remain challenging to obtain beyond the local universe. With the advent of new state-of-the-art millimeter wave high angular resolution instruments such as NIKA2, installed on the IRAM 30m telescope \citep[The New IRAM KIDs Array 2, $< 20$ arcsec resolution at 150 and 260 GHz,][]{Calvo2016,Catalano2016,NIKA2017}, or MUSTANG2, on the Green Bank Telescope \citep[The MUltiplexed Squid Tes Array at Ninety Gigahertzh 2, $\sim 8$ arcsec at 90 GHz,][]{Dicker2014}, the use of tSZ data to study the inner structure of clusters is about to enter a new era. Alternatively, the use of interferometers such as ALMA (Atacama Large Millimeter/submillimeter Array) have already shown the huge potential of such observations to map the tSZ effect at unprecedented angular resolutions \citep{Kitayama2016,Basu2016}.

The NIKA2 pathfinder, NIKA \citep{Monfardini2011,Catalano2014}, has already been used to image galaxy clusters at high angular resolution, including deep observations \citep{Adam2014,Adam2015,Adam2016a,Adam2016b,Ruppin2016}. In this paper, we apply filtering methods to the NIKA tSZ maps in order to detect and study pressure sub-structures in the ICM of six clusters of galaxies at $0.45 \leq z \leq 0.89$. The same filtering methods are applied to toy models and to synthetic tSZ maps extracted from the RHAPSODY-G hydrodynamical simulations \citep{Wu2013,Hahn2017} to provide a better interpretation of the observed structures. We also study the noise propagation through the filters and quantify possible systematic effects arising from contaminating point sources and data processing. The detection of sub-structures allows us to infer the presence of ongoing merger activity of the targets and show the huge potential of future instrument for investigating cluster formation in distant clusters.

This paper is organized as follow. Section \ref{sec:Pressure_substructures_detection} describes the filtering algorithms. In Section \ref{sec:Application_to_toy_models} and \ref{sec:Application_to_hydrodynamical_simulations}, we apply the filtering procedure to toy models and to the RHAPSODY-G simulations. We investigate possible systematic effects and study the noise properties in Section \ref{sec:Systematics_and_noise_properties}. Finally, the filtering algorithms are applied to the NIKA cluster sample in Section \ref{sec:Application_to_the_NIKA_clusters_sample} and we discuss their implications in Section \ref{sec:discussions}. Summary and conclusions are provided in Section \ref{sec:Summary_and_conclusions}. Throughout this paper, we assume a flat $\Lambda$CDM cosmology according to \textit{Planck} results \citep{Planck2016XIII} with $H_0 = 67.8$ km s$^{-1}$ Mpc$^{-1}$, $\Omega_{\rm M} = 0.308$, and $\Omega_{\Lambda} = 0.692$.

\section{Detection of pressure sub-structures}\label{sec:Pressure_substructures_detection}
\subsection{The thermal Sunyaev-Zel'dovich effect}
The tSZ effect consists in the spectral distortion of the black-body spectrum of the cosmic microwave background (CMB) radiation. Its frequency dependence is given by \citep{birkinshaw1999}
\begin{equation}
	f(x, T_e) = \frac{x^4 e^x}{\left(e^x-1\right)^2} \left(x \ \mathrm{coth}\left(\frac{x}{2}\right) - 4\right) \left(1 + \delta_{\rm tSZ}(x, T_e) \right),
	\label{eq:sz_f_x}
\end{equation}
where $x = \frac{h \nu}{k_{\mathrm{B}} T_{\mathrm{CMB}}}$ is the dimensionless frequency, $h$ the Planck constant, $k_{\mathrm{B}}$ the Boltzmann constant, $\nu$ the observation frequency, and $T_{\mathrm{CMB}}$ the temperature of the CMB. The term $\delta_{\rm tSZ}(x,T_e)$ corresponds to relativistic corrections, which depend on the observing frequency and the electron temperature $T_e$ \citep[see, e.g.][]{Itoh2003}. The induced change in intensity with respect to the primary CMB intensity, $I_0$, can be expressed as
\begin{equation}
	\frac{\Delta I_{\rm tSZ}}{I_0} = y \ f(x, T_e) \ ,
\label{eq:deltaI}
\end{equation}
where $y$ is the Compton parameter, which measures the integrated electronic pressure, $P_{e}$, along the line-of-sight, $d\ell$, written as
   \begin{equation}
	y = \frac{\sigma_{\mathrm{T}}}{m_{\mathrm{e}} c^2} \int P_{e} \ d\ell.
	\label{eq:y_compton}
   \end{equation}
The parameter $\sigma_{\mathrm{T}}$ is the Thomson cross-section, $m_{\mathrm{e}}$ is the electron rest mass, and $c$ the speed of light. In this paper, we are using the NIKA 150 GHz data only, frequency at which the tSZ signal is close to its maximum decrement. The NIKA maps thus provide a direct measurement of the line-of-sight integrated electron pressure assuming relativistic effects are small.

\subsection{NIKA data}\label{sec:NIKA_Data}
NIKA has been used to map six clusters of galaxies using the tSZ effect at 150 and 260 GHz between November 2012 and February 2015. This sample contains both well known objects with various multi-wavelength coverage and Planck discovered sources, at intermediate and high redshifts. The names, coordinates, and main properties of the clusters are listed in Table \ref{tab:cluster_summary}, and by design, the sample contains a wide variety of cluster morphologies, spanning redshifts between 0.45 and 0.89. It was dedicated to provide pilot observations to test the feasibility of the NIKA2 \citep{NIKA2017} tSZ Large Program, which is under preparation \citep{Comis2016,Mayet2017}. 

The main steps of the data reduction are described in \cite{Adam2014,Adam2015}. The reduction of the raw NIKA data affects the reconstructed astrophysical sky signal by filtering out structures on large scales. We compute the angular transfer function of the reduction, as described in \cite{Adam2015} and use it to deconvolve the data to compensate for filtering effects. The absolute zero level of the brightness on the map remains unconstrained when observing clusters with NIKA, and it cannot be corrected using the transfer function of the processing. This corresponds to the transfer function being zero at angular wavenumber $k = 0$. In this paper, we use the NIKA 150 GHz tSZ maps only, deconvolved from the transfer function except for the beam smoothing (see Section \ref{sec:Transfer_function_filtering} and Appendix \ref{sec:Transfer_function_deconvolution}). The overall calibration uncertainty is estimated to be about 10\% depending on the observation campaign (see Table \ref{tab:cluster_summary}), including the brightness temperature model of our primary calibrator, the NIKA bandpass uncertainties, the opacity correction, and the stability of the instrument \citep{Catalano2014}. All clusters are contaminated by foreground and/or background galaxies that appear as point sources in the maps and they are removed according to the procedure described in \cite{Adam2015} and \cite{Adam2016a}. 

\begin{figure}[h]
\centering
\resizebox{0.5\textwidth}{!} {
\begin{tabular}{l}
\includegraphics[trim=0cm 0.7cm 0cm 0cm, clip=true, scale=1]{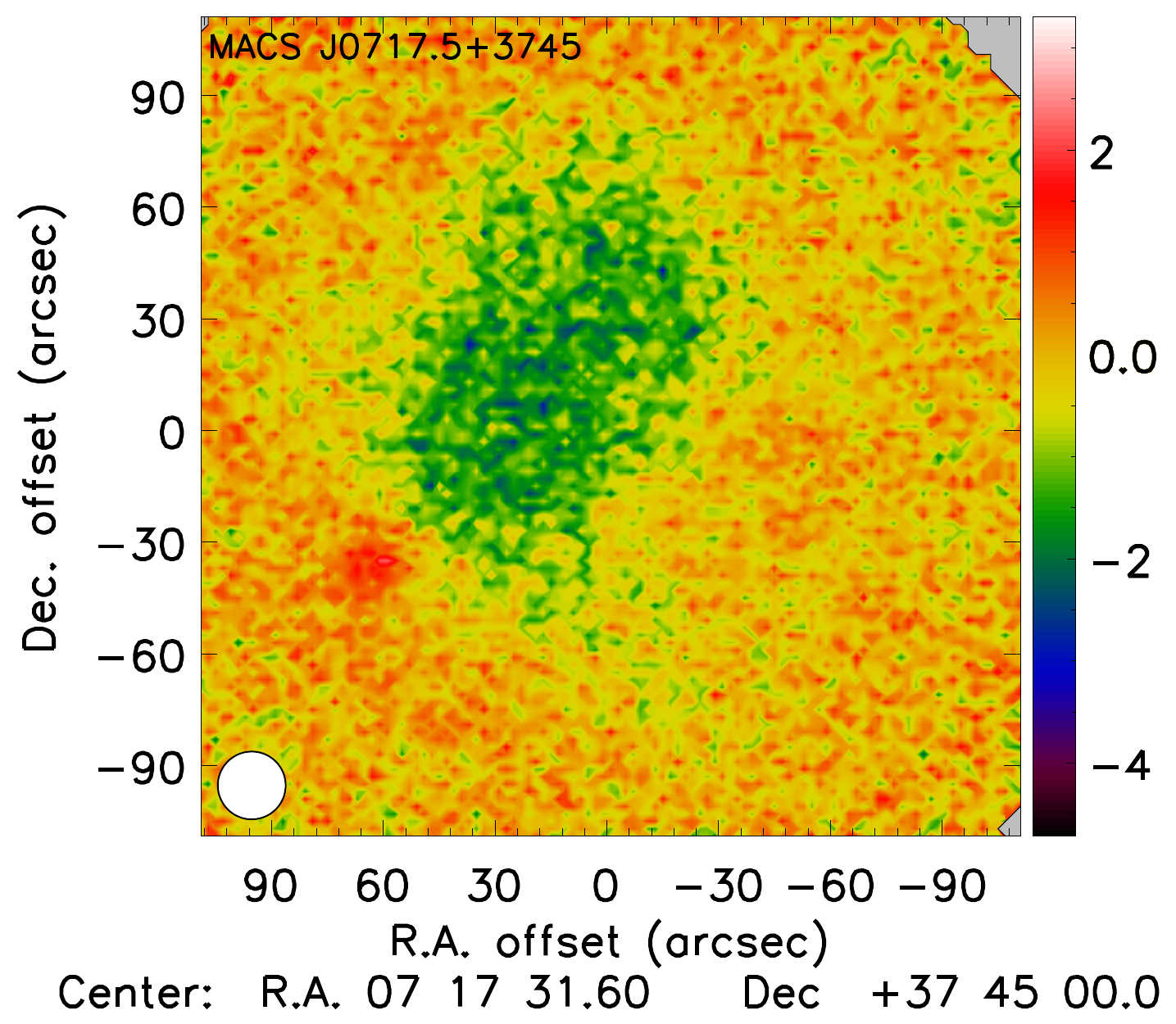} 
\put(-60,350){\makebox(0,0){\rotatebox{0}{\LARGE mJy/beam}}}
\end{tabular}}
\caption{\footnotesize{Example of the raw surface brightness image of \mbox{MACS~J0717.5+3745}, prior to any filter application (e.g. smoothing), point source removal and deconvolution. The data is dominated by noise on small scales. The cluster diffuse emission is nonetheless visible as a negative decrement, as well as a radio foreground galaxy (positive) on the southeast. The white circle on the bottom left corner provide the NIKA beam FWHM.}}
\label{fig:raw_image_macsj0717}
\end{figure}

The NIKA maps, $M$, are produced on 2 arcsec $\times$ 2 arcsec pixel grids to ensure proper Nyquist sampling, with respect to the 18.2 arcsec FWHM beam. On small scales, the data are dominated by Gaussian noise. The peak signal-to-noise, after smoothing the maps at a 22 arcsec FWHM resolution, depends strongly on the observing time and weather conditions, ranging from $\sim 4 \sigma$ to about $20 \sigma$ for the deepest observations. As an example, Figure \ref{fig:raw_image_macsj0717} provides the raw image of one of the NIKA cluster, \mbox{MACS~J0717.5+3745}, to illustrate the nature of the data.

\begin{table*}[]
\caption{\footnotesize{Summary of the main properties of the NIKA cluster sample, ordered by increasing redshift.}}
\begin{center}
\resizebox{\textwidth}{!} {
\begin{tabular}{c|c|c|c|c|c||c|c|c|c}
\hline
\hline
Name & $z$ & kpc/arcsec & $M_{500}$ & $Y_{500}$ & Comments & Configuration & Calibration uncertainty & On source time & Central rms$^{(b)}$ \\
 &  & & ($10^{14}$ M$_{\odot}$)& ($10^{-3}$arcmin$^2$) & & & & (hour) & (mJy/beam) \\
\hline
RX~J1347.5-1145 & 0.452 & 5.9 & 11.0 & 1.40 & Cool-core+merger & Nov. 2012, preliminary instrument & 15\% & 5.8 & 1.2 \\ 
MACS~J1423.8+2404 & 0.545 & 6.6 & 4.9 $^{(a)}$ & 0.47 $^{(a)}$ & Relaxed elliptical cool-core & Feb. 2014, Open Pool & 7\% &1.5 & 0.35 \\ 
MACS~J0717.5+3745 & 0.546 & 6.6 & 11.5 & 1.74 & Multiple merger, strong kSZ & Jan./Feb. 2014 \& 2015, Open Pool & 7\% & 13.1 & 0.10 \\ 
PSZ1~G046.13+30.75 & 0.569 & 6.7 & 6.4 & 0.70 & Disturbed & Nov. 2015, Open Pool & 9\% & 6.0 &  0.32\\ 
PSZ1~G045.85+57.71 & 0.611 & 6.9 & 7.0 & 0.92 & Elliptical cool-core & Nov. 2015, Open Pool & 9\% & 6.4 & 0.17 \\ 
CL~J1226.9+3332 &  0.888 & 8.0 & 5.7 & 0.90 & Disturbed core & Feb. 2014, Open Pool & 7\% & 7.8 & 0.17 \\ 
\hline
\end{tabular}
}
\end{center}
{\small {\bf Notes.} The masses are extracted from the \textit{Planck} catalog \citep{PlanckXXVII2015}. The integrated Compton parameter, $Y_{500}$, is computed by normalizing $Y_{5R_{500}}$ (extracted from the \textit{Planck} catalog) by a factor of 1.79, assuming a Universal pressure profile \citep{Arnaud2010}. $^{(a)}$ From \cite{Adam2016a}, as this cluster is not in the \textit{Planck} catalog. $^{(b)}$ At the 22 arcsec FWHM effective angular resolution.}
\label{tab:cluster_summary}
\end{table*}

\subsection{Algorithms}\label{sec:Algorithms}
In order to reveal pressure sub-structures, we make use of two algorithms, namely a Gaussian Gradient Magnitude \citep[GGM, see also][]{Roediger2013,Sanders2016b} filter and a Difference of Gaussians \citep[DoG, similar to unsharp-masking, as used also in X-ray analysis, see e.g.][]{Fabian2003} filter. They allow us to identify strong pressure gradients or discontinuities, and pressure peaks at specific scales, respectively.

\subsubsection{Gaussian gradient magnitude filter}
Because the NIKA data are dominated by noise on small scales, we convolve the maps with a Gaussian kernel, $G_{\theta_0}$, of FWHM $\theta_0$, to reduce it. We then compute the magnitude of the gradient of the maps as 
\begin{equation}
	M_{\rm GGM} = \sqrt{\left(\mathcal{D}_{\rm R.A.} \ast \left[G_{\theta_0} \ast M\right]\right)^2 + \left(\mathcal{D}_{\rm Dec.} \ast \left[G_{\theta_0} \ast M\right]\right)^2},
	\label{eq:GGM_filter}
\end{equation}
where the convolution kernel $\mathcal{D}_{\rm R.A., Dec.}$ along the R.A. and Dec. axis, are respectively given by
\begin{equation}
	\mathcal{D}_{\rm R.A.}= \mathcal{D}_{\rm Dec.}^{\rm T} = \frac{1}{8 \Delta \theta}
	\begin{pmatrix}
	-1 & 0 & 1\\
	-2 & 0 & 2\\
	-1 & 0 & 1
	\end{pmatrix},
	\label{eq:GGM_kernel}
\end{equation}
where $\Delta \theta$ is the map pixel size. The resulting quantity, $M_{\rm GGM}$, is therefore a measurement of the projected pressure gradient on scale $\theta_0$. It is expressed in units of surface brightness per arcmin, and can be converted to physical units of keV/cm$^3$ Mpc per Mpc accounting for the angular diameter distance of the cluster and the conversion coefficients from Compton parameter to Jy/beam given in \cite{Adam2016b}.

The direction of the gradient is obtained by computing its orientation on the sky, given by the angle
\begin{equation}
	\Psi = {\rm atan} \left( \frac{\mathcal{D}_{\rm Dec.} \ast \left[G_{\theta_0} \ast M\right]}{\mathcal{D}_{\rm R.A.} \ast \left[G_{\theta_0} \ast M\right]} \right).
	\label{eq:GGM_angle}
\end{equation}

\subsubsection{Difference of Gaussian filter}
In order to extract pressure sub-structures on a specific scale, we compute the difference of the NIKA maps convolved respectively with Gaussian kernels of FWHM $\theta_1$ and $\theta_2$:
\begin{equation}
	M_{\rm DoG} = G_{\theta_1} \ast M - G_{\theta_2} \ast M.
	\label{eq:DoG_filter}
\end{equation}
The filter thus removes signal on scales larger than $\theta_2$ and on scale smaller than $\theta_1$, allowing us to search for structures in a narrow range of angular scales. The resulting map is homogeneous to the input map, i.e. it is sensitive to the projected pressure along the line-of-sight in the selected range of angular scales.

\subsubsection{Baseline filtering parameters}\label{sec:Baseline_filtering_parameters}
The NIKA beam FWHM is 18.2 arcsec at 150 GHz \citep{Catalano2014}. On small scales, the noise dominates over the signal and therefore, the smallest scale at which we can search for sub-structures is slightly below, but of the order of the beam FWHM, depending on the signal-to-noise of the data. As the data reduction attenuates the signal on scales larger than $\sim 2$ arcmin \citep[see][and Section \ref{sec:Systematics_and_noise_properties}]{Adam2015}, the signal is also dominated by noise on large scales. Consequently the spatial dynamics accessible with NIKA data is about one order of magnitude, between 15 and 150 arcsec (limited by the beam and signal processing, respectively). Here, we aim at detecting pressure discontinuities and compression regions in distant clusters ($z>0.4$), which requires to probe the smallest scales available. If present, sub-components are expected to have typical angular size smaller than that of the main clusters ($\sim$ 1--2 arcmin) by a factor of a few. Therefore, we focus on the following baseline parameters for our filters: $\theta_0 = 15$ arcsec, and $\left(\theta_1, \theta_2\right) = \left(15, 45\right)$ arcsec. In practice, it is possible to increase the signal-to-noise ratio of the filtered maps by increasing the filters scales, at the cost of washing out the signal arising from the sub-structures we aim to detect. In Appendix \ref{sec:Impact_of_the_filter_parameters}, we show how the extracted signal changes according to the filter parameters.

\section{Application to cluster toy models}\label{sec:Application_to_toy_models}
The filtering algorithms are first tested on toy models in order to highlight the signature of the filtered signal in idealized physical situations. The toy models that we consider are: a spherical generalized Navarro Frenk and White (gNFW) pressure profile \citep{Nagai2007}, a bimodal pre-merging system, a main core plus an extension, and a pressure profile that includes a shock, respectively. We apply the filters using the baseline parameters discussed in Section \ref{sec:Baseline_filtering_parameters}.

\subsection{Construction of toy models}
The construction of the toy models is described below. The clusters redshift are $z=0.6$ because it corresponds to the mean redshift of the NIKA sample (see Section \ref{sec:NIKA_Data} and Table \ref{tab:cluster_summary}), but we also test other redshifts between $z=0.4$ and $z=1$.
\begin{enumerate}
\item {\bf Spherical gNFW profile:} one of the most common and simple description of clusters pressure is that of a spherically symmetric radial distribution following a gNFW \citep{Nagai2007} profile, and before using the filtered maps to investigate disturbances in the ICM, it is important to understand the response of the filters to perfectly spherical cluster signals. We thus simulate such a cluster using the best fit slopes and concentration parameters obtained by \cite{Planck2013V} and assuming a characteristic radius $R_{500} = 1000$ kpc (comparable to NIKA clusters). The pressure is integrated along the line of sight, to produce an azimuthally symmetric Compton parameter map (Eq. \ref{eq:y_compton}), which we convert into surface brightness in the 150 GHz NIKA band (Eq. \ref{eq:sz_f_x}). We also use best fit parameter profiles of cool-core and disturbed clusters from \cite{Arnaud2010} to test changes in the steepness of the profile.
\item {\bf Bimodal pre-merger system:} clusters can form via the merging of sub-clusters and groups. We model such system using our gNFW model to simulate a pair of identical clusters, separated by 100 arcsec. We add a pressure bar component between the two clusters to mimic the adiabatic compression caused by the merger. The bar has an arbitrary projected size of $R_{500}/5$ along the merger direction and $R_{500}/1.5$ in the perpendicular direction. The tSZ amplitude of the bar is 0.3 times that of the gNFW components peaks.
\item {\bf Main core plus extension:} it is common to find clusters made of a main core plus an extension that arises from both post-merger and pre-merger disturbances in the ICM. We model this system by a main gNFW cluster, to which we add a second gNFW sub-cluster with $R_{500} = 300$ kpc, normalized to a peak of 1/3 that of the main core, and located 50 arcsec away.
\item {\bf Shocks:} merging events can cause discontinuities in the ICM gas, such as shocks in the pressure. We thus simulate a radially symmetric cluster including a pressure discontinuity in the profile. We use the Rankine-Hugoniot jump conditions \citep[see e.g.][for a review]{Sarazin2002}, given by $\frac{P_1}{P_2} = \frac{2 \gamma_{\rm ad}}{\gamma_{\rm ad}+1} \mathcal{M}^2 + \frac{\gamma_{\rm ad}-1}{\gamma_{\rm ad}+1}$, with $P_1$ and $P_2$ the pressure before and after the shock and $\gamma_{\rm ad} = 5/3$ the adiabatic index. We use a Mach number $\mathcal{M} = 3$ as observed in typical shocks \citep[e.g.][]{Markevitch2007}. The pressure profile before and after the jump is described by a power law with index 0 in the inner part (to avoid confusion with the signal that would arise from a steep core, as in the spherical gNFW profile), and index -2 in the outer part, respectively. The pressure is integrated along the line of sight and the resulting Compton parameter profile is used to generate the surface brightness. We consider a shock moving outward, with respect to the cluster center.
\end{enumerate}
The toy model images are finally convolved with the NIKA beam and normalized to have a typical peak surface brightness of -1 mJy/beam. This corresponds to a Compton parameter of about $10^{-4}$ and it is also the typical scale of the NIKA clusters.

\subsection{Application of the filters}
\begin{figure}[h]
\centering
\resizebox{0.5\textwidth}{!} {
\begin{tabular}{lll}
\includegraphics[trim=0cm 2.2cm 0cm 0cm, clip=true, scale=1]{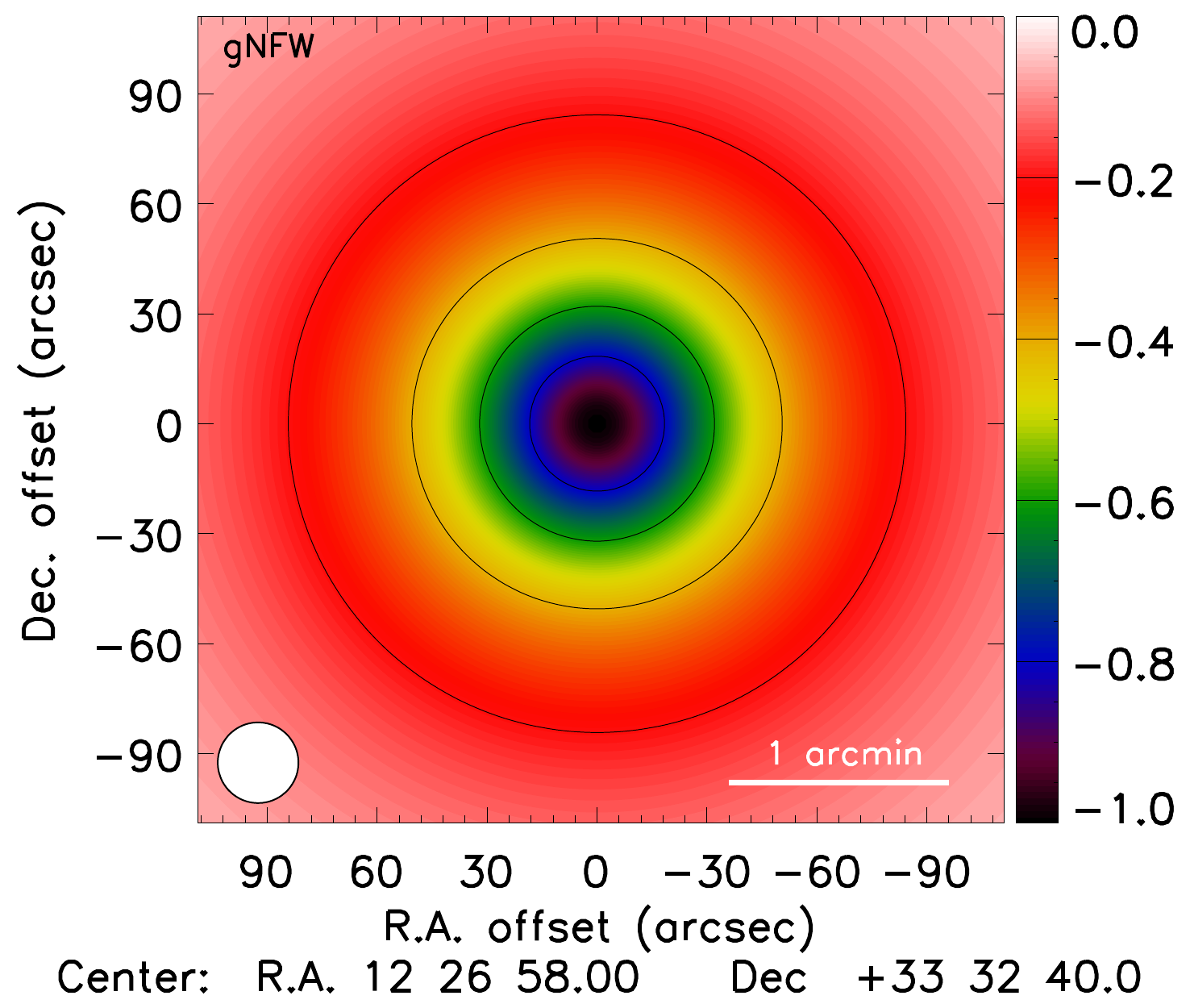} 
\put(-60,310){\makebox(0,0){\rotatebox{0}{\LARGE mJy/beam}}} & 
\includegraphics[trim=2.3cm 2.2cm 0cm 0cm, clip=true, scale=1]{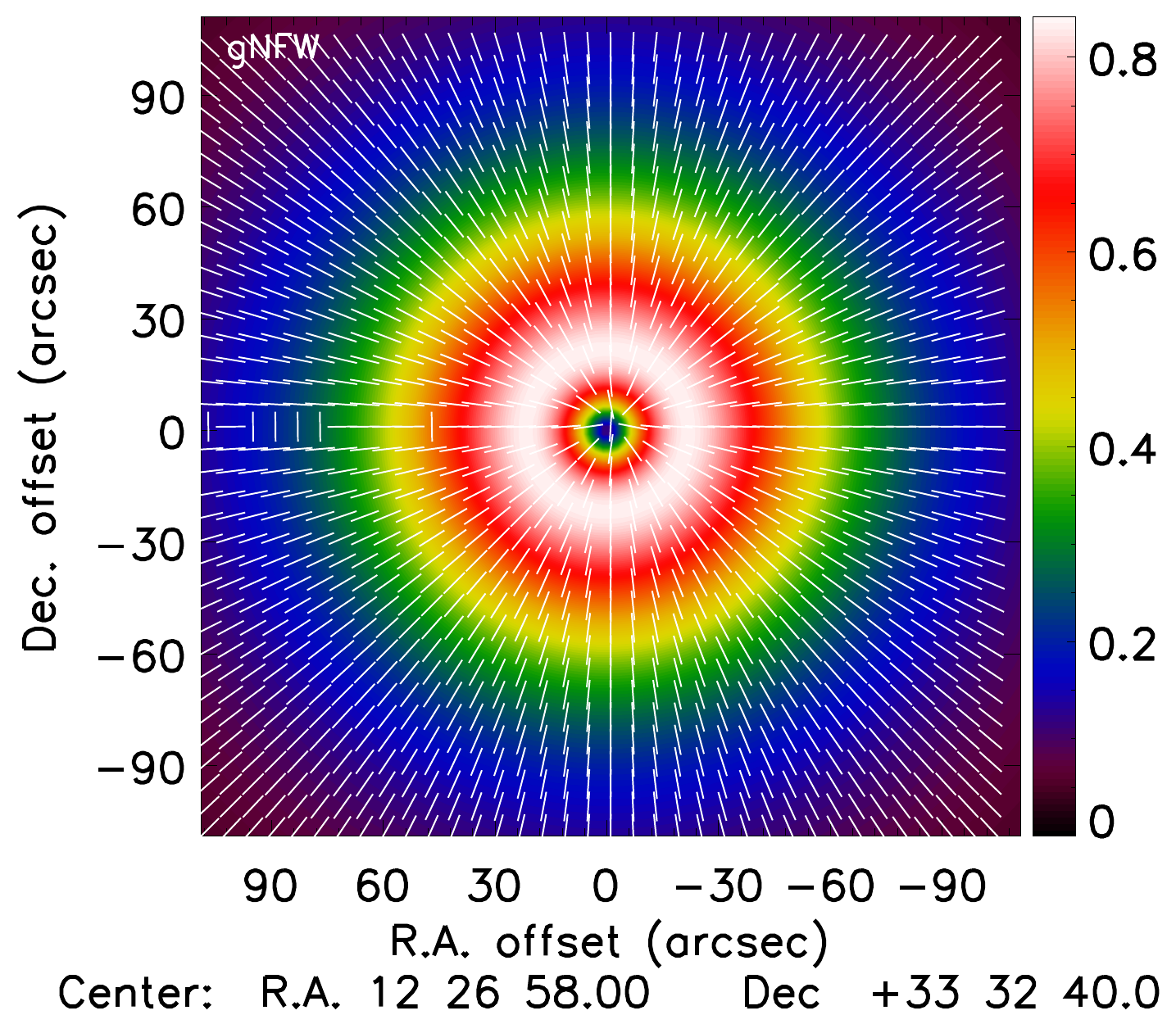} 
\put(-60,310){\makebox(0,0){\rotatebox{0}{\LARGE mJy/beam/arcmin}}} & 
\includegraphics[trim=2.3cm 2.2cm 0cm 0cm, clip=true, scale=1]{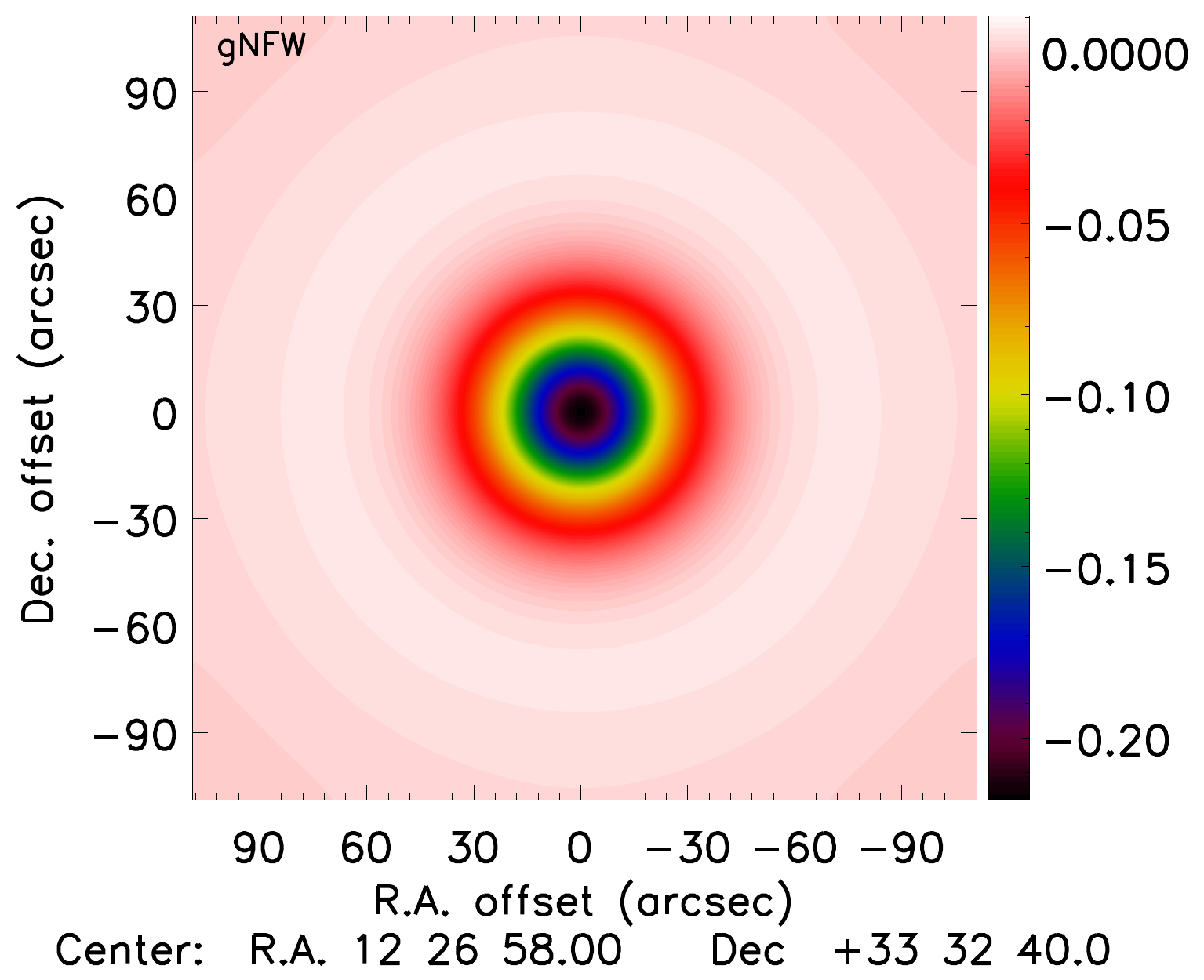} 
\put(-60,310){\makebox(0,0){\rotatebox{0}{\LARGE mJy/beam}}} \\
\includegraphics[trim=0cm 2.2cm 0cm 0cm, clip=true, scale=1]{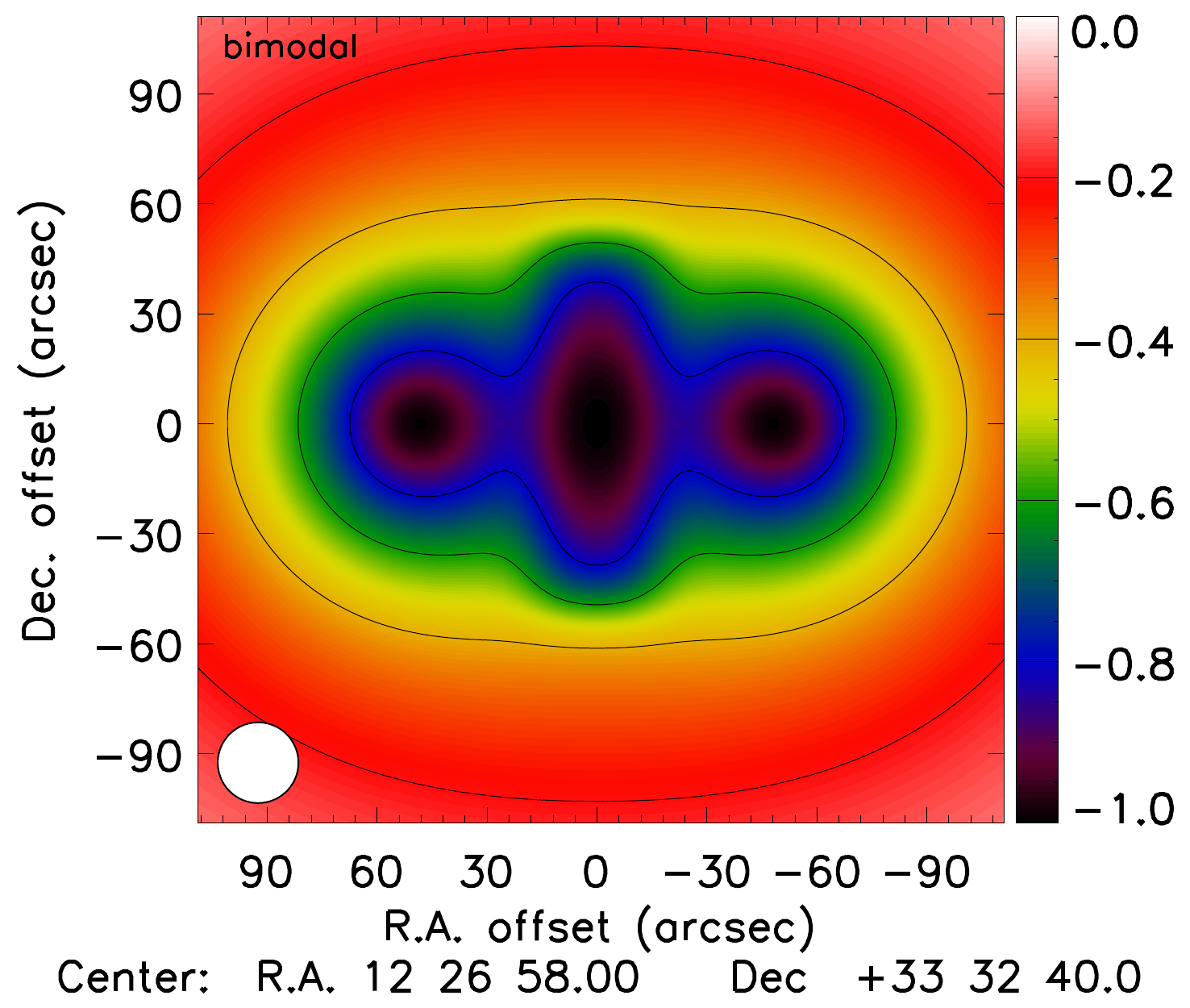} &
\includegraphics[trim=2.3cm 2.2cm 0cm 0cm, clip=true, scale=1]{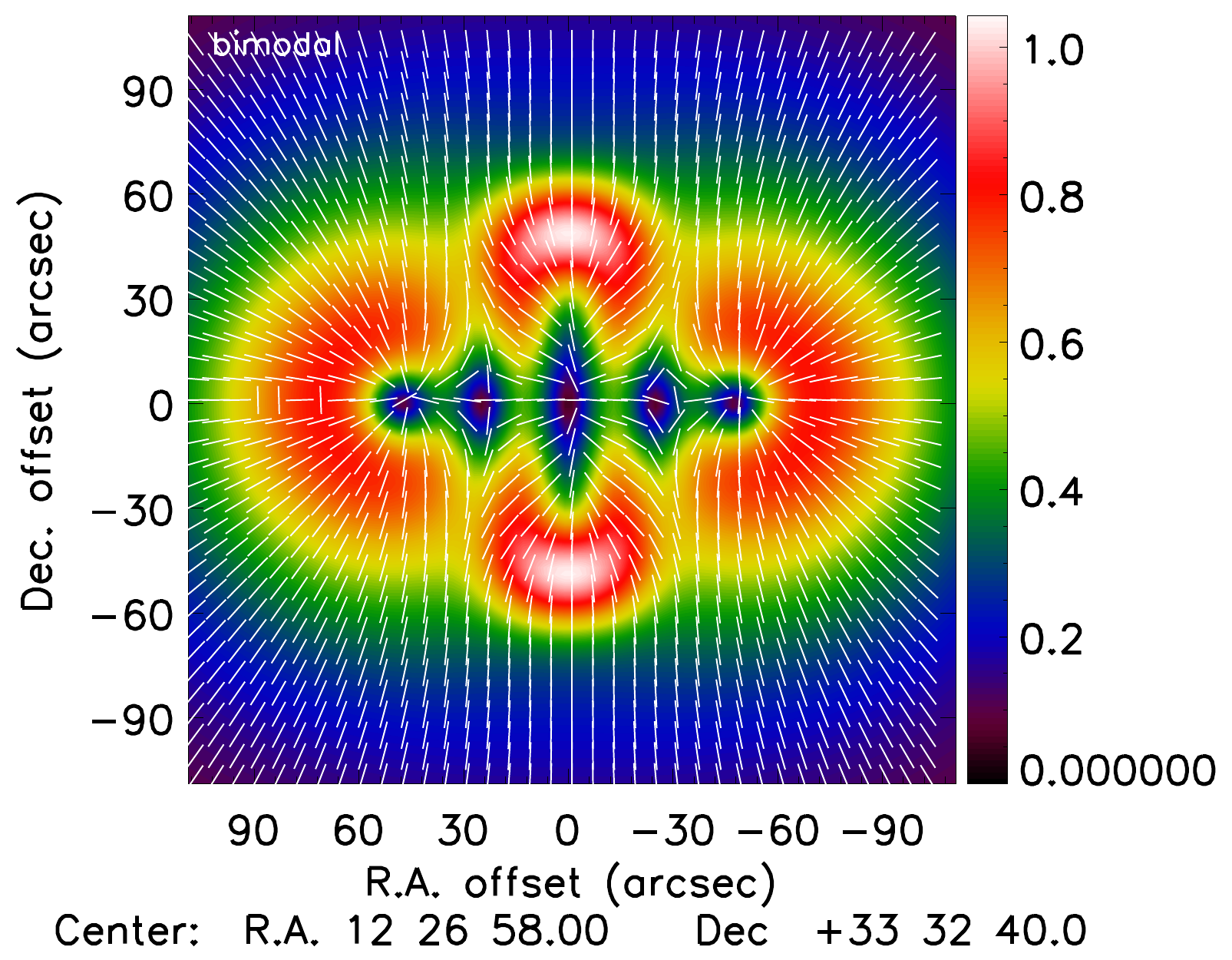} &
\includegraphics[trim=2.3cm 2.2cm 0cm 0cm, clip=true, scale=1]{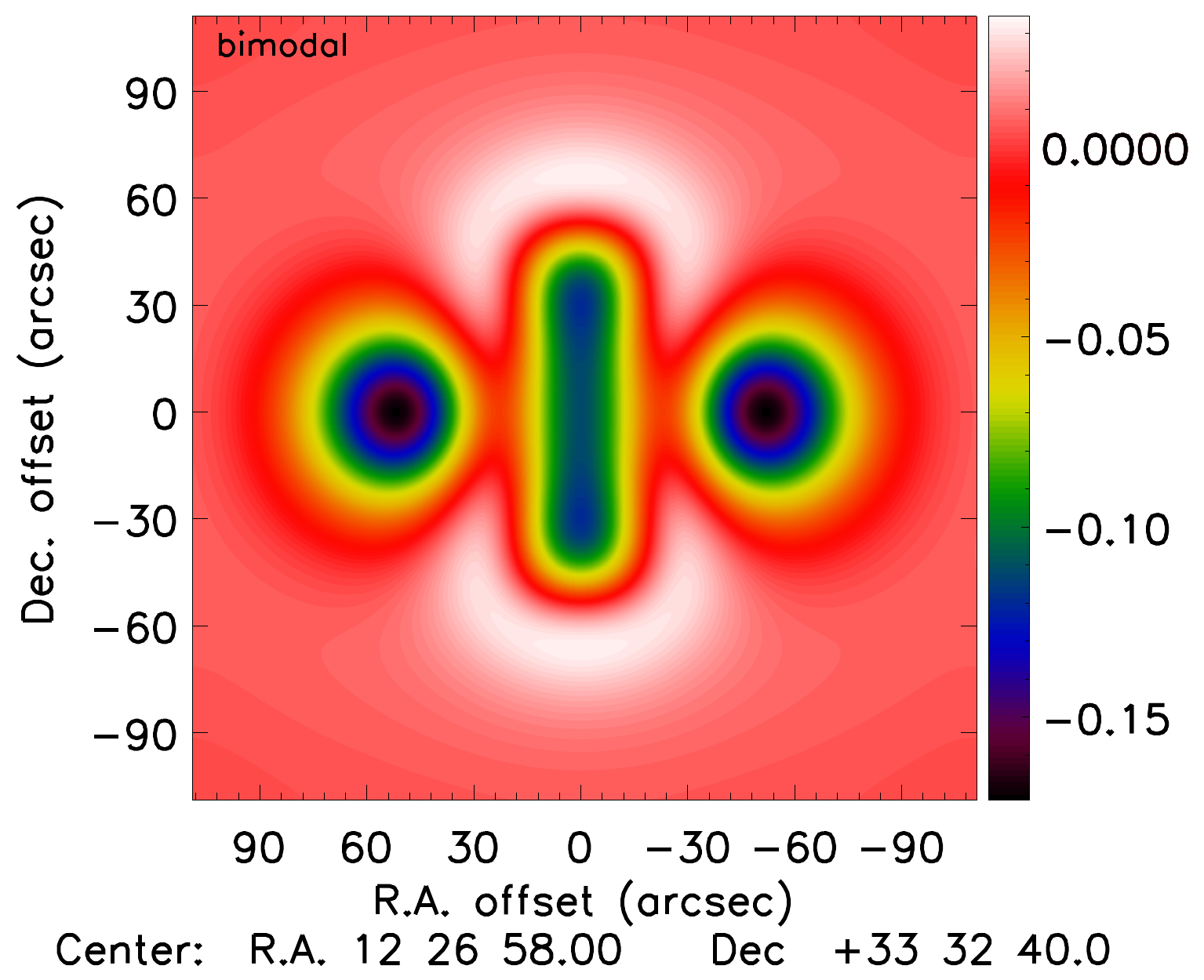} \\
\includegraphics[trim=0cm 0.7cm 0cm 0cm, clip=true, scale=1]{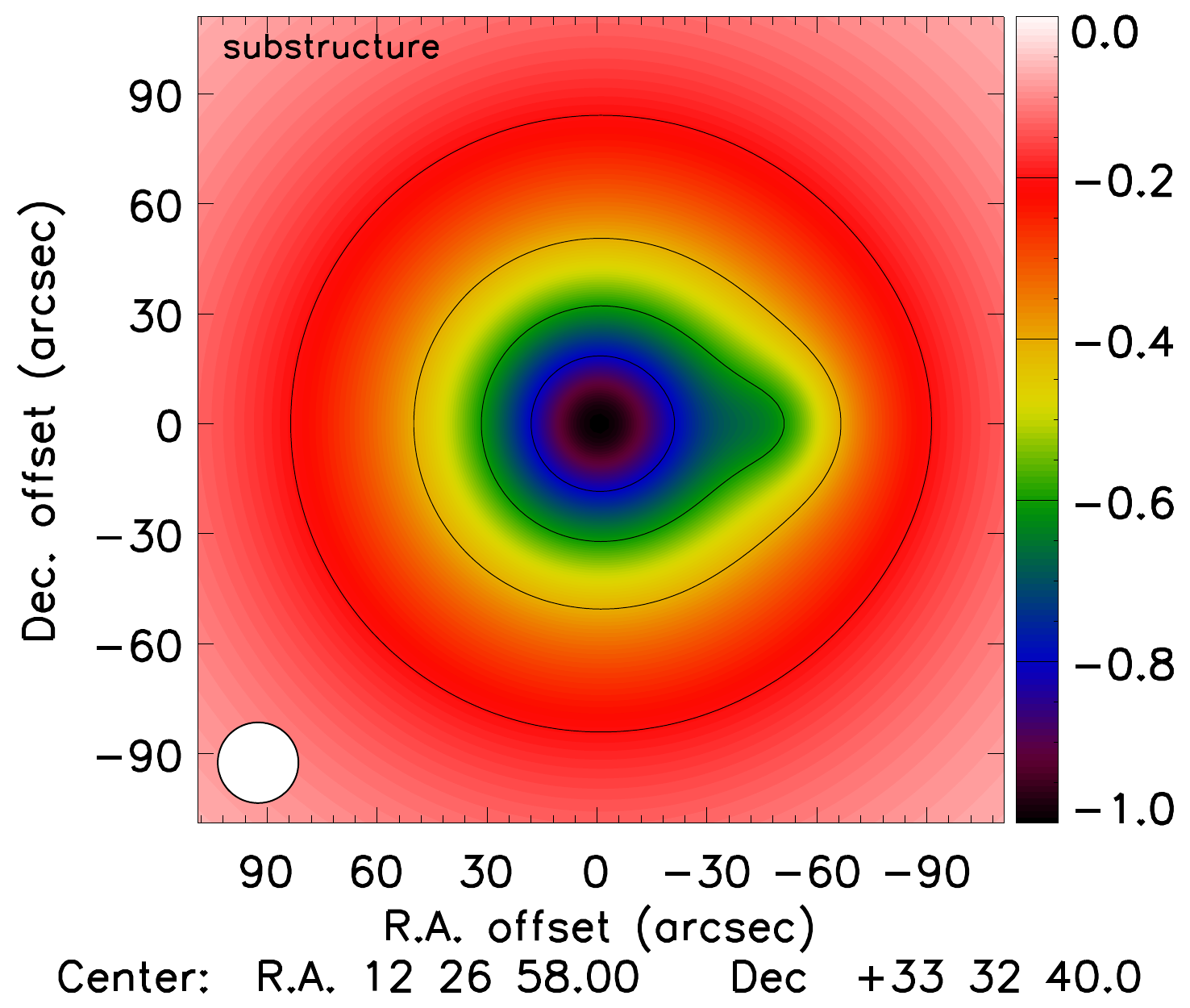} &
\includegraphics[trim=2.3cm 0.7cm 0cm 0cm, clip=true, scale=1]{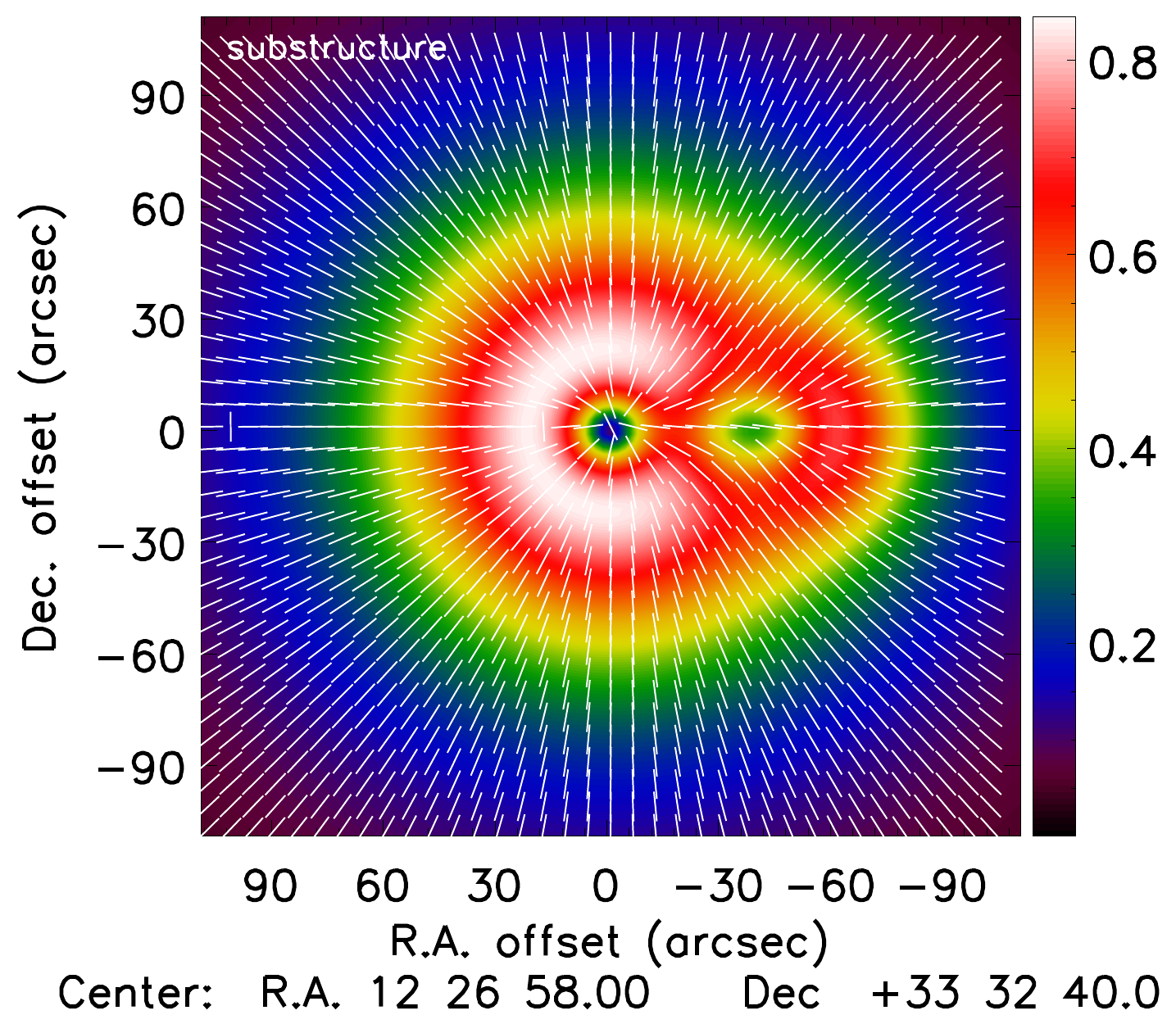} &
\includegraphics[trim=2.3cm 0.7cm 0cm 0cm, clip=true, scale=1]{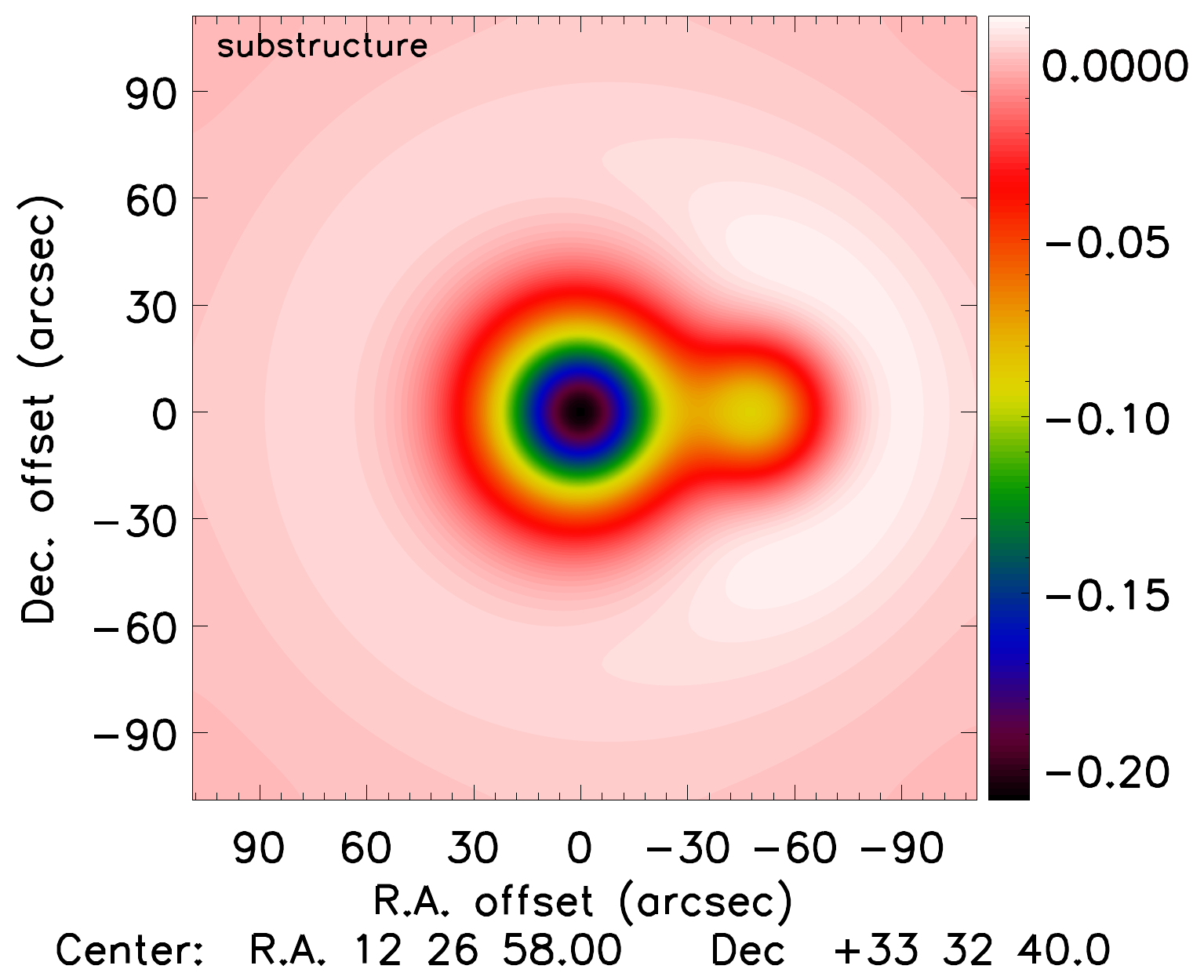} 
\end{tabular}}
\caption{\footnotesize{Surface brightness and GGM and DoG response to the tSZ signal expected for a spherically symmetric gNFW pressure profile (top), a bimodal cluster plus a pressure bar (middle), and a main core plus an extension (bottom).
{\bf Left:} tSZ simulated surface brightness. The white circle on the bottom left provides the beam FWHM.
{\bf Middle:} GGM filtered maps with $\theta_0 = 15"$. The white vectors represent the direction of the gradient, $\Psi$. 
{\bf Right:} DoG filtered maps with $\theta_1 = 15"$ and $\theta_2 = 45"$.}}
\label{fig:test_filter_gNFW_and_bimodal}
\end{figure}

\begin{figure*}[h]
\centering
\includegraphics[trim=0cm 0cm 0cm 0cm, clip=true, width=0.49\textwidth]{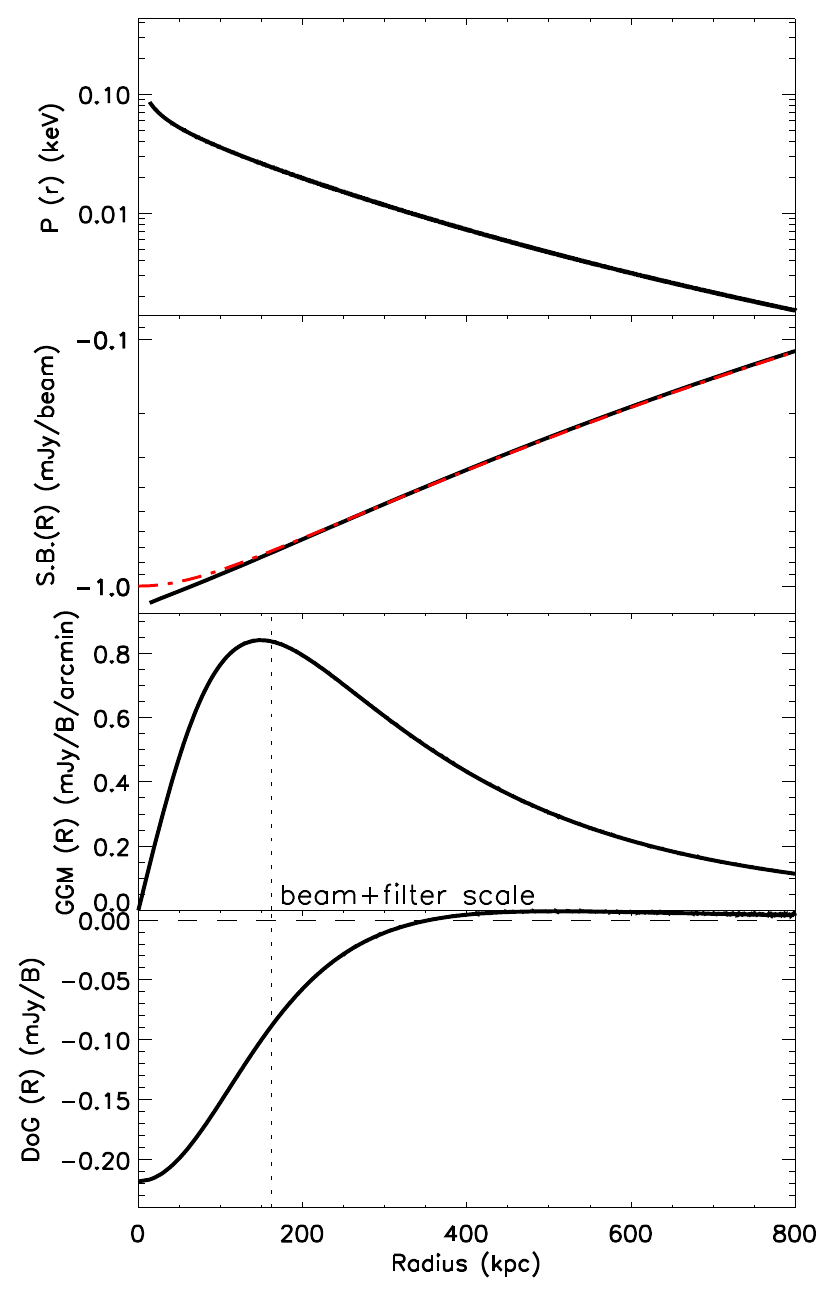}
\includegraphics[trim=0cm 0cm 0cm 0cm, clip=true, width=0.49\textwidth]{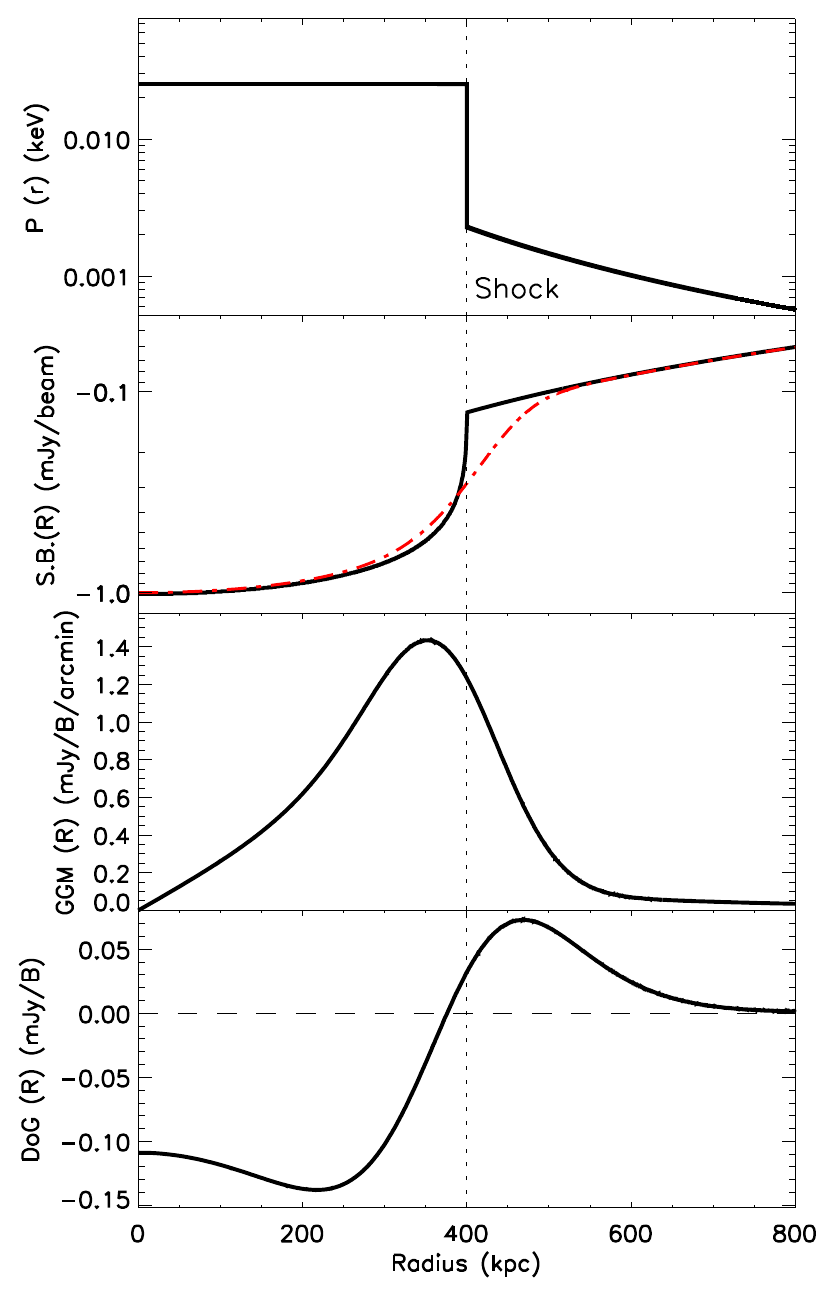}
\caption{\footnotesize{{\bf Left:} GGM and DoG response to a gNFW radial pressure profile. The vertical dashed line provide the physical size corresponding to the 15 arcsec filter plus the NIKA beam size, at the cluster redshift. 
{\bf Right:} GGM and DoG response to a shock propagating outward, in a radially symmetrical way, with Mach number $\mathcal{M} = 3$. The shock position is given by the vertical dashed line. 
From top to bottom, we show the pressure, the surface brightness, the GGM and the DoG radial profile. In the surface brightness case, the thick black lines provide the raw profile while the thin red dashed-line accounts for beam smoothing. The pressure depends of the physical radius in 3D, while the other quantities depend on the projected radius, in 2D.}}
\label{fig:test_filter_GNFW_and_shock}
\end{figure*}

We first consider the gNFW model (see the maps of Figure \ref{fig:test_filter_gNFW_and_bimodal}, top row). Because the pressure distribution is spherically symmetric, the surface brightness and the filtered maps are azimuthally symmetric (see the profiles of Figure \ref{fig:test_filter_GNFW_and_shock}, left panel). The surface brightness profiles are peaked toward the center. We thus observe a ring with a radius that matches the filter plus beam size on the GGM map and profile. This means that at the angular scales recovered by NIKA and the redshift we consider, we are not able to observe a flattening in the profile in the core (if it exists), and the profile surface brightness appears always steeper as we get closer to the center. We observe a ring radius larger than the filter plus beam size, i.e. we detect a flattening in the core, only when reaching redshifts lower than 0.5, in the case of the disturbed clusters pressure profile \citep[i.e. the flatter that we consider here][]{Arnaud2010}. On larger scales, the GGM map smoothly vanishes. The null of the GGM map coincides with the cluster center due to smoothing. The typical scale of the gradient peak is 0.8--1.2 mJy/beam/arcmin, for a surface brightness normalized to -1 mJy/beam at the maximum decrement. The DoG maps allow us to extract the core signal, whose amplitude is typically 20--30\% of that of the peak at our scales.

The middle row of Figure \ref{fig:test_filter_gNFW_and_bimodal} provides the maps of the bimodal cluster. The GGM maps is elongated along the merger axis. The strong gradients on the north and south are due to the bar, that produces an artificial edge on the surface brightness (before filter and beam smoothing), and we also observe the gradient resulting of the gNFW cores on the east and west sides. The DoG map allows us to clearly define the different components that would not necessarily be obvious on the surface brightness map in particular in the presence of noise (see also Section \ref{sec:Systematics_and_noise_properties}).

The case of the main core plus extension is given on the bottom row of Figure \ref{fig:test_filter_gNFW_and_bimodal}. The GGM map presents two minima that correspond to the two cores and the gradient is reduced between the two sub-clusters, with respect to the single gNFW profile. Similarly to the bimodal case, the DoG map allow us to clearly identify the two components and their locations.

Figure \ref{fig:test_filter_GNFW_and_shock} (right panel) presents the toy model and the corresponding filter response as a function of radius, in the case of the shock, similarly to the signal that one would extract along cones in the direction of the shock. It provides the pressure profile ($P$, as a function of the physical radius, $r$, in 3D), the surface brightness profile (as a function of the projected radius $R$, in 2D), and the filtered map profiles, ${\rm GGM}(R)$ and ${\rm DoG}(R)$, respectively. The shock location is $r_{\rm shock} = 400$ kpc. Because of line of sight integration, the surface brightness profile does not present a trivial discontinuity. Instead, it progressively vanishes before the shock. The sharp discontinuities and peaks are then degraded due to instrumental beam smoothing. The GGM filter allow us to pick up the strong gradients in the map, and is thus well designed for highlighting shocks. It presents a peak associated to the shock, but it is not coincident with it (due to smoothing). The DoG filtered maps allows us to highlight compact sub-structures and the filter is efficient to pick up the cluster core. For such a shock and filter parameters, the GGM amplitude typically reaches 1-2 mJy/beam/arcmin, depending on the distance from the center via the amplitude of the surface brightness at the shock location. The DoG amplitude reaches -0.1-0.5 mJy/beam.

The different toy models show the expected signal after application of the filters under simplistic assumptions, but they correspond to typical features that one can expect in the NIKA sample. It is clear that the filtered maps provide a new way to look at our data, in particular in terms of morphological studies. At the angular scales probed by NIKA and the redshifts we consider (i.e., that of the NIKA sample), we can however see some limitations when pushing the extraction of the features to smaller angular scales. As a typical example, Figure \ref{fig:test_filter_GNFW_and_shock} (right panel) shows that a shock will be hardly distinguishable from a ramp at redshift above 0.5. Similarly, the cluster core, even if purely spherical, will also show up at amplitudes similar to that of the signal arising from merging events that we want to extract. 

The most accessible features correspond to scales of a few tens of arcsec, i.e. a few hundred kpc. After this preparatory exercise on toy models, we will next test the performances of the filters under controlled but realistic situations using cosmological hydrodynamic simulations.

\section{Application to hydrodynamical simulations}\label{sec:Application_to_hydrodynamical_simulations}
\begin{table*}[]
\caption{\footnotesize{Summary of the properties of the RHAPSODY-G selected cluster sample.}}
\begin{center}
\begin{tabular}{c|c|c|c|c|c}
\hline
\hline
Name & $z$ & kpc/arcsec & $M_{500}$ ($10^{14}$ M$_{\odot}$)& $Y_{500}$ ($10^{-3}$arcmin$^2$) & Comments \\
\hline
RG361\_00188 & 0.61 & 6.9 & 3.8 & 0.23 & Very relaxed, elongated along the line-of-sight \\ 
RG474\_00172 & 0.90 & 8.0 & 4.1 & 0.20 & Major merger after the first crossing of the two main cores \\ 
RG377\_00181 & 0.54 & 6.5 & 3.7 & 0.24 & Multiple merger \\
RG448\_00211 & 0.40 & 5.5 & 3.5 & 0.29 & Pressure shell caused by the central AGN \\ 
RG474\_00235 & 0.39 & 5.4 & 12.7 & 2.26 & Major merger, mostly along the line-of-sight \\ 
\hline
\end{tabular}
\end{center}
\label{tab:rhapsody_summary}
\end{table*}

\begin{figure*}[h]
\resizebox{\textwidth}{!} {
\begin{tabular}{llll}
\includegraphics[trim=0cm 2.2cm 0cm 0cm, clip=true, scale=1]{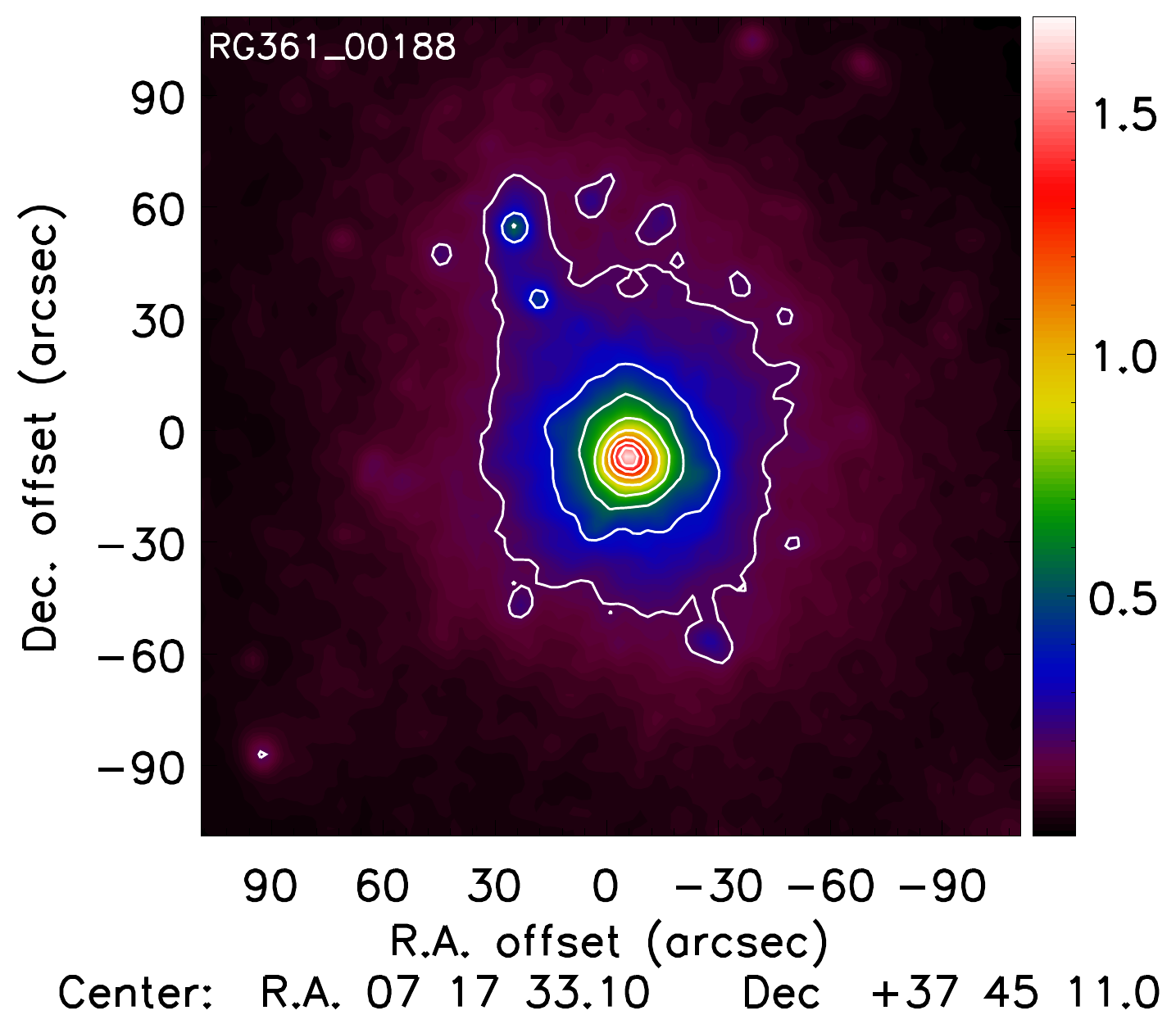} 
\put(-60,310){\makebox(0,0){\rotatebox{0}{\LARGE a.d.u.}}} & 
\includegraphics[trim=2.3cm 2.2cm 0cm 0cm, clip=true, scale=1]{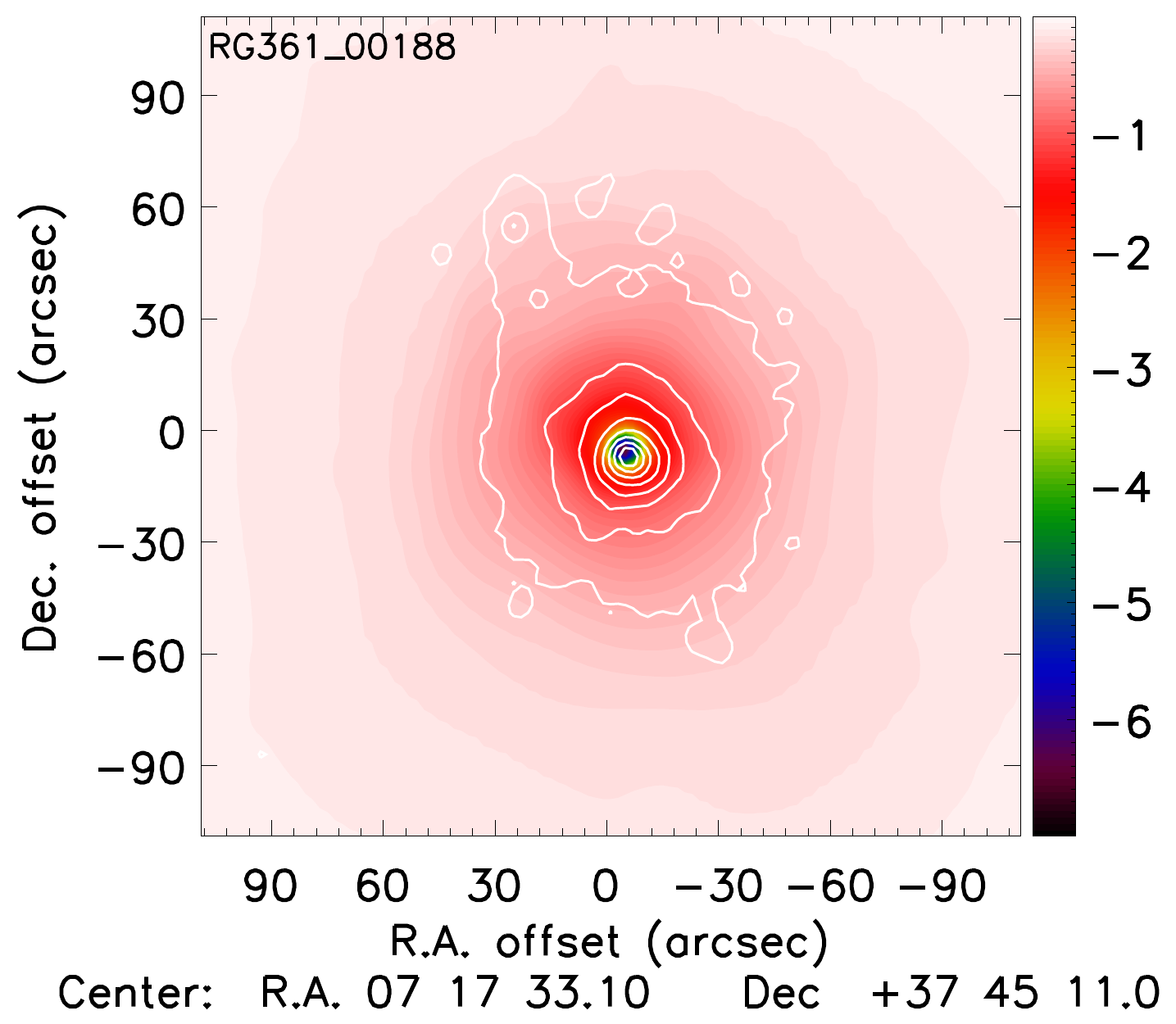} 
\put(-60,310){\makebox(0,0){\rotatebox{0}{\LARGE mJy/beam}}} & 
\includegraphics[trim=2.3cm 2.2cm 0cm 0cm, clip=true, scale=1]{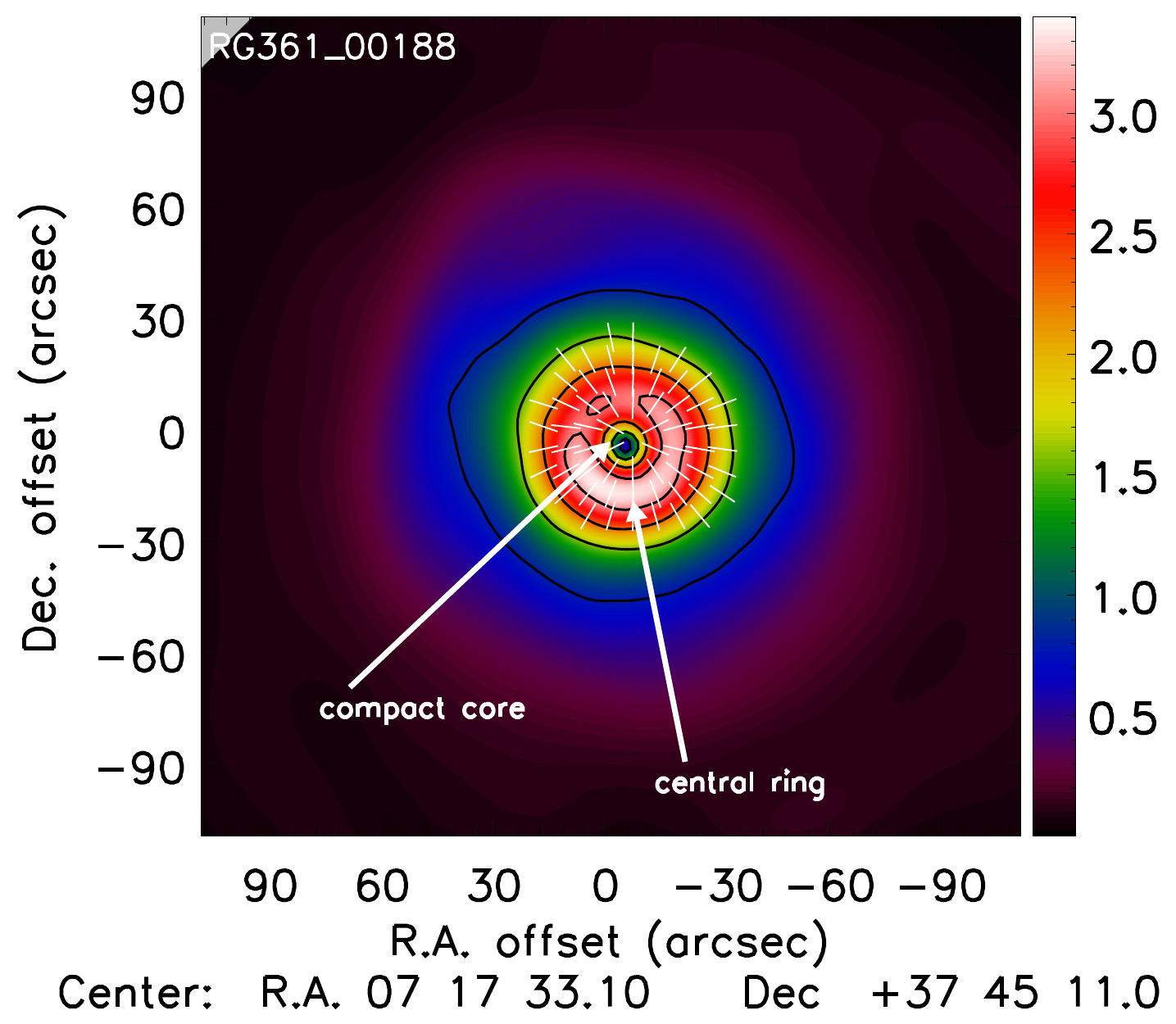} 
\put(-60,310){\makebox(0,0){\rotatebox{0}{\LARGE mJy/beam/arcmin}}} & 
\includegraphics[trim=2.3cm 2.2cm 0cm 0cm, clip=true, scale=1]{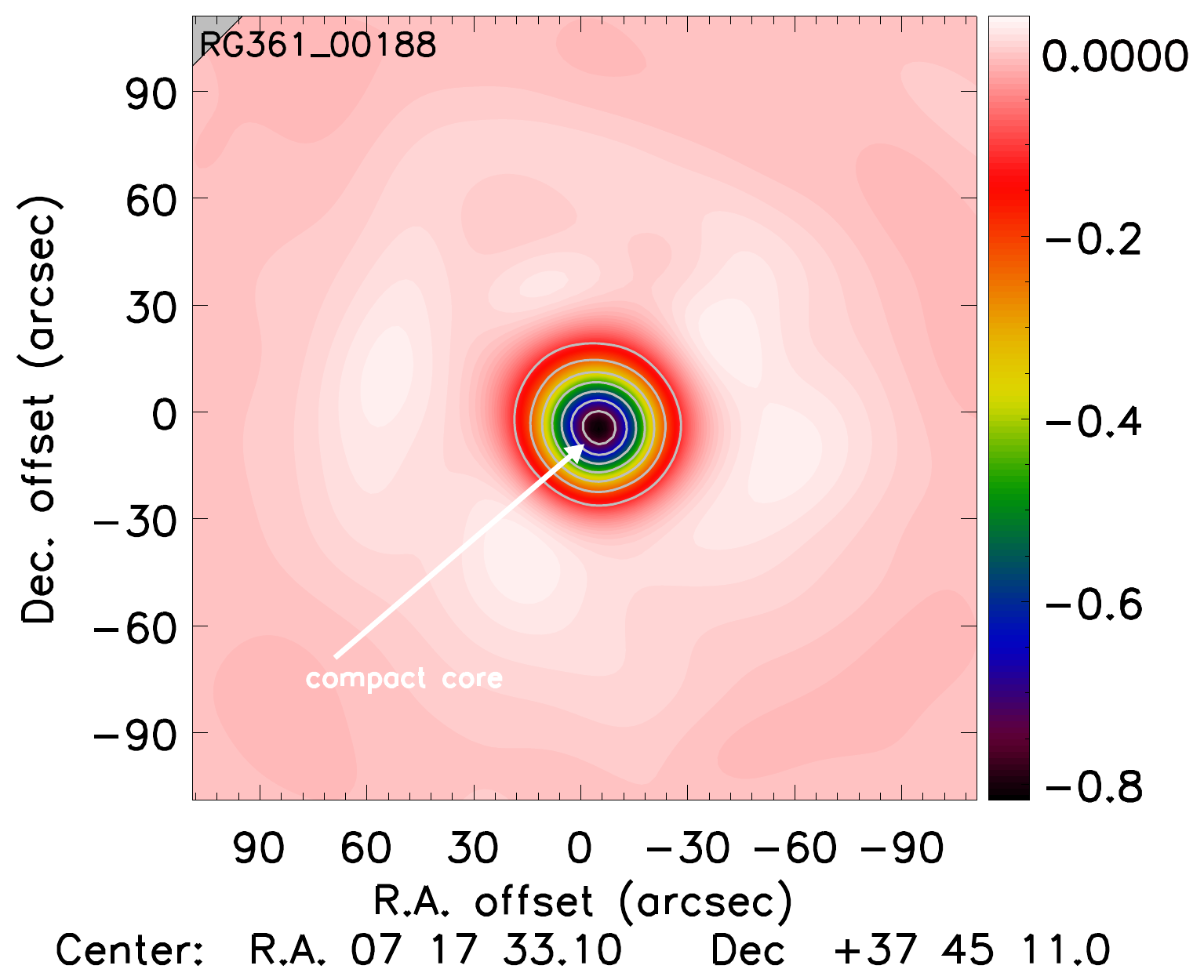} 
\put(-60,310){\makebox(0,0){\rotatebox{0}{\LARGE mJy/beam}}} \\
\includegraphics[trim=0cm 2.2cm 0cm 0cm, clip=true, scale=1]{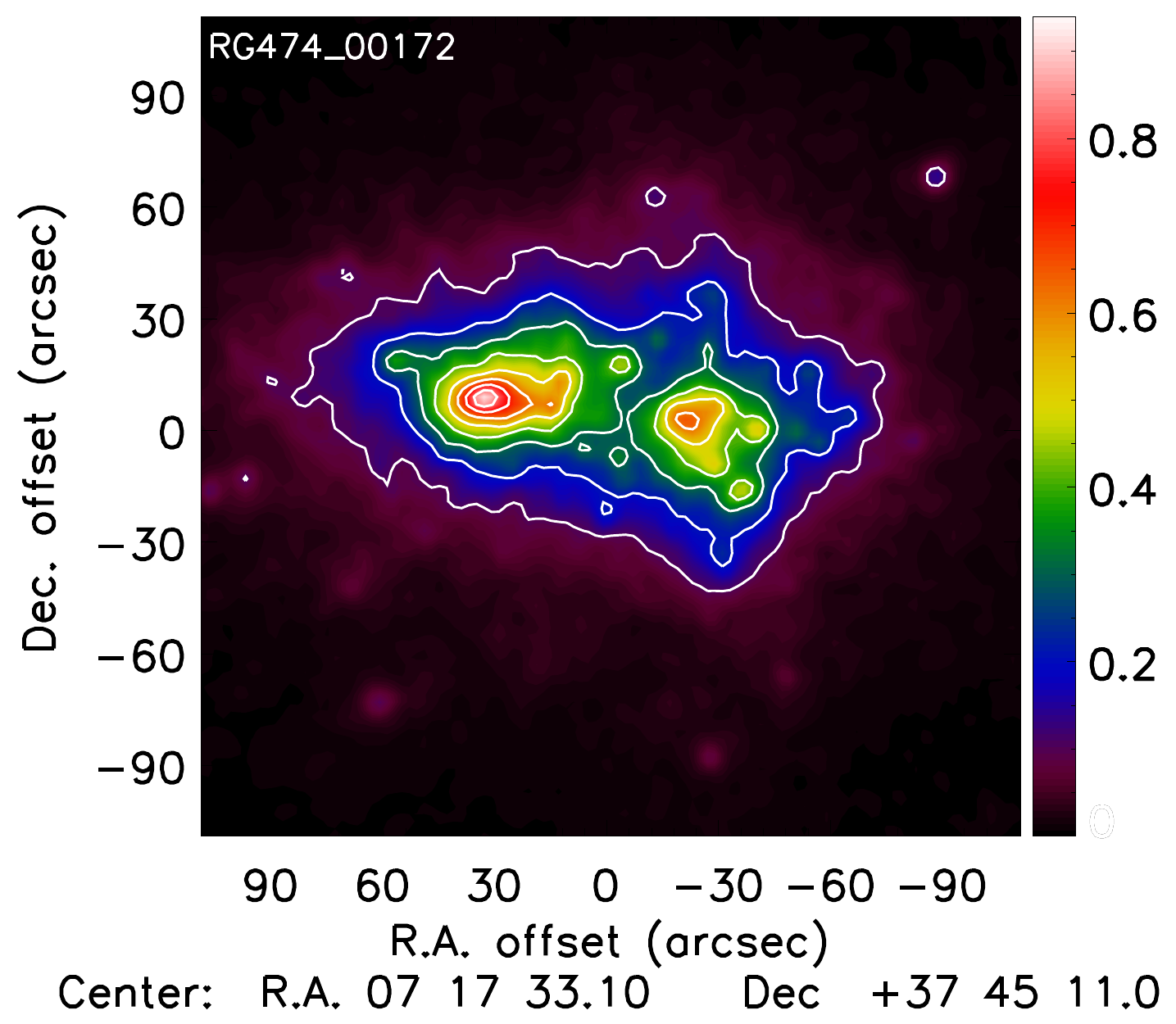} & 
\includegraphics[trim=2.3cm 2.2cm 0cm 0cm, clip=true, scale=1]{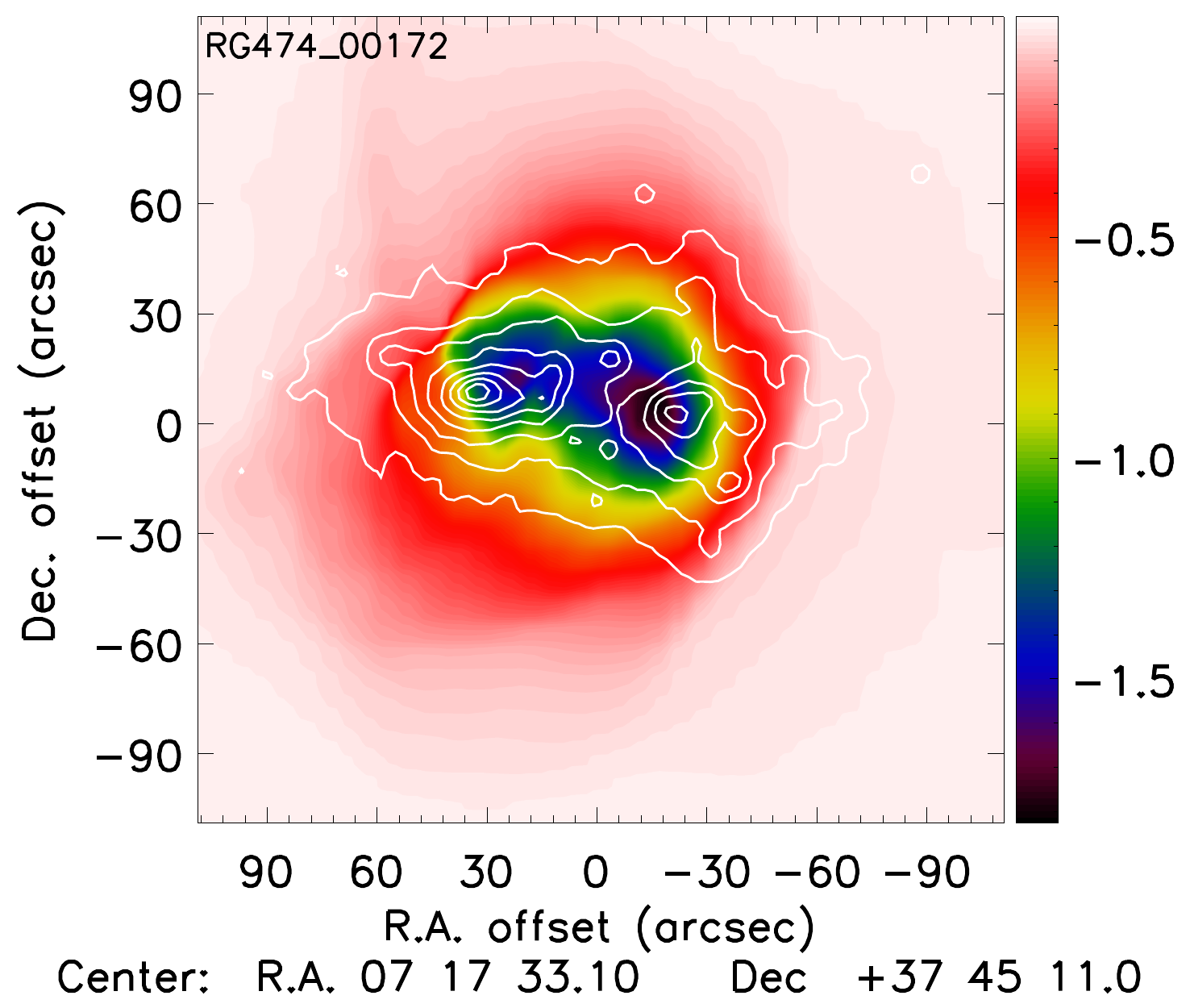} & 
\includegraphics[trim=2.3cm 2.2cm 0cm 0cm, clip=true, scale=1]{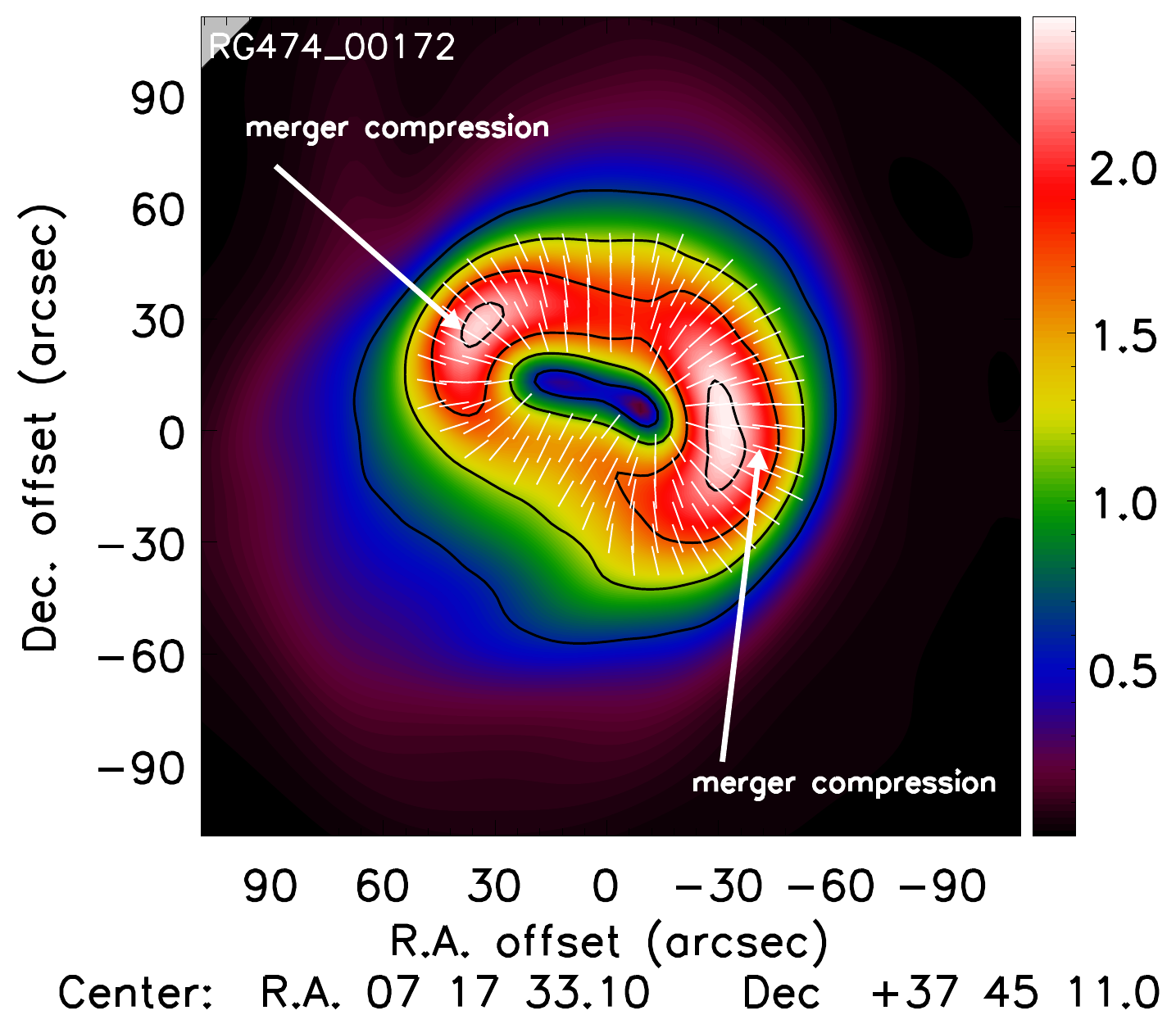} & 
\includegraphics[trim=2.3cm 2.2cm 0cm 0cm, clip=true, scale=1]{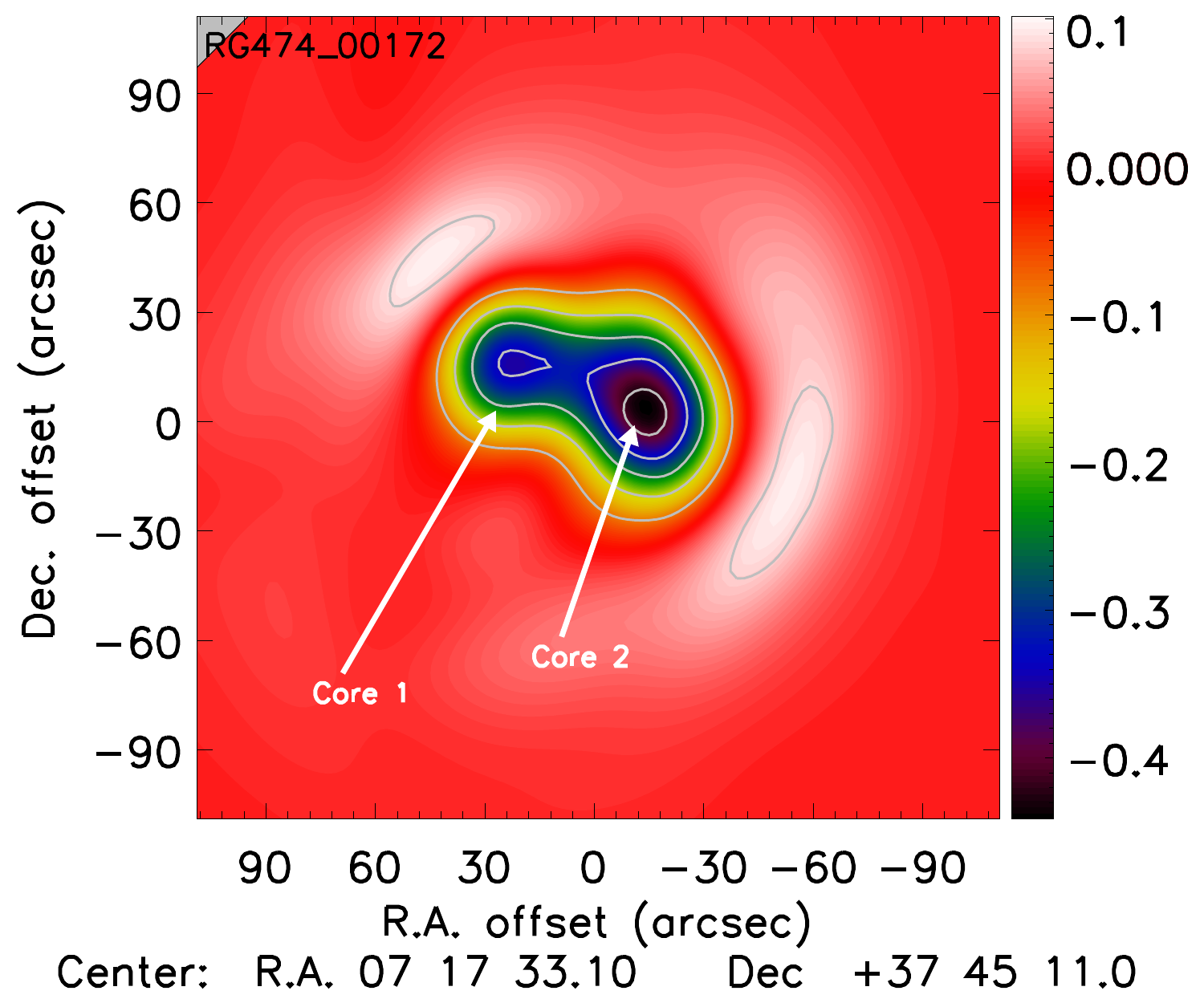} \\
\includegraphics[trim=0cm 2.2cm 0cm 0cm, clip=true, scale=1]{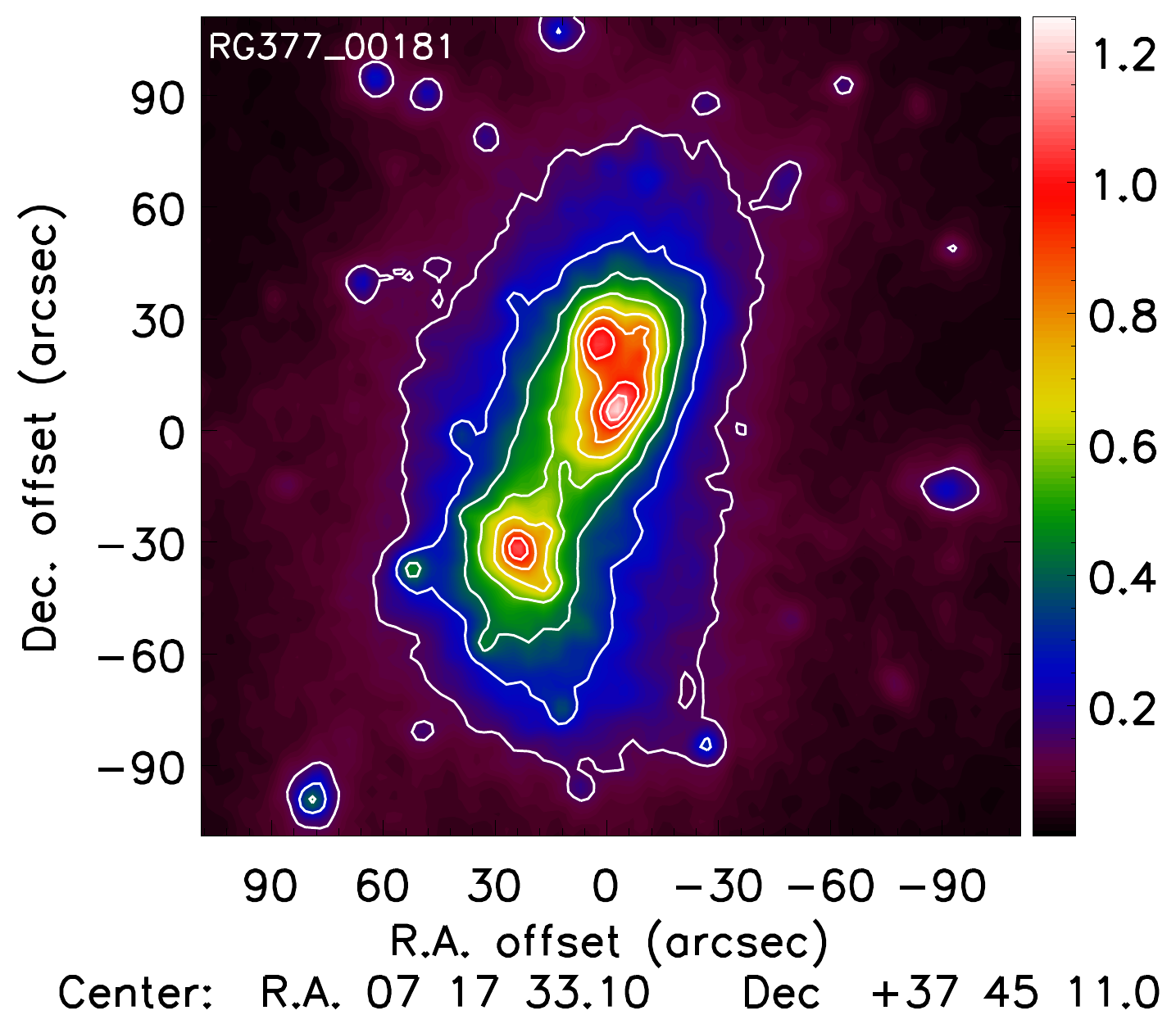} & 
\includegraphics[trim=2.3cm 2.2cm 0cm 0cm, clip=true, scale=1]{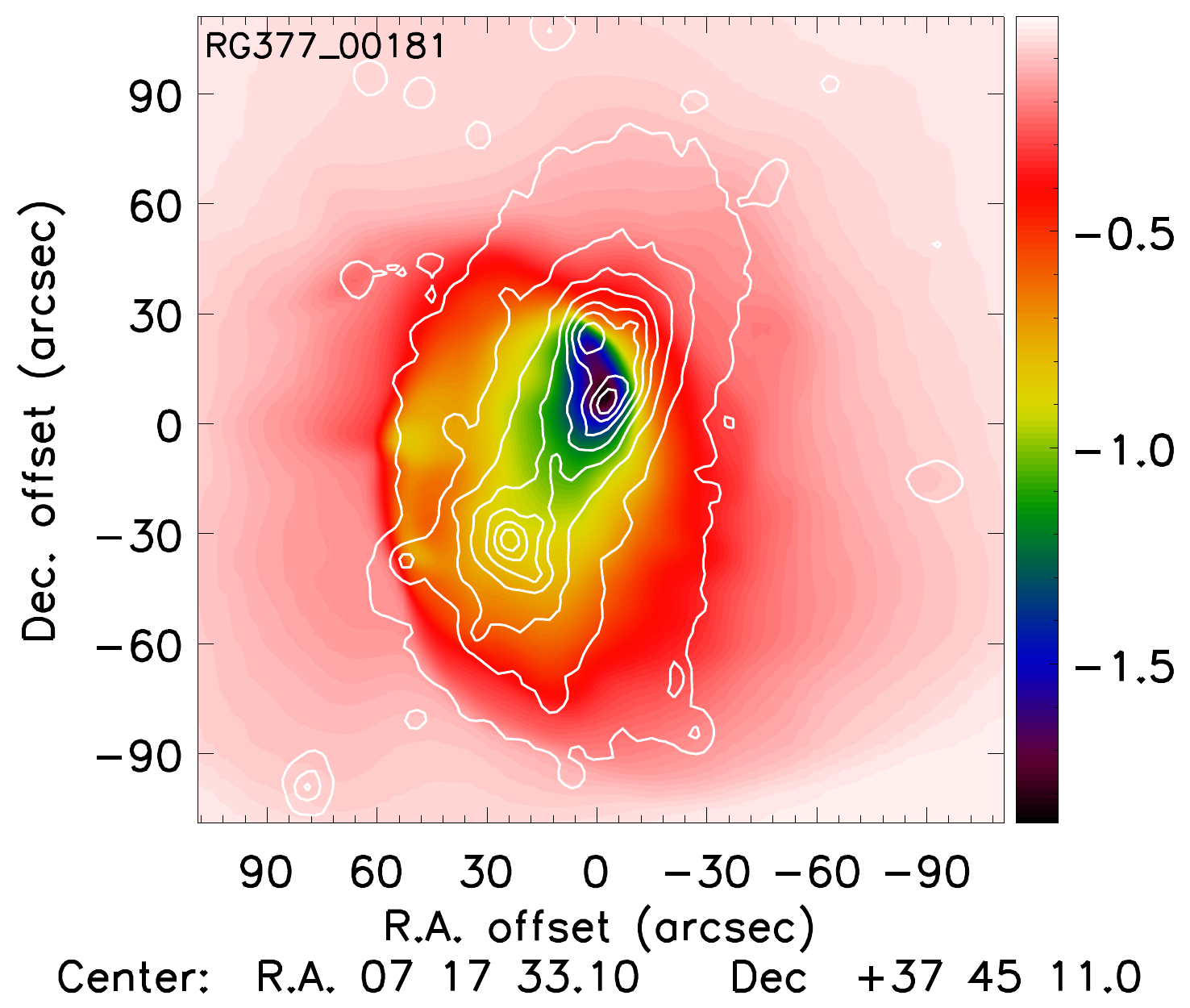} & 
\includegraphics[trim=2.3cm 2.2cm 0cm 0cm, clip=true, scale=1]{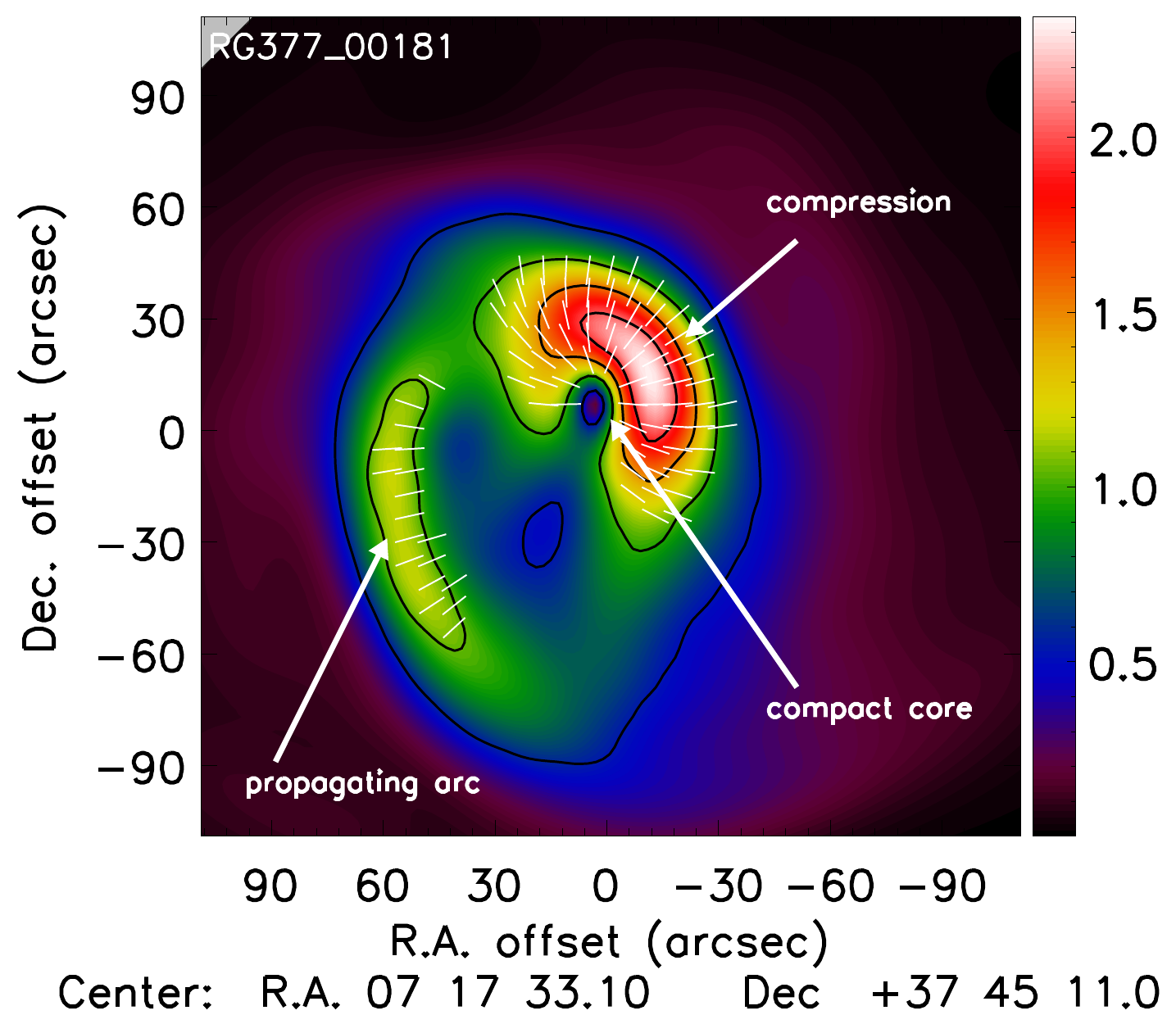} & 
\includegraphics[trim=2.3cm 2.2cm 0cm 0cm, clip=true, scale=1]{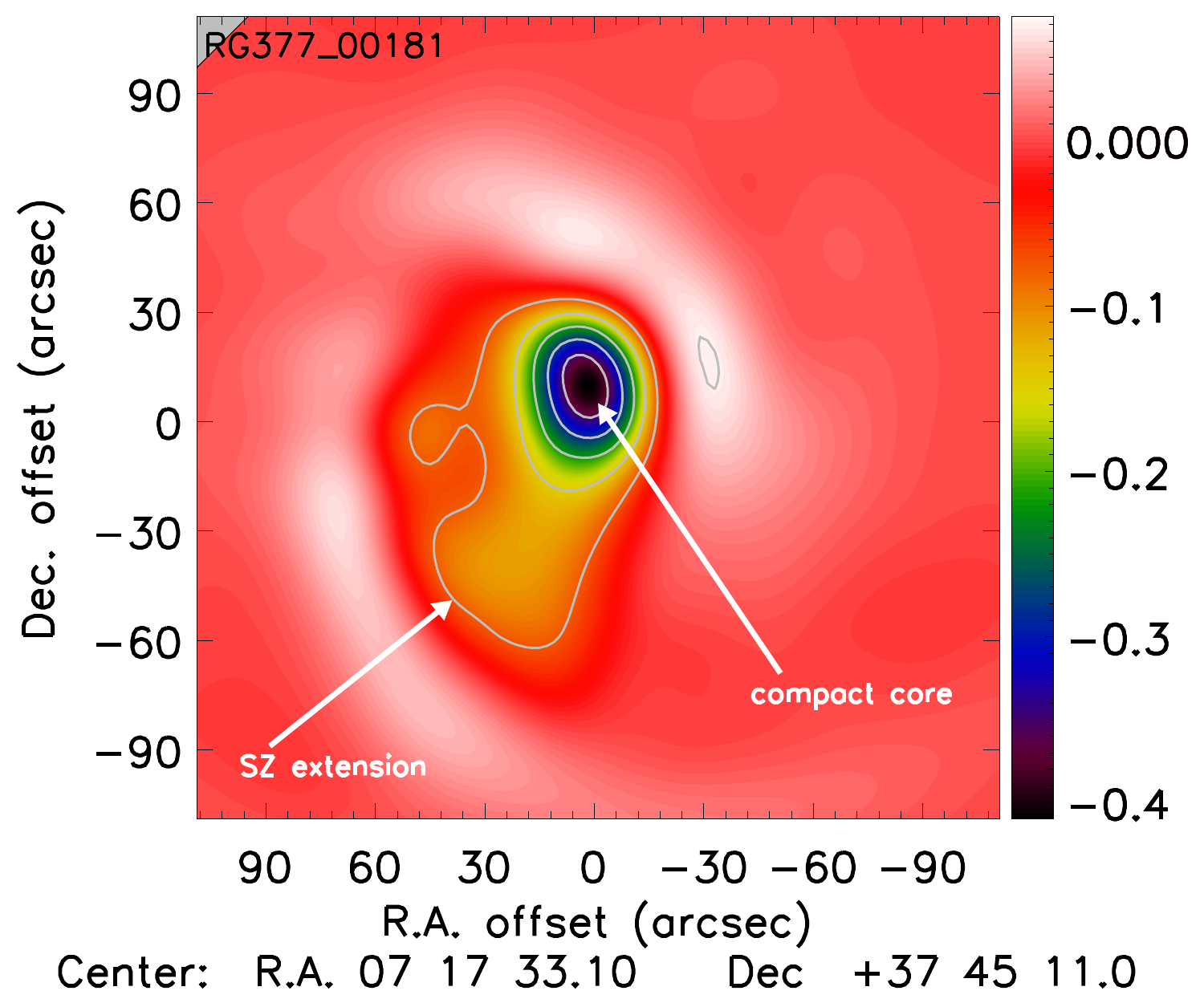} \\
\includegraphics[trim=0cm 2.2cm 0.0cm 0cm, clip=true, scale=1]{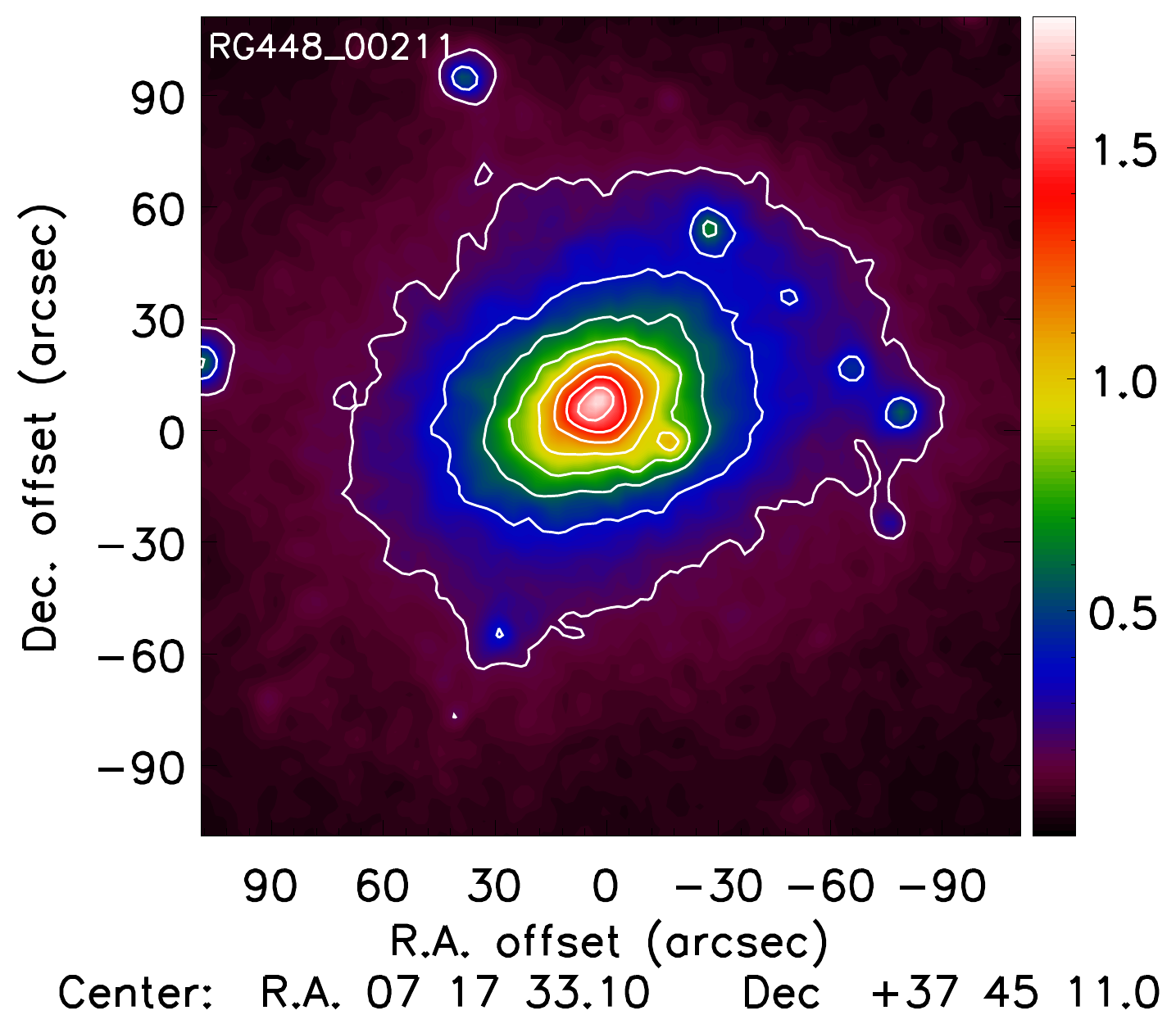} & 
\includegraphics[trim=2.3cm 2.2cm 0.0cm 0cm, clip=true, scale=1]{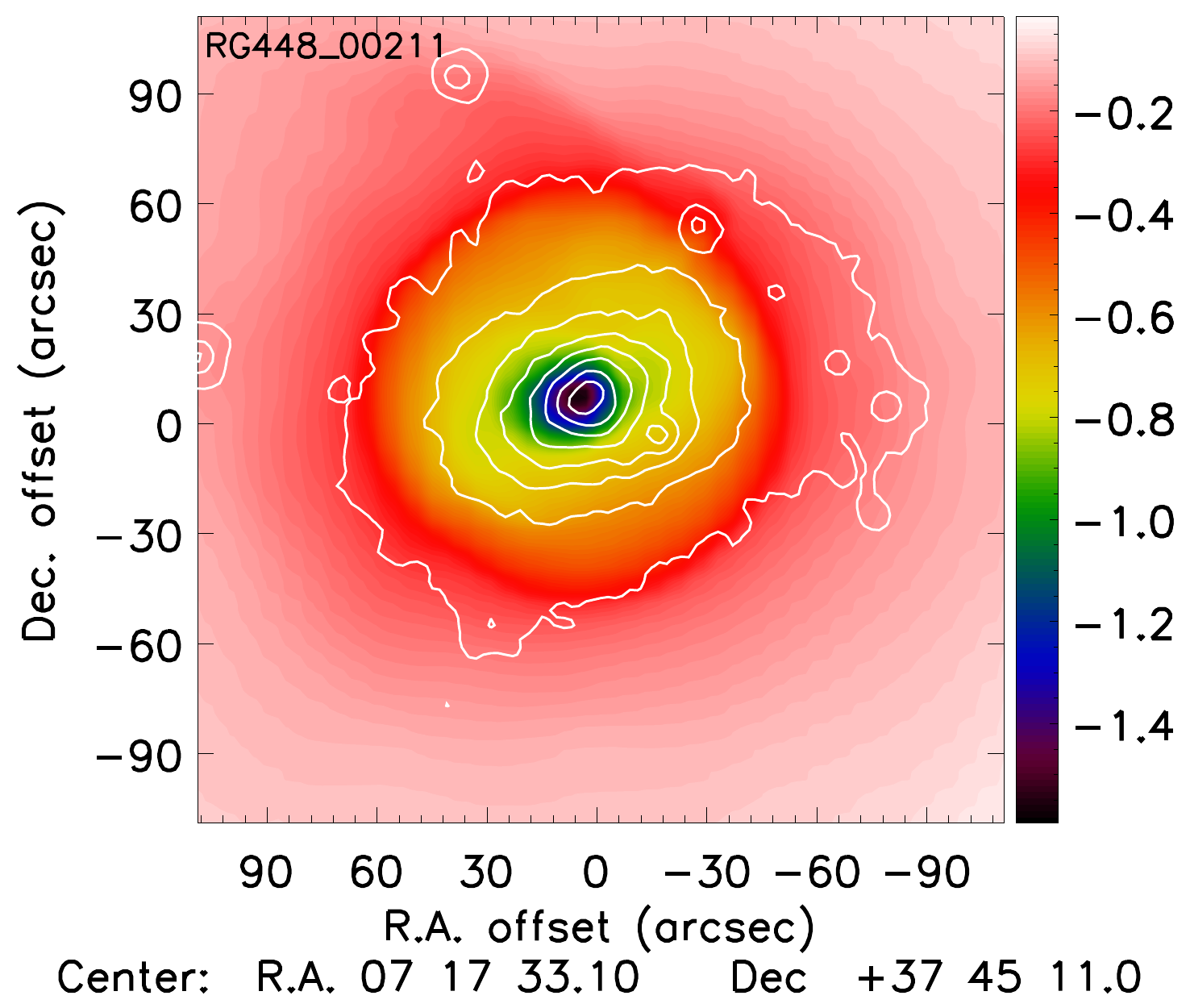} & 
\includegraphics[trim=2.3cm 2.2cm 0.0cm 0cm, clip=true, scale=1]{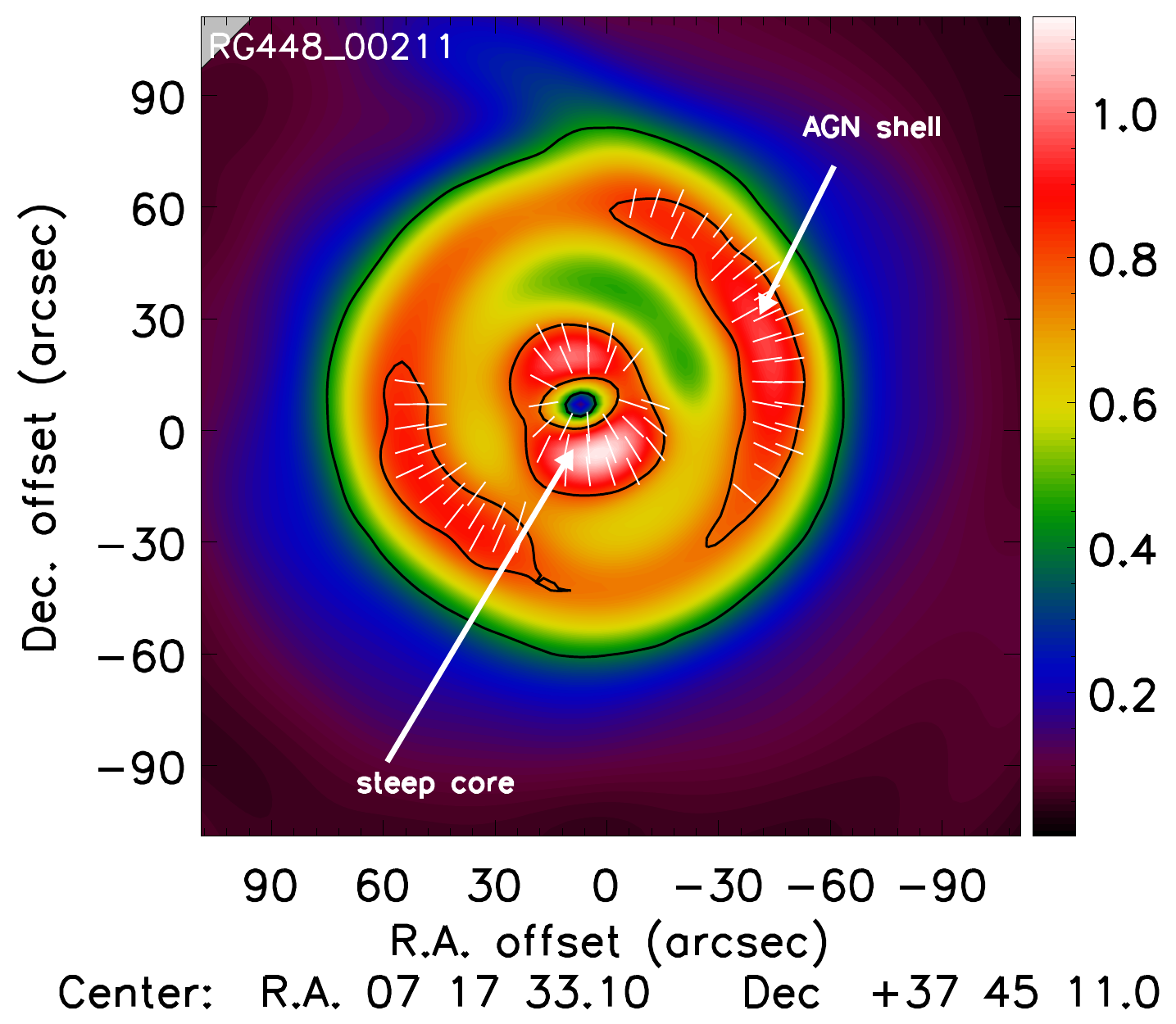} & 
\includegraphics[trim=2.3cm 2.2cm 0.0cm 0cm, clip=true, scale=1]{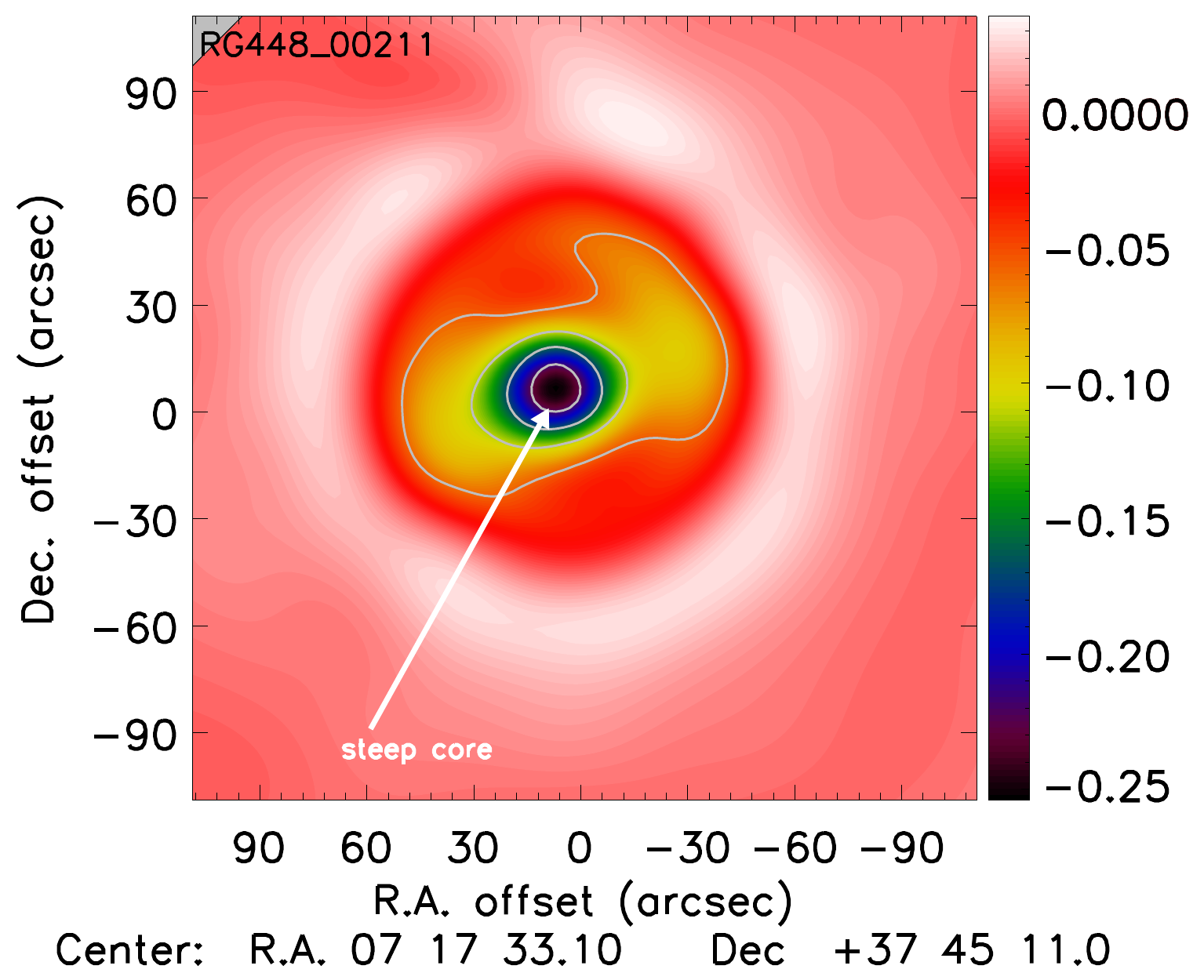} \\
\includegraphics[trim=0cm 0.7cm 0.0cm 0cm, clip=true, scale=1]{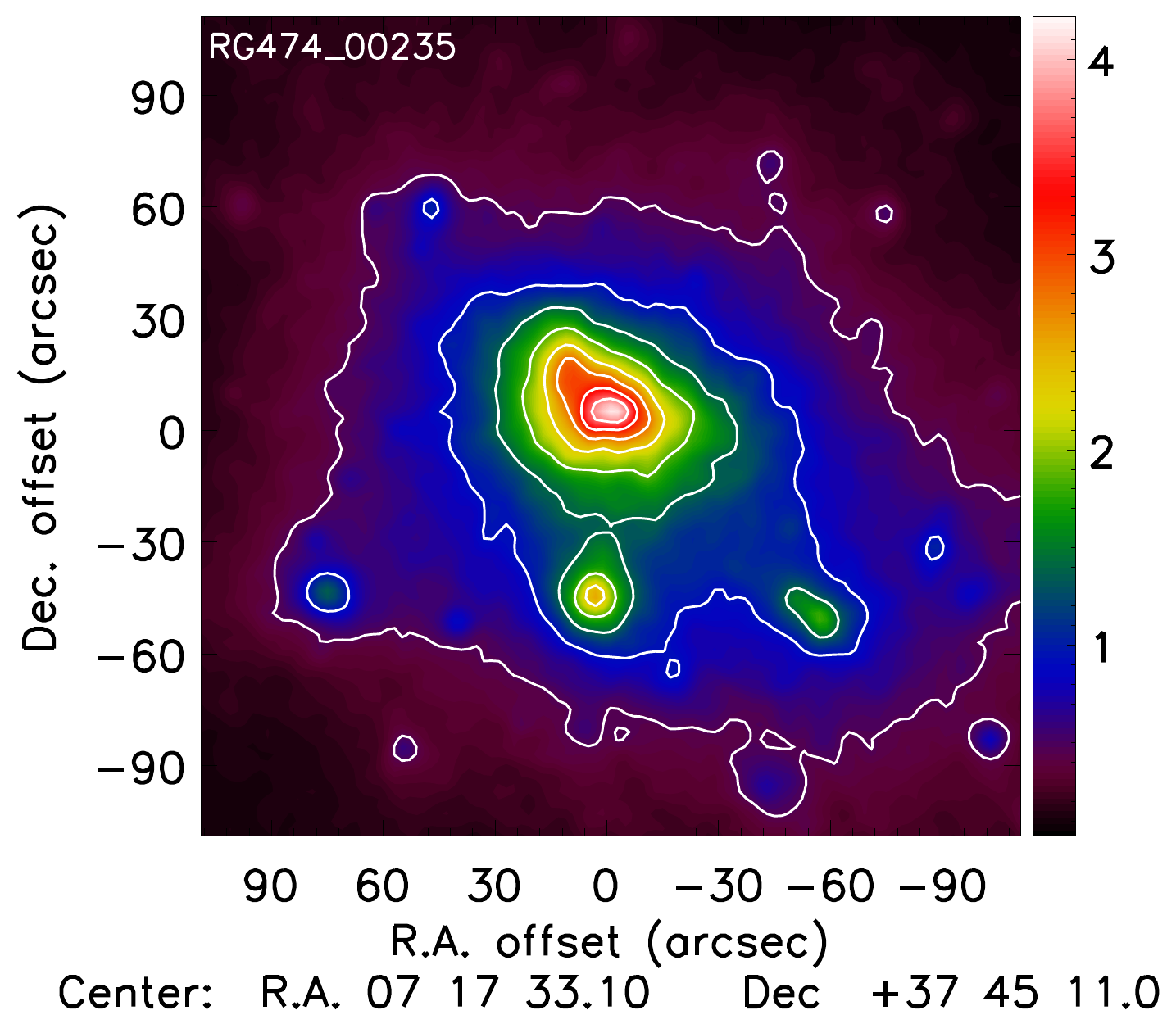} & 
\includegraphics[trim=2.3cm 0.7cm 0.0cm 0cm, clip=true, scale=1]{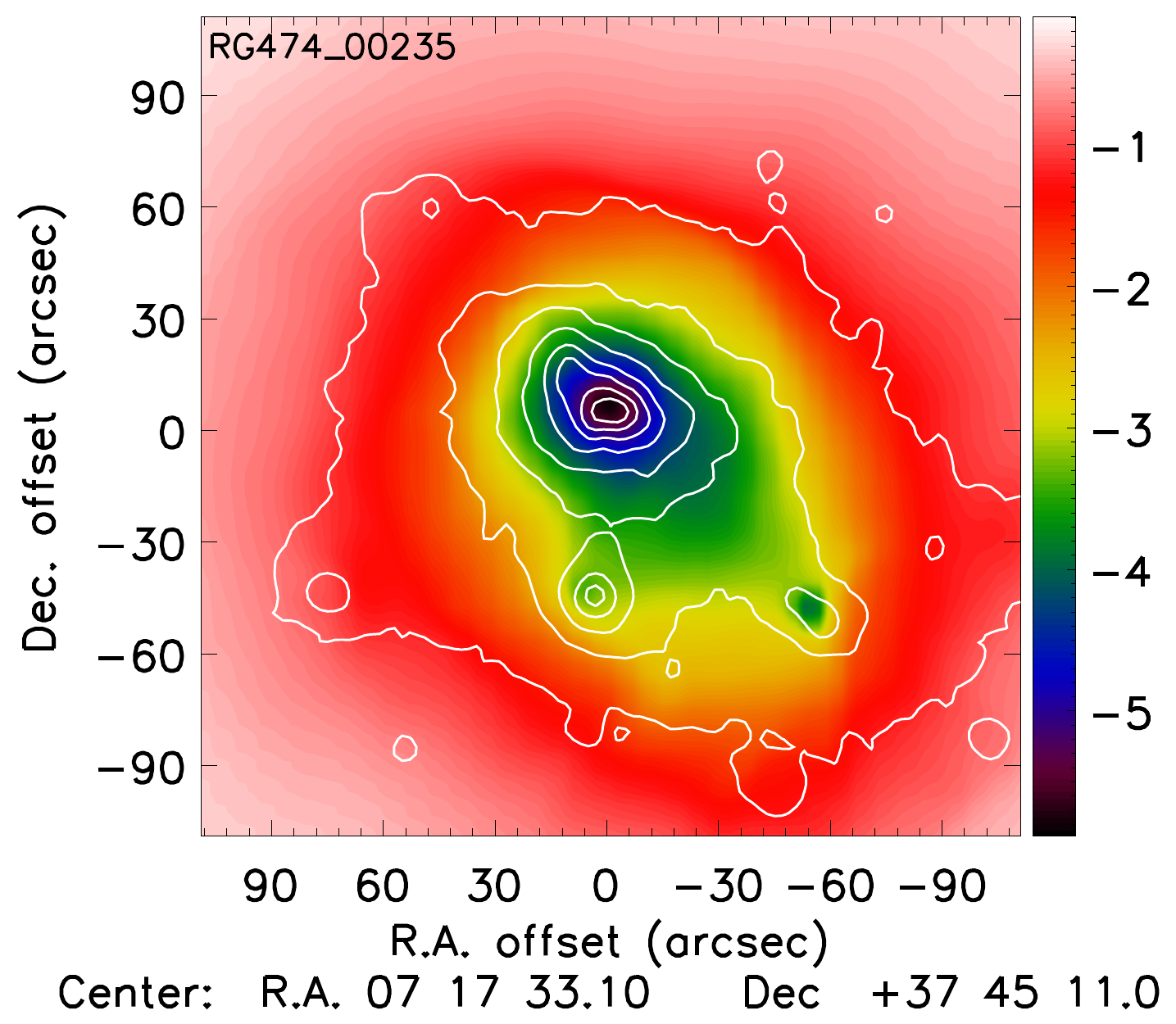} & 
\includegraphics[trim=2.3cm 0.7cm 0.0cm 0cm, clip=true, scale=1]{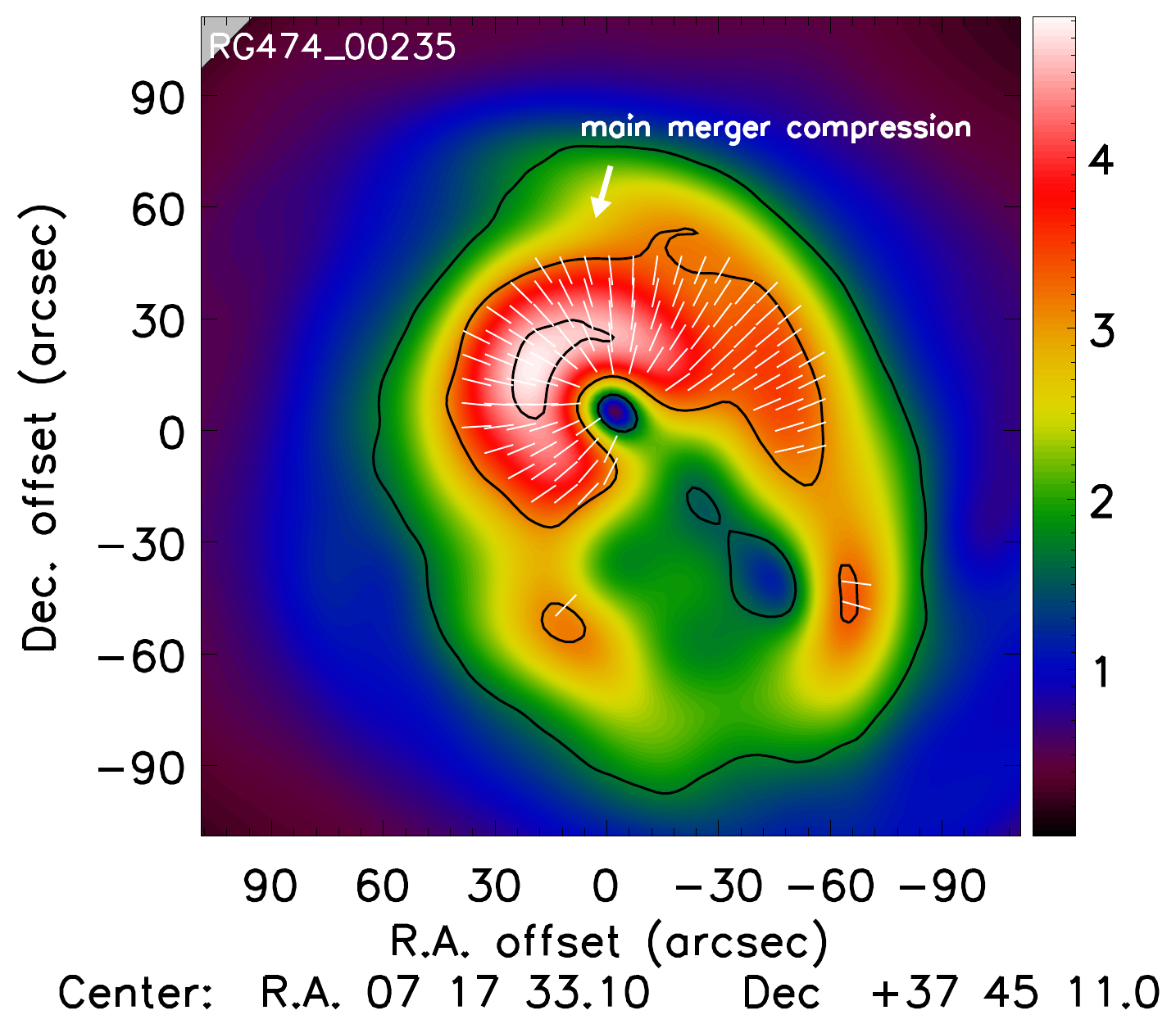} & 
\includegraphics[trim=2.3cm 0.7cm 0.0cm 0cm, clip=true, scale=1]{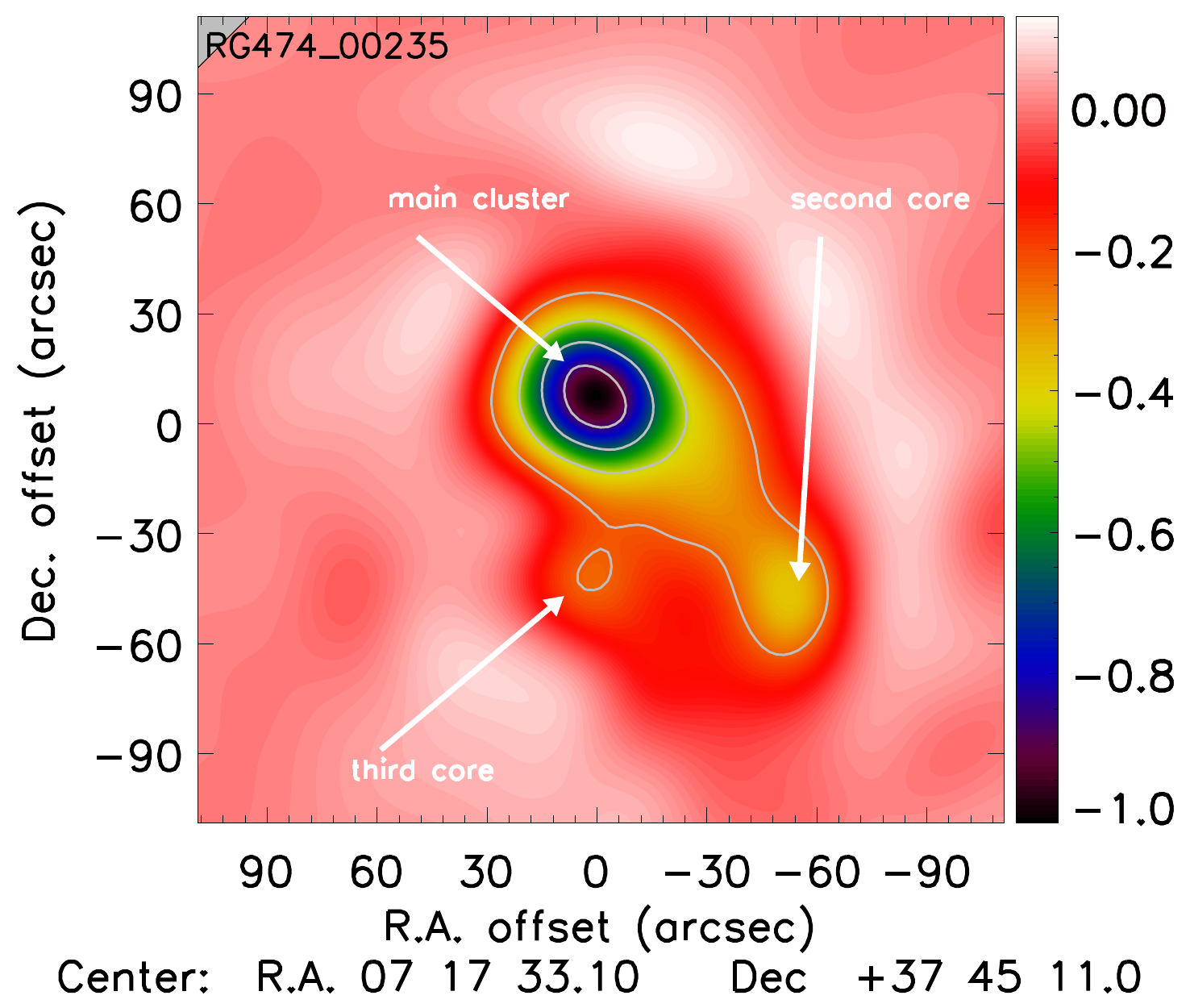} \\
\end{tabular}}
\caption{\footnotesize{Application of the filtering algorithms to clusters from the RHAPSODY-G suite of simulations. From top to bottom, the clusters are RG361\_00188, RG474\_00172, RG377\_00181, RG448\_00211, and RG474\_00235 (see also Table \ref{tab:rhapsody_summary} for the main properties of the sample). {\bf Left:} projected dark matter density. Contours are linearly spaced. Units are arbitrary. {\bf Middle left:} input raw tSZ surface brightness maps with dark matter map contours overlaid. Units are mJy/beam. {\bf Middle right:} GGM filtered maps with $\theta_0 = 15"$. The white vectors represent the direction of the gradient, $\Psi$. Units are mJy/beam/arcmin. {\bf Right:} DoG filtered maps with $\theta_1 = 15"$ and $\theta_2 = 45"$. Units are mJy/beam. The contours provide the true signal-to-noise expected for the simulation of the observation of these sources done in Section \ref{sec:Systematics_and_noise_properties}. They are given by steps of $2 \sigma$ excluding 0 (see Section \ref{sec:Systematics_and_noise_properties} for further discussions about the signal-to-noise estimates).}}
\label{fig:RG_cluster_sample}
\end{figure*}

Before performing our analysis on the NIKA data, we test the behavior of the filters described in Section \ref{sec:Pressure_substructures_detection} by using realistic simulated cluster data. Such simulations also aid in the interpretation of the observed structures in real data, in addition to the toy models described in Section \ref{sec:Application_to_toy_models}. To do so, we use synthetic tSZ images extracted from the RHAPSODY-G simulations for different cluster configurations.

\subsection{Extraction of tSZ maps from the RHAPSODY-G simulations}
The RHAPSODY-G simulations \citep{Wu2013,Hahn2017} is a suite of cosmological hydrodynamics adaptive mesh refinement zoom simulations of ten massive clusters. The simulations include cooling and sub-resolution models for star formation and supermassive black hole feedback. 

Boxes of co-moving volume $\left(5 \ {\rm Mpc}\right)^3$ were used to extract Compton parameter maps around the clusters by integrating Eq \ref{eq:y_compton} along a fixed line-of-sight direction, using the gas pressure in each grid cell. These maps were converted into surface brightness images, in the NIKA 150 GHz bandpasses \citep[see the coefficient provided in][]{Adam2016b}. We neglect relativistic corrections, as they can only reduce the surface brightness by up to 8\% for the hottest clusters at our frequency \citep{Itoh2003}. The maps were produced for a large subset of the available simulation snapshots (about 150 per cluster), tracing the evolution of the clusters from redshift $z \sim 1.5$ to $z=0$. The knowledge of the cluster formation history from the simulation data is a key point for the interpretation of the maps.

\subsection{Selection of a RHAPSODY-G sub-sample}\label{sec:Selection_of_a_RHAPSODY-G_sub-sample}
A sub-sample of clusters was selected to assess the sub-structure detection in a quantitative way and to investigate the behavior of the filters in details (see Section \ref{sec:Systematics_and_noise_properties} for the end-to-end processing of the simulations). The clusters were selected to match the redshift range of the NIKA sample (see Section \ref{sec:NIKA_Data}) and to explore various cluster configurations. We thus restricted the RHAPSODY-G snapshots to the redshift range $0.4 \lesssim z \lesssim 1$ and selected five distinct snapshots (see Figure \ref{fig:RG_cluster_sample}), by visual inspection, for different clusters\footnote{The 3-digit number after "RG" refers to the respective RHAPSODY-G halo, while the second 5-digit number refers to the snapshot number of the analyzed map.}:
\begin{enumerate}
\item {\bf RG361\_00188:} a relaxed spherical cluster at redshift $z = 0.61$, with minimal merging activity since $z \gtrsim 1$. The cluster is slightly elongated along the line-of-sight.
\item {\bf RG474\_00172:} a major merger at $z = 0.90$. The two main sub-clusters already passed through each other, the closest encounter happening at $z \sim 0.95$. The merger axis is close to the plane of the sky. 
\item {\bf RG377\_00181:} a triple-merger with two nearly equal mass objects and a smaller sub-group at $z = 0.54$. The two main sub-clusters have already experienced a first encounter at $z \sim 0.59$, while the smaller group is spatially coincident (aligned) with one of the two main clusters.
\item {\bf RG448\_00211:} a relaxed, slightly elliptical cluster, exhibiting strong AGN feedback at $z = 0.40$. This produces a pressure shell around the main core.
\item {\bf RG474\_00235:} a very massive cluster, corresponding to the evolved version of RG474\_00172, at $z = 0.39$. This cluster was selected mainly because it is relatively nearby and very massive, such that its spatial extent fits better that of the NIKA clusters, allowing us to validate our procedure in such cases. It is undergoing a major merger, mostly oriented along the line-of-sight.
\end{enumerate}
We note that the RHAPSODY-G cluster masses match well those of the NIKA sample at redshift zero, but they are lower by a factor of a few at the considered redshifts (excepted for RG474\_00235, which reaches a mass similar to the most massive NIKA cluster). Therefore the simulated surface brightnesses are also lower by a factor of $\sim 2$. The summary of the main properties of the four RHAPSODY-G selected clusters is provided in Table \ref{tab:rhapsody_summary}.

\subsection{Filter application to raw RHAPSODY-G tSZ images}
We first apply the filters defined in Section \ref{sec:Pressure_substructures_detection} to the raw RHAPSODY-G surface brightness images (i.e. without including noise or observational artifacts). The filtering parameters are set according to our baseline choice (Section \ref{sec:Baseline_filtering_parameters}), as optimized for the NIKA sample. Figure \ref{fig:RG_cluster_sample} provides images of the input surface brightness of the RHAPSODY-G selected sample, and the corresponding filtered maps together with the projected dark matter density distribution. 

RG361\_00188 (relaxed) presents a very compact core and is azimuthally symmetric. The tSZ signal matches well that of the dark matter. As the steepness of the surface brightness profile increases toward the center, the corresponding GGM map is null at the cluster peak and exhibits a quasi perfect ring with radius matching the filter scale (see also the gNFW model of Figure \ref{fig:test_filter_gNFW_and_bimodal}), with a small excess in the south. The DoG map presents a single compact core component. The RG474\_00172 and RG377\_00181 (major and multiple mergers, respectively) tSZ maps show that the two clusters are both clearly non-spherical. The corresponding dark matter maps reflect the presence of multiple sub-clusters and the dark matter distribution deviates significantly from that of the tSZ signal. The GGM maps provide additional information by highlighting regions of strong gas compression. In the case of the major merger, RG474\_00172, we observe a strongly elongated ring with two main pressure gradients in the east and west regions, comparable to the merger toy model of Figure \ref{fig:test_filter_gNFW_and_bimodal} without the bar component. Discontinuities are observed in the simulation at these locations, but they are too close to the core to be distinguished from it considering our baseline scales, and the signal we observe results from the sum of the shocks and the gNFW-like steep cores, as an overall compression. In the case of the multiple merger, RG377\_00181, a strong gradient is visible in the west region, corresponding to the main core, and a $\sim 70$ arcsec (460 kpc) long arc propagates through the ICM on the southeast sector, corresponding to another sub-cluster moving within the ICM of the main cluster and causing a shock. The DoG maps present a double peak structure, allowing us to quantitatively identify the two sub-clusters in the case of RG474\_00172. In contrast, one only finds a main core plus a weak extension in the case of RG377\_00181 (comparable to the main core plus extension toy model of Figure \ref{fig:test_filter_gNFW_and_bimodal}). RG448\_00211 appears to be relatively relaxed and azimuthally symmetric, in agreement with the dark matter distribution. However, the GGM map reveals, in addition to the steep core, the presence of a pressure shell responsible for a strong gradient ring extending 45 arcsec (250 kpc) away from the peak. This is due to the feedback from the central AGN onto the ICM, and possibly an artifact of the specific implementation of AGN feedback as a thermal blastwave in these simulations. We also observe a plateau extending to 45 arcsec (250 kpc) radius on the DoG filtered map and we note that the core signal is slightly elongated with a main axis inclined by about 20 degrees with respect to the R.A. axis. RG448\_00211 presents a structure that compares well to RG377\_00181, both in the dark matter and the gas distribution. However, no shock front can clearly be identify, probably due to projection effects since the merger axis is mainly along the line of sight in this case. In addition, both the amplitude and the angular size of RG448\_00211 are much larger than RG377\_00181, because it is located at lower redshift, and because its mass is about 3.5 times larger. RG448\_00211 is mainly used to validate our analysis, even in the case of very bright and very extended clusters (see Appendix \ref{sec:Transfer_function_deconvolution}).

The morphologies seen in the filtered RHAPSODY-G sub-sample provide templates that we can compare to the structures observed in the real NIKA data. In practice, we performed this analysis using all available RHAPSODY-G snapshots and projected the data along three line-of-sight directions. Therefore, we dispose of a much larger variety of configurations, even if the four selected clusters already qualitatively sample most of the structures visible in the simulation, according to the selection of the sub-sample. These templates will allow us to better understand the physics at play in the NIKA clusters, in particular in merging systems. When using the tSZ signal as a tracer of the overall matter distribution \citep[e.g.][]{Adam2015,Adam2016a,Ruppin2016}, the application of such filters could also help to interpret scatters and biases between the total mass distribution and the tSZ signal, as such features reflect spatial misalignment between the dark matter distribution and the hot gas.

In summary, the GGM and the DoG filters allow for the detection of strong gradient in the surface brightness (seen as ridges in the GGM maps), and tSZ peaks in the RHAPSODY-G images, respectively. They provide templates for interpreting features observed in the NIKA data in Section \ref{sec:Application_to_the_NIKA_clusters_sample}. These features correspond to shocks and gas compression regions induced by merging events, and the main cores of the merging sub-clusters, respectively. In the case of isolated spherical clusters, GGM features are also observed, showing up as rings around the tSZ peak.

\section{Systematic effects and noise properties}\label{sec:Systematics_and_noise_properties}
In addition to the tSZ sub-structures, NIKA data are affected by several systematic effects that can potentially alter the reconstructed filtered images. In this Section we test the response of the filters described in Section \ref{sec:Pressure_substructures_detection} to potential bouncing artifacts due to the transfer function of the NIKA processing, spatially correlated noise, and point source residuals.

\begin{figure*}[h]
\centering
\resizebox{0.75\textwidth}{!} {
\begin{tabular}{lll}
\includegraphics[trim=0cm 2.2cm 0cm 0cm, clip=true, scale=1]{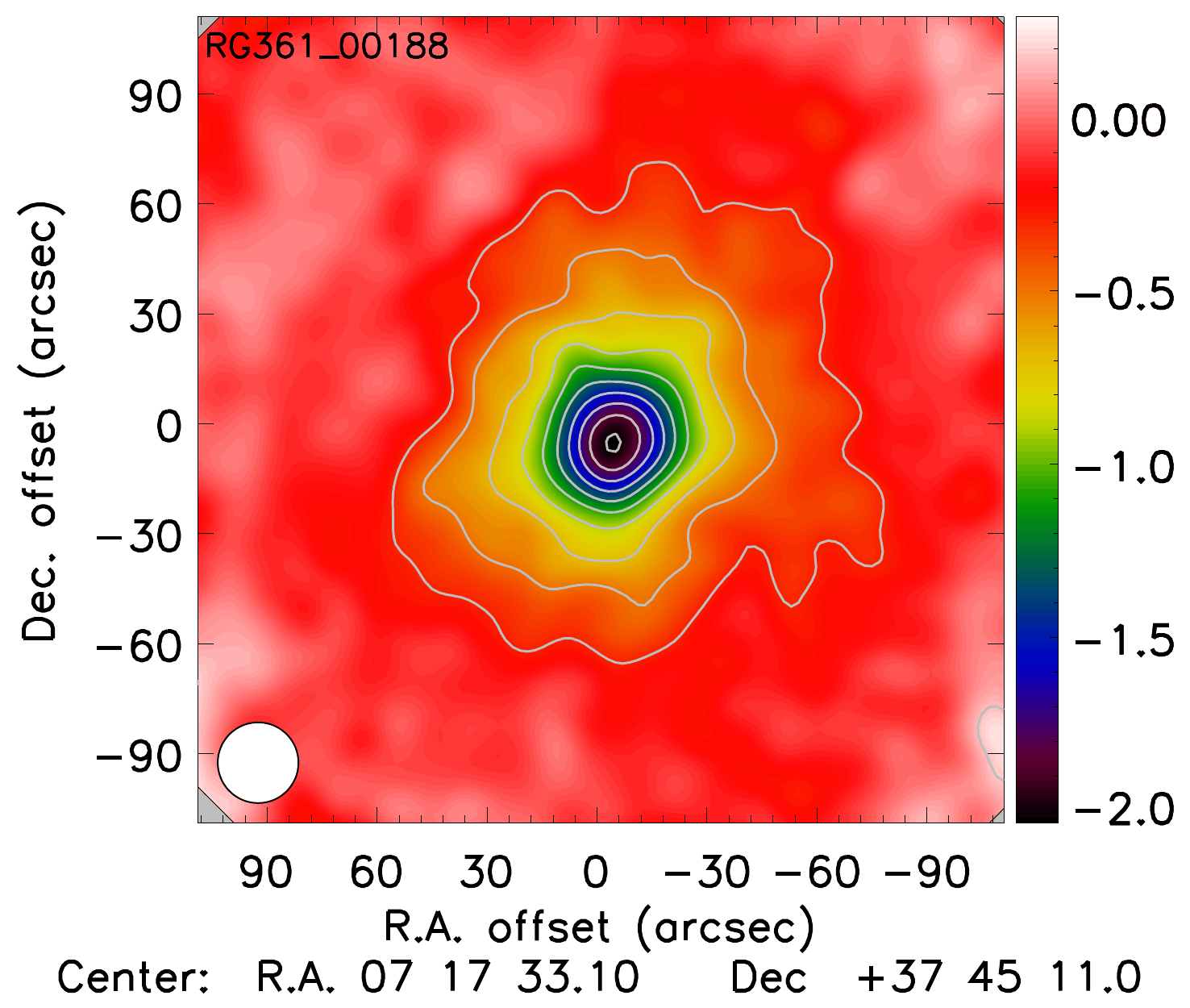} 
\put(-60,310){\makebox(0,0){\rotatebox{0}{\LARGE mJy/beam}}} & 
\includegraphics[trim=2.3cm 2.2cm 0cm 0cm, clip=true, scale=1]{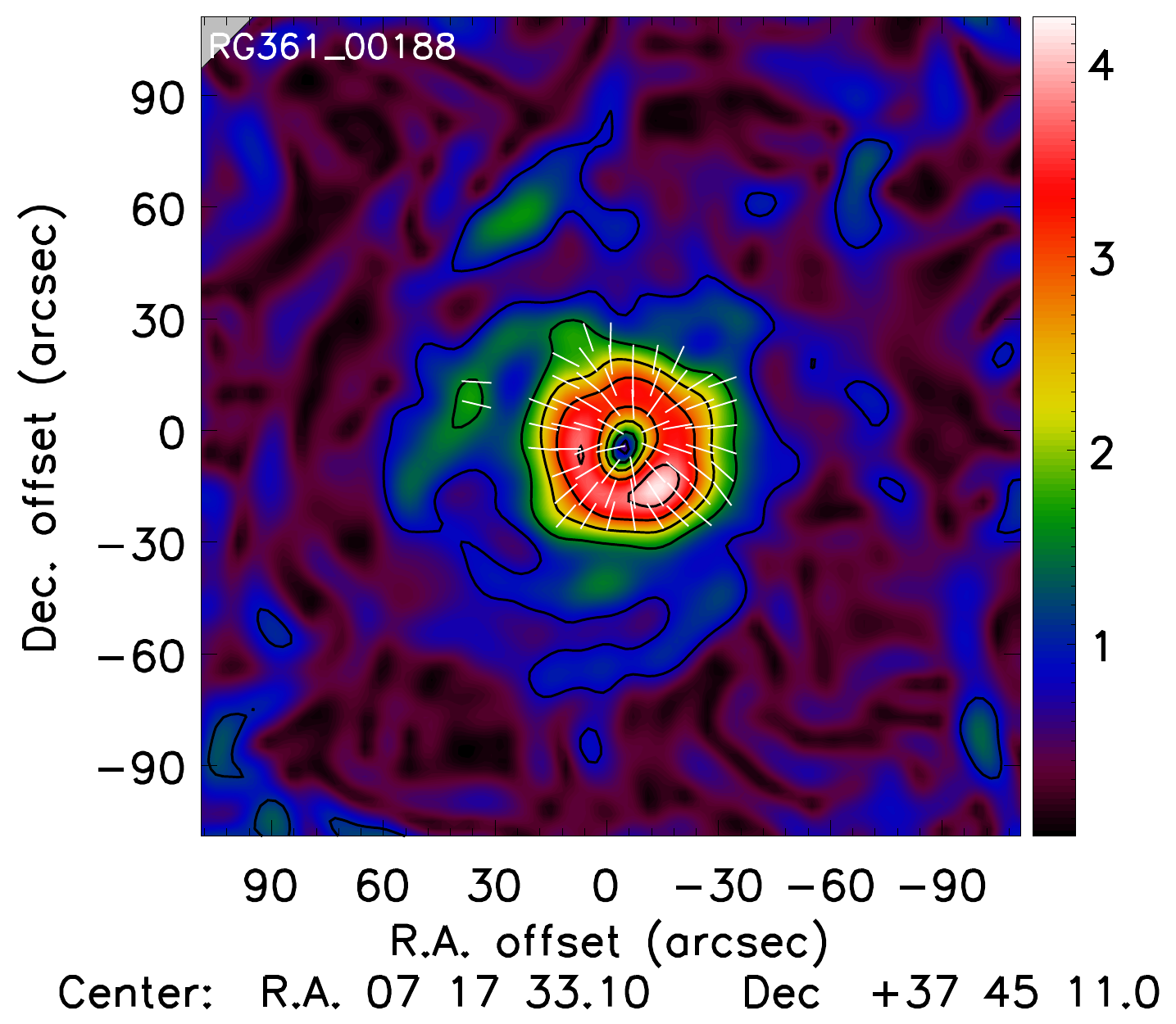} 
\put(-60,310){\makebox(0,0){\rotatebox{0}{\LARGE mJy/beam/arcmin}}} & 
\includegraphics[trim=2.3cm 2.2cm 0cm 0cm, clip=true, scale=1]{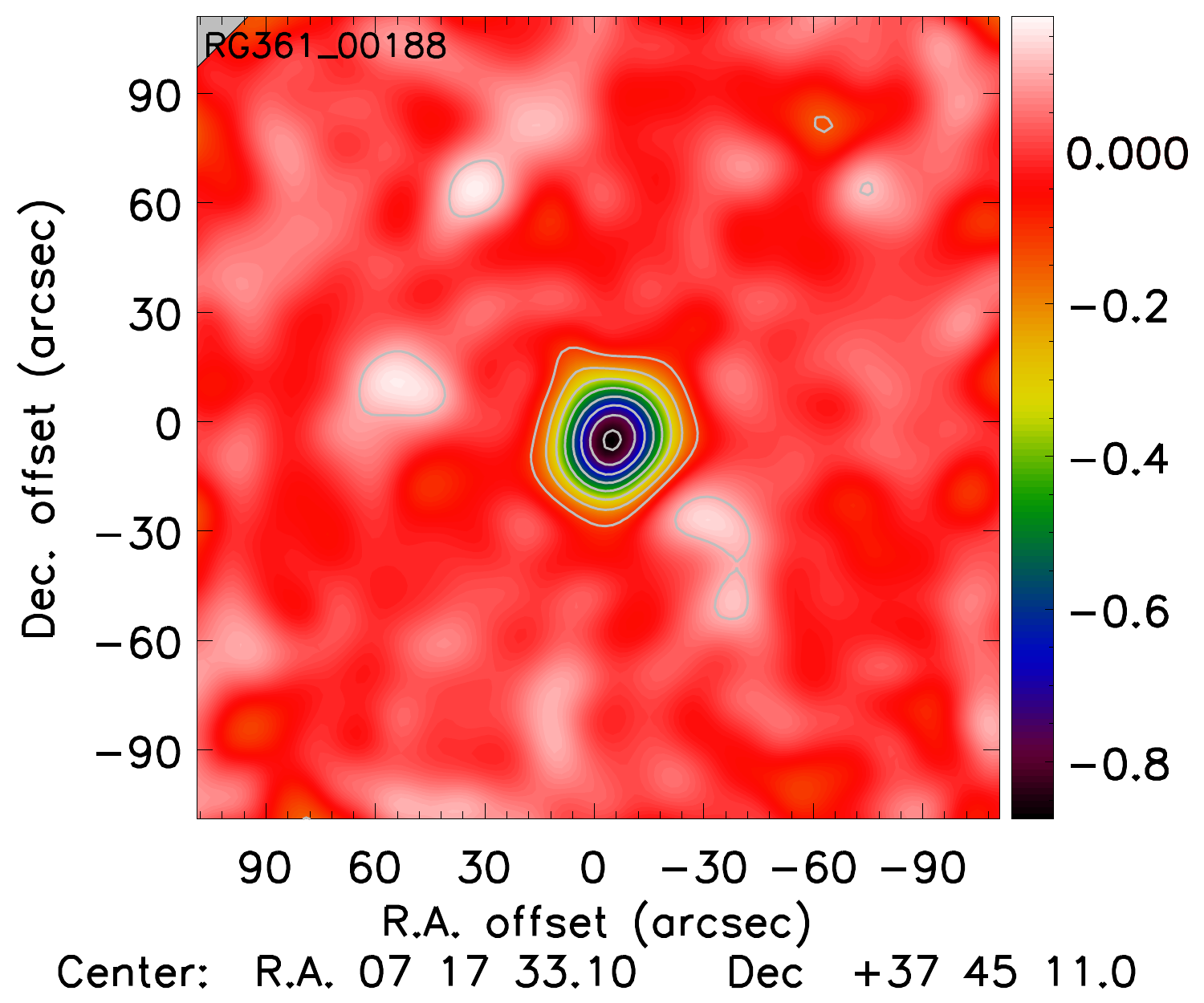} 
\put(-60,310){\makebox(0,0){\rotatebox{0}{\LARGE mJy/beam}}} \\
\includegraphics[trim=0cm 2.2cm 0cm 0cm, clip=true, scale=1]{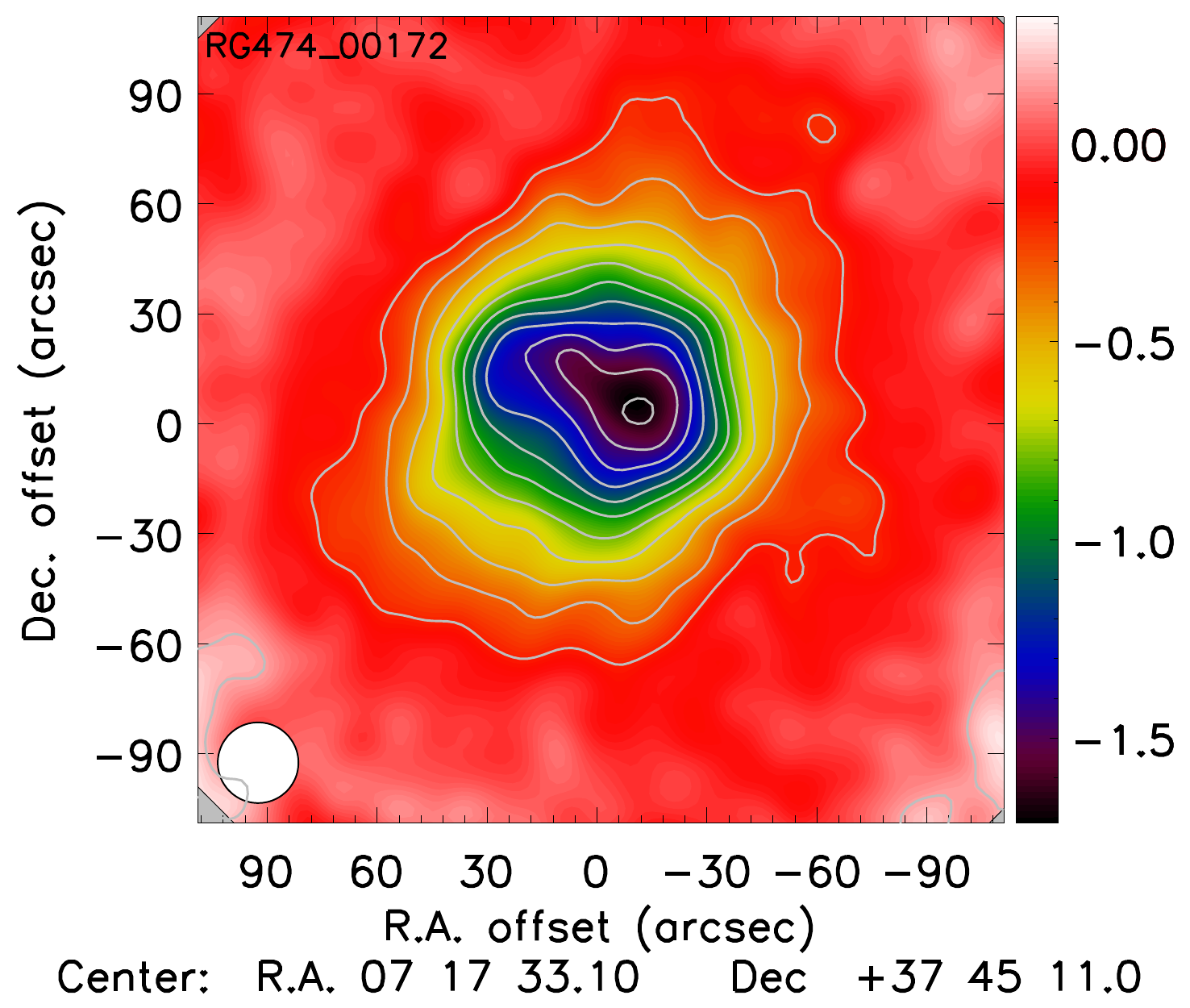} & 
\includegraphics[trim=2.3cm 2.2cm 0cm 0cm, clip=true, scale=1]{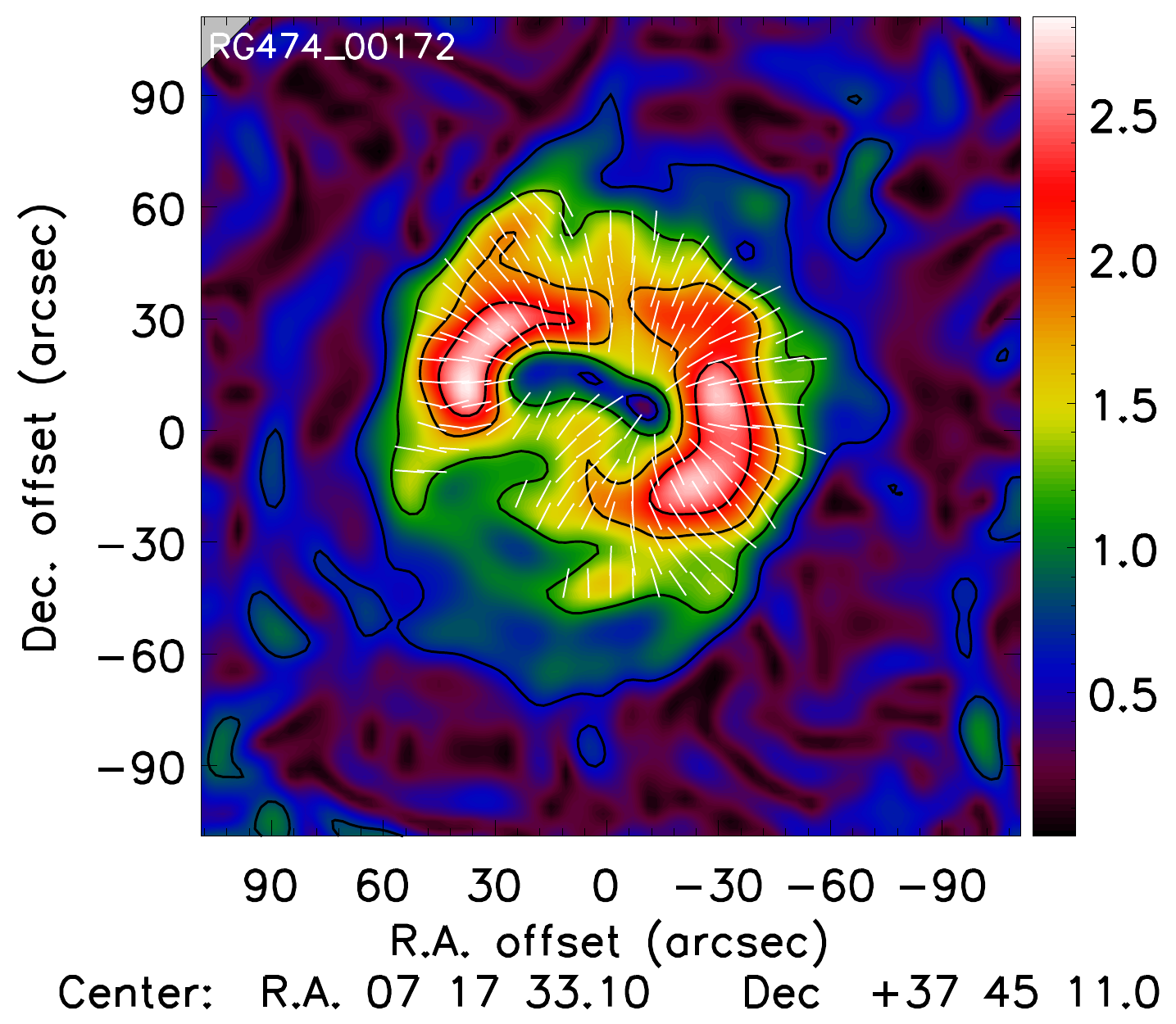} & 
\includegraphics[trim=2.3cm 2.2cm 0cm 0cm, clip=true, scale=1]{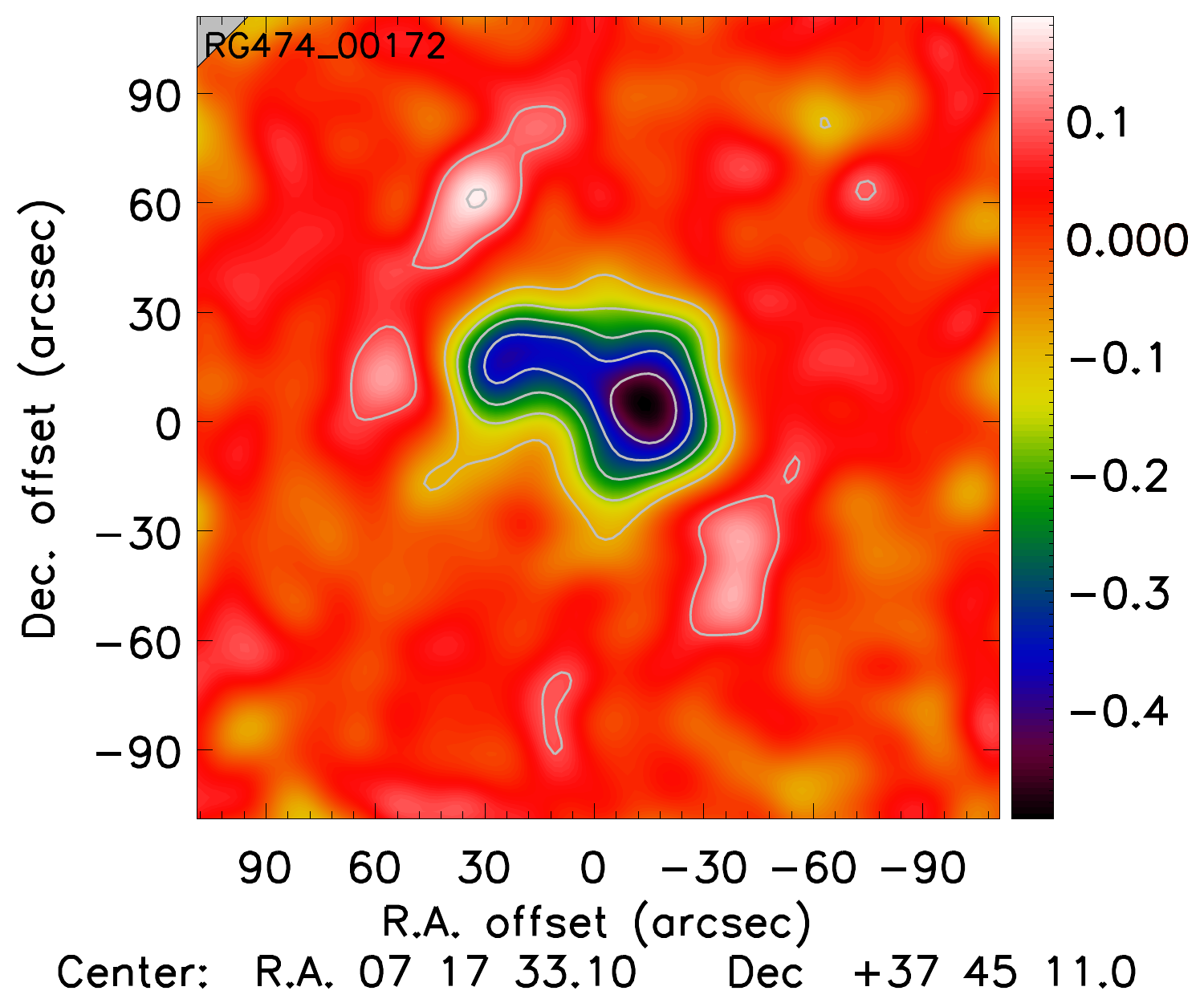} \\
\includegraphics[trim=0cm 2.2cm 0cm 0cm, clip=true, scale=1]{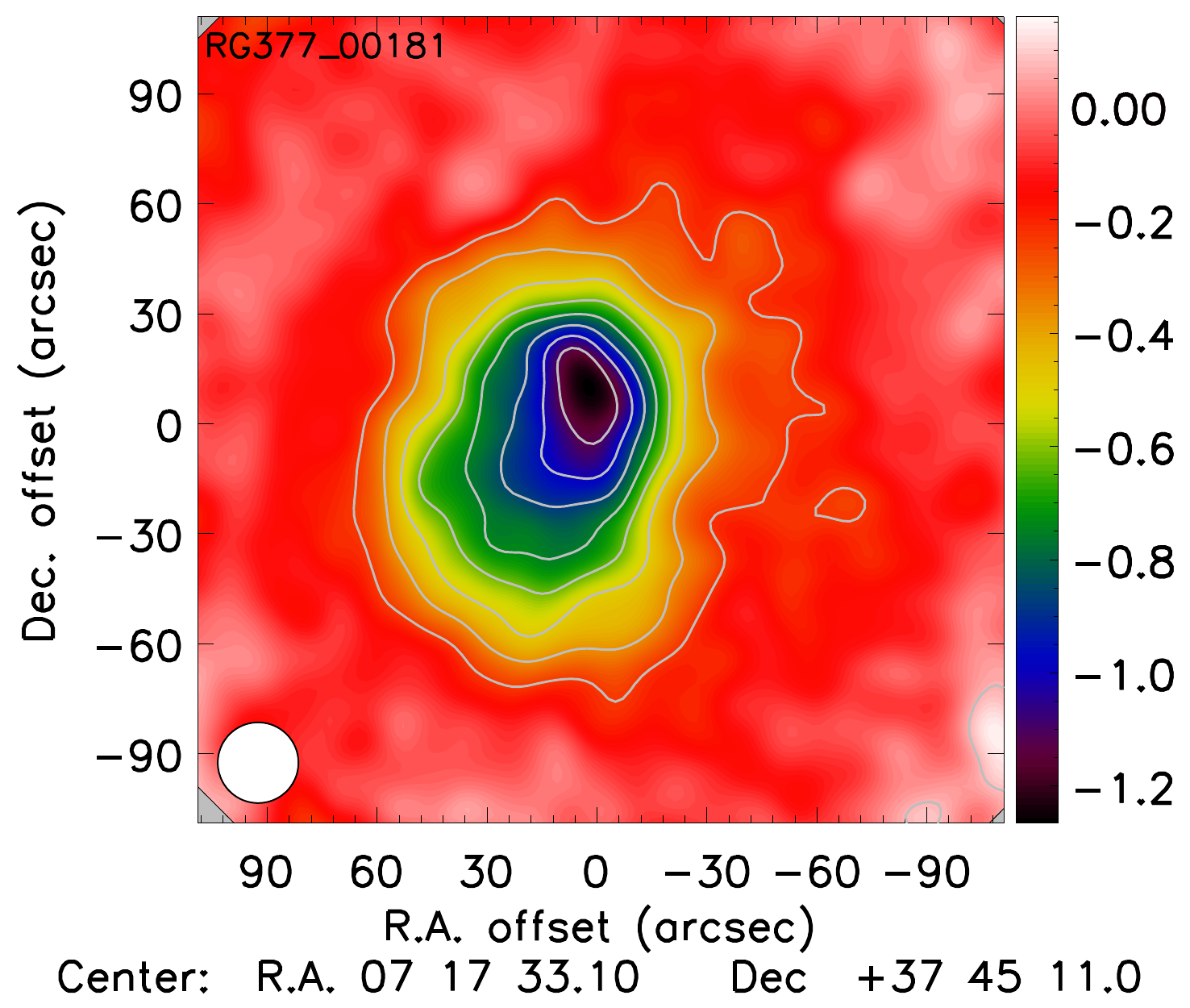} & 
\includegraphics[trim=2.3cm 2.2cm 0cm 0cm, clip=true, scale=1]{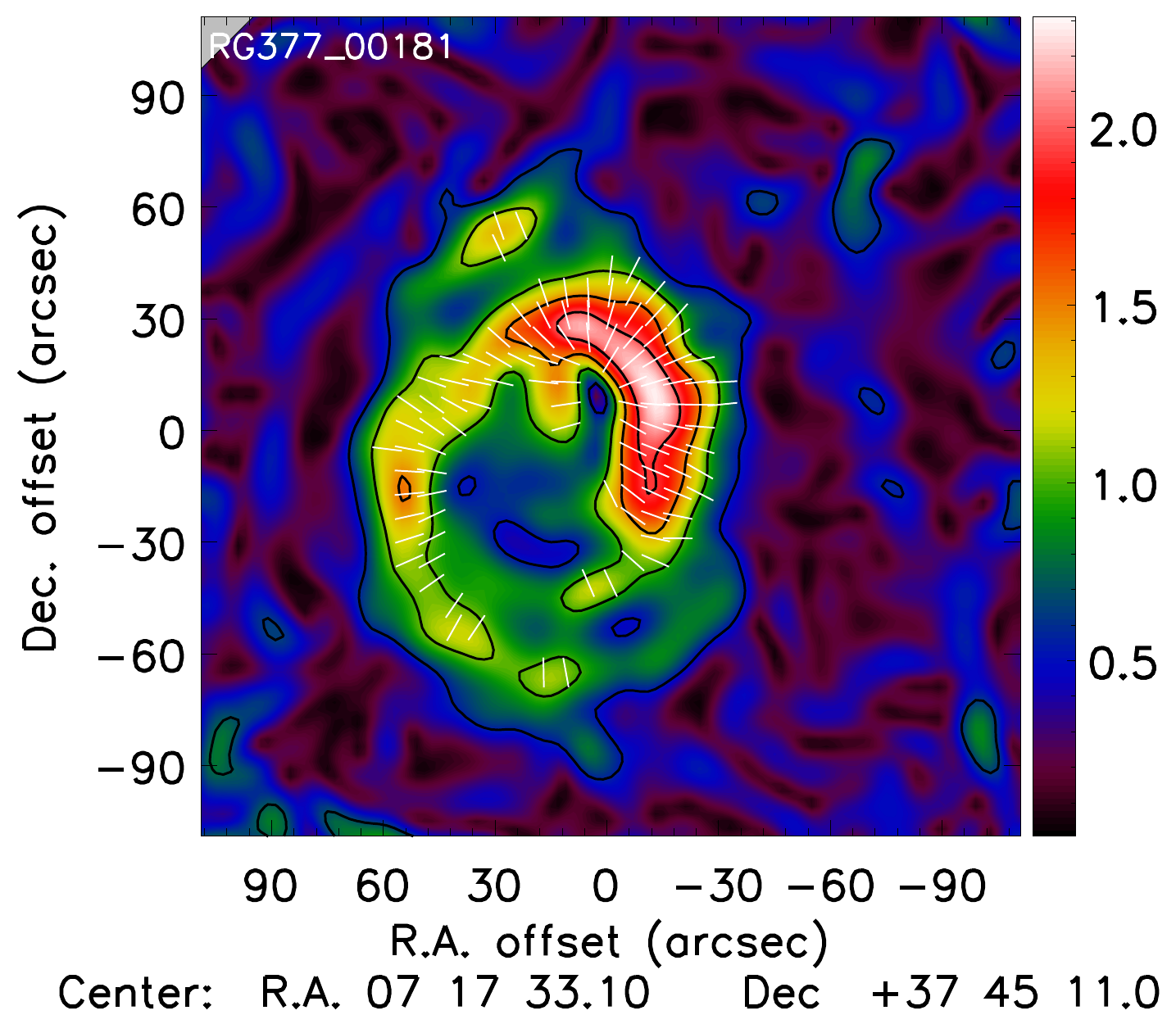} & 
\includegraphics[trim=2.3cm 2.2cm 0cm 0cm, clip=true, scale=1]{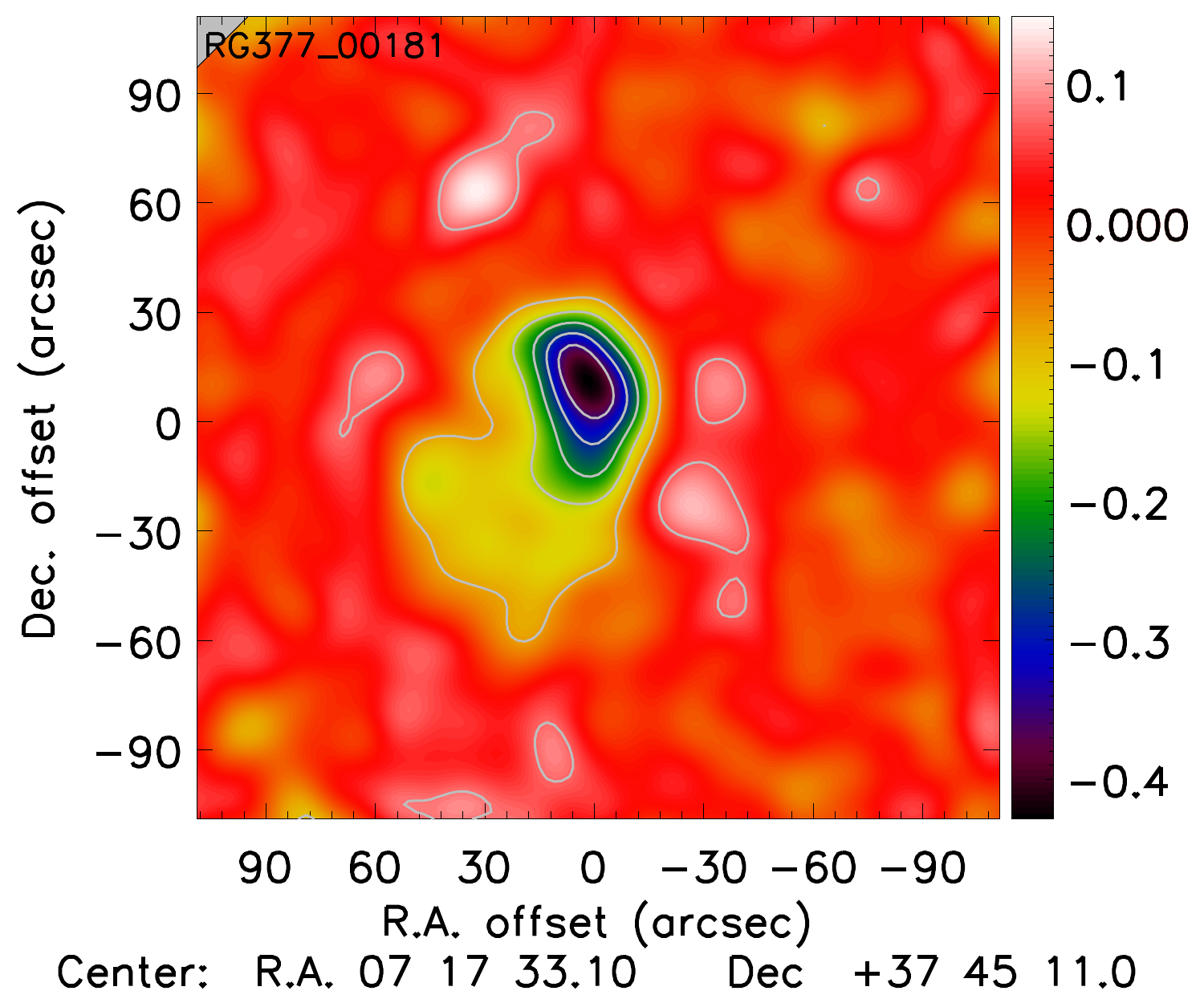} \\
\includegraphics[trim=0cm 2.2cm 0cm 0cm, clip=true, scale=1]{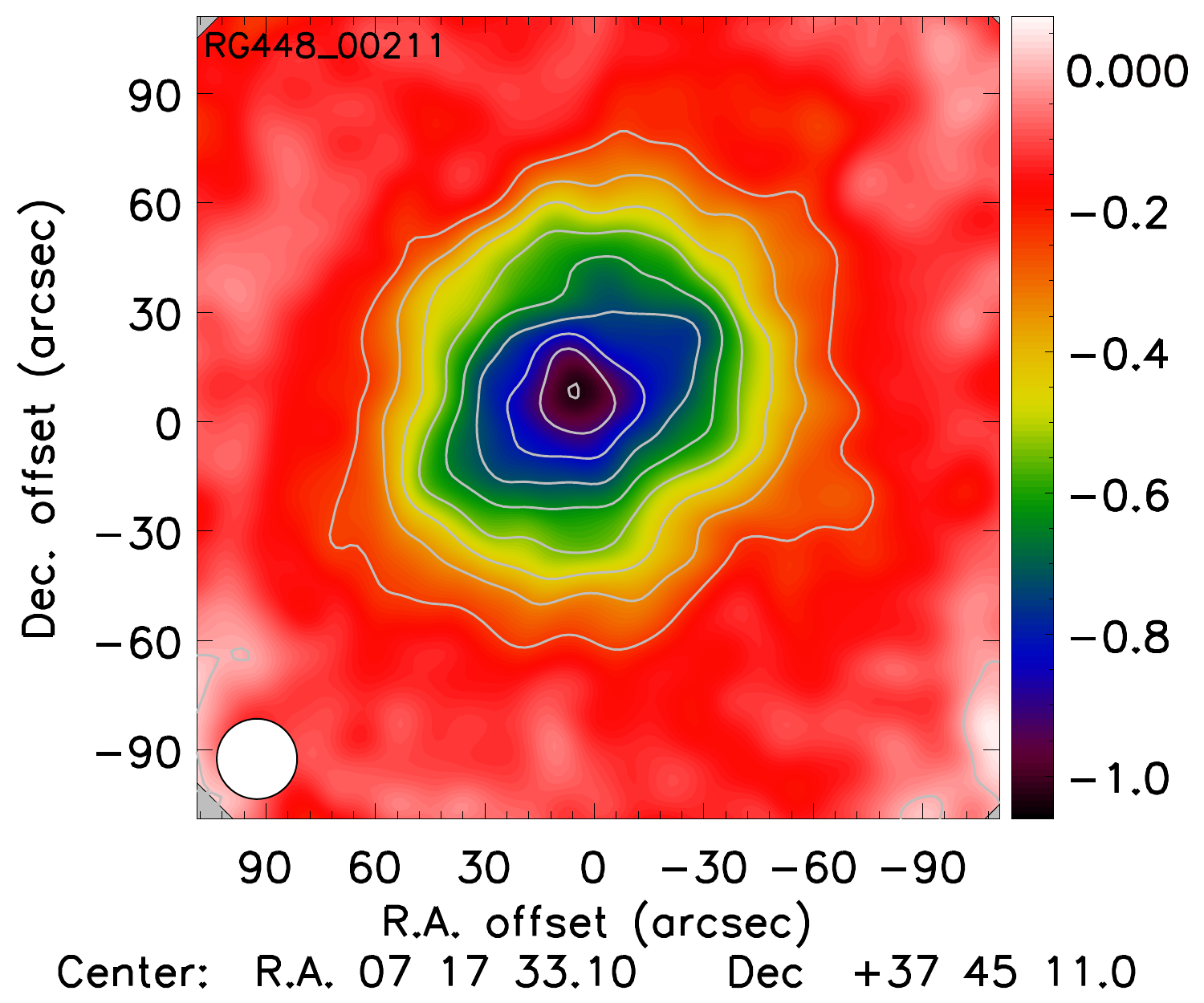} & 
\includegraphics[trim=2.3cm 2.2cm 0cm 0cm, clip=true, scale=1]{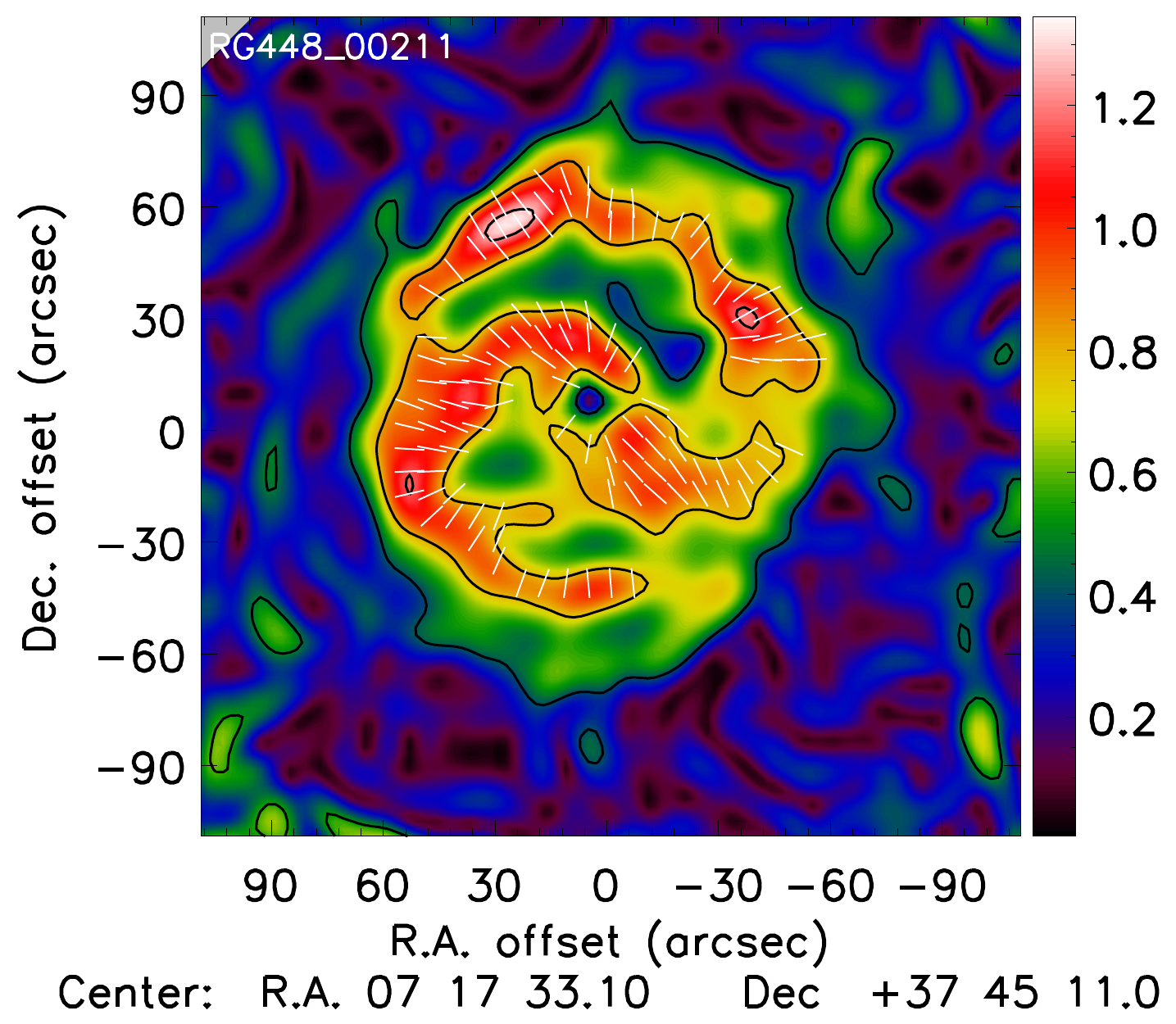} & 
\includegraphics[trim=2.3cm 2.2cm 0cm 0cm, clip=true, scale=1]{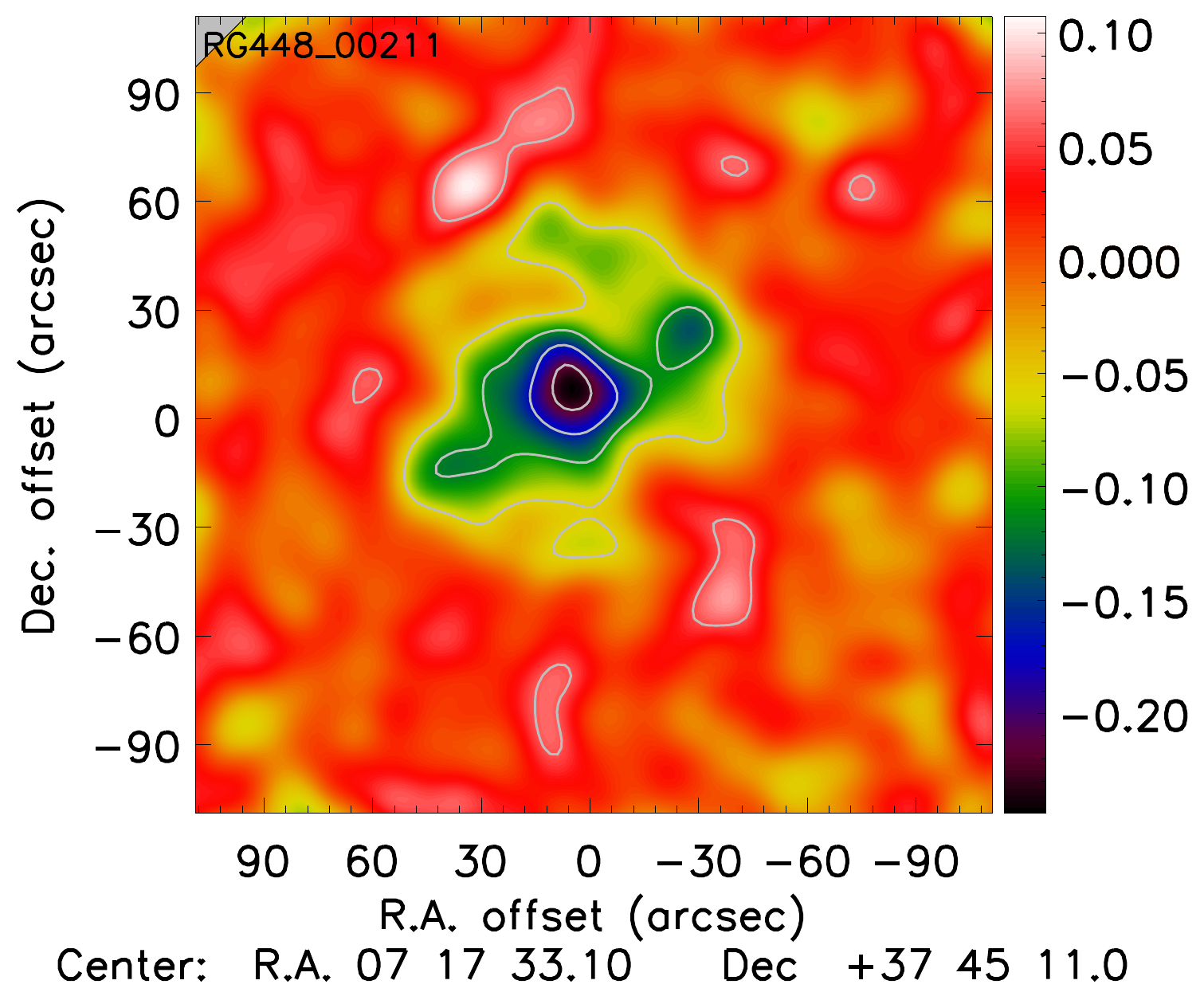} \\
\includegraphics[trim=0cm 0.7cm 0cm 0cm, clip=true, scale=1]{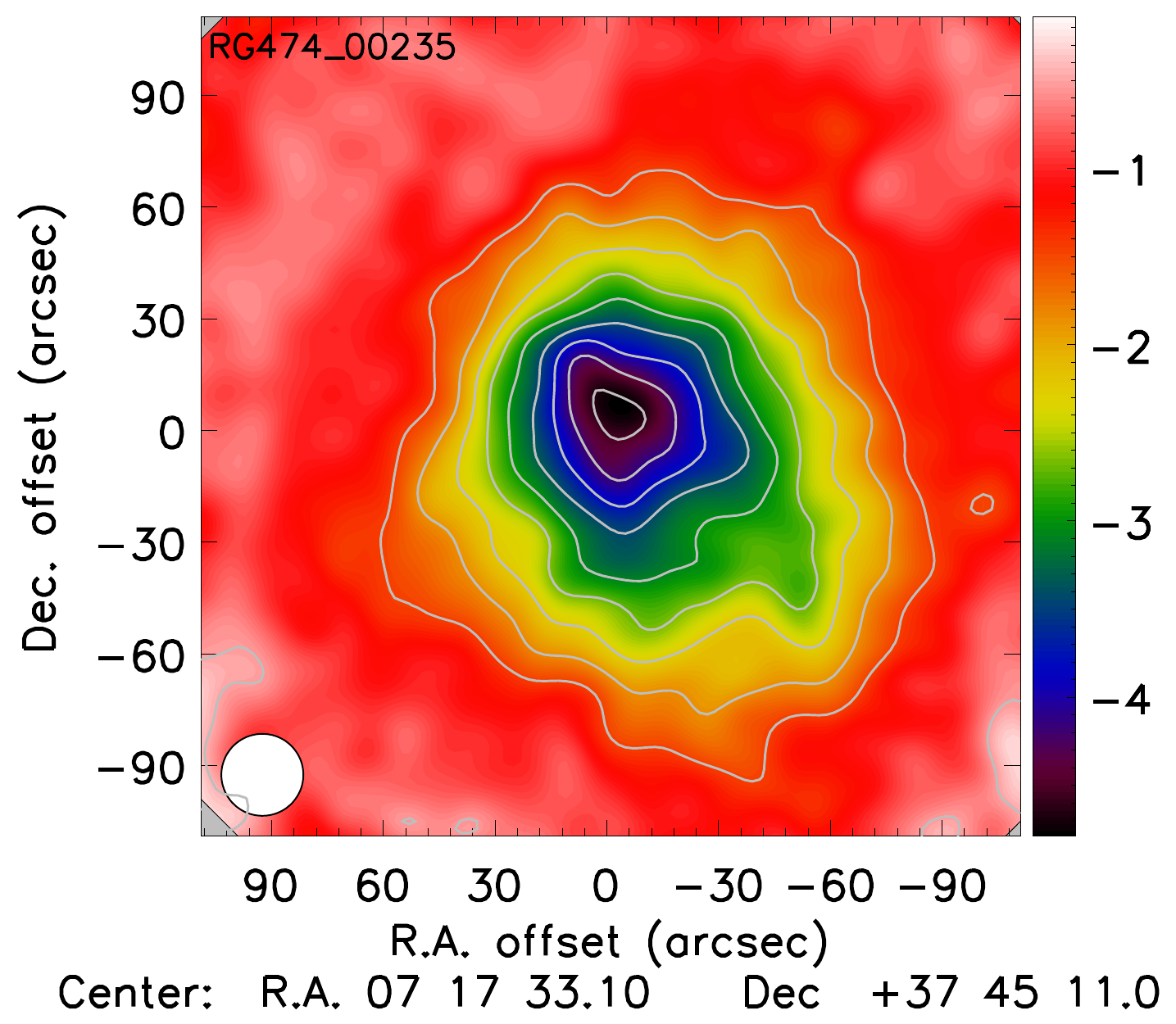} & 
\includegraphics[trim=2.3cm 0.7cm 0cm 0cm, clip=true, scale=1]{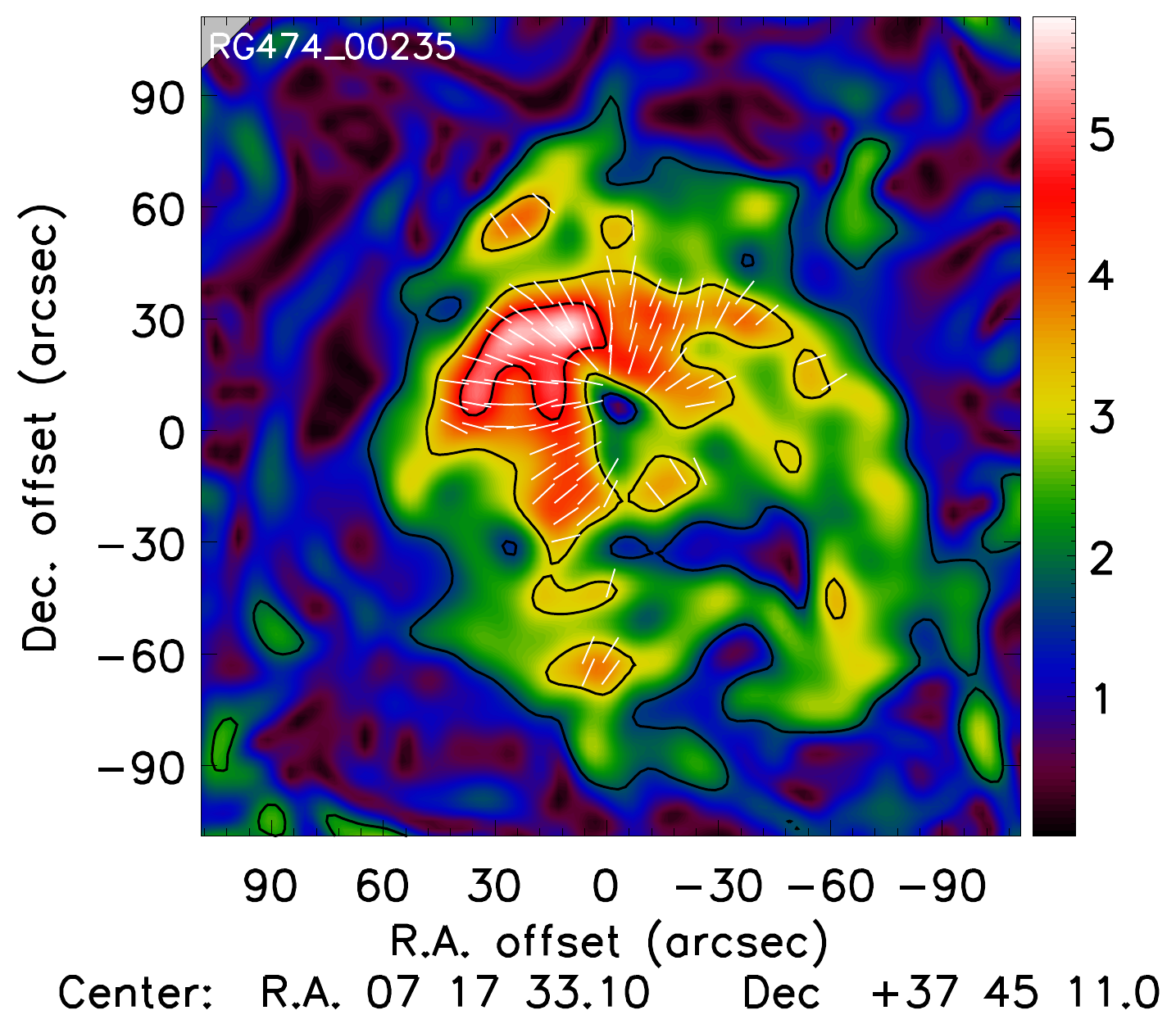} & 
\includegraphics[trim=2.3cm 0.7cm 0cm 0cm, clip=true, scale=1]{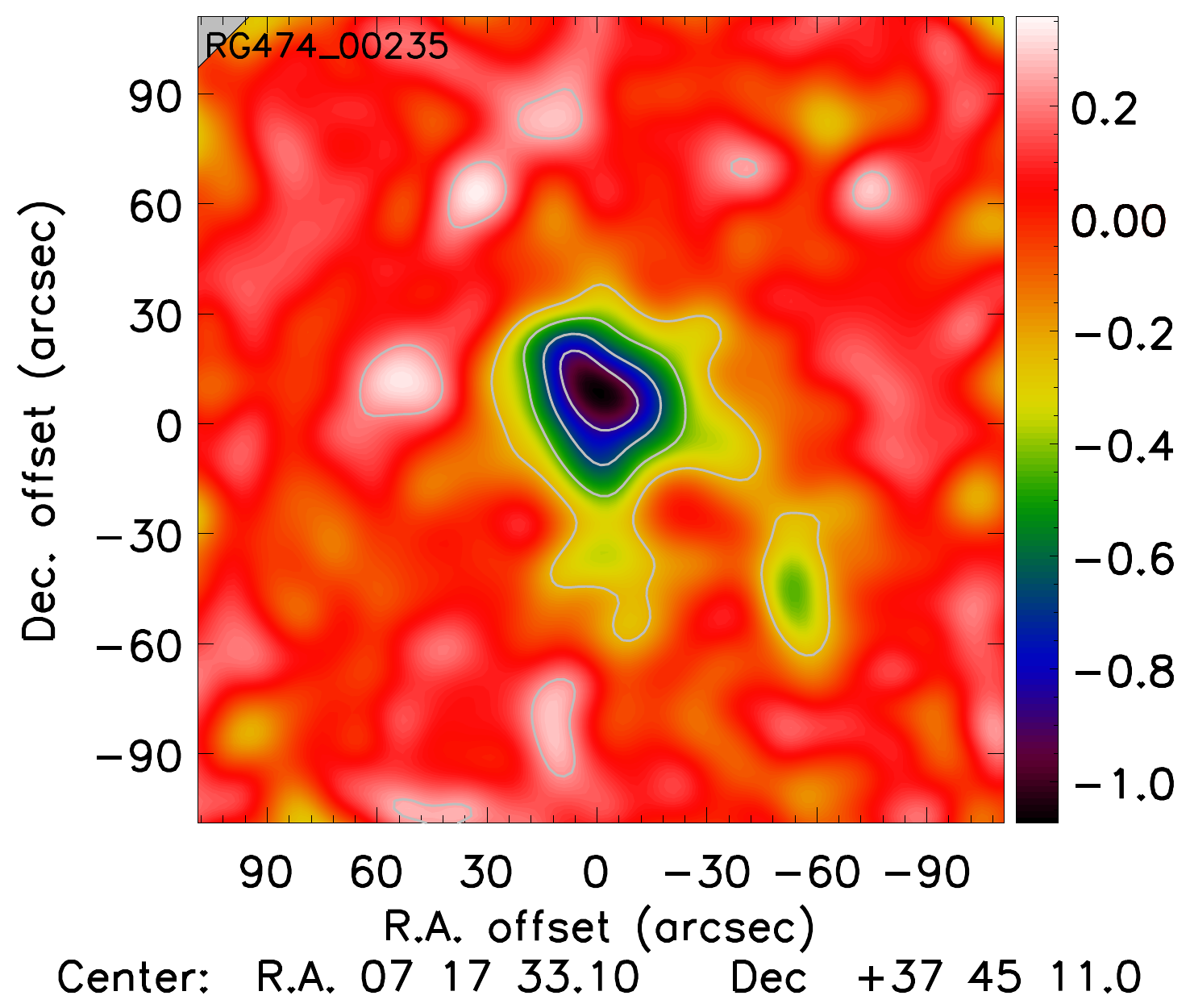} \\
\end{tabular}}
\caption{\footnotesize{Same as figure \ref{fig:RG_cluster_sample} in the case of end-to-end processing of the simulated RHAPSODY-G sub-sample: surface brightness maps (left), GGM filtered maps (middle) and DoG filtered maps (right). From top to bottom, the clusters are RG361\_00188, RG474\_00172, RG377\_00181 and RG448\_00211 (see also Table \ref{tab:rhapsody_summary} for the main properties of the sample). The surface brightness images are deconvolved from the transfer function (but not from the beam smoothing) and the signal-to-noise contours are estimated as in the case of real data (separated by steps of $2 \sigma$).}}
\label{fig:RG_cluster_sample_proc}
\end{figure*}

\subsection{NIKA processing of the RHAPSODY-G sub-sample}
While the DoG filter is linear and preserves the gaussian nature of the noise, this is not the case for the GGM filter. Therefore, in order to validate the behavior of the filter, we use end-to-end realistic simulations of the observations and post-processing of the RHAPSODY-G sub-sample, including all the artifacts present in the real data, but for which the true input signal is known.

To obtain simulated observed surface brightness images, we make use of the real raw telescope time ordered data (TOD) collected for the cluster \mbox{MACS~J0717.5+3745} (the one with the largest observing time, see Section \ref{sec:Application_to_the_NIKA_clusters_sample}), in which we inject the signal corresponding to the considered RHAPSODY-G tSZ maps, according to the telescope scanning strategy. Prior to adding the simulated signal to the data, half of the TOD scans are multiplied by $-1$ in order to cancel the astrophysical signal present in the real data, on the final co-added map. We aim at obtaining simulations corresponding to the highest signal-to-noise NIKA clusters. Therefore, before adding the simulated signal to the TOD, the noise is rescaled by the ratio of the RHAPSODY-G clusters peak surface brightness to the one of the deconvolved map of \mbox{MACS~J0717.5+3745} (typically a factor of $\sim 2$, see Section \ref{sec:Selection_of_a_RHAPSODY-G_sub-sample}). We note that this high signal-to-noise case also allows us to check the behavior of the filters in the low signal-to-noise regime. Indeed, the signal-to-noise is maximal at the peak, but smoothly vanishes towards the outskirts. Nevertheless, we also consider the case where the signal-to-noise peaks is 4, at 22~arcsec resolution (lowest signal-to-noise NIKA cluster, \mbox{PSZ1~G046.13+30.75}), to validate our procedure in the cluster center in this regime (Appendix \ref{sec:Transfer_function_deconvolution}). After including the simulated signal, the TOD are then processed similarly to the one of the real NIKA clusters \citep[see][for more details]{Adam2015}.

Finally, we obtain realistic maps of the RHAPSODY-G sub-sample, as if they were truly observed using NIKA, with a signal-to-noise peak of about 20, at 22~arcsec resolution. The maps, once deconvolved from the processing transfer function (see Section \ref{sec:Transfer_function_filtering}), are displayed in Figure~\ref{fig:RG_cluster_sample_proc}, together with their GGM and DoG filtered maps obtained using our baseline parameter values (cf. Section~\ref{sec:Baseline_filtering_parameters}). These maps can be compared to that of Figure~\ref{fig:RG_cluster_sample} to validate our procedure, and they are used in the following sub-sections to quantify the properties of the filtered maps in the presence of systematic effects and noise.

\subsection{Transfer function filtering}\label{sec:Transfer_function_filtering}
By construction, the zero level surface brightness of the NIKA maps is not defined. However, this does not affect the results presented in this paper since both the GGM and DoG filtering procedures are insensitive to zero level effects. In addition, the NIKA processing transfer function reduces the signal on scales that are larger than the instrument field of view ($\sim 2$ arcmin). After deconvolution, small differences can remain between the deconvolved image and the original input image, due in particular to anisotropies in the scanning strategy and uncertainties in the estimated transfer function. However, these differences are mostly prominent on scales larger than the NIKA field of view, where filtering effects are important \citep{Adam2015}. By comparing the NIKA-like processed RHAPSODY-G sub-sample to the non-processed maps, we find that these differences have a negligible impact on the filter application with respect to the noise (see also figures \ref{fig:RG_cluster_sample} and \ref{fig:RG_cluster_sample_proc}). In appendix \ref{sec:Transfer_function_deconvolution}, we provide more details on the deconvolution and its impact on the results.

\subsection{Noise spatial correlations and propagation through the filters}\label{sec:Noise_spatial_correlations_and_propagation_through_the_filters}
\begin{figure*}[h]
\centering
\includegraphics[trim=0cm 0cm 0cm 0cm, clip=true, totalheight=4.4cm]{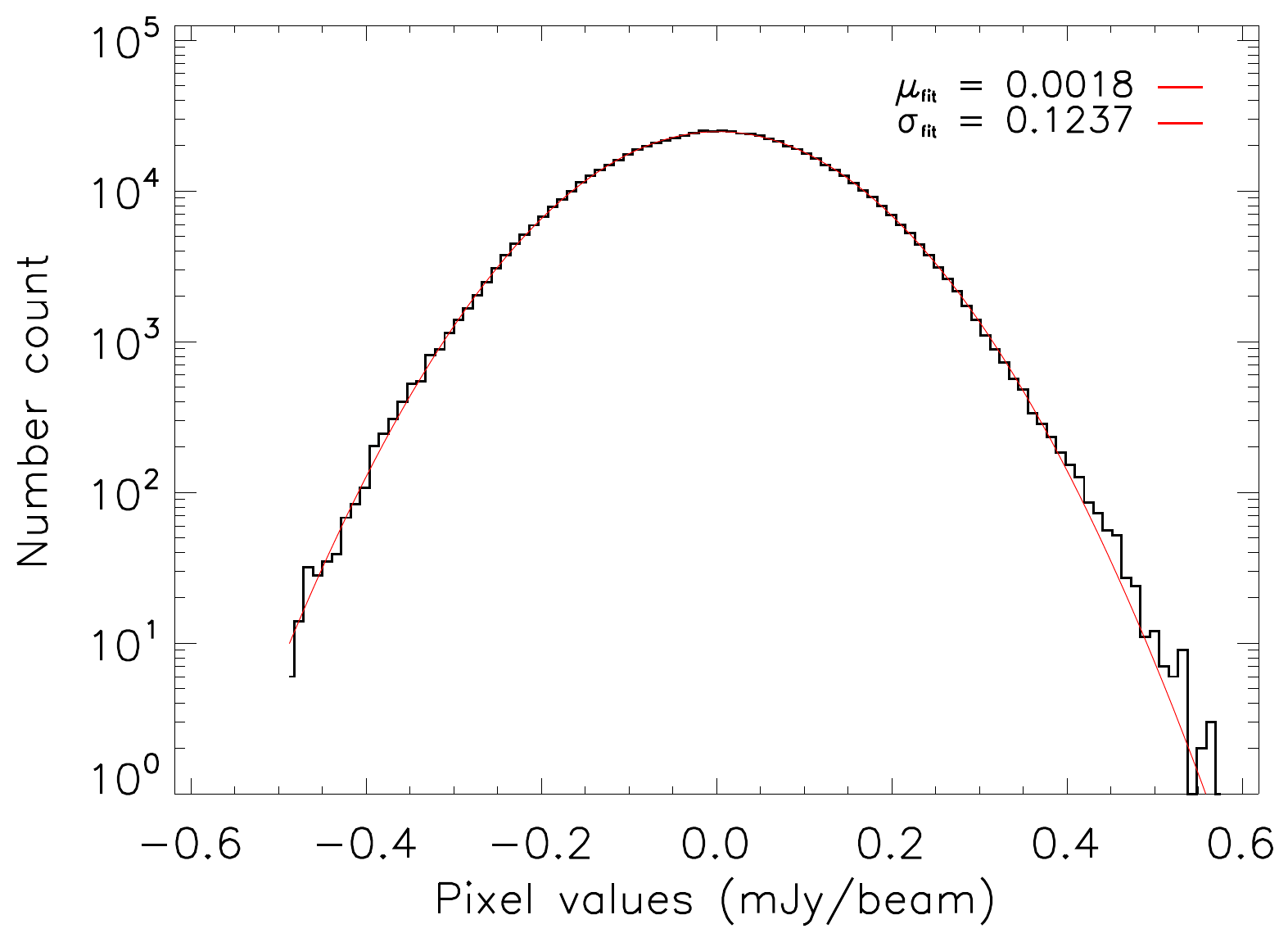}
\includegraphics[trim=0cm 0cm 0cm 0cm, clip=true, totalheight=4.4cm]{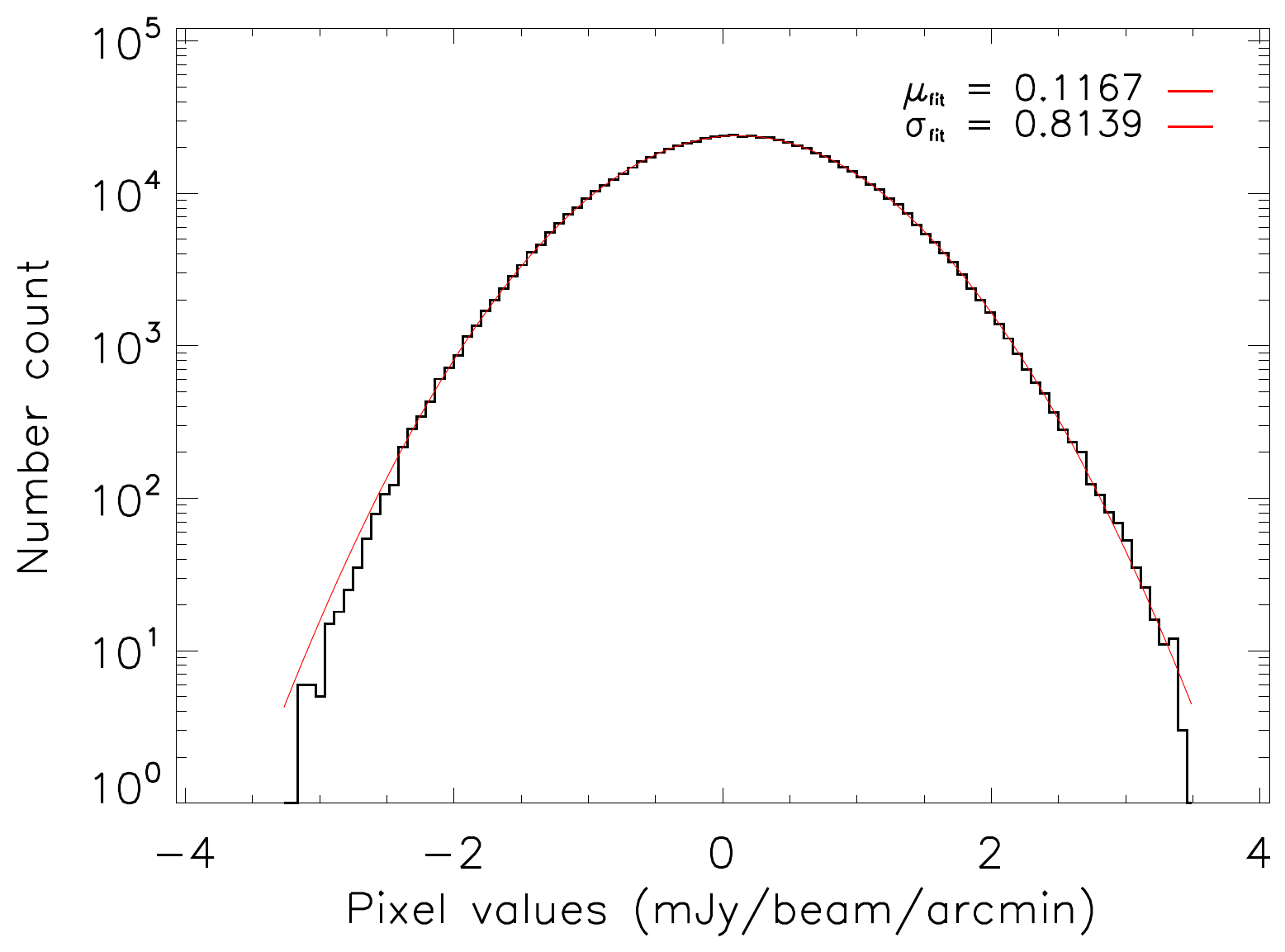}
\includegraphics[trim=0cm 0cm 0cm 0cm, clip=true, totalheight=4.4cm]{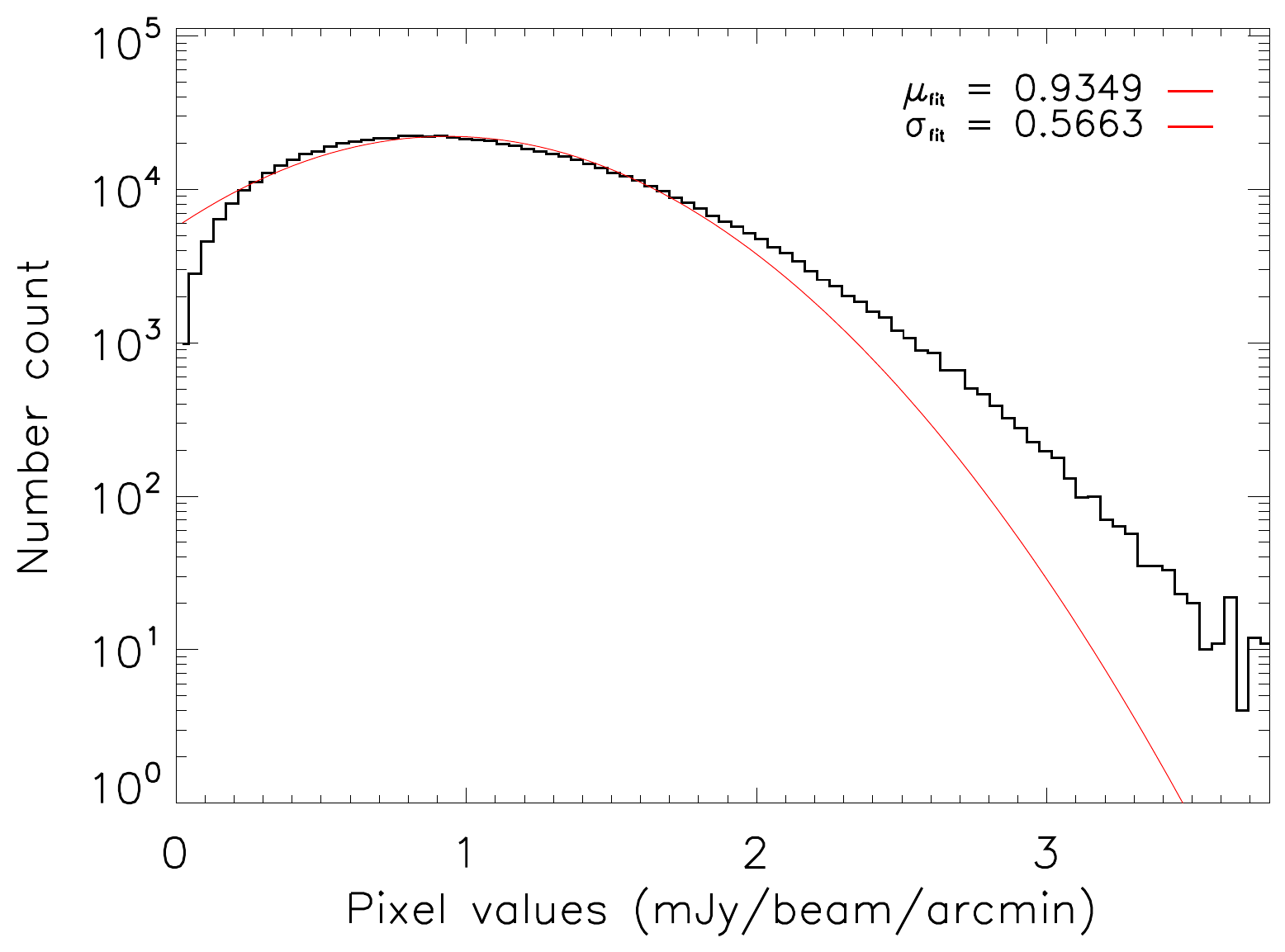}
\caption{\footnotesize{Noise distribution of all the pixels enclosed within 30 arcsec radius from the cluster center, in the case of \mbox{CL~J1226.9+3332}, for our baseline filter parameters. The red curves provides the best Gaussian fit to the histograms. {\bf Left:} DoG noise histogram. {\bf Middle:} GGM noise histogram in the case of $\hat{S}$ set to the observed signal of \mbox{CL~J1226.9+3332}. {\bf Right:} GGM noise histogram in the case where $\hat{S} = 0$ in Eq \ref{eq:GGM_filter_noise}.}}
\label{fig:noise_statistics1}
\end{figure*}

\begin{figure}[h]
\centering
\includegraphics[width=0.5\textwidth]{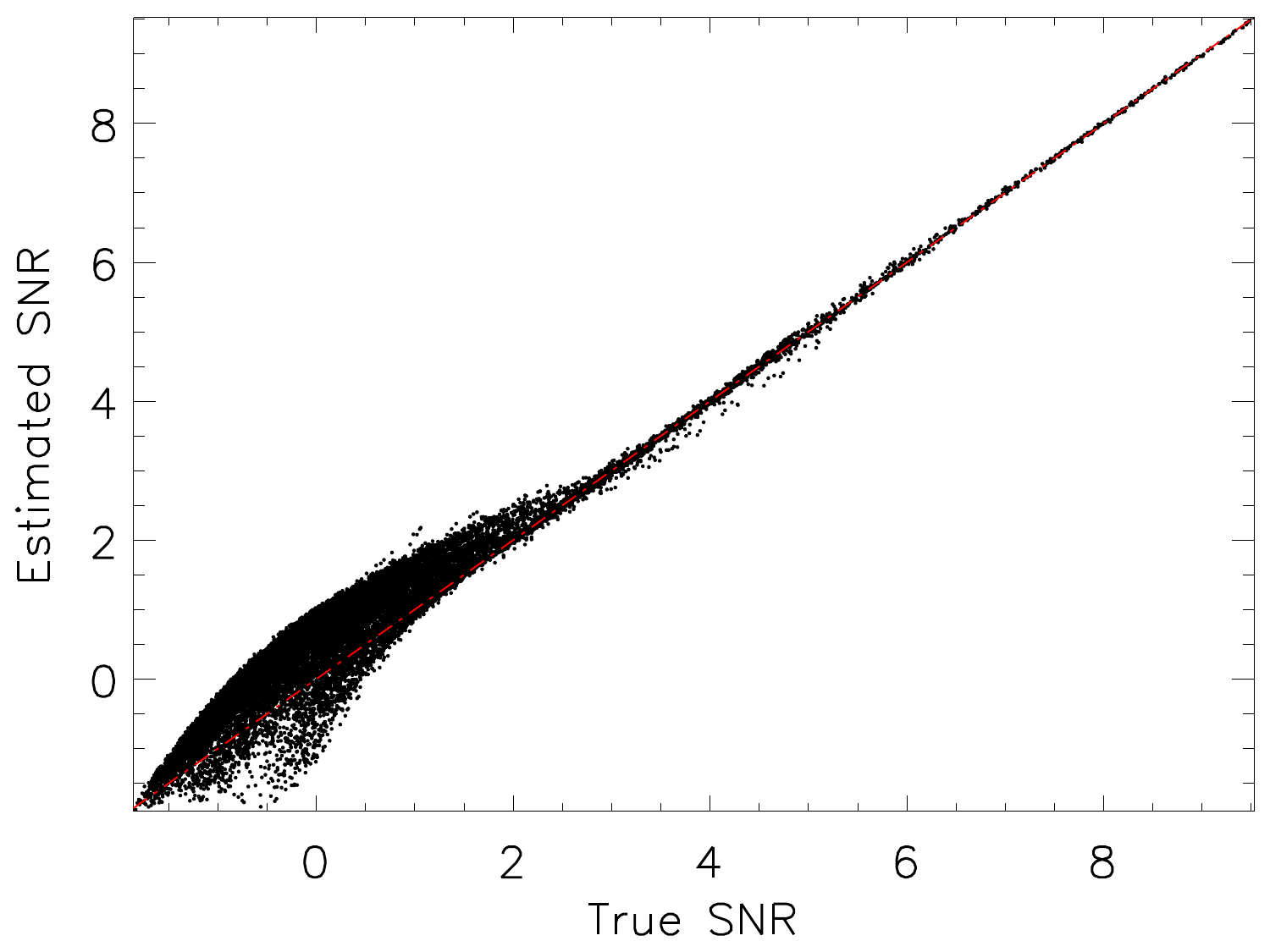}
\caption{\footnotesize{Comparison of the estimated GGM filtered maps signal-to-noise ratio as a function of the true signal-to-noise, using the RHAPSODY-G sub-sample. The estimated signal-to-noise is computed using the NIKA processed RHAPSODY-G sub-sample (Eq \ref{eq:GGM_filter_snr}). The red dashed line represent the one to one correspondence. Each point corresponds to a pixel of the map.}}
\label{fig:noise_statistics2}
\end{figure}

The NIKA noise is Gaussian and spatially correlated \citep{Adam2016a} but its behavior does not necessarily propagate in a trivial way under the application of the filters, especially in the case of the GGM filter, which is non-linear. In order to estimate the significance of the structures that we observe in Section~\ref{sec:Application_to_the_NIKA_clusters_sample}, we use noise realizations of the surface brightness maps generated as described in \cite{Adam2016a}. They account for the map pixel-to-pixel correlated noise induced by residual atmospheric and electronic noise, the inhomogeneities of the noise due to the scanning strategy, and the contribution of the cosmic infrared background, which is non-negligible for the deepest observations. For each cluster, we produce a set of 1000 surface brightness noise realizations, $N^{(i)}$, which are deconvolved from the transfer function associated to the NIKA data processing similarly to their corresponding cluster data and used to estimate the noise in the filtered maps using the following procedure. 

As the DoG filter is linear and preserves the noise Gaussianity, we process these surface-brightness-noise-only Monte Carlo realizations through the same filter as the data. Then, we estimate the respective DoG filtered maps standard deviation by taking the root-mean-square of all the processed Monte Carlo realizations, for each pixel of the map. The DoG signal-to-noise ratio are straightforwardly obtained by normalizing the filtered cluster data by the standard deviation maps, and follow Gaussian statistics.

According to the definition of Eq \ref{eq:GGM_filter}, the noise of the GGM filtered maps depends on the surface brightness signal itself because of the non-linearity of the filter. Therefore, we use the observed signal as an estimate of the true signal, $\hat{S}$, to which we add the surface brightness noise realizations, prior to processing them through the GGM filter. The estimate of the filtered map noise contribution, from each Monte Carlo realization $i$, is then given by
\begin{equation}
	N^{(i)}_{\rm GGM} = \left(\hat{S} + N^{(i)}\right)_{\rm GGM} -  \hat{S}_{\rm GGM},
	\label{eq:GGM_filter_noise}
\end{equation}
where the index GGM corresponds to the filter application defined by Eq \ref{eq:GGM_filter}. For each pixel of the map, we compute $\mu_{\rm GGM}$ and $\sigma_{\rm GGM}$, the mean and the standard deviation of the noise realizations $N^{(i)}_{\rm GGM}$ per map pixel, respectively. Accounting for the non-zero mean of the filtered map noise, the effective signal-to-noise ratio is defined as 
\begin{equation}
{\rm SNR}_{\rm GGM} = \frac{\hat{S}_{\rm GGM} - \mu_{\rm GGM}}{\sigma_{\rm GGM}}.
	\label{eq:GGM_filter_snr}
\end{equation}
In practice, the estimated surface brightness signal is the sum of the true signal, $S$, and the noise, $N$, as $\hat{S} = S + N$. In the low surface brightness signal-to-noise regions, $\hat{S}$ is therefore dominated by the noise, which can bias low the value of $\mu_{\rm GGM}$. In the case of the RHAPSODY-G sub-sample, we also dispose of the true input surface brightness signal, $\hat{S} = S$, and we use it to compute the true signal-to-noise, to quantify the reliability of our signal-to-noise estimation procedure, as discussed below.

The noise distribution is illustrated in Figure \ref{fig:noise_statistics1} for both the GGM and the DoG filters. The data of \mbox{CL~J1226.9+3332} are used as an example, in the case of our baseline filter parameters (Section \ref{sec:Baseline_filtering_parameters}). The histograms account for the pixels that are enclosed within 30~arcsec from the cluster center, where the noise is homogeneous. In the case of the GGM filter, we consider both the possibility in which the cluster signal, $\hat{S}$, is set to the measured signal from \mbox{CL~J1226.9+3332}, and where it is set to zero in Eq \ref{eq:GGM_filter_noise}, to illustrate the noise behavior in low signal-to-noise regions. As we can see, the DoG noise statistics (Figure \ref{fig:noise_statistics1}, left panel) is described by a Gaussian distribution with a mean compatible with zero, by construction. The GGM noise distribution is well described by a Gaussian distribution in signal dominated regions (middle panel). However, as the signal vanishes, such as in the external regions, the noise becomes non Gaussian, its mean increases and the distribution gets positively skewed. Despite the fact that the non-zero mean of the noise is corrected for when computing the significance of the data, we note that the noise is boosted by up to 15\% (in the wing of the distribution), in the extreme case of zero signal, with respect to what one would expect for Gaussian noise (Figure \ref{fig:noise_statistics1}, red line).

The validation of the signal-to-noise reconstruction of the GGM filtered maps is illustrated in Figure~\ref{fig:noise_statistics2}. It shows the estimated signal-to-noise as a function of the true signal-to-noise, computed using the RHAPSODY-G sub-sample for which both $\hat{S}$ and $S$ are available. We can observe that in the low signal-to-noise regime, the estimated signal-to-noise is overall biased high, due to the fact that $\mu_{\rm GGM}$ is biased low. At higher signal-to-noise ($\gtrsim 3$), the estimated signal-to-noise converges to the true signal-to-noise within a few percent. Our procedure is thus valid in this regime.

Finally, we use the end-to-end processed and filtered maps from a RHAPSODY-G sub-sample, shown in Figure \ref{fig:RG_cluster_sample_proc}, to check that the structures we recover are reliable. Indeed, above 3, the signal-to-noise of the recovered features is consistent with what we expect in the case of the raw maps of Figure \ref{fig:RG_cluster_sample} when using the true signal-to-noise.

\subsection{Point sources residuals}\label{sec:Point_sources_residuals}
\begin{figure}[h]
\resizebox{0.5\textwidth}{!} {
\begin{tabular}{lll}
\includegraphics[trim=0cm 2.2cm 0cm 0cm, clip=true, scale=1]{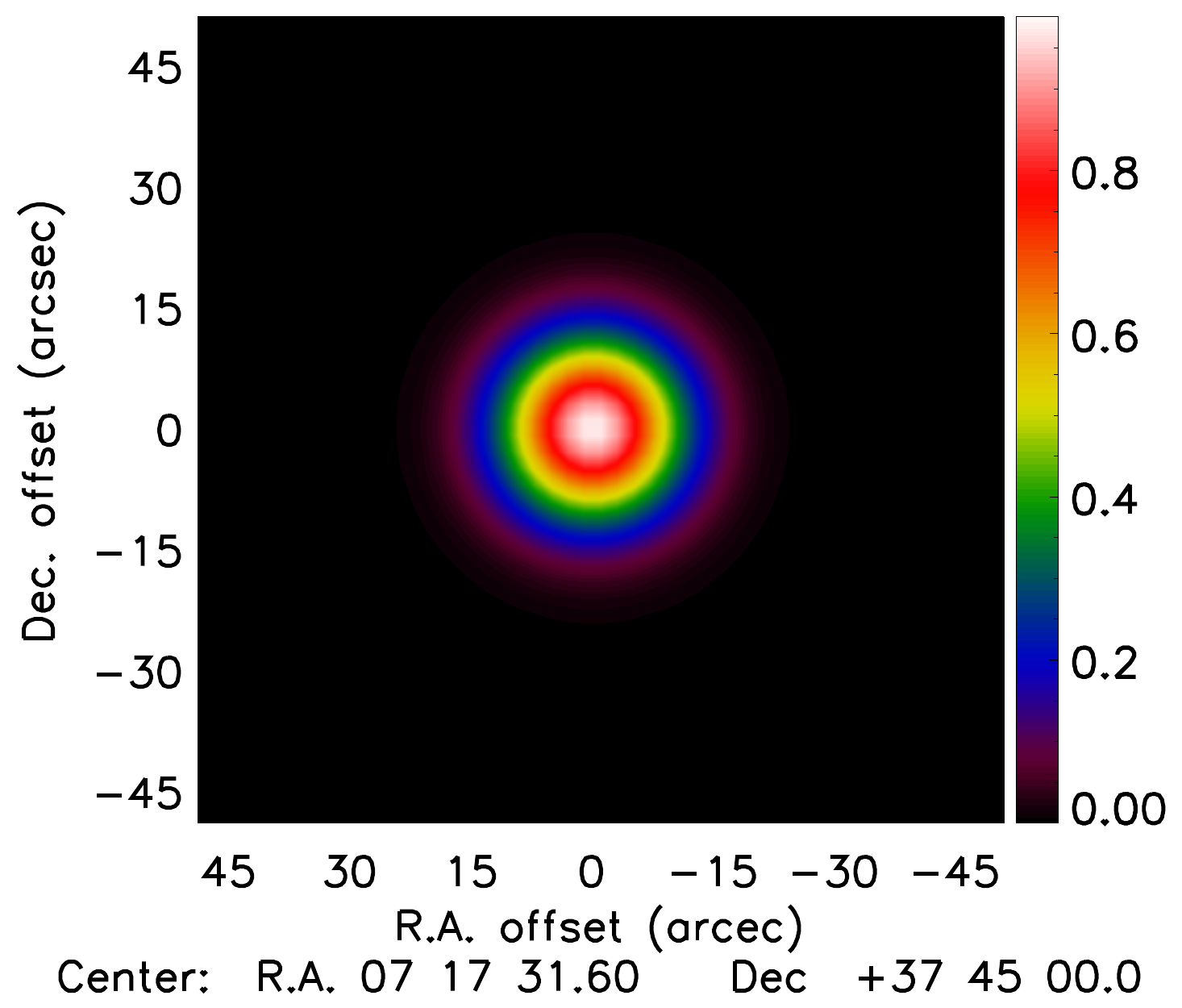} 
\put(-60,310){\makebox(0,0){\rotatebox{0}{\LARGE mJy/beam}}} & 
\includegraphics[trim=2.3cm 2.2cm 0cm 0cm, clip=true, scale=1]{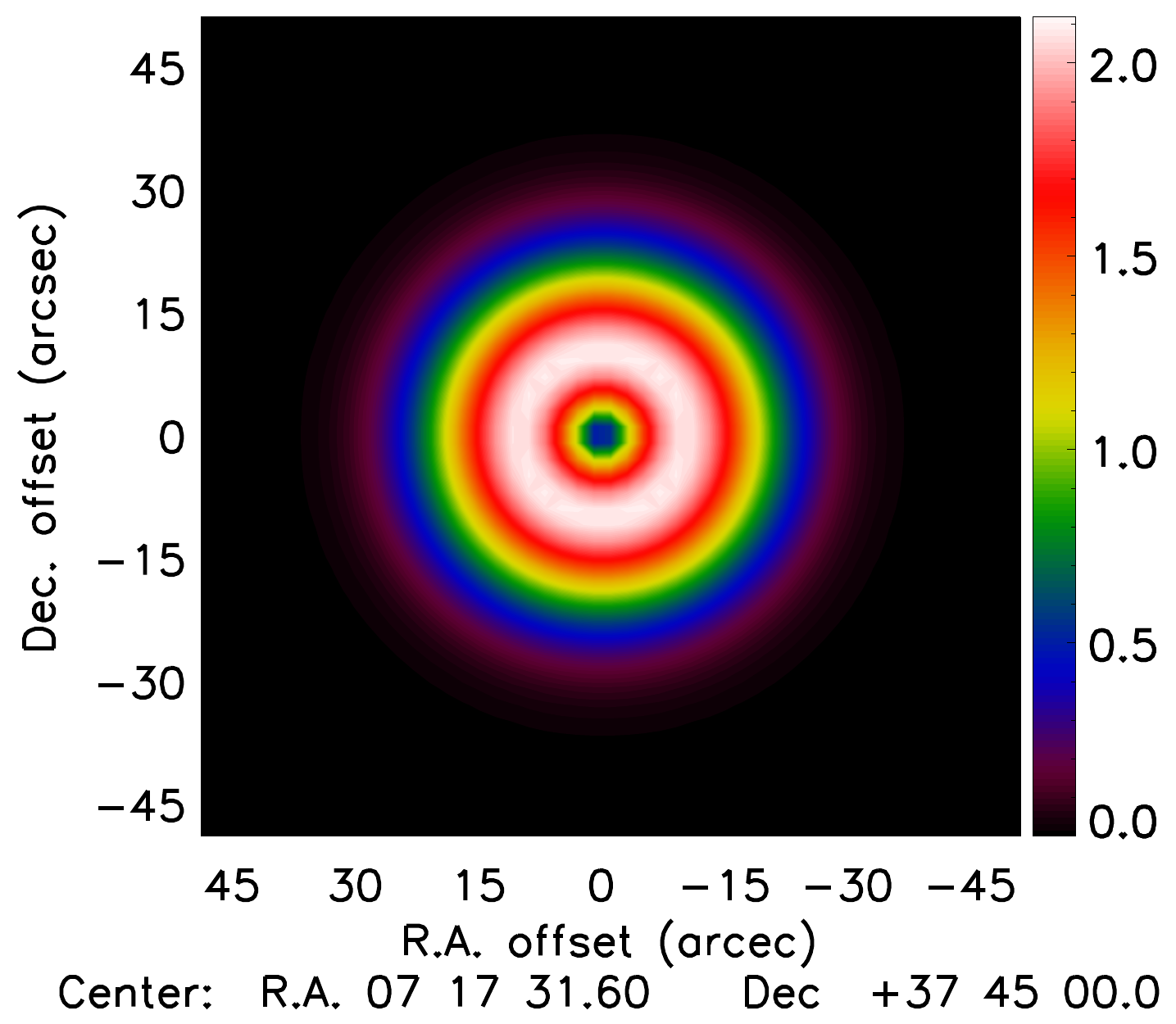} 
\put(-60,310){\makebox(0,0){\rotatebox{0}{\LARGE mJy/beam/arcmin}}} & 
\includegraphics[trim=2.3cm 2.2cm 0cm 0cm, clip=true, scale=1]{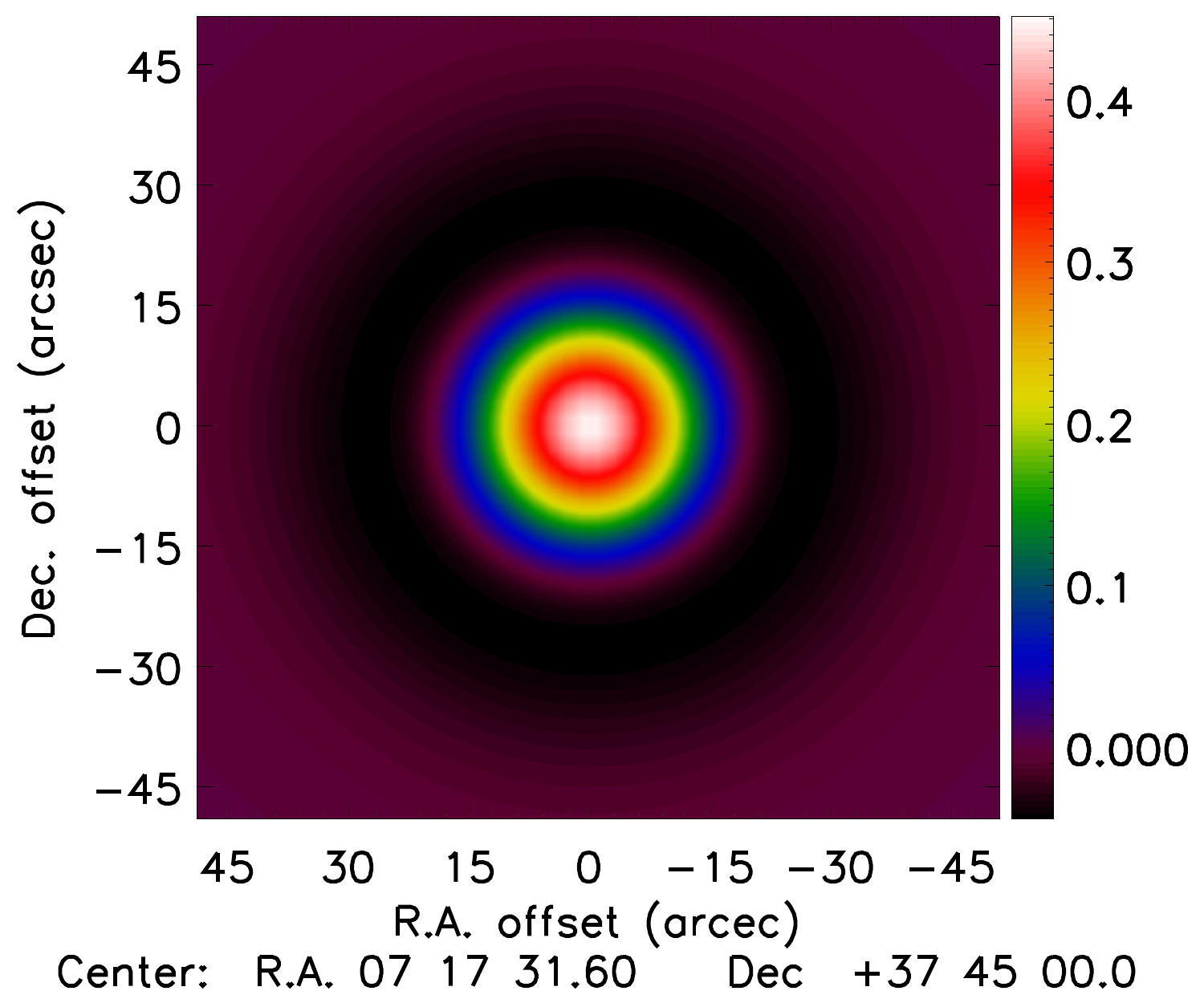} 
\put(-60,310){\makebox(0,0){\rotatebox{0}{\LARGE mJy/beam}}} \\
\includegraphics[trim=0cm 0.7cm 0cm 0cm, clip=true, scale=1]{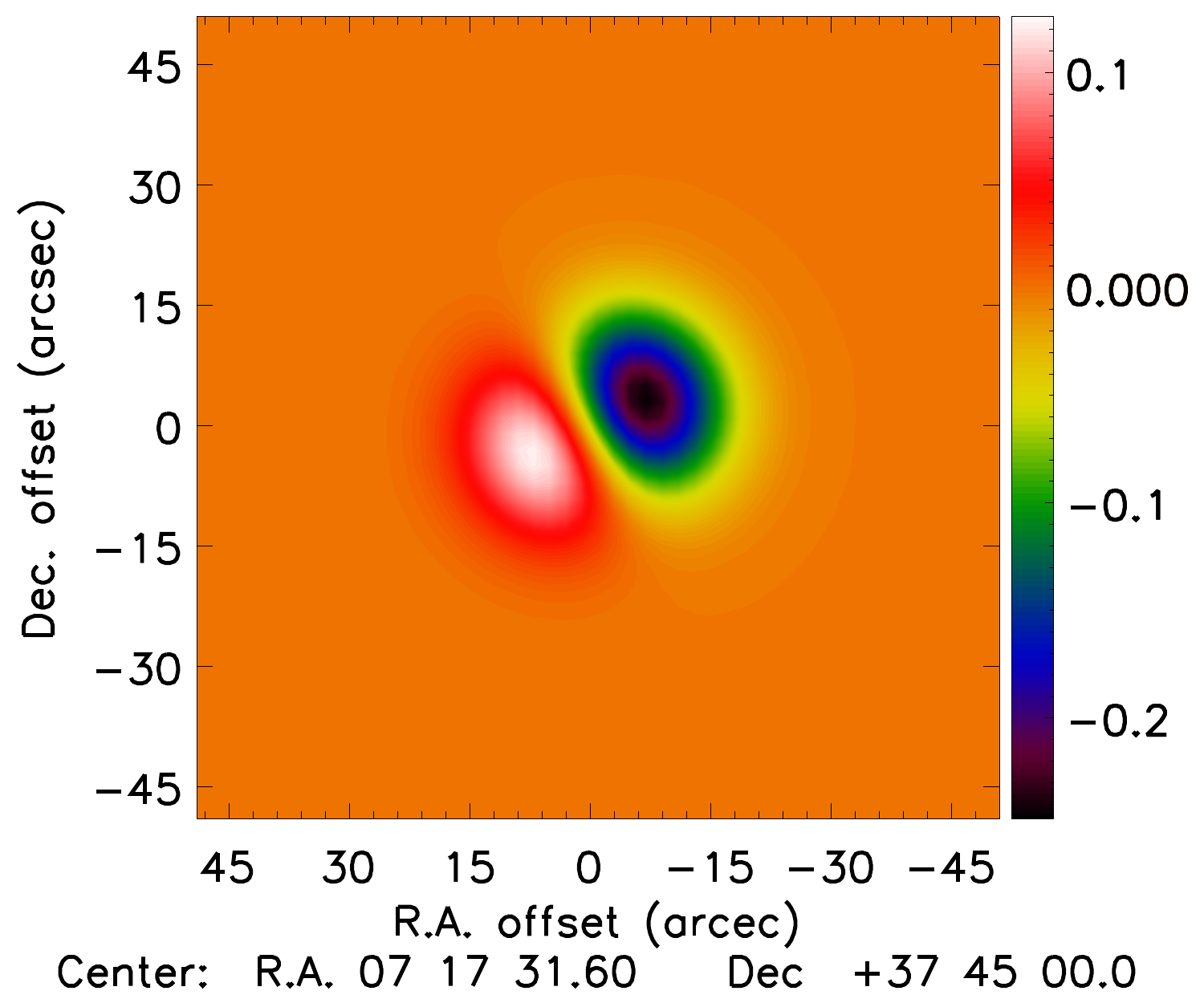} &
\includegraphics[trim=2.3cm 0.7cm 0cm 0cm, clip=true, scale=1]{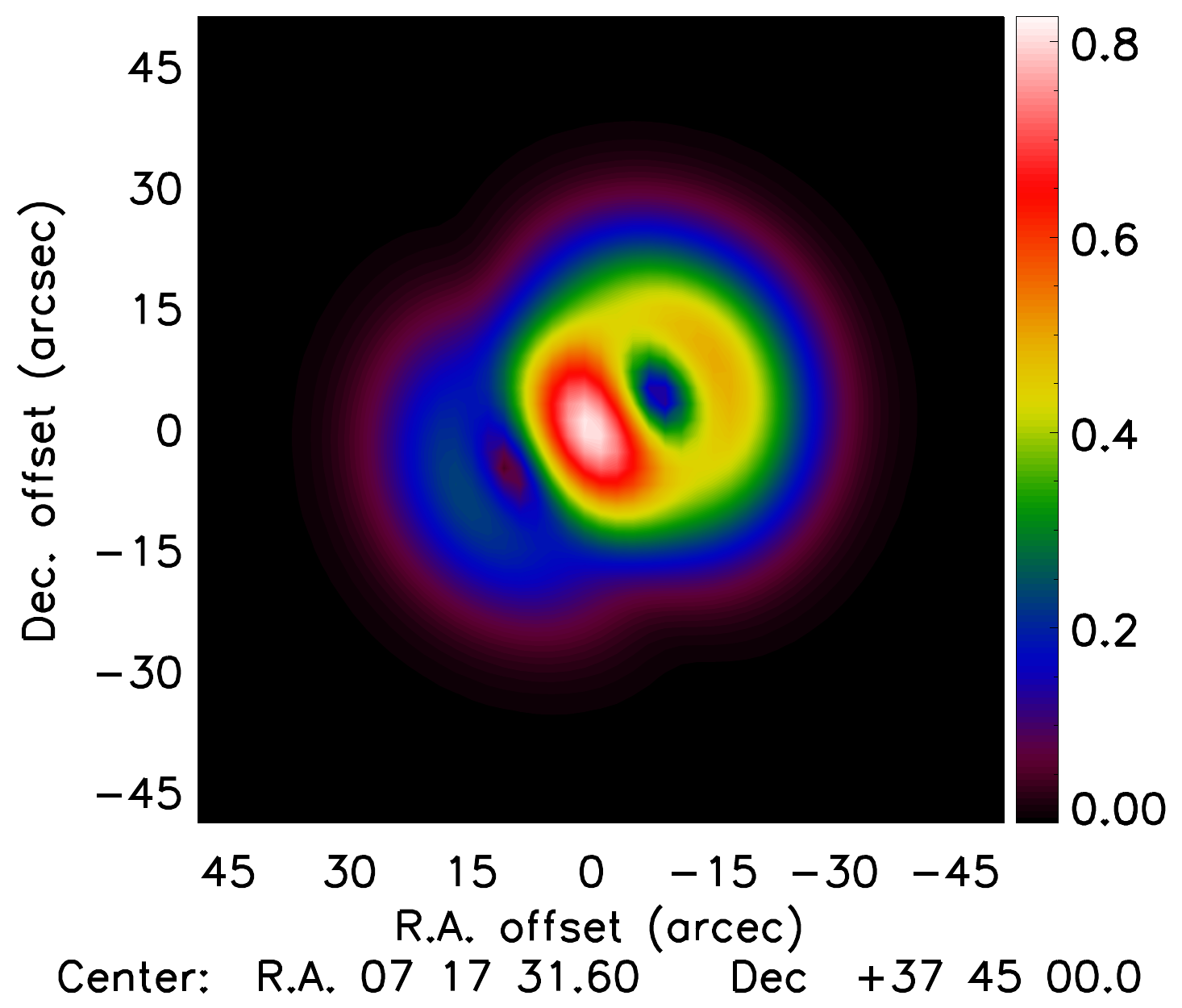} & 
\includegraphics[trim=2.3cm 0.7cm 0cm 0cm, clip=true, scale=1]{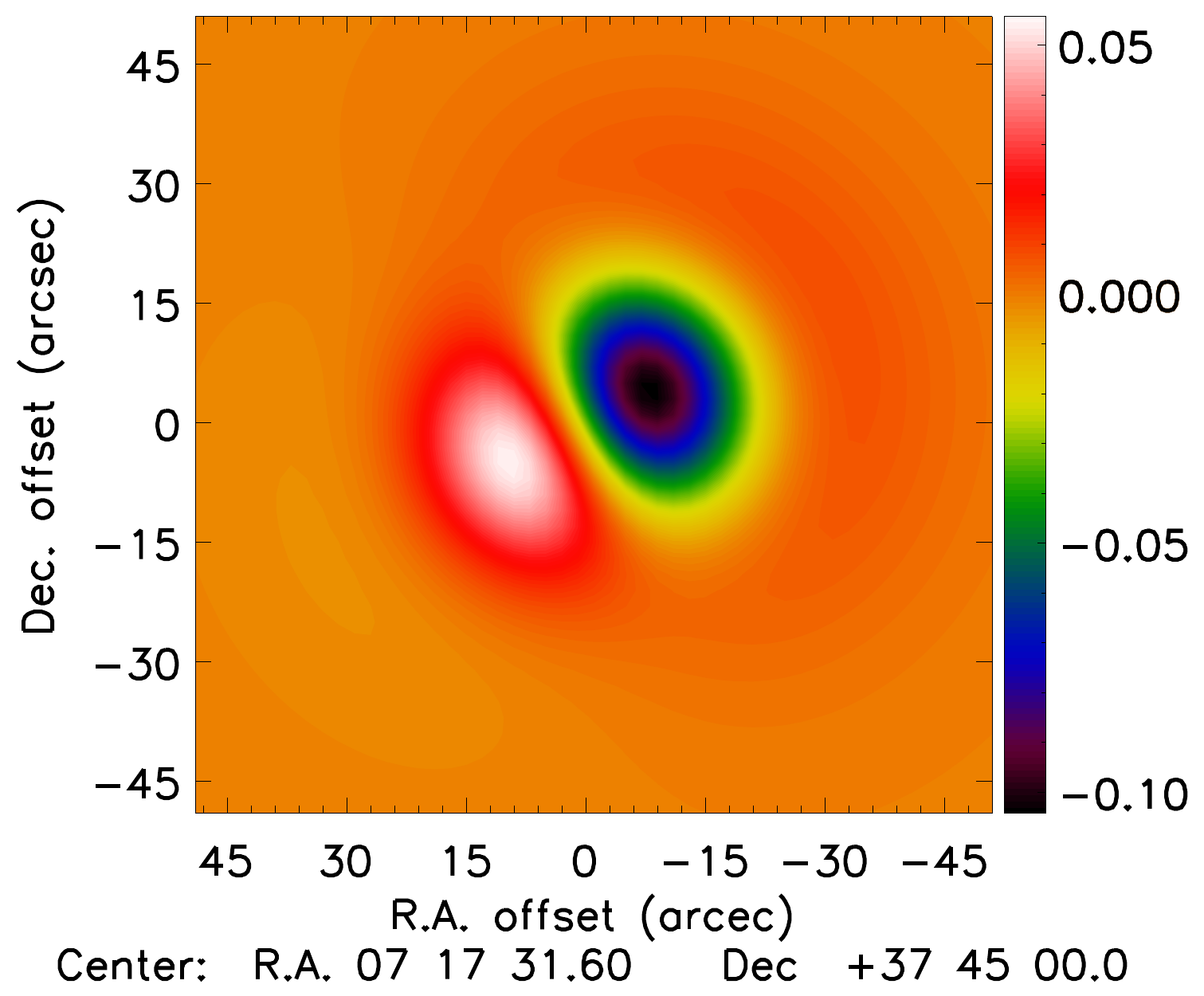} 
\end{tabular}}
\caption{\footnotesize{Application of the filters to point source residuals in case 1 (no source subtraction, top) and case 2 (source improperly subtracted, bottom). The input point source flux has been normalized to one. {\bf Left}: input surface brightness map. {\bf Middle}: GGM filtered map. {\bf Right}: DoG filtered map.}}
\label{fig:Point_source_maps}
\end{figure}
Foreground, background, or cluster-member galaxies can show up as point sources in tSZ images, which constitutes a well known potential bias. In the case of NIKA, these sources are subtracted either by fitting their flux together with the tSZ signal \citep{Adam2015}, or by modeling their Spectral Energy Distribution (SED) and fitting them to multi-wavelength photometric data to extrapolate their flux in the respective NIKA bands \citep[see the method detailed in][]{Adam2016a}. Even if such procedures provide a first good estimate of the contaminant, they are limited by the accuracy of the model to describe the data, the limited available extra data, and miscentering arising from limited accuracy in the source coordinate or telescope pointing precision. At 150 GHz, the contaminant sources are mostly radio galaxies, but infrared galaxies can also have a significant contribution.

We estimate the impact of residual point sources by considering the following approach in two upper limit cases:
\begin{enumerate} 
\item Point sources below the detection threshold that are not subtracted: we simulate point sources with a flux of $3 \sigma$ times that of the noise rms, and assume that it is not subtracted from the tSZ map. 
\item Point sources that are poorly subtracted: we assume a given degree of contamination by simulating a 3 mJy point source (the typical flux of the brightest radio sources observed in the NIKA clusters at 150 GHz), which we subtract assuming a flux that is off by 10\%, and including a miscentering offset of 3 arcsec \citep[the NIKA pointing accuracy for one scan][]{Catalano2014}.
\end{enumerate}
As the GGM filter is not linear, the point source contamination induced by the application of the filter depends not only on the point sources themselves, but also on the local environment of the source. Therefore, we test both the case where the sources are simulated in the outskirts of the clusters, where the noise dominates, and within a few arcsec of the cluster core, where the signal dominates.

Using our baseline filter parameters, we find that the difference between the point-source-contaminated and the tSZ-only maps can reach deviations of up to $4 \sigma$, both for the GGM and the DoG filters. As the signal-to-noise ratio of the simulated clusters is set to match the highest signal-to-noise NIKA clusters, it represents an upper limit of the contamination. In addition, the biases remain local on the maps and exhibit specific structures that are distinct from the ones that appear in the case of interacting clusters (ring-like shaped, azimuthally symmetric, or dipole-like at the scale of the filter, see figure \ref{fig:RG_cluster_sample}). Nonetheless, point source residuals could mimic compact tSZ cores, at the scale of the beam. As we increase the filter scales, the signal arising from point source residuals quickly becomes insignificant. The signatures of point source residuals are illustrated in figure \ref{fig:Point_source_maps}. Due to the GGM non linearity, the corresponding extra signal arising from point sources can slightly deviate from the one shown in figure \ref{fig:Point_source_maps}.

We conclude that undetected or mismodeled point sources are an unavoidable potential bias. They can affect the interpretation of the signal at a level of a few $\sigma$, but they cannot lead to a mis-interpretation of the overall signal observed in tSZ maps as the contamination remains local. Other artifacts are expected to produce negligible signal in our filter data. Despite the non-linearity of the GGM filter, we have developed a method to estimate the statistical significance of the corresponding signal that is accurate above signal-to-noise of three.

\section{Application to the NIKA clusters sample}\label{sec:Application_to_the_NIKA_clusters_sample}
\begin{figure*}[p]
\centering
\resizebox{0.75\textwidth}{!} {
\begin{tabular}{lll}
\includegraphics[trim=0cm 2.2cm 0cm 0cm, clip=true, scale=1]{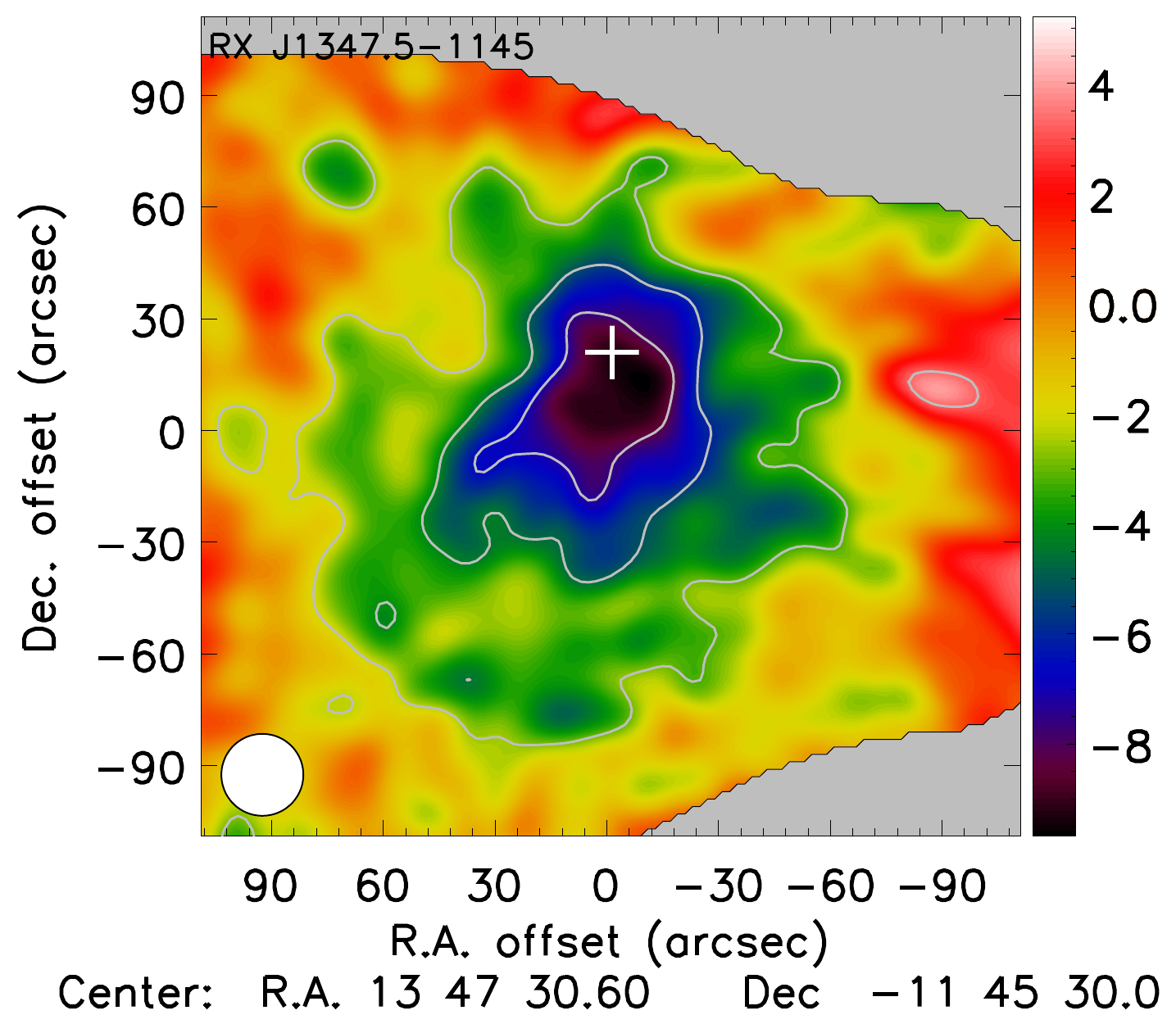} 
\put(-60,310){\makebox(0,0){\rotatebox{0}{\LARGE mJy/beam}}} & 
\includegraphics[trim=2.3cm 2.2cm 0cm 0cm, clip=true, scale=1]{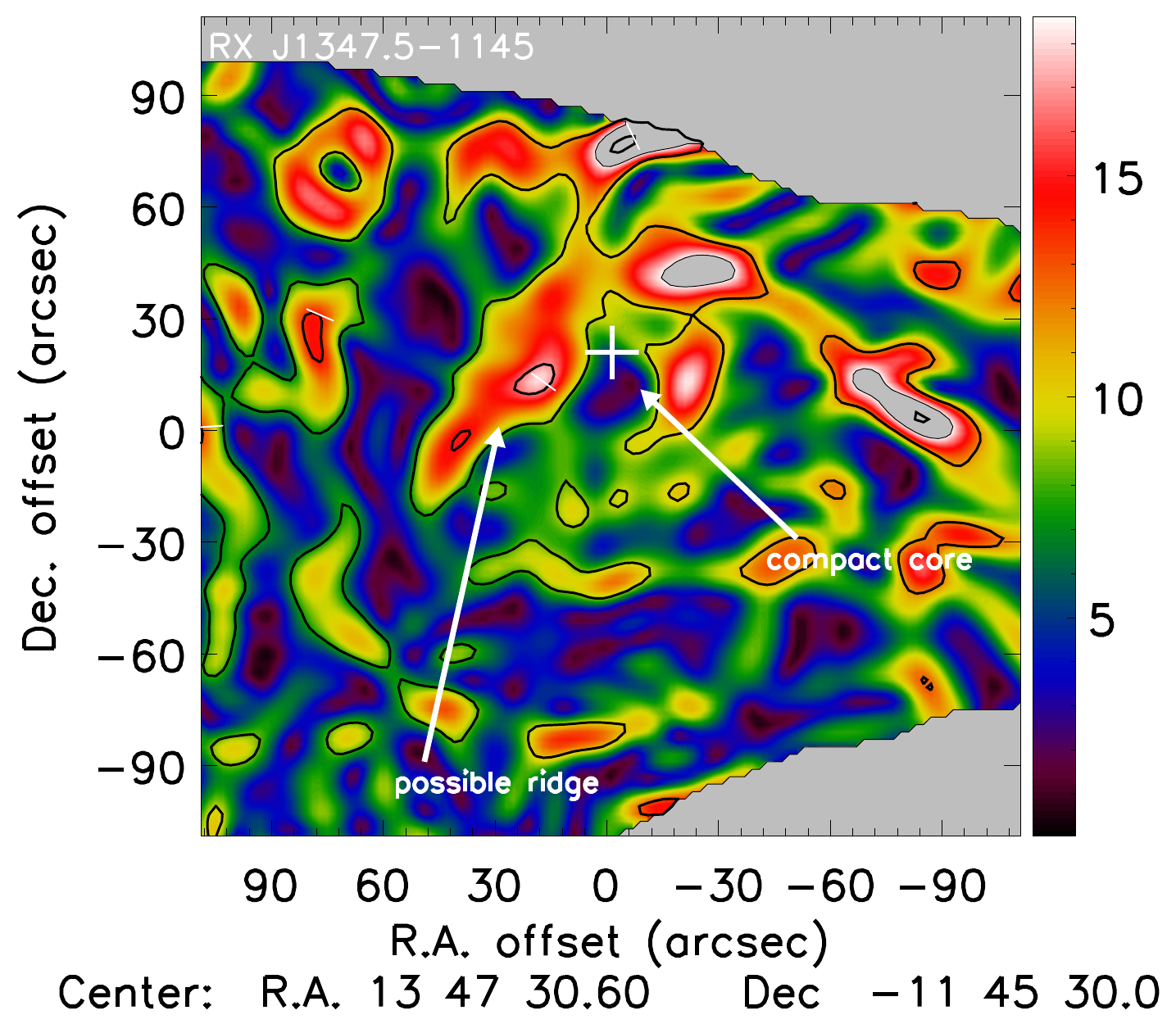} 
\put(-60,310){\makebox(0,0){\rotatebox{0}{\LARGE mJy/beam/arcmin}}} & 
\includegraphics[trim=2.3cm 2.2cm 0cm 0cm, clip=true, scale=1]{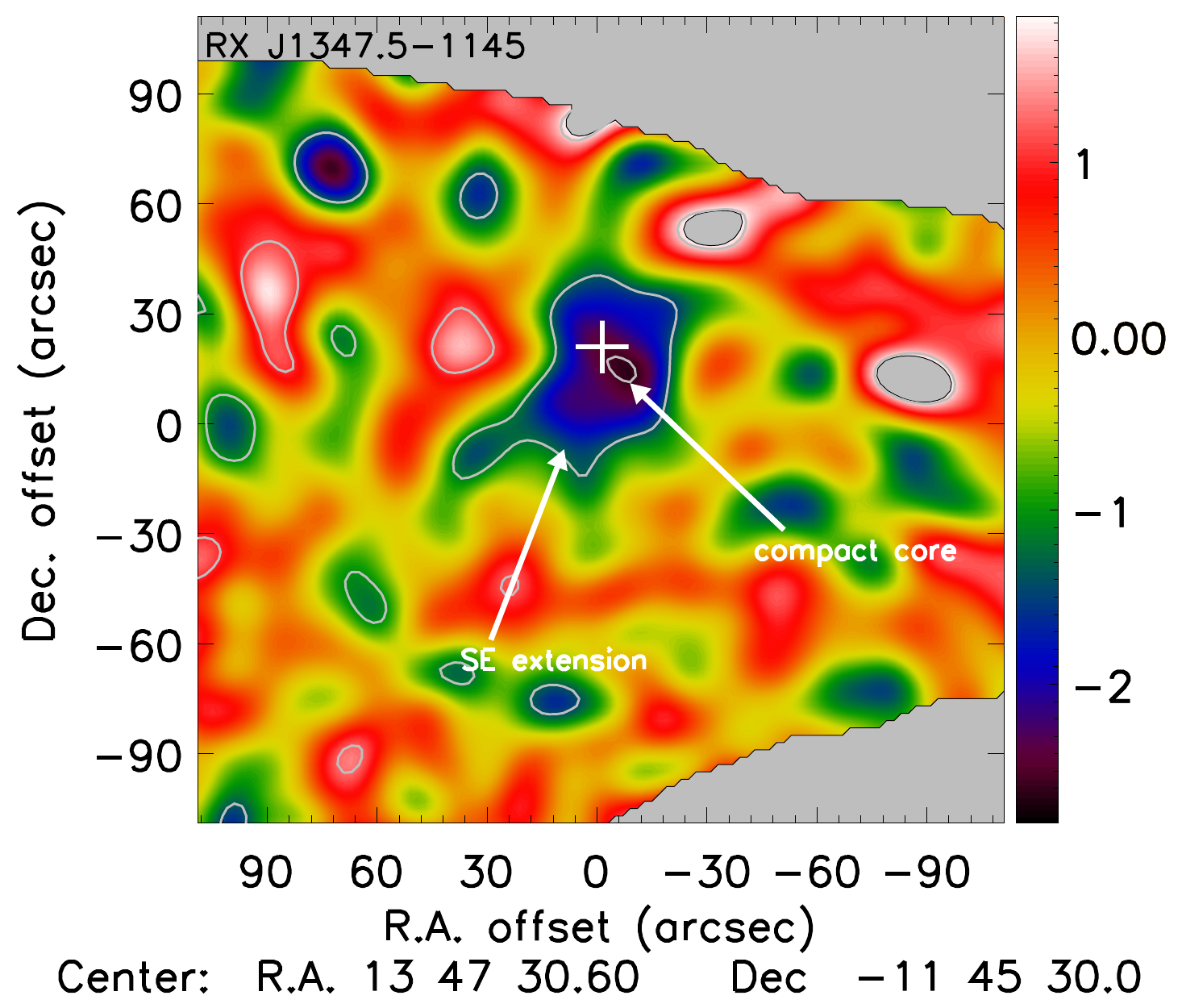} 
\put(-60,310){\makebox(0,0){\rotatebox{0}{\LARGE mJy/beam}}} \\
\includegraphics[trim=0cm 2.2cm 0cm 0cm, clip=true, scale=1]{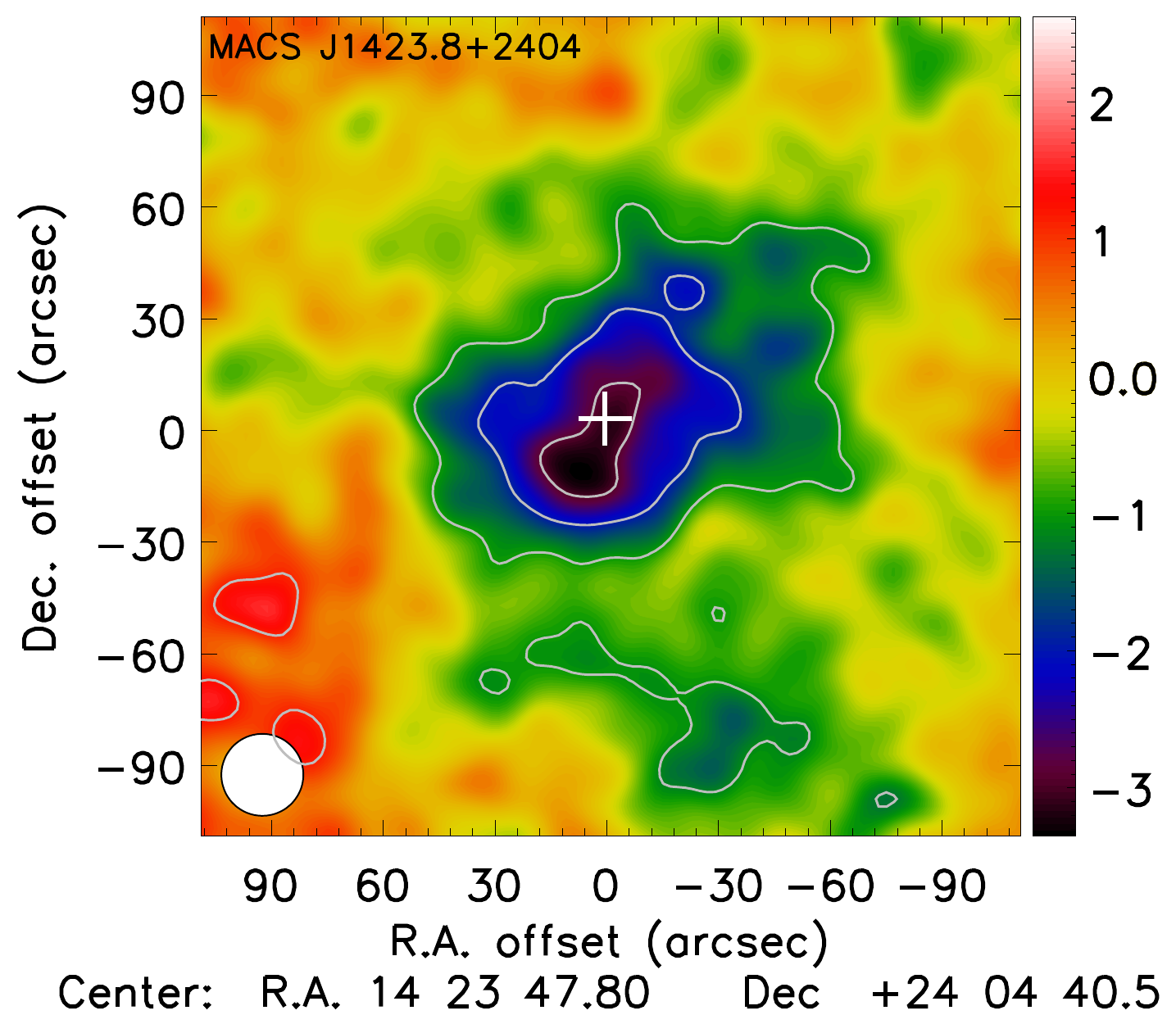} & 
\includegraphics[trim=2.3cm 2.2cm 0cm 0cm, clip=true, scale=1]{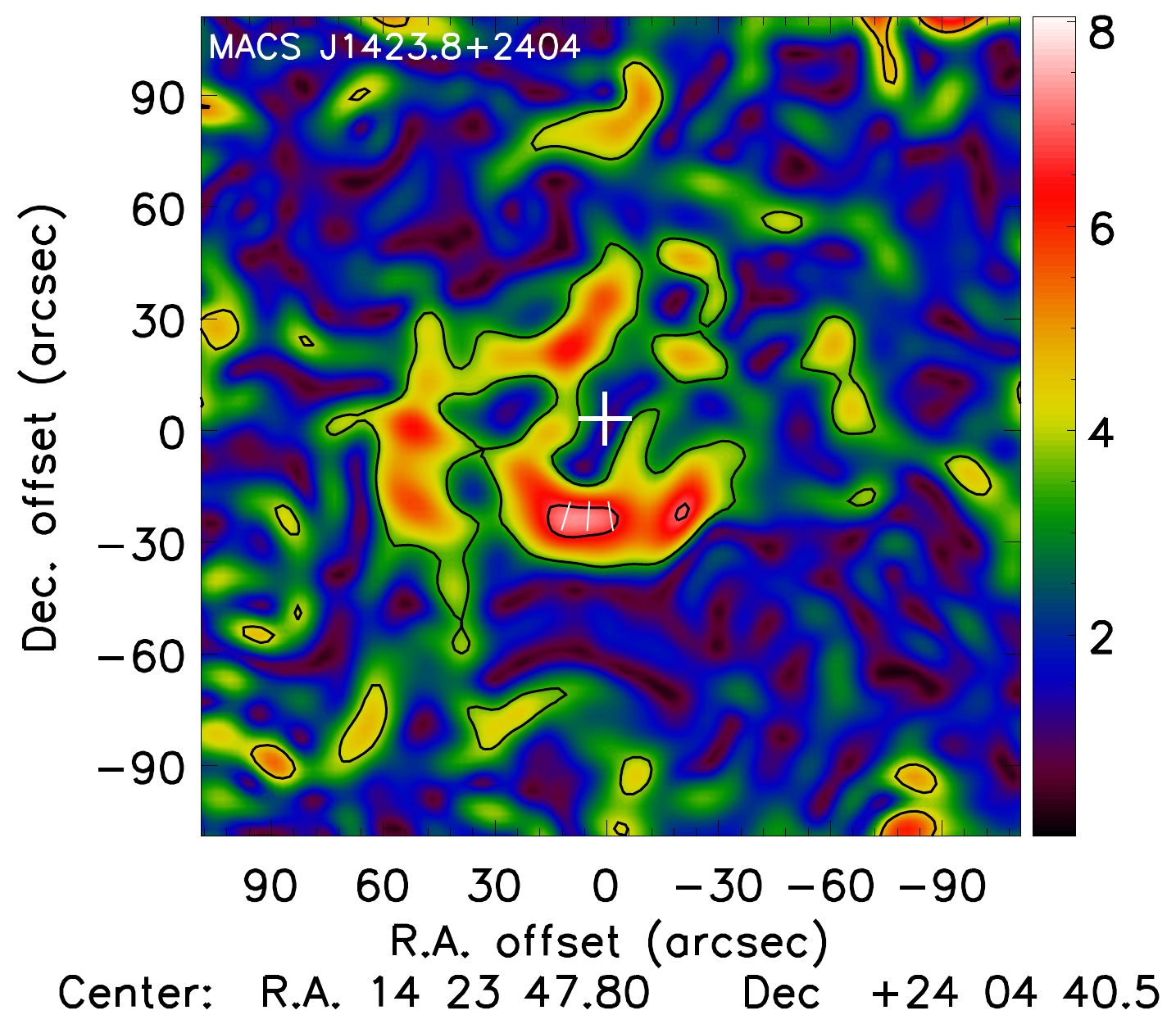} & 
\includegraphics[trim=2.3cm 2.2cm 0cm 0cm, clip=true, scale=1]{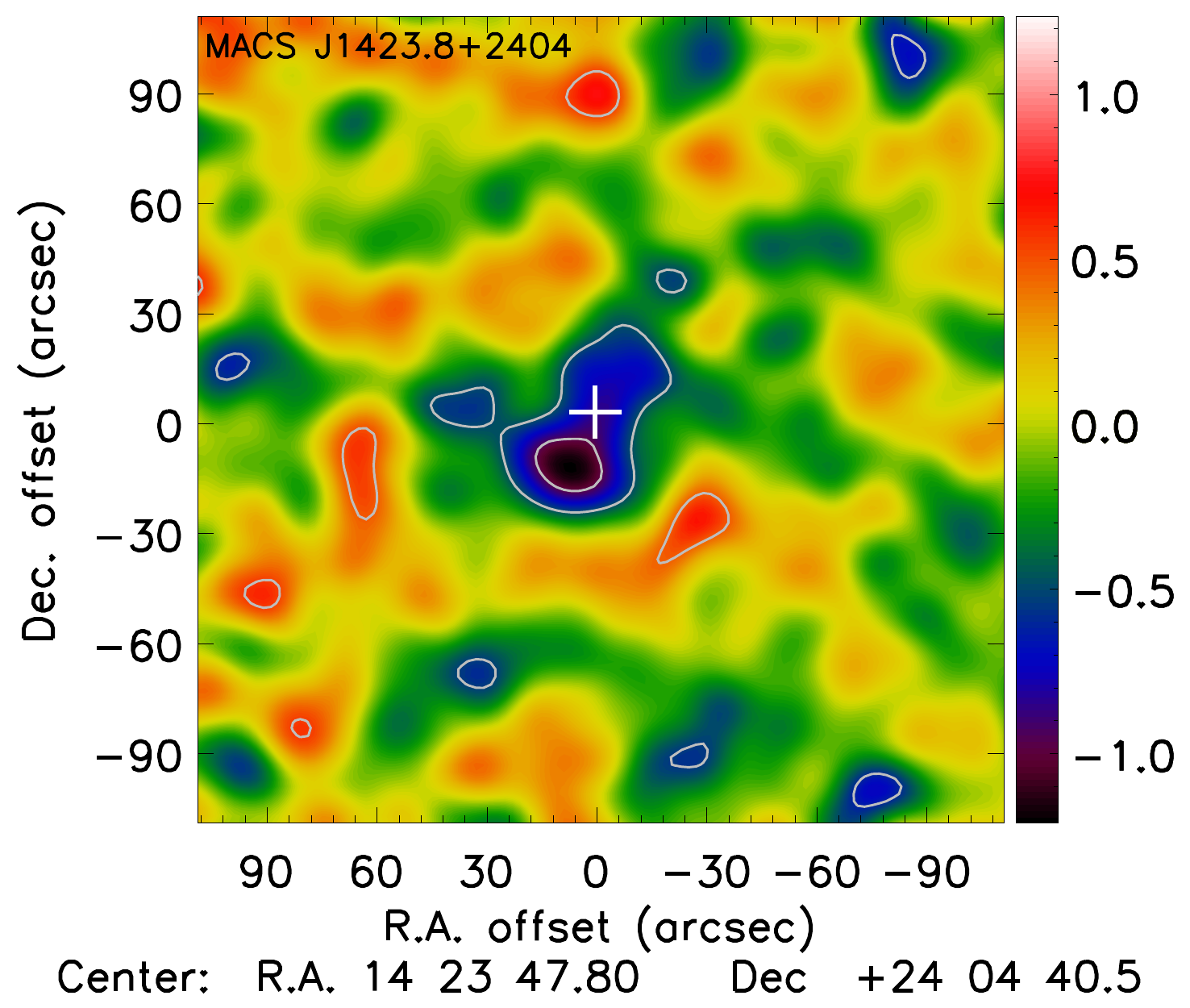} \\
\includegraphics[trim=0cm 2.2cm 0cm 0cm, clip=true, scale=1]{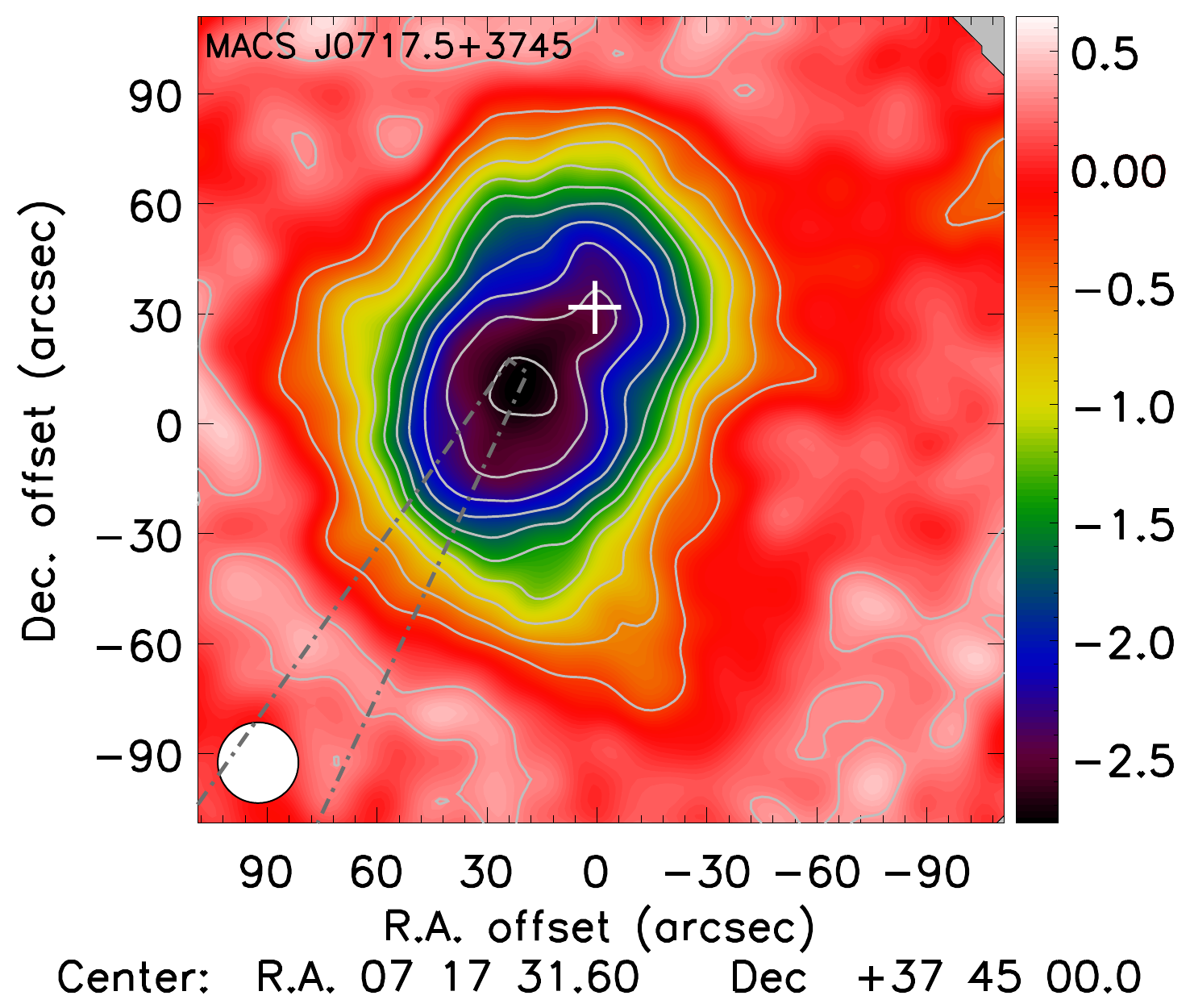} & 
\includegraphics[trim=2.3cm 2.2cm 0cm 0cm, clip=true, scale=1]{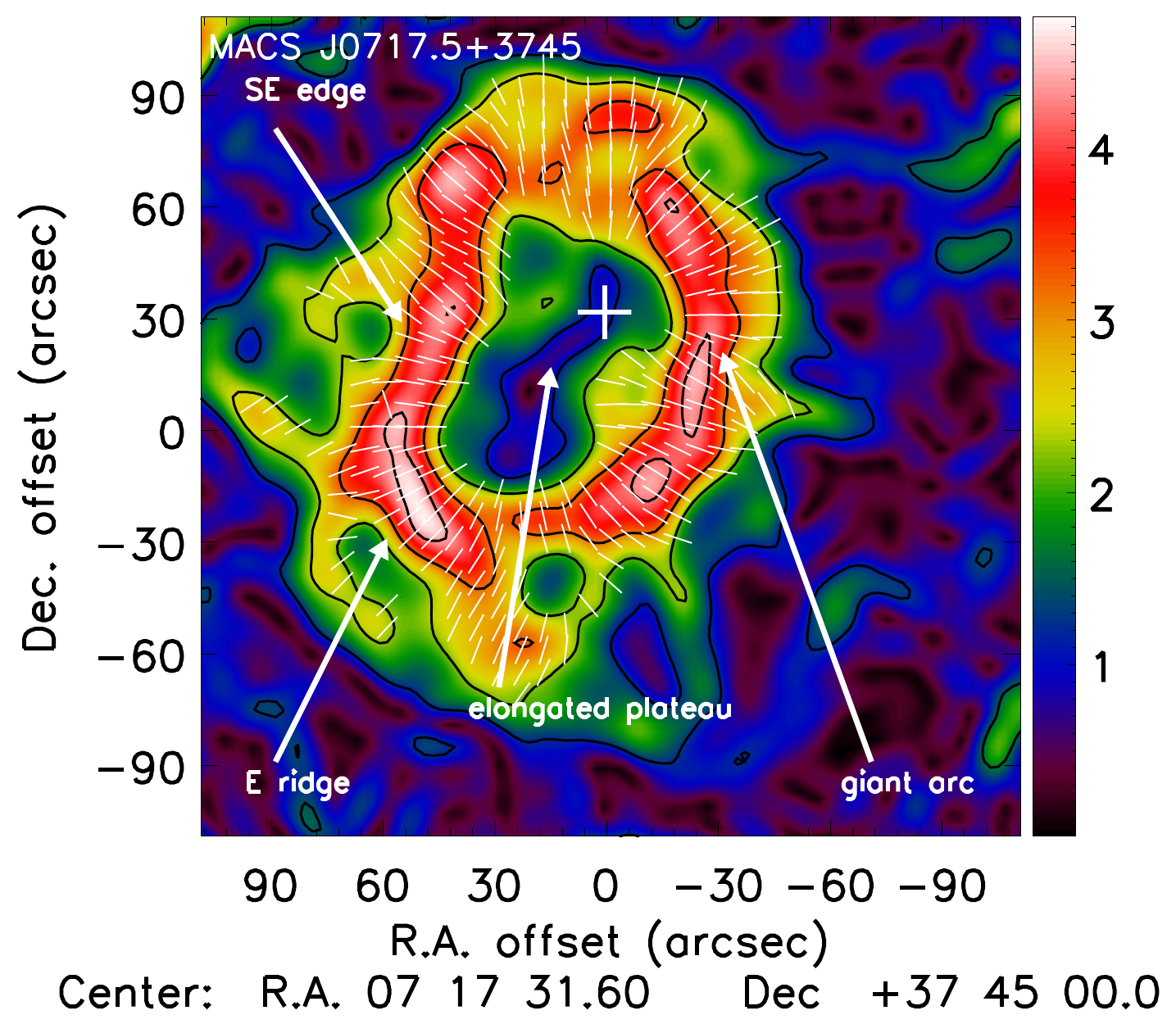} & 
\includegraphics[trim=2.3cm 2.2cm 0cm 0cm, clip=true, scale=1]{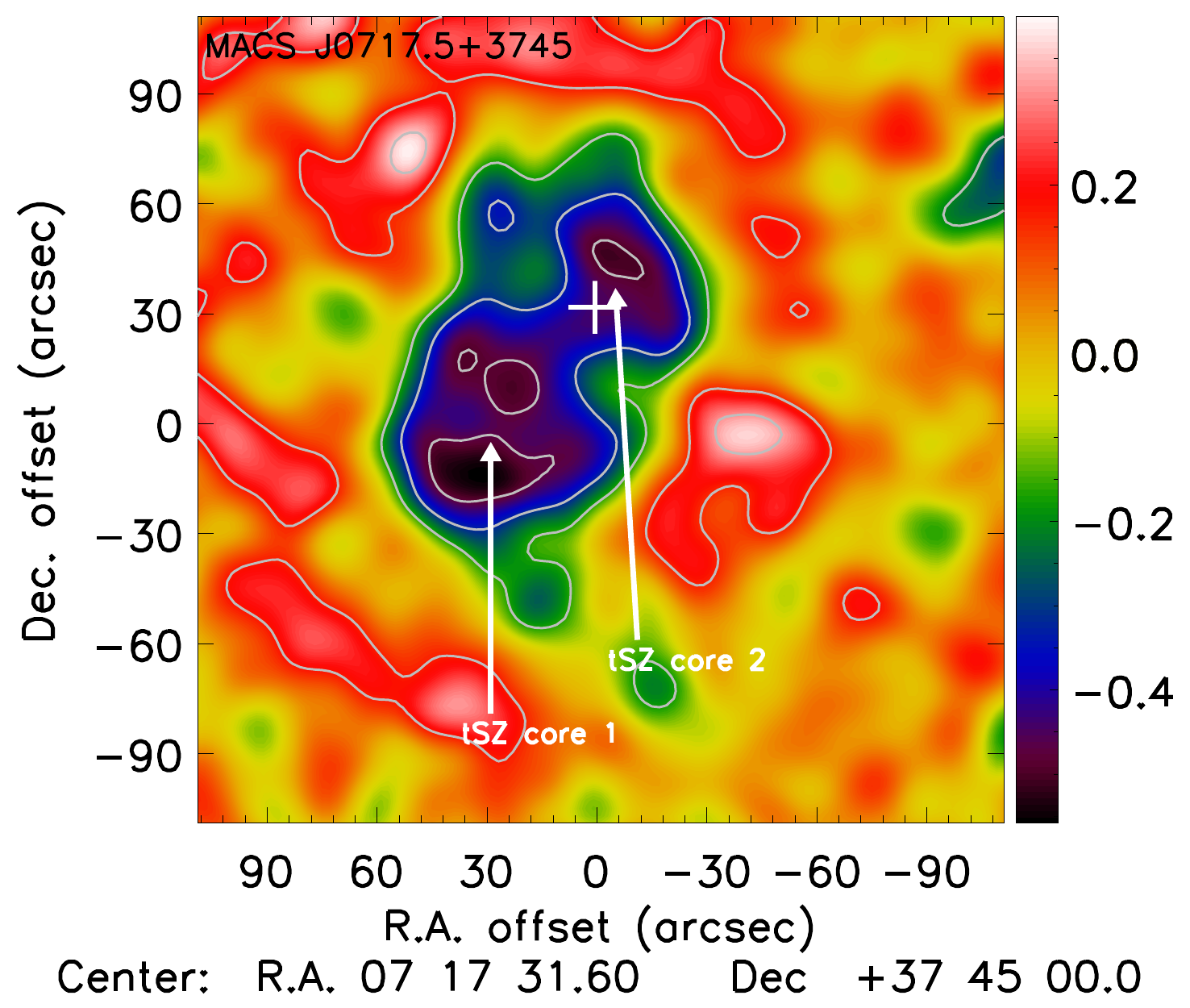} \\
\includegraphics[trim=0cm 2.2cm 0cm 0cm, clip=true, scale=1]{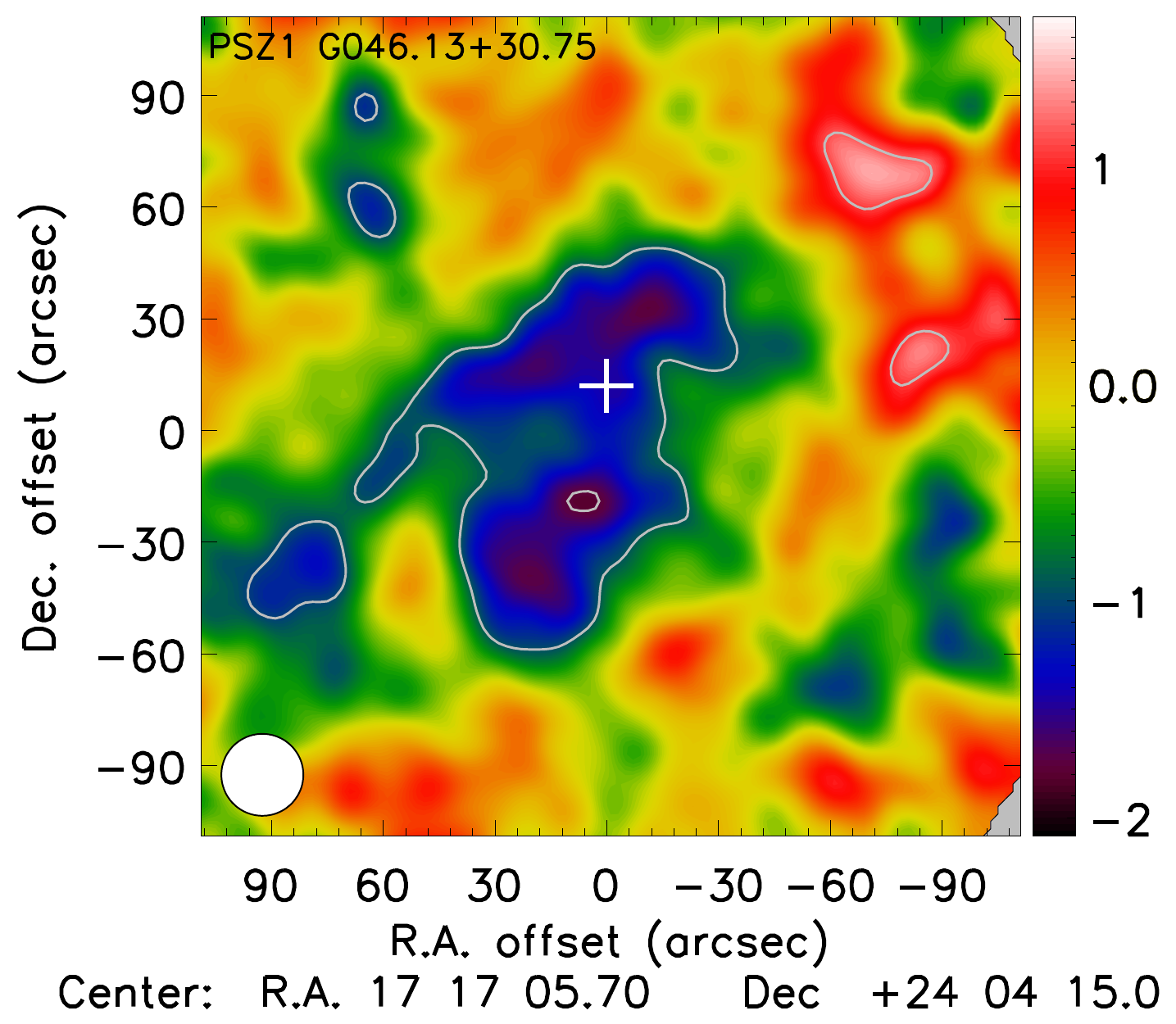} & 
\includegraphics[trim=2.3cm 2.2cm 0cm 0cm, clip=true, scale=1]{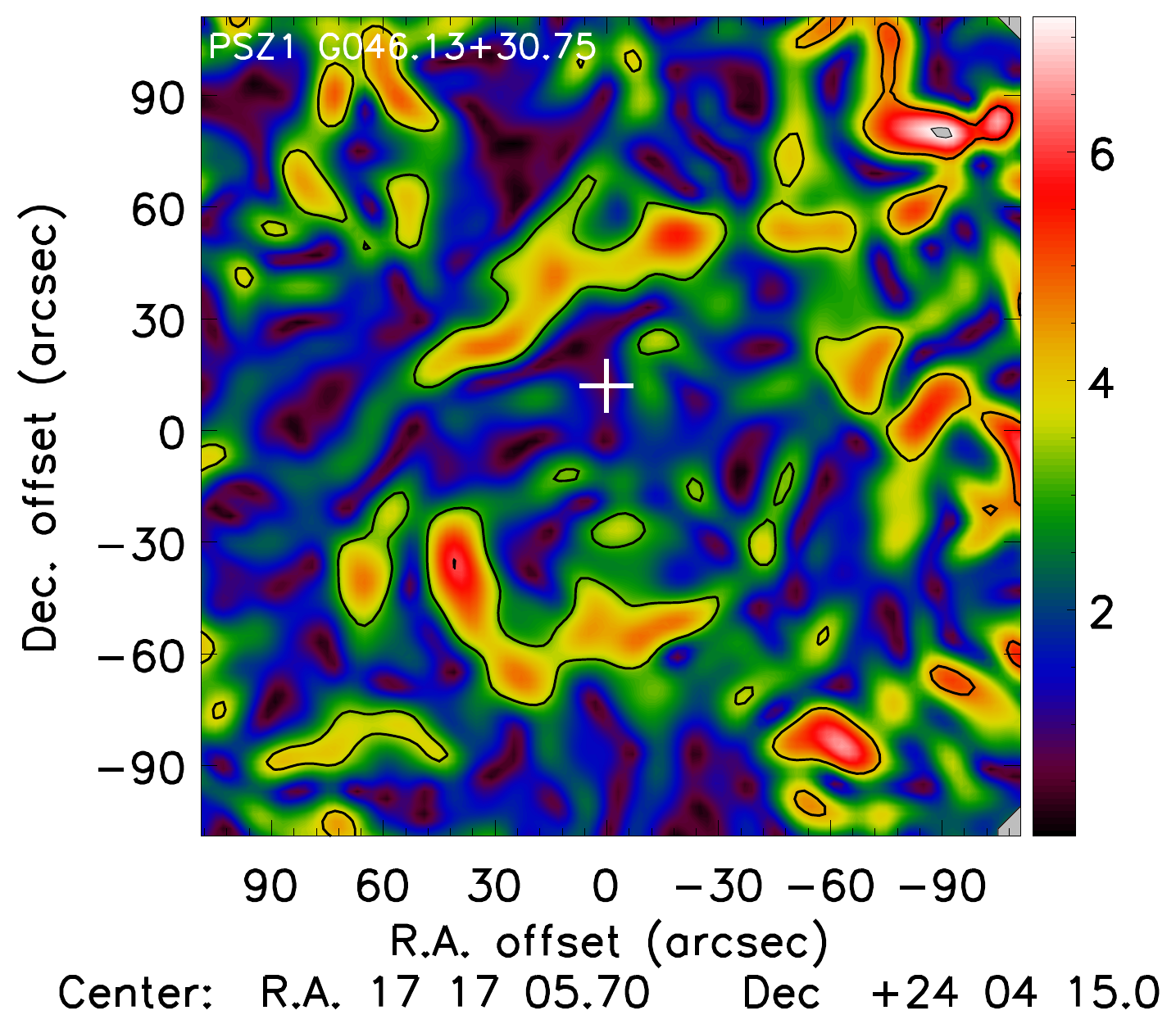} & 
\includegraphics[trim=2.3cm 2.2cm 0cm 0cm, clip=true, scale=1]{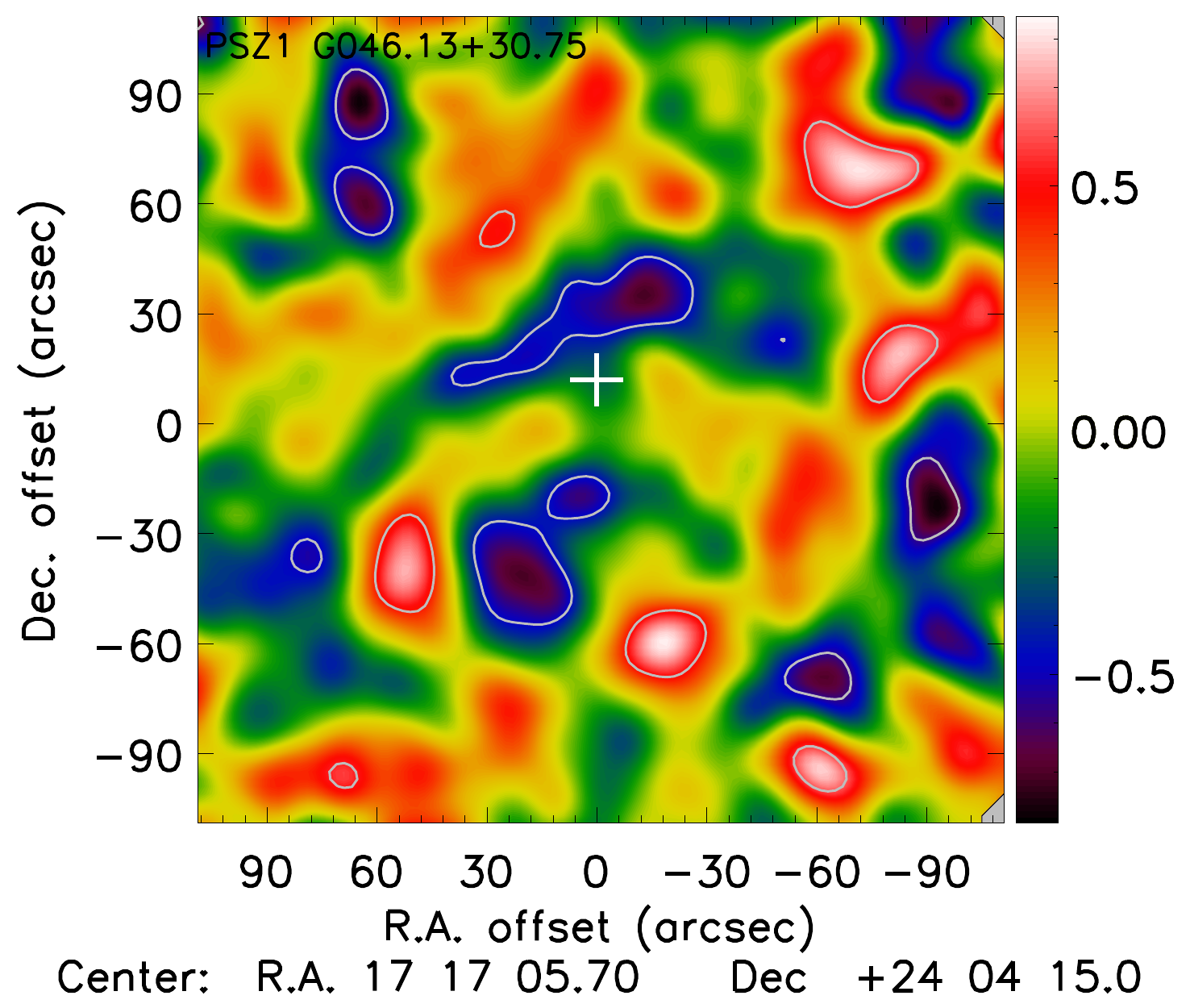} \\
\includegraphics[trim=0cm 2.2cm 0cm 0cm, clip=true, scale=1]{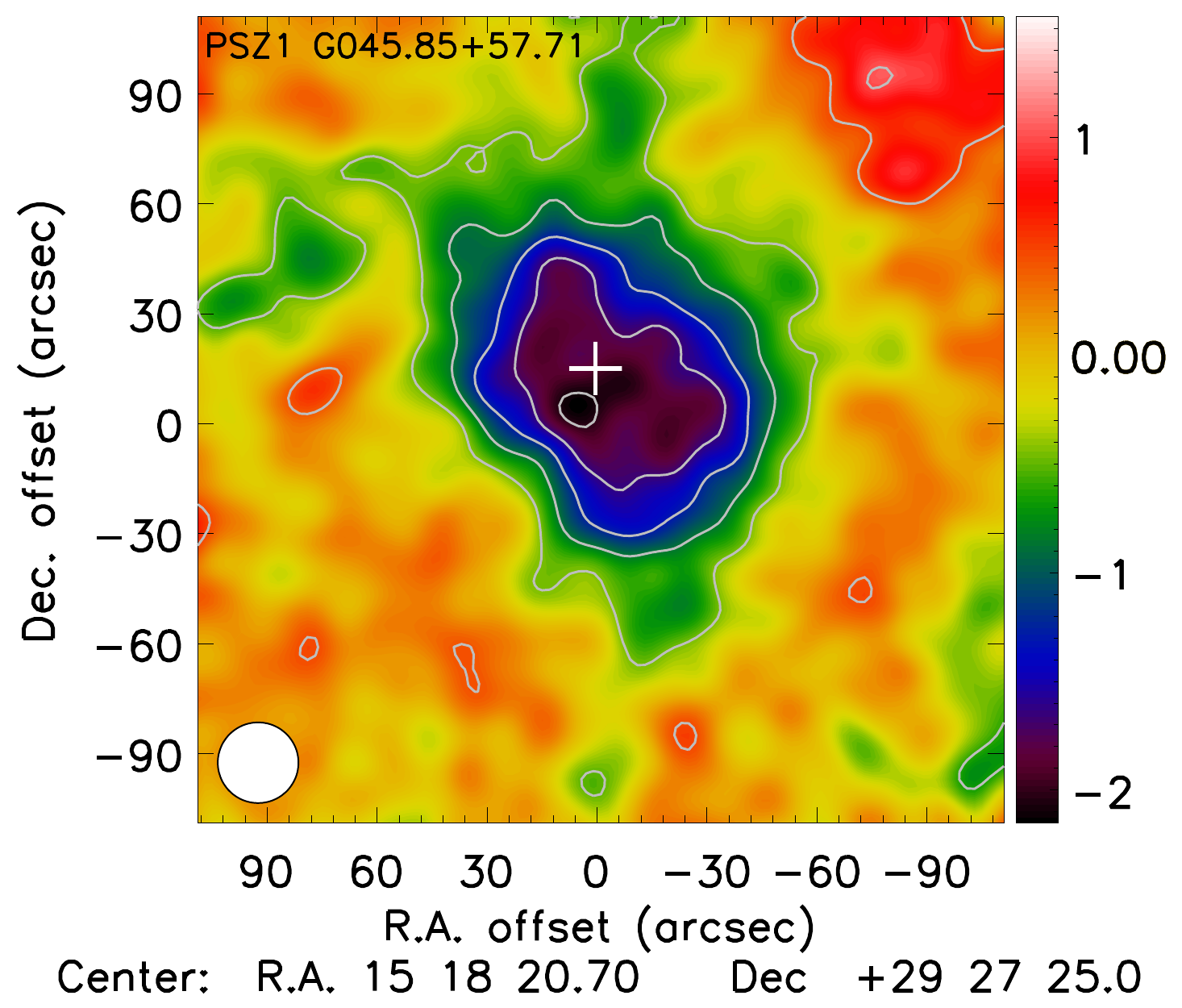} & 
\includegraphics[trim=2.3cm 2.2cm 0cm 0cm, clip=true, scale=1]{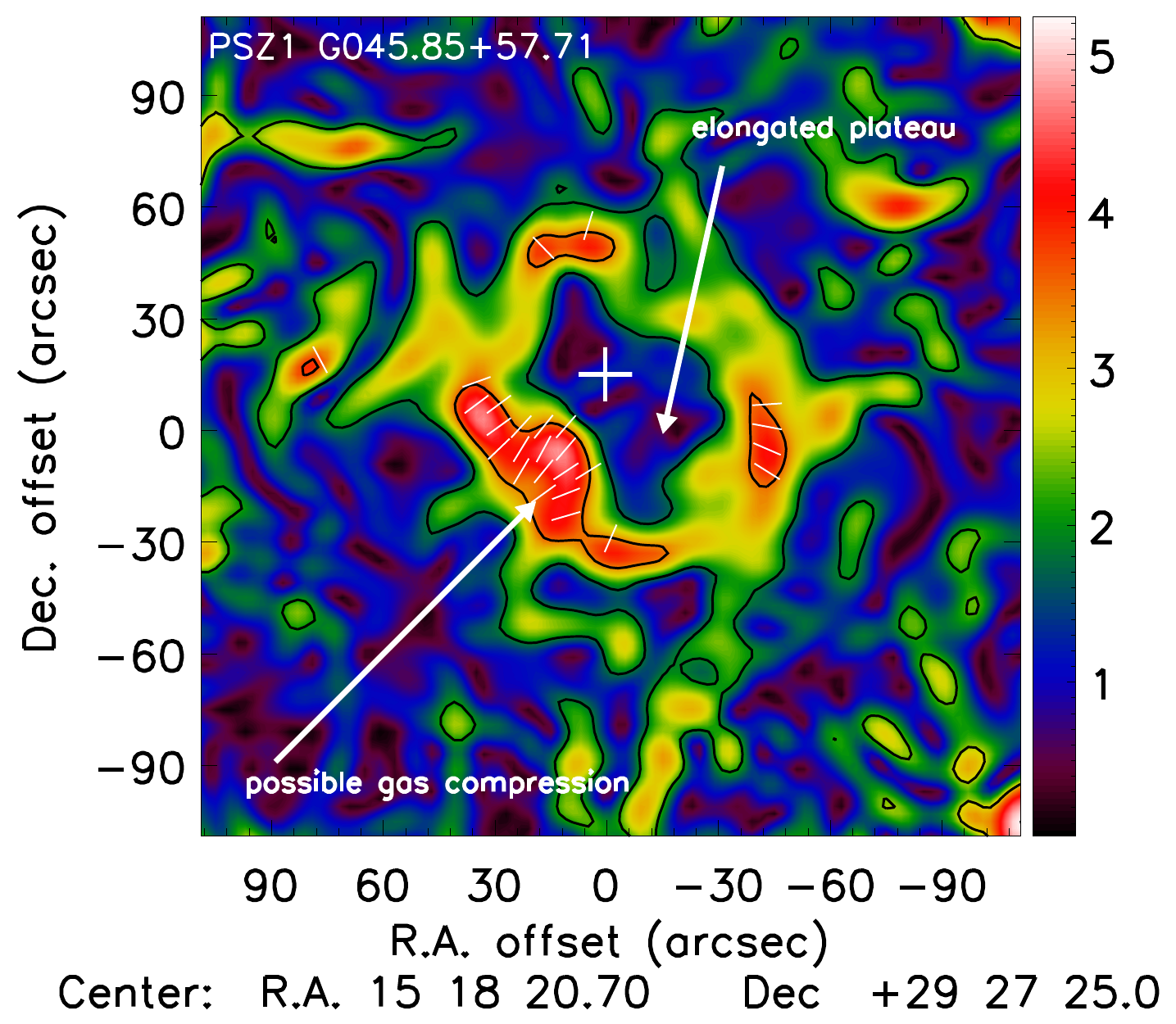} & 
\includegraphics[trim=2.3cm 2.2cm 0cm 0cm, clip=true, scale=1]{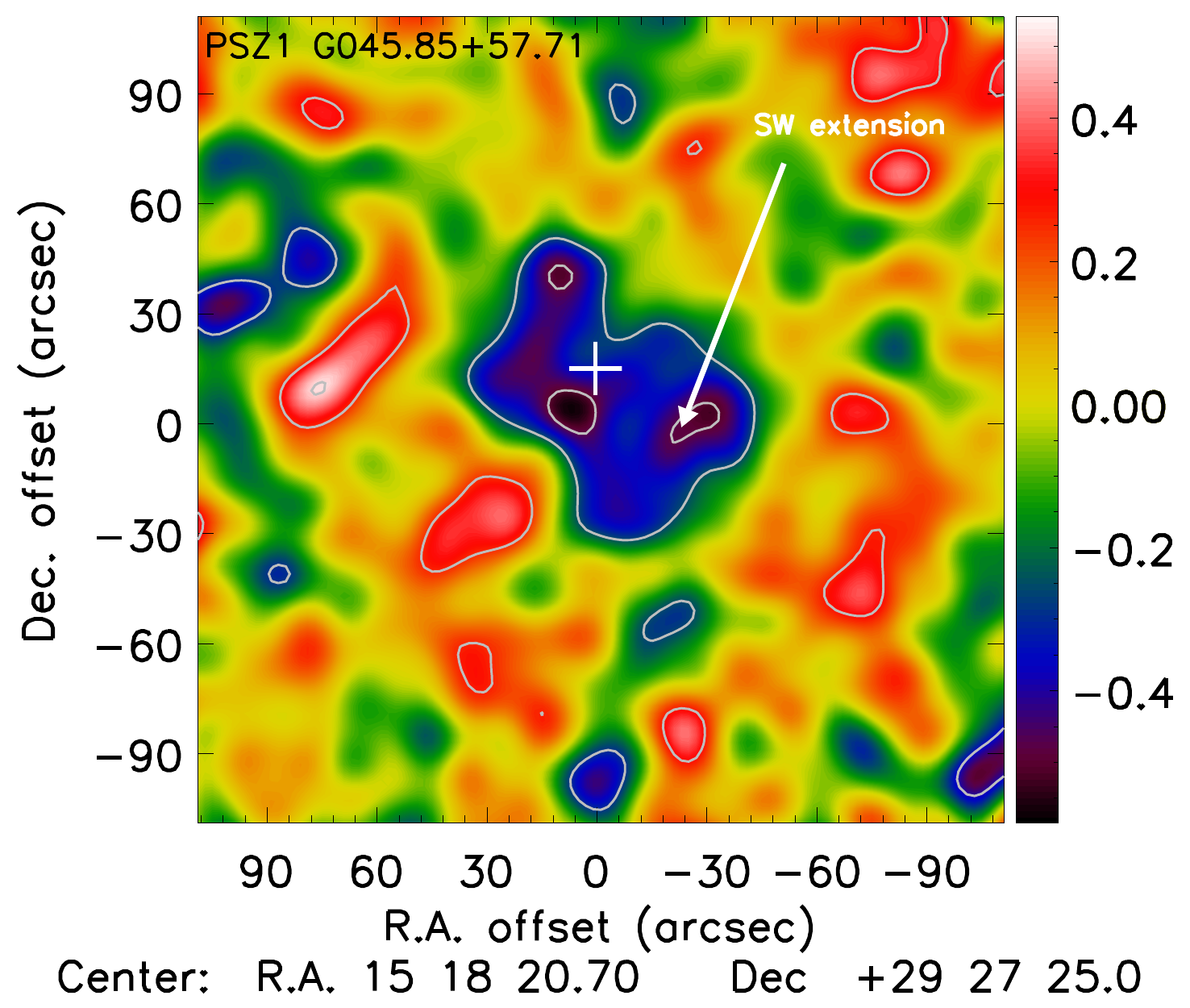} \\
\includegraphics[trim=0cm 0.7cm 0cm 0cm, clip=true, scale=1]{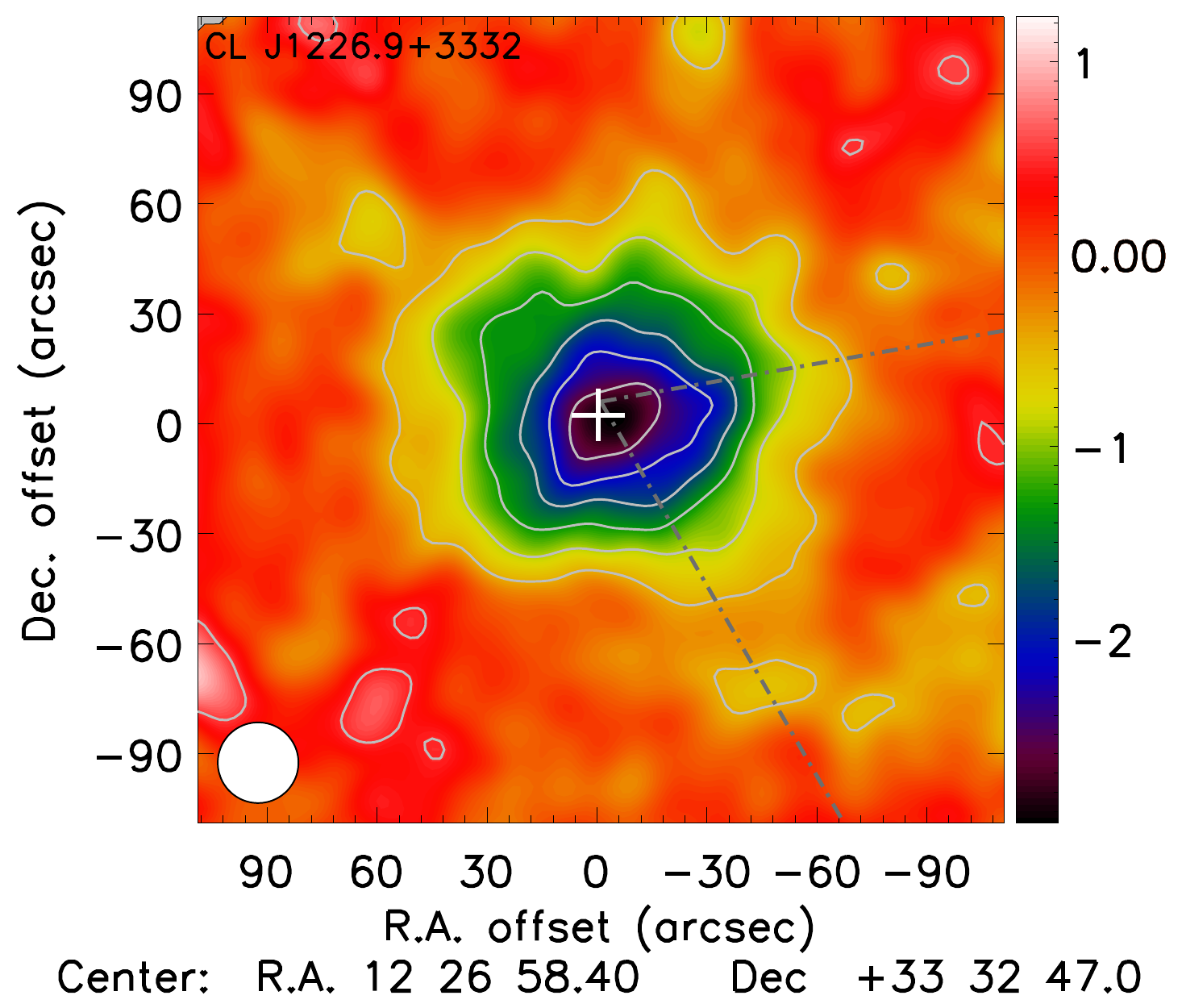} & 
\includegraphics[trim=2.3cm 0.7cm 0cm 0cm, clip=true, scale=1]{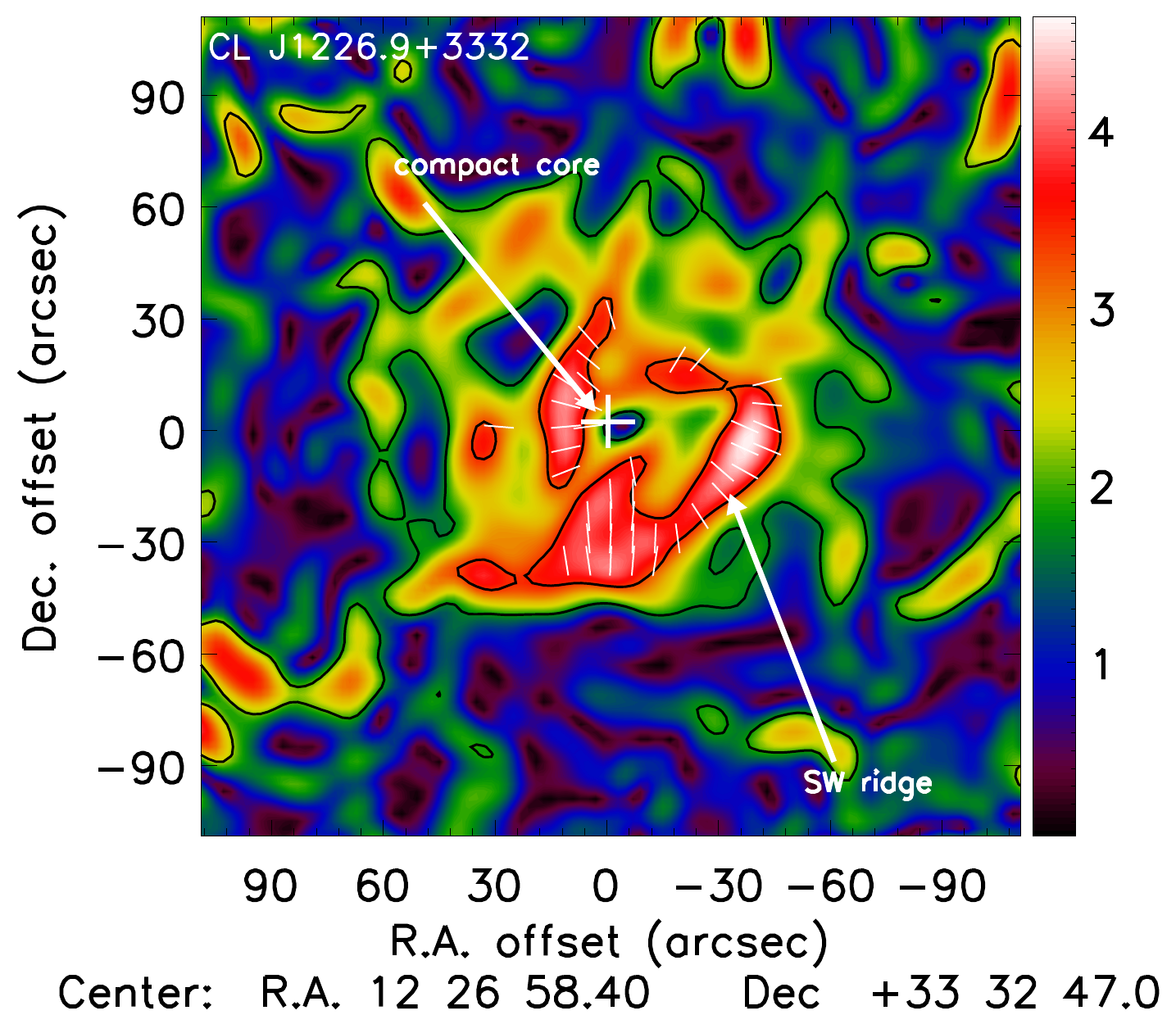} & 
\includegraphics[trim=2.3cm 0.7cm 0cm 0cm, clip=true, scale=1]{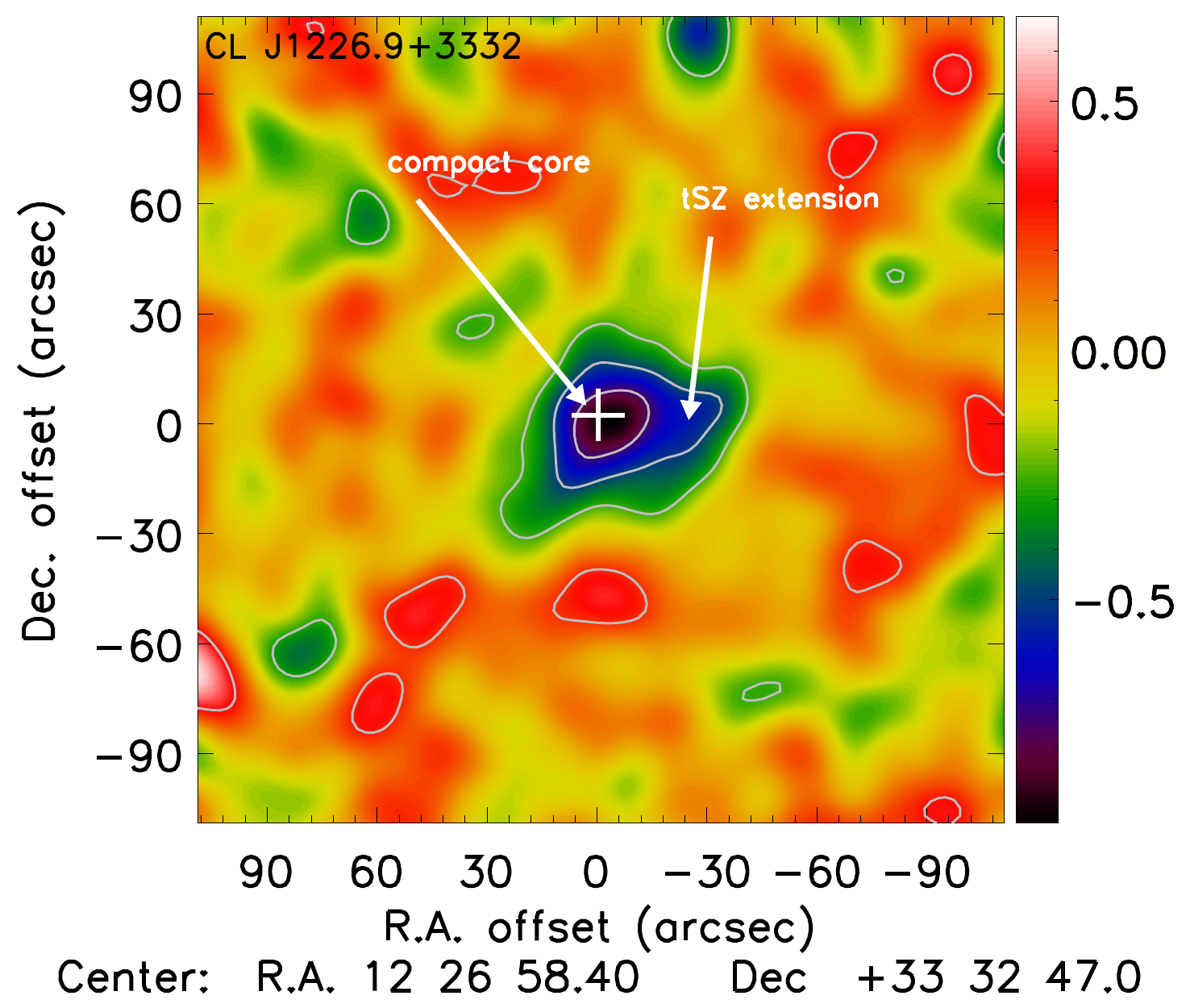}
\end{tabular}}
\caption{\footnotesize{Redshift-ordered maps of the NIKA cluster sample at 150 GHz. 
{\bf Left:} Surface brightness images. Each map has been smoothed to an effective angular resolution of 22 arcsec FWHM, as shown by the bottom left white circle. Point sources have been subtracted and the data have been deconvolved from the transfer function, but the zero level absolute brightness remains unconstrained. 
{\bf Middle:} GGM filtered maps.
{\bf Right:} DoG filtered maps.
In all cases, contours provide the signal-to-noise ratio, starting at $\pm 2 \sigma$ and increasing by $2 \sigma$ steps. The white crosses provide the main X-ray peak (see Table \ref{tab:xray_peak}). In the case of \mbox{MACS~J0717.5+3745}, the kSZ contribution is not accounted for in the map presented here (see Appendix \ref{sec:Impact_of_kSZ} and Section \ref{sec:MACSJ0717} for more details). In the case of \mbox{CL~J1226.9+3332} and \mbox{MACS~J0717.5+3745}, the dashed lines correspond to the cone used to define the GGM ridge region when extracting profiles in Figure \ref{fig:profile_CLJ1227} and \ref{fig:profile_MACSJ0717}.}}
\label{fig:NIKA_cluster_sample}
\end{figure*}

\begin{table}[]
\caption{\footnotesize{Coordinates of the main X-ray peak.}}
\begin{center}
\resizebox{0.5\textwidth}{!} {
\begin{tabular}{c|c|c|c}
\hline
\hline
Name & R.A. & Dec. & Reference \\
\hline
RX~J1347.5-1145 & 13:47:30.575 & -11:45:10.08 & \textit{Chandra} (ObsID3592) \\ 
MACS~J1423.8+2404 & 14:23:47.908 & +24:04:42.69 & \textit{Chandra} (ObsID4195) \\ 
MACS~J0717.5+3745 & 07:17:31.740 & +37:45:30.73 & \textit{Chandra} (ObsID4200) \\ 
PSZ1~G046.13+30.75 & 17:17:05.780 & +24:04:26.00 & \textit{XMM-Newton} (ObsID0693661401) \\ 
PSZ1~G045.85+57.71 & 15:18:20.811 & +29:27:39.10 & \textit{XMM-Newton} (ObsID0693661101) \\ 
CL~J1226.9+3332 & 12:26:58.454 & +33:32:48.36 & \textit{Chandra} (ObsID5014) \\ 
\hline
\end{tabular}
}
\end{center}
\label{tab:xray_peak}
\end{table}

The NIKA surface brightness maps at 150 GHz are provided in Figure \ref{fig:NIKA_cluster_sample} together with the filter algorithms results. They are deconvolved from the processing transfer function, smoothed to an effective resolution of 22 arcsec FWHM and cleaned from point source contamination. The three clusters \mbox{RX~J1347.5-1145}, \mbox{MACS~J1423.8+2404} and \mbox{PSZ1~G046.13+30.75} have been imaged with NIKA. Despite the relatively low significance of this sub-sample (less than 10 $\sigma$ per beam at the 22 arcsec resolution), we apply the GGM and DoG filter to the maps. We observe hints for internal structures, but only at low significance level, as briefly discussed below. The three other NIKA clusters, \mbox{MACS~J0717.5+3745}, \mbox{PSZ1~G045.85+57.71} and \mbox{CL~J1226.9+3332}, present a peak significance larger than 10 $\sigma$ at the 22 arcsec resolution, with a maximum of more than $18 \sigma$ for \mbox{MACS~J0717.5+3745} (see figure \ref{fig:NIKA_cluster_sample}). This is comparable to the RHAPSODY-G processed tSZ maps of Figure \ref{fig:RG_cluster_sample_proc} and we thus expect to be able to find sub-structure features at a similar significance within these clusters. The GGM and DoG filtered maps of the NIKA clusters are presented in Figure \ref{fig:NIKA_cluster_sample} for our baseline filter parameters. In the following sub-sections, we discuss the individual cluster analysis.

\subsection{RX~J1347.5-1145}\label{sec:RXJ1347.5-1145}
\mbox{RX~J1347.5-1145} is a massive cluster at redshift 0.45, and one of the most luminous X-ray cluster. Being an extremely bright tSZ source, it has already been observed using the tSZ effect by several groups \citep[e.g.][]{Komatsu1999,Pointecouteau1999,Kitayama2004,Mason2010,Plagge2012,Adam2014,Sayers2016,Kitayama2016}. \mbox{RX~J1347.5-1145} presents a dense core, but also shows an extension toward the southeast, which is associated with the merging of a sub-cluster. The tSZ peak is aligned with the southeast extension, corresponding to the overpressure caused by the merger in this region. This cluster was observed with an early setup of the NIKA instrument \citep[bandpass, sensitivity, calibration procedure, see][for more details]{Adam2014}. In addition, the scanning strategy was not isotropic due to a mistake in the control software, leading to possible uncontrolled systematics in the filtered map. Therefore, the application of the filtering algorithms on the NIKA map of \mbox{RX~J1347.5-1145} is meant to remain qualitative in this paper. The GGM filtered map presents a $\gtrsim 2 \sigma$ ridge ($\gtrsim 4 \sigma$ when using $\theta_0 = 20$ arcsec), inclined by 40 degrees with respect to the Dec. axis, about 30 arcsec (178 kpc) east from the X-ray center. It could be associated to the eastern shock front seen in \textit{Chandra} data by \cite{Kreisch2016}, but deeper tSZ observation are necessary to confirm it. The DoG filtered map shows that the tSZ pressure in the cluster core is elongated from SE to NW. Large scales being filtered out, the morphology of the map compares well with that of MUSTANG \citep{Mason2010} and ALMA \citep{Kitayama2016} that are sensitive to smaller scales and intrinsically filter out large scales. The tSZ SE extension that we observe, and which is tracking a hot spot in \mbox{RX~J1347.5-1145}, is coincident with that identified by NOBA \citep[Nobeyama Bolometer Array][]{Kitayama2004} and DIABOLO \citep[on the IRAM 30m telescope][]{Pointecouteau1999,Pointecouteau2001} observations. Nevertheless, the exact location of the tSZ peak is not clear due to the uncertainties in the flux of a radio source located in the brightest cluster galaxy.

\subsection{MACS~J1423.8+2404}
\mbox{MACS~J1423.8+2404} is a very relaxed system and a typical cool-core, at redshift 0.55. The corresponding NIKA observations and data reduction are detailed in \cite{Adam2016a}, including a detailed description of the cluster. The peak signal-to-noise ratio of the NIKA map reaches about 6 at the 22 arcsec resolution. The GGM and DoG maps do not allow us to identify sub-structures that deviate from that expected for a spherically symmetric object, within the noise fluctuations whose magnitude is large. A small GGM excess ($\gtrsim 4 \sigma$), with respect to gNFW expectation (see Figure \ref{fig:test_filter_gNFW_and_bimodal}), is visible in the south, but point source contamination (AGN and IR) is strong near the cluster center, and this could be due to mis-modeling of the contamination \citep[see][and the discussion of Section \ref{sec:Point_sources_residuals}]{Adam2016a}.

\subsection{MACS~J0717.5+3745}\label{sec:MACSJ0717}
\begin{figure}[h]
\center
\includegraphics[trim=5cm 0cm 8cm 2cm, clip=true, width=0.5\textwidth]{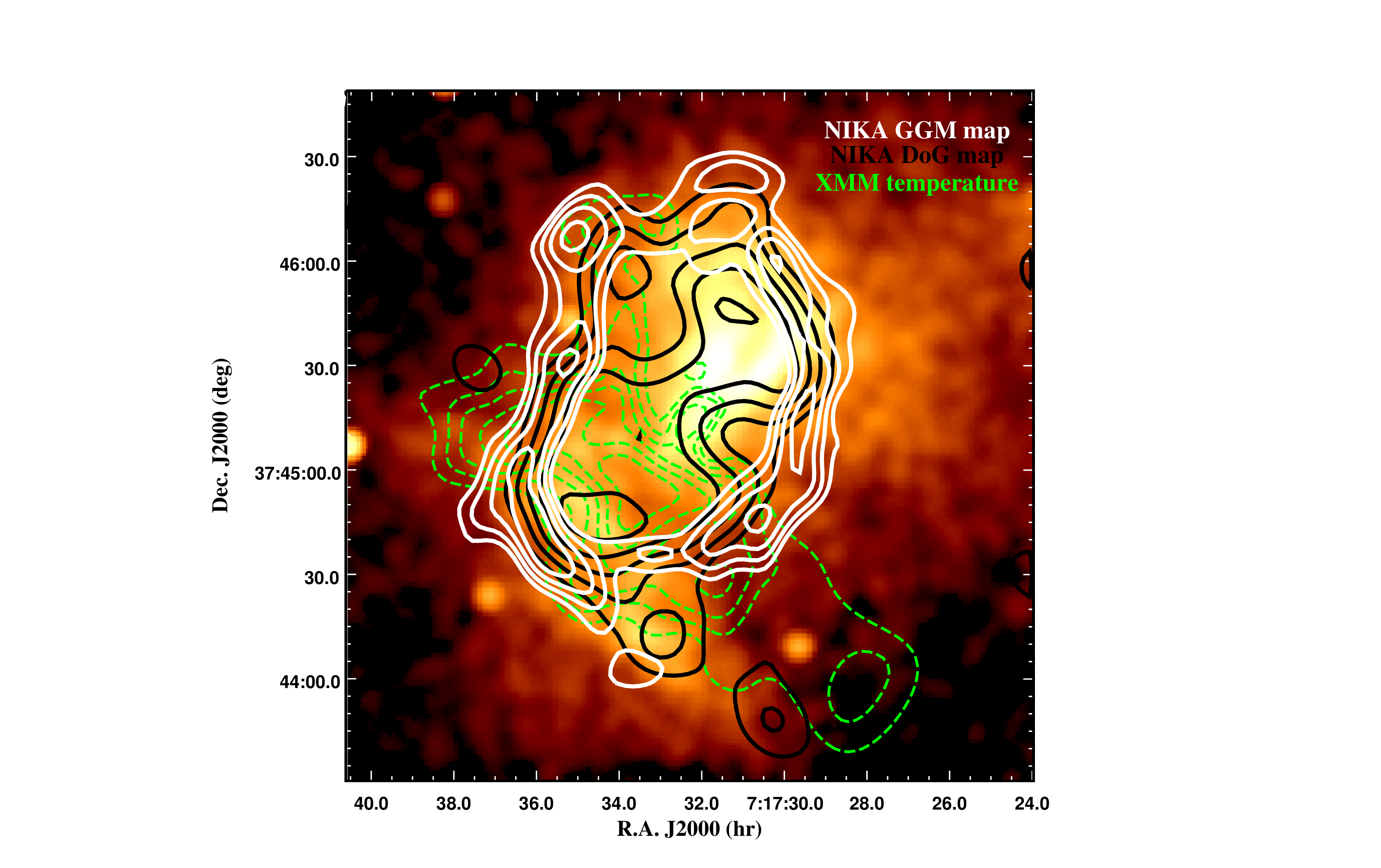} 
\caption{\footnotesize{Multi-probe view of the cluster \mbox{MACS~J0717.5+3745}: \textit{Chandra} X-ray surface brightness image (ObsID4200), plotted on a square root scale (thus proportional to the projected density) with \textit{XMM-Newton} temperature \citep[green contours from][]{Adam2017} and the DoG (black contours) and GGM (white contours) maps overlaid.}}
\label{fig:MACSJ0717_multiL}
\end{figure}

\mbox{MACS~J0717.5+3745} is one of the most disturbed clusters known to date. It is made of at least four interacting sub-clusters \citep{Ma2009}, and is thus an excellent target to search for compression regions, discontinuities and secondary peaks in tSZ maps. In addition to the systematic uncertainties discussed in Section \ref{sec:Systematics_and_noise_properties}, \mbox{MACS~J0717.5+3745} contains a significant amount of kinetic Sunyaev-Zel'dovich \citep[kSZ,][]{Sunyaev1980}. Because of large uncertainties in the estimates of this contaminant, it is only possible to test the impact of the kSZ signal on our result by comparing the recovered signal in the cases with and without kSZ correction, as detailed in Appendix \ref{sec:Impact_of_kSZ}. Since \mbox{MACS~J0717.5+3745} disposes of excellent X-ray data, we also provide a multi-wavelength view of the cluster together with our filtered maps in Figure \ref{fig:MACSJ0717_multiL}.

\begin{figure}[h]
\center
\includegraphics[trim=0cm 0cm 0cm 0cm, clip=true, width=0.5\textwidth]{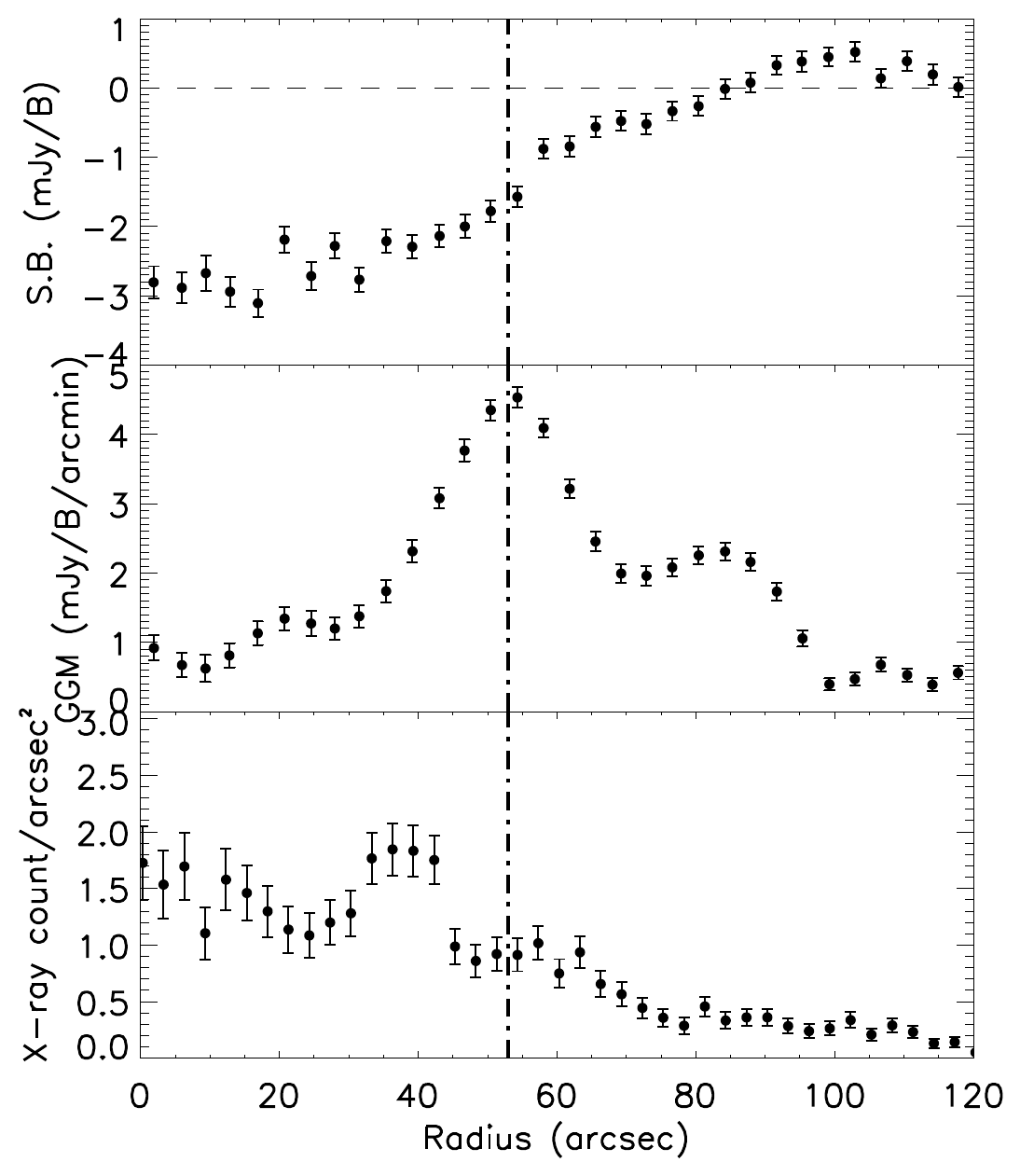} 
\caption{\footnotesize{Profile of \mbox{MACS~J0717.5+3745} in the region defined to include the strongest GGM ridge (east ridge, see Figure \ref{fig:NIKA_cluster_sample}, as the dark gray dashed line on the surface brightness image). The dashed-dotted line correspond to the GGM ridge location. {\bf Top:} surface brightness profile. {\bf Middle:} GGM profile. {\bf Bottom:} \textit{Chandra} photon count per unit area profile.}}
\label{fig:profile_MACSJ0717}
\end{figure}

The disturbed dynamical state of \mbox{MACS~J0717.5+3745} is obvious, as can be observed in Figure \ref{fig:NIKA_cluster_sample} and \ref{fig:MACSJ0717_multiL}. The overall structure of the GGM and the DoG maps are the same for the case with and without kSZ, but the relative amplitude of the structures strongly depends on the correction (see Appendix \ref{sec:Impact_of_kSZ}). Therefore, the kSZ contaminant only allows us a qualitative estimate of the gas inner structure. The GGM maps present two strong ridges on the east and southeast regions, of $\sim 1$ arcmin (395 kpc) each (${\rm SNR}_{\rm GGM} > 8$). In addition, we observe a large arc ($\sim 90$ arcsec) in the west sector that surrounds the X-ray main core. The southeast GGM ridge is spatially coincident with the southeast edge of the hot temperature bar caused by adiabatic compression (see Figure \ref{fig:MACSJ0717_multiL}), and could trace a shock in this region (see also the comparison to X-ray in Figure \ref{fig:profile_MACSJ0717}, and the corresponding discussion below). The eastern GGM ridge is nearly aligned with a radio relic \citep[see, e.g.][]{vanWeeren2017}, and both could be sourced by a shock caused by the ongoing merger. It also corresponds to a secondary vertical structure in the temperature map. The null of the gradient (i.e. flat surface brightness or constant projected ICM pressure) extends across all the brightest regions of the cluster in the case without kSZ correction, while it presents two nulls for the kSZ cleaned version of the map. In both cases, the main X-ray peak is aligned with a null region of the gradient, but we note that the the X-ray structure is itself complex and presents several peaks. Both DoG maps highlight the presence of two main peaks in the pressure distribution (at $> 4 \sigma$). This was also observed by \cite{Mroczkowski2012} using MUSTANG data. One of the peaks coincides with the main X-ray peak while the other one is coincident with the most massive sub-cluster, as identified from a strong lensing reconstruction \citep[e.g.][]{Limousin2015}. It is also the location of the hottest region that shows adiabatically compressed gas from the merger of the two main clusters. Weaker extensions in the DoG map, such as the one extending to the southwest, coincide with secondary X-ray peaks and allow us to identify sub-structures also in the pressure distribution.

In Figure \ref{fig:profile_MACSJ0717}, we focus on the southeast ridge (the brightest one) by providing the surface brightness and GGM profiles in a slice of the cluster that encloses the ridge (see the gray dashed line in Figure \ref{fig:NIKA_cluster_sample}). We also compare these profiles to the X-ray photon count, as measured using \textit{Chandra} data. We can see that the most prominent GGM feature traces well the steepest surface brightness change in the profile, which is possibly related to a discontinuity in the pressure. It also corresponds to a jump in the density, as probed by the X-ray photon counts, which is expected in the case of a shock. We note that the latter is offset by about 7 arcsec with respect to the GGM expectation. However, this could be the result of the very different resolution of \textit{Chandra} and NIKA, respectively 0.5 and 18.2 arcsec, or projection effect in such a complex system.

The GGM and DoG filtered maps of \mbox{MACS~J0717.5+3745} compare well to multiple major mergers as identified in the RHAPSODY-G simulation. They provide further insight into the gas structure of the cluster, with respect to the surface brightness image, by allowing us to identify regions of compressed gas. They are consistent with a major merger of two main sub-clusters and that of one (or more) additional sub-cluster likely to be smaller. Nevertheless, projection effects limit our interpretation of the data, in particular in the case of such a complex object. Using the filtered maps together with multi-wavelength data might in the future provide a better understanding of the ongoing merger scenario.

\mbox{MACS~J0717.5+3745} is among the most violent mergers in the Universe, in which kSZ contamination is expected to be large \citep{Mroczkowski2012,Sayers2013,Adam2016b}, and where the gas can reach a temperature of up to 25 keV \citep[e.g.][]{Adam2017}. As shown in Appendix \ref{sec:Impact_of_kSZ}, kSZ contamination can significantly alter the relative brightness and location of the tSZ peaks we identify using the DoG filter. Similarly. the location of the identified ridges in the GGM map can change significantly when considering the kSZ signal. While \mbox{MACS~J0717.5+3745} is certainly an extreme case (the only single cluster in which kSZ has been clearly detected so far), we note that the kSZ signal potentially affects all the tSZ maps to some extent, and thus the filtered maps. Unfortunately, the current sensitivity to kSZ imaging is not yet sufficient to extract accurate kSZ corrections for all clusters in our sample. By using maps of the ICM temperature \citep[e.g.][]{Adam2017}, it is possible to account for relativistic corrections to the tSZ spectrum in the NIKA band. In the hottest region of the cluster ($\sim 25$ keV), the correction reaches 13\% \citep[see also Table 1 in][]{Adam2016b}, and the relative amplitude of the correction is 8\% over the cluster extension. By comparing the filtered map with and without considering the correction, we find that only the amplitude of the signal can change significantly (by typically 10\%), but the relative spatial variations are not significant with respect to the noise. Since \mbox{MACS~J0717.5+3745} is the most disturbed and the hottest cluster in our sample, it is thus justified to neglect relativistic corrections for our purpose. Accounting for them would require high resolution mapping of the temperature, which are generally difficult to obtain at high redshift.

\subsection{PSZ1~G046.13+30.75}
\mbox{PSZ1~G046.13+30.75} is a \textit{Planck}-discovered cluster \citep{PlanckXXIX2014}, at redshift 0.55, and was observed during the same campaign as another NIKA cluster, \mbox{PSZ1~G045.85+57.71} \citep[see][for more details]{Ruppin2016}. The observing weather conditions were unstable with a high opacity and the cluster significance barely reaches $4 \sigma$ per 22 arcsec FWHM beam at the peak on the NIKA map. The GGM and DoG filtered maps are consistent with noise.

\subsection{PSZ1~G045.85+57.71}
\mbox{PSZ1~G045.85+57.71} is a \textit{Planck}-discovered cluster \citep{PlanckXXIX2014}, at redshift 0.61, and was observed during the same campaign as \mbox{PSZ1~G046.13+30.75} \citep[see][for more details]{Ruppin2016}. It was classified as a cool-core cluster based on its temperature and entropy profile, both fully consistent with the \rexcess\ cool-core sub-sample \citep{Bohringer2007,Arnaud2010,Pratt2010}. Its morphology is highly elliptical, with a main axis oriented about 45 degrees with respect to the R.A. axis, as can be observed in Figure \ref{fig:NIKA_cluster_sample}, but \cite{Ruppin2016} do not find evidence for merging activity. The tSZ peak is consistent with that of the X-ray, and the tSZ elongation is mostly prominent toward the southwest.

The GGM map of \mbox{PSZ1~G045.85+57.71} is that of an elongated ring following the morphology of the cluster. On the southeast region, we observe a stronger pressure gradient extending over about 45 arcsec (312 kpc) and reaching ${\rm SNR}_{\rm GGM} = 5.8$. In contrast to the RHAPSODY-G relaxed cluster RG361\_00188, the GGM map shows that the projected pressure is approximately constant (i.e., null of the gradient, or constant surface brightness), over a wide area from the X-ray core to the southwest extension. As \mbox{PSZ1~G045.85+57.71} is a cool-core cluster, with a dense X-ray core, this indicates that the temperature rises in the southwest sector. The DoG map does not show the presence of any strong core and is consistent with the projected pressure being relatively constant from the X-ray core to the southwest extension. As a consequence of the lack of signal on small scales, the DoG map significance is relatively low, reaching only $4 \sigma$ at the peak. The main peak is located within a few arcsec of the X-ray core, but two secondary peaks are visible towards the southwest and on the north.

The GGM and DoG images of \mbox{PSZ1~G045.85+57.71} are consistent with that of a merging sub-group falling from the southwest to the main cluster. It would be responsible for the local temperature rise and the gas overpressure on the southwest. Such scenarios are commonly observed in the RHAPSODY-G clusters. In contrast, relaxed RHAPSODY-G systems do not present similar features in their filtered maps, which reinforces the interpretation proposed above. Nevertheless, we note that the signal-to-noise ratio remains limited in the case of this cluster, and any stronger conclusions would require deeper data.

\subsection{CL~J1226.9+3332}
\mbox{CL~J1226.9+3332} is a hot and massive high redshift cluster, at $z=0.89$. It was discovered by ROSAT \citep{Ebeling2001} and has been the object of several multi-wavelength studies since then. In particular, \cite{Maughan2007} identified a temperature excess in the southwest region from \textit{XMM-Newton} and \textit{Chandra} X-ray observations, in agreement with a tSZ sub-structure observed by MUSTANG \citep{Korngut2011}. This extension seen in the gas distribution is attributed to the merging of a smaller cluster that is visible from Hubble lensing data \citep{Jee2009}. NIKA was used to map the tSZ signal toward \mbox{CL~J1226.9+3332} as discussed in \cite{Adam2015}, who also found an extension in the southwest region when subtracting a spherical model fitted to their data.

The large scale morphology of the NIKA map presented in Figure \ref{fig:NIKA_cluster_sample} shows that \mbox{CL~J1226.9+3332} is overall spherical and we observe a good match between the X-ray and tSZ peaks. This is confirmed by the GGM filtered map, where the location of the null of the gradient (the tSZ main peak), at the tSZ peak, is in excellent agreement with the X-ray peaks. The gradient is strong around the core, indicating the presence of a compact pressure structure. However, we observe a strong GGM ridge in the southwest region (${\rm SNR}_{\rm GGM} = 5.7$), extending over about 45 arcsec (360 kpc). The DoG map agrees with \mbox{CL~J1226.9+3332} being dominated by a compact core ($-7.7 \sigma$) that matches the X-ray peak. It also shows an extension towards the west ($> -4 \sigma$).

\begin{figure}[h]
\center
\includegraphics[trim=0cm 0cm 0cm 0cm, clip=true, width=0.5\textwidth]{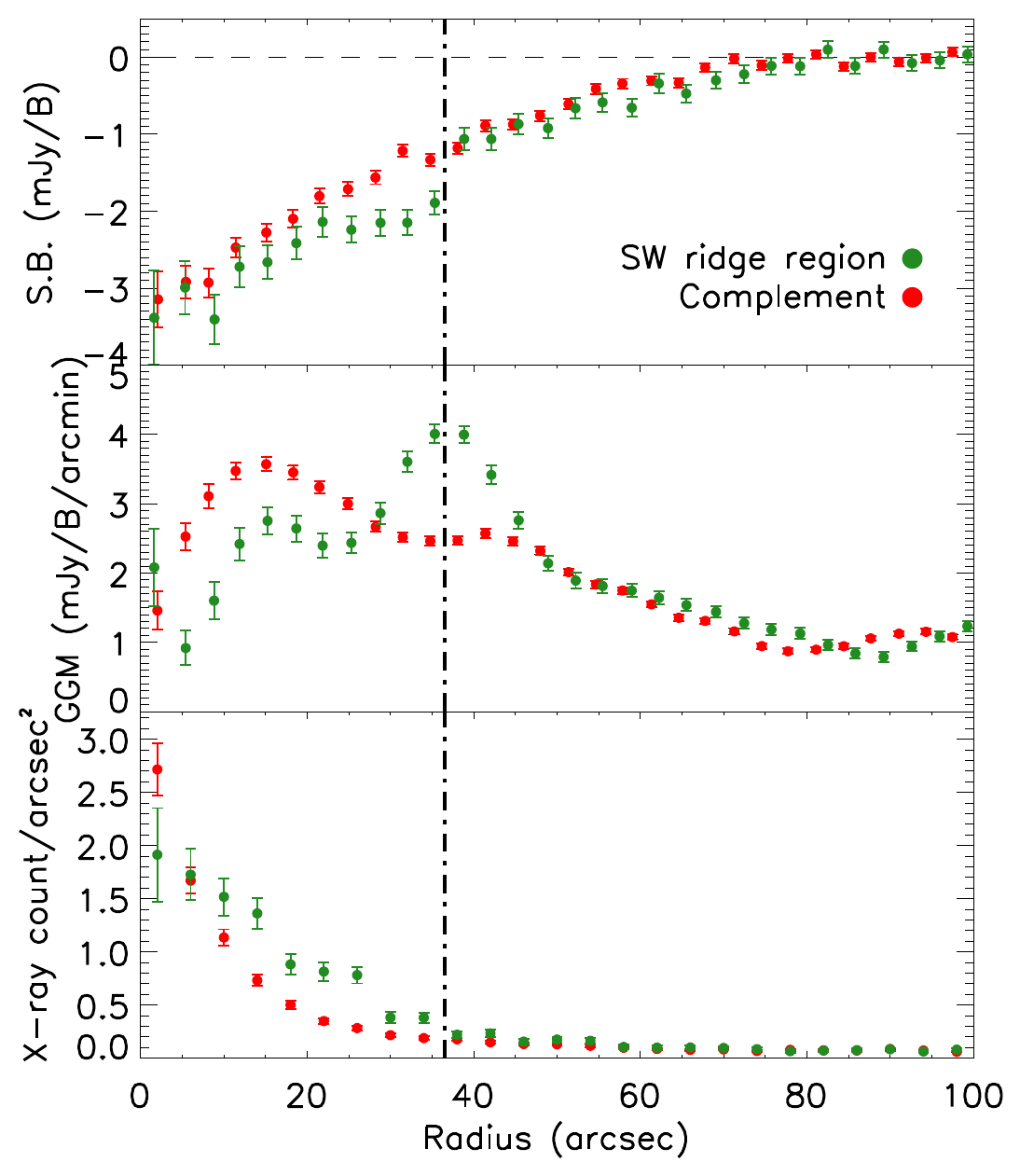} 
\caption{\footnotesize{Comparison of the profile of \mbox{CL~J1226.9+3332} in the region of the GGM ridge (green) to its complement (red). The GGM ridge region is define in Figure \ref{fig:NIKA_cluster_sample}, as the black dashed line on the surface brightness image. The dashed-dotted line correspond to the GGM ridge location. {\bf Top:} surface brightness profile. {\bf Middle:} GGM profile. {\bf Bottom:} \textit{Chandra} photon count per unit area profile.}}
\label{fig:profile_CLJ1227}
\end{figure}

In Figure \ref{fig:profile_CLJ1227}, we provide the surface brightness and GGM profiles of \mbox{CL~J1226.9+3332} in the region of the GGM ridge and in the complementary region. The region is defined by the cone shown on the surface brightness image of Figure \ref{fig:NIKA_cluster_sample}. It is designed to enclose the tSZ extension and to point in a direction perpendicular to the GGM ridge. We also compare the profiles to \textit{Chandra} X-ray photon counts, extracted in the same region, as a proxy for the gas density.

The signal observed in the GGM ridge region is in excellent agreement with that expected from a shock, as in Figure \ref{fig:test_filter_GNFW_and_shock} (right panel). The GGM profile peaks at $\theta = 37.5$ arcsec and we observe a discontinuity at the same radius in the surface brightness. On the contrary, the signal extracted from the complementary region is in good agreement with that of a relaxed spherical cluster. As expected in this case, the GGM signal peak is located at a radius that corresponds to the filter size (see also Figure \ref{fig:test_filter_gNFW_and_bimodal} and Section \ref{sec:Application_to_toy_models}) and the GGM profile smoothly vanishes as the radius increases. The X-ray photon count profiles also show an excess of signal in the region of the GGM ridge (at radii between 10 and 35 arcsec from the center) with respect to the rest of the cluster. A possible discontinuity, which is also expected in the gas density in the case of a shock, is seen in the X-ray photon count profile, but it is located at a radius about 5 arcsec lower than the GGM peak, as it was the case in Figure \ref{fig:profile_MACSJ0717} for \mbox{MACS~J0717.5+3745}. While the presence of a compression region is clear, confirmation of the presence of a shock would require deeper observations and a more thorough analysis. We note that presence of the GGM ridge could affect the pressure profile extraction as done in \cite{Romero2017}. It could be responsible for the preferred large value of the $\alpha$ gNFW parameter, which describes the width of the transition between the core and the outskirt.

The GGM and DoG analysis of \mbox{CL~J1226.9+3332} is fully consistent with that of a shock occurring within the merger scenario discussed above \citep[see also][for more details]{Adam2015}. \textit{Chandra} data also seems to support this hypothesis. The structure observed in the signal matches well what is seen in the case of the RHAPSODY-G multiple merger RG377\_00181. Nonetheless, the strength of the tSZ extension with respect to the main cluster is larger in the case of \mbox{CL~J1226.9+3332} and the GGM ridge we observe is in a more compact configuration, closer to the core.

\section{Discussions and perspectives}\label{sec:discussions}
Cosmological studies using tSZ observations of clusters of galaxies generally rely on mass estimates based on various hypotheses, such as the hydrostatic equilibrium of the ICM or the 3D structure for lensing masses, and the self-similarity of the galaxy cluster population. These assumptions enable a calibration of the scaling law linking the integrated Compton parameter to the cluster total mass \citep[see, e.g,][]{Arnaud2007,Arnaud2010,Planck2013V}, and using a universal pressure profile to compute the tSZ power spectrum \citep[e.g.][]{Komatsu2002}. However, deviations from these hypotheses have already been identified and may lead to biases on cosmological parameters estimated using galaxy clusters \citep[see, e.g.][]{Ichikawa2013,McDonald2014,Planck2014XX,Planck2015XXIV,Planck2016XXII}.

Deviations from hydrostatic equilibrium are expected for unrelaxed clusters where the non-thermal pressure content due to turbulence and bulk flows contributes significantly to the equilibrium of the gas in the gravitational potential \citep[e.g.][]{Siegel2016}. Such effects suppress power principally on large scales in the tSZ power spectrum whereas AGN feedback has a significant impact on small scales \citep[e.g.][]{Shaw2010}. Furthermore, the bias affecting the estimate of the cluster mass under the hydrostatic equilibrium assumption in unrelaxed clusters is expected to be higher than for relaxed clusters. The complex inner structure that can arise in the case of disturbed clusters can also affect lensing mass estimates. In this context, dynamical state indicators such as the features identified by the GGM and DoG filters are powerful tools to identify unrelaxed gas structures, which can help to understand the scatter and offset in the scaling relations.

While the tSZ surface brightness map of a cluster subject to AGN feedbacks can display an apparent relaxed morphology (see the cluster RG448\_00211 in Figure \ref{fig:RG_cluster_sample}), the GGM map of this system reveals a characteristic ring feature which is not observed for a fully relaxed cluster (see the cluster RG361\_00188 in the same figure). The GGM map can then be used as a dynamical indicator to identify unrelaxed ICM regions in cluster outskirts and could therefore be linked to the amplitude of the hydrostatic bias.

Information on the dynamical state of the ICM is also given by the comparison of the distinct projected distributions of gas and dark matter. In the case of relaxed systems such as RG361\_00188, the gas distribution follows the one of the dark matter density, contrarily to major mergers or clusters with disturbed outskirts. Although the tSZ surface brightness map does not enable to constrain the dark matter density distribution, the gas compression direction provided by both the GGM and DoG maps can give information on the projected shape of the underlying gravitational potential. For example, while the apparent ICM morphology of the cluster RG448\_00211, in Figure \ref{fig:RG_cluster_sample}, is azimuthally symmetric, the DoG map of this cluster displays an elliptical morphology with a major axis direction matching the one of the dark matter density distribution. The features in the DoG filter maps can therefore be used as a morphology indicator to identify potential deviations between the gas and the dark matter distribution.

Constraining the distribution of the cluster merger rate in large ranges of halo mass and redshift is of key importance to improve our understanding of the formation of large-scale structures \citep[e.g.][]{Fakhouri2010}. The GGM map of the head-on merger RG474\_00172 (see Figure \ref{fig:RG_cluster_sample_proc}) displays a characteristic bimodal distribution giving the locations of high gas compression. Such information is not provided by the tSZ surface brightness map obtained after the end-to-end processing of the RHAPSODY-G simulated cluster. The GGM filter could therefore be used to improve on the identification of mergers by discriminating relaxed elongated ICM and major mergers. 

The gas compression intensity, given by the amplitude of the gradient in the GGM maps, could also be compared to radio relic observations \citep[e.g.][]{vanWeeren2010}. The identification of a correlation between the two observables would enable to link the amplitude of the gradient in the GGM maps to the Mach number. Such information would greatly enhance the understanding of structure assembly especially when combined with the gas line of sight velocity provided by kSZ effect observations.

\section{Summary and conclusions}\label{sec:Summary_and_conclusions}
The tSZ effect provides a direct observable for the ICM pressure in galaxy clusters. In order to search for pressure sub-structures, which are related to the formation history of the clusters, we have applied filtering methods to resolved NIKA tSZ maps of six galaxy clusters at intermediate and high redshift. The same methods were applied to toy models and to the RHAPSODY-G hydrodynamical simulations in order to better interpret our results. Additionally, a careful investigation of the impact of possible systematic effects was performed. We considered contamination from compact sources, the filtering due to the NIKA processing, and the propagation of the spatially correlated noise.

While half of the clusters in the NIKA sample do not present sufficiently high signal-to-noise ratio to enable the identification of internal pressure structures with confidence, the three others show significant features that are not visible in the raw tSZ maps. The observed structures show signatures that are similar to the ones observed in the RHAPSODY-G simulations where they can be clearly attributed to compressions and shocks in the hot gas due to merging events. The potential point source residual contamination is unlikely to cause such structures because its signature is different from the one we observe, and expected to be subdominant. Similarly, we find that the effects of the NIKA processing are not significant at the scales of the observed structures and are subdominant given the available signal-to-noise ratio. Finally, we account for spatial correlations of the noise in order to assess the significance of the detections.

The three highest signal-to-noise clusters in our sample, \mbox{MACS~J0717.5+3745} at $z=0.54$, \mbox{PSZ1~G045.85+57.71} at $z=0.61$ and \mbox{CL~J1226.9+3332} at $z=0.89$, were investigated in more depth. \mbox{MACS~J0717.5+3745} is clearly disturbed at all scales probed by NIKA. Three strong pressure gradient features are observed in the northwest, east, and southeast sectors, using the GGM filter. We also identified two main peaks in the pressure distribution from the DoG filtered map, which are associated with sub-groups seen at other wavelengths. \mbox{PSZ1~G045.85+57.71} is elongated on large scales, but we do not observe significant sub-clumps. A GGM ridge is observed in the southeast of the cluster, potentially indicating ongoing merging activity. \mbox{CL~J1226.9+3332} appears to be relaxed on large scales, but the GGM filtered map shows a $\sim 45$ arcsec (360 kpc) long GGM ridge in the west region associated with an elongation of the gas on small scales as observed in the DoG filtered map. Its signature is in excellent agreement with that of a shock.

The combined high angular resolution and high sensitivity of current tSZ data has now reached a state where the detailed study of the ICM structure of galaxy clusters is possible up to intermediate and high redshifts. In the case of our sample, significant sub-structures start to be visible for clusters with a peak signal-to-noise of $\gtrsim 10$, per $\sim 22$ arcsec beam.

Our analysis shows that it is possible to explore the details of cluster assembly in distant clusters using deep tSZ imaging. New generation instruments, such as NIKA2 \citep{Calvo2016,Catalano2016,NIKA2017} at the IRAM 30m telescope, are expected to provide unprecedented high sensitivity tSZ data at high angular resolution. NIKA2 observations \citep[such as the ones of the tSZ large program,][]{Comis2016,Mayet2017} should therefore allow for in-depth investigations of the inner structure of the ICM pressure of a cosmologically representative cluster sample. Currently, the sample that we used in this paper is limited to redshifts larger than 0.45, at which the 18 arcsec FWHM NIKA resolution at 150 GHz corresponds to about 100 kpc. With an instantaneous field of view of 6.5 arcmin, NIKA2 will also allow us to observe lower redshift clusters more easily, and thus enable the access to smaller physical scales at similar integration time. At intermediate and high redshift, the mapping speed should increase by a factor of up to $\sim 10$ from NIKA to NIKA2, thus providing very high signal-to-noise images in very short time.

\begin{acknowledgements}
We are thankful to the anonymous referee for useful comments that helped improve the quality of the paper.
We would like to thank the IRAM staff for their support during the campaigns. 
The NIKA dilution cryostat has been designed and built at the Institut N\'eel. In particular, we acknowledge the crucial contribution of the Cryogenics Group, and  in particular Gregory Garde, Henri Rodenas, Jean Paul Leggeri, Philippe Camus. 
This work has been partially funded by the Foundation Nanoscience Grenoble, the LabEx FOCUS ANR-11-LABX-0013 and the ANR under the contracts "MKIDS", "NIKA" and ANR-15-CE31-0017. 
This work has benefited from the support of the European Research Council Advanced Grants ORISTARS and M2C under the European Union's Seventh Framework Programme (Grant Agreement nos. 291294 and 340519).
We acknowledge fundings from the ENIGMASS French LabEx (B. C. and F. R.), the CNES post-doctoral fellowship program (R. A.),  the CNES doctoral fellowship program (A. R.) and the FOCUS French LabEx doctoral fellowship program (A. R.).
R.A. acknowledges support from Spanish Ministerio de Econom\'ia and Competitividad (MINECO) through grant number AYA2015-66211-C2-2.
E. P. acknowledges the support of the French Agence Nationale de la Recherche under grant ANR-11-BS56-015.
O.H. acknowledges funding from the European Research Council (ERC) under the European Union's Horizon 2020 research and innovation programme (grant agreement No 679145, project “COSMO\_SIMS”).
D.M. acknowledges support from the Swiss National Science Foundation (SNSF) through the SNSF Early.Postdoc and Advanced.Postdoc Mobility Fellowships. 
\end{acknowledgements}

\bibliography{biblio_ED}

\appendix
\section{Impact of the filter parameters}\label{sec:Impact_of_the_filter_parameters}
\begin{figure}[h]
\centering
\resizebox{0.5\textwidth}{!} {
\begin{tabular}{ll}
\includegraphics[trim=0cm 2.2cm 0cm 0cm, clip=true, scale=1]{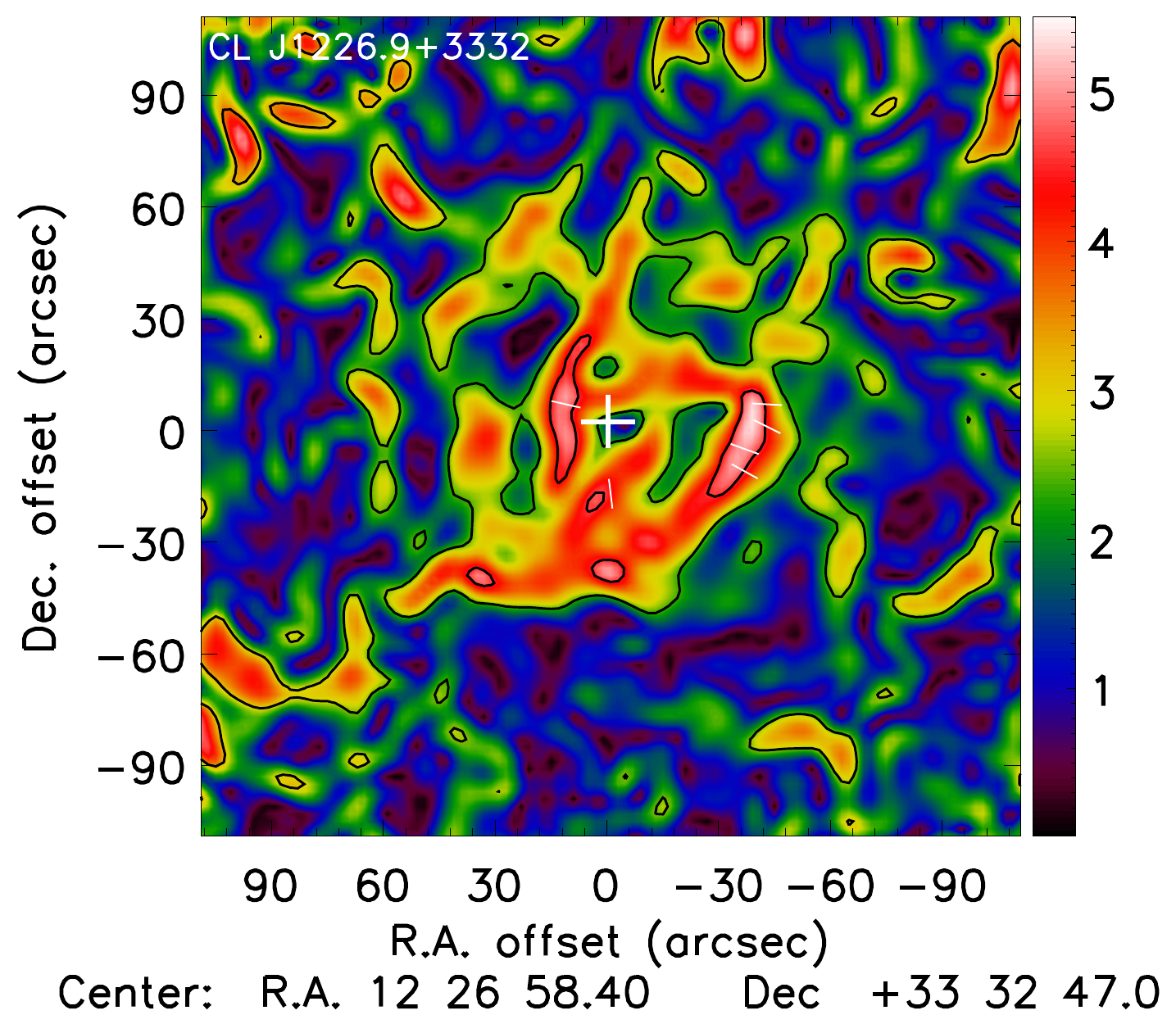} 
\put(-60,310){\makebox(0,0){\rotatebox{0}{\LARGE mJy/beam/arcmin}}} & 
\includegraphics[trim=2.3cm 2.2cm 0cm 0cm, clip=true, scale=1]{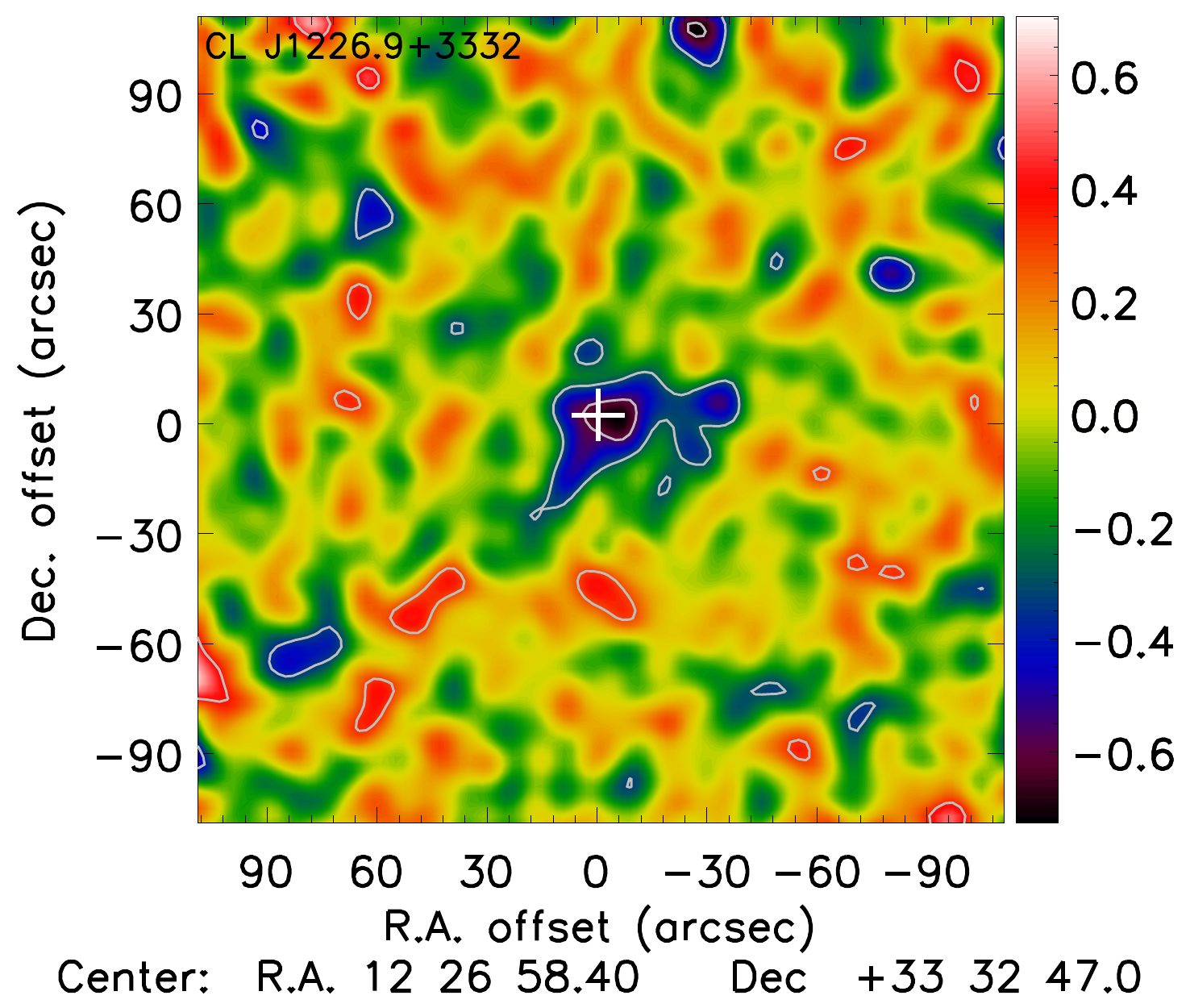} 
\put(-60,310){\makebox(0,0){\rotatebox{0}{\LARGE mJy/beam}}} \\
\includegraphics[trim=0cm 2.2cm 0cm 0cm, clip=true, scale=1]{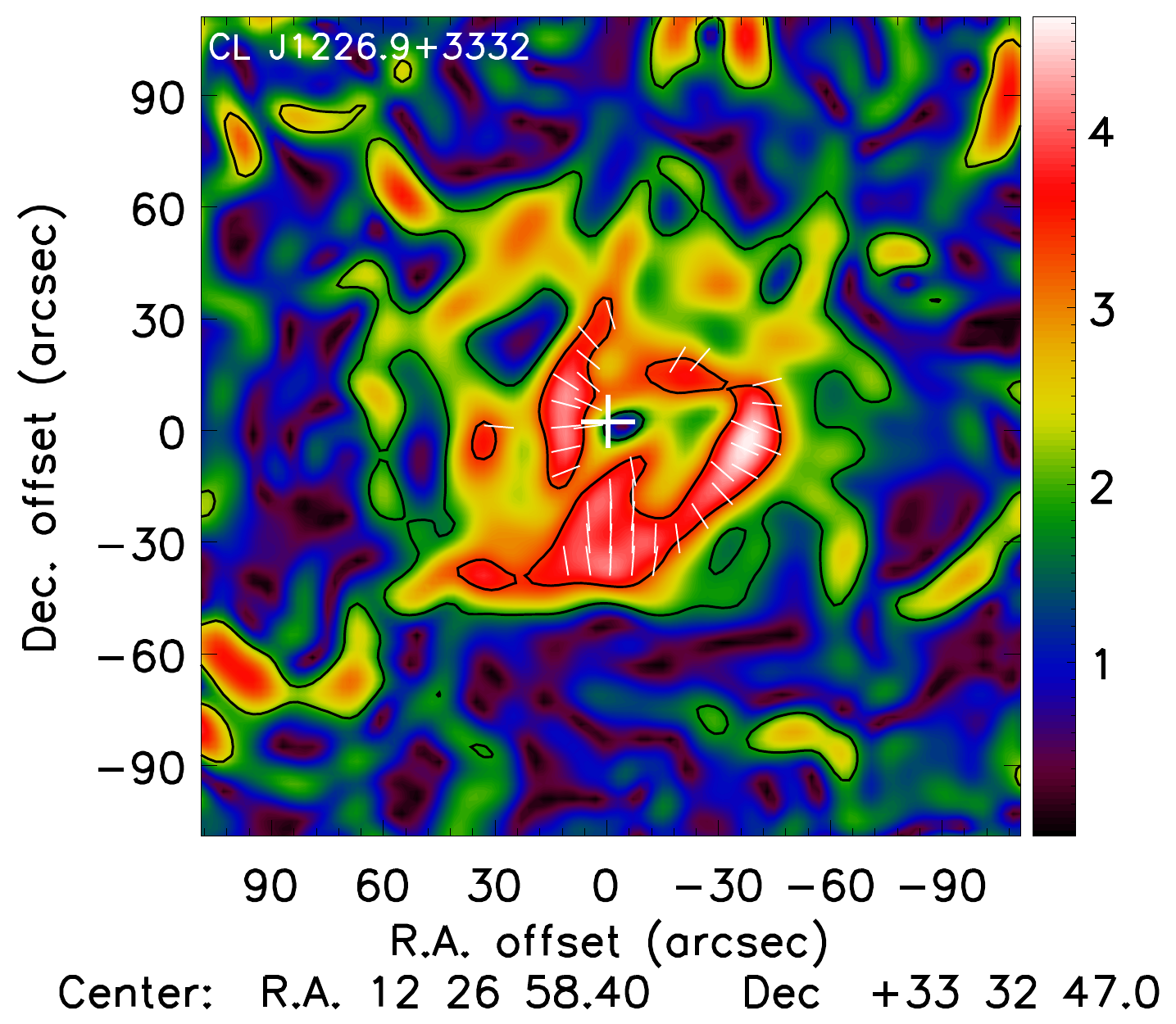} & 
\includegraphics[trim=2.3cm 2.2cm 0cm 0cm, clip=true, scale=1]{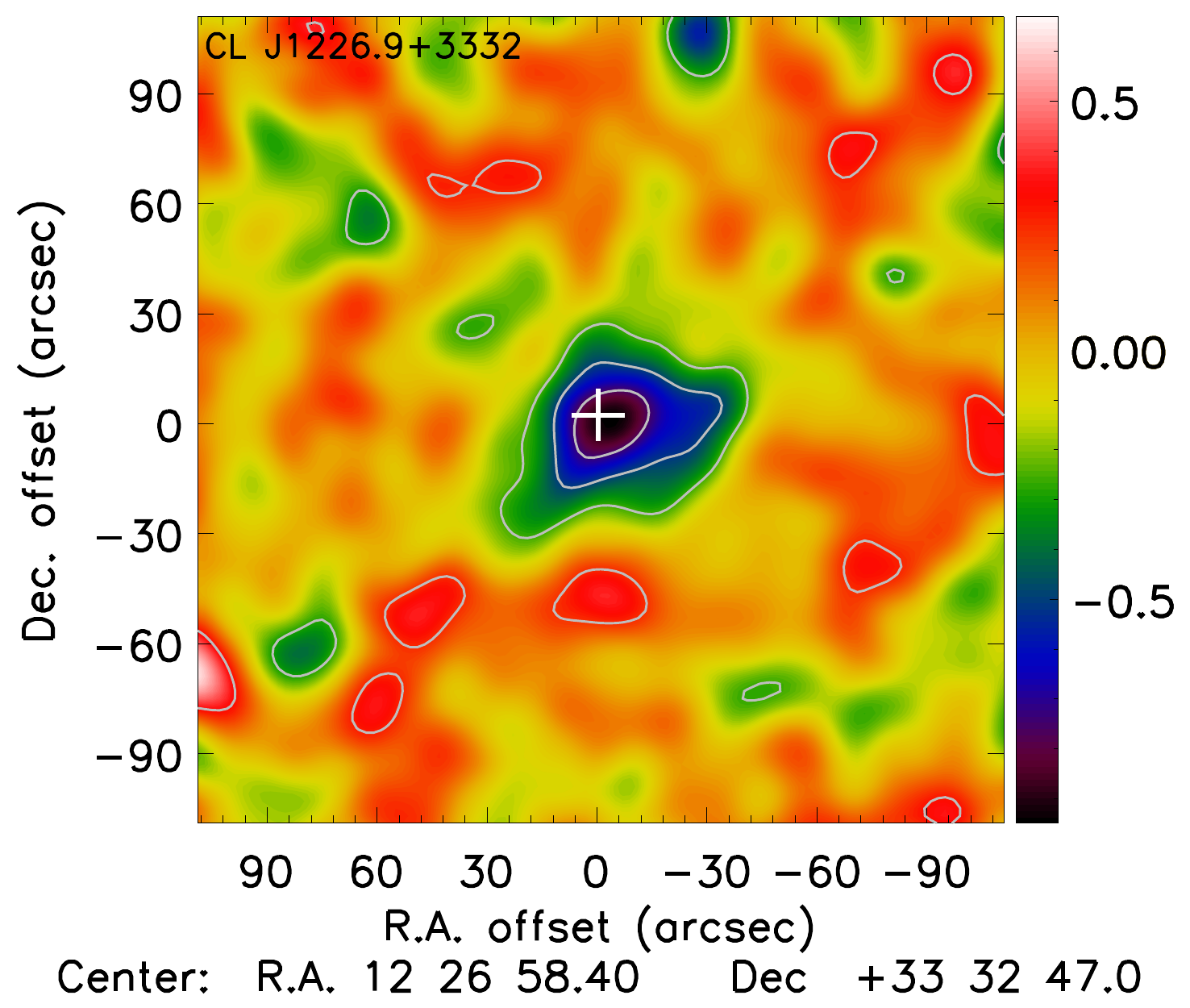} \\
\includegraphics[trim=0cm 2.2cm 0cm 0cm, clip=true, scale=1]{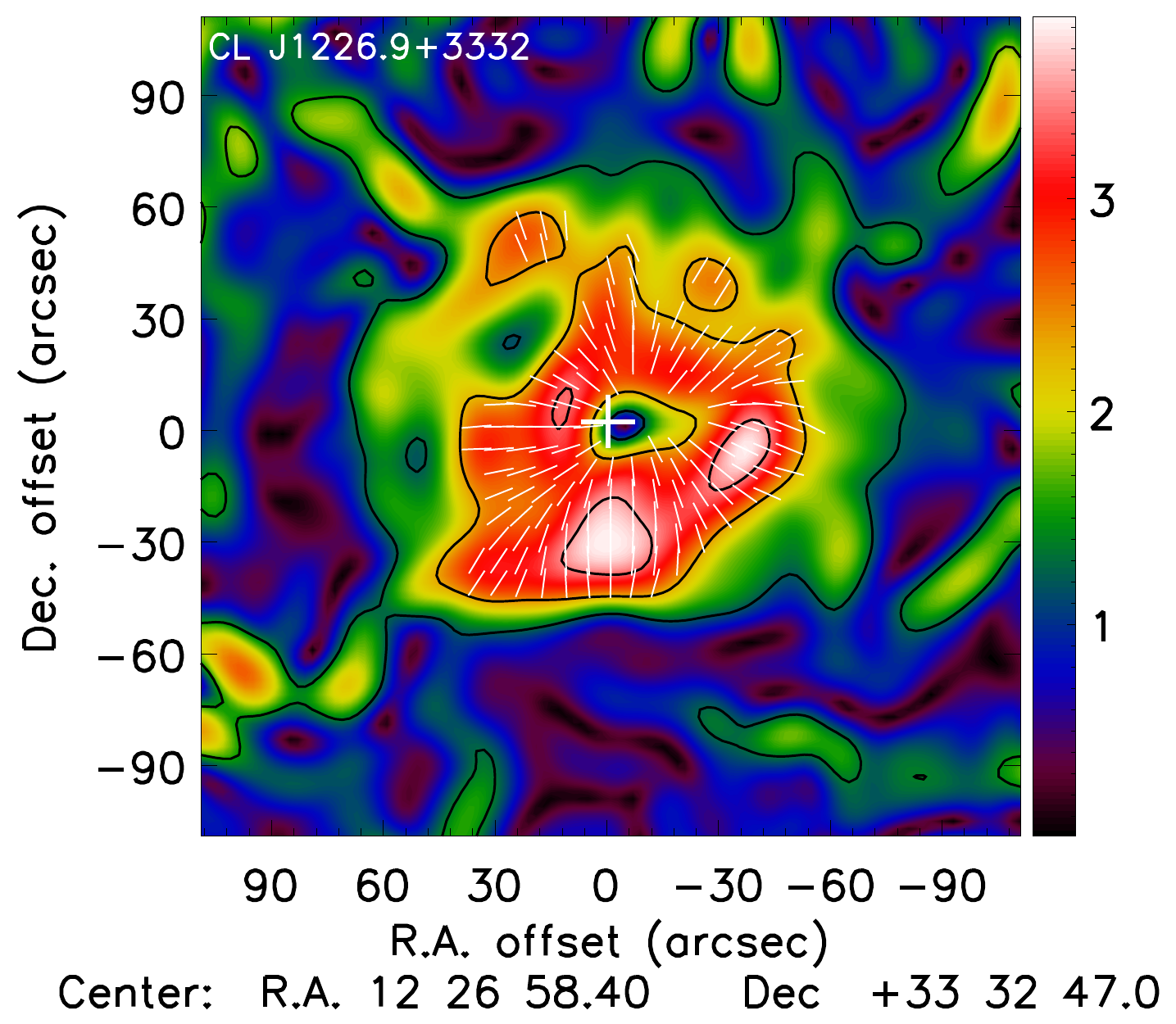} & 
\includegraphics[trim=2.3cm 2.2cm 0cm 0cm, clip=true, scale=1]{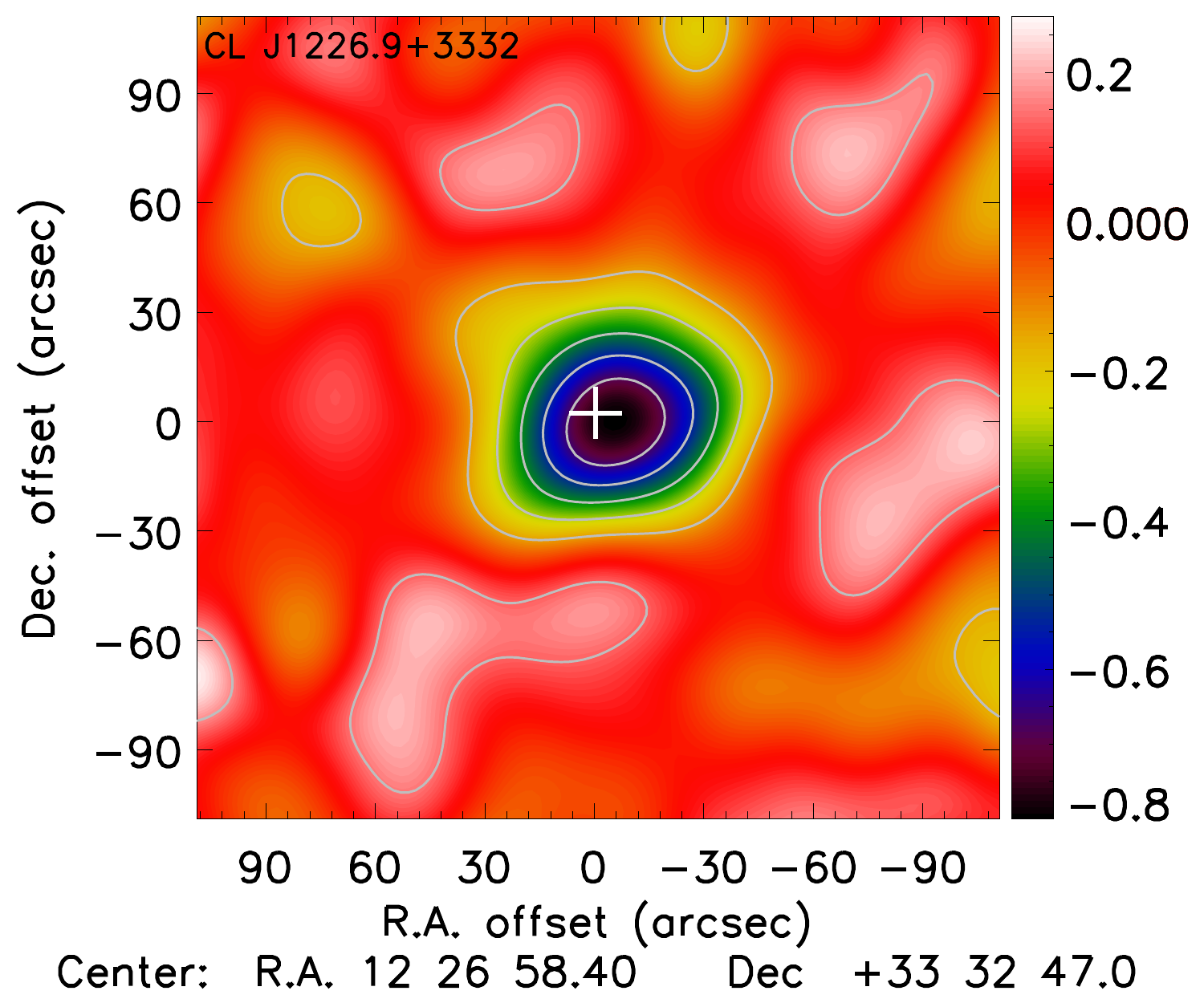} \\
\includegraphics[trim=0cm 0.7cm 0cm 0cm, clip=true, scale=1]{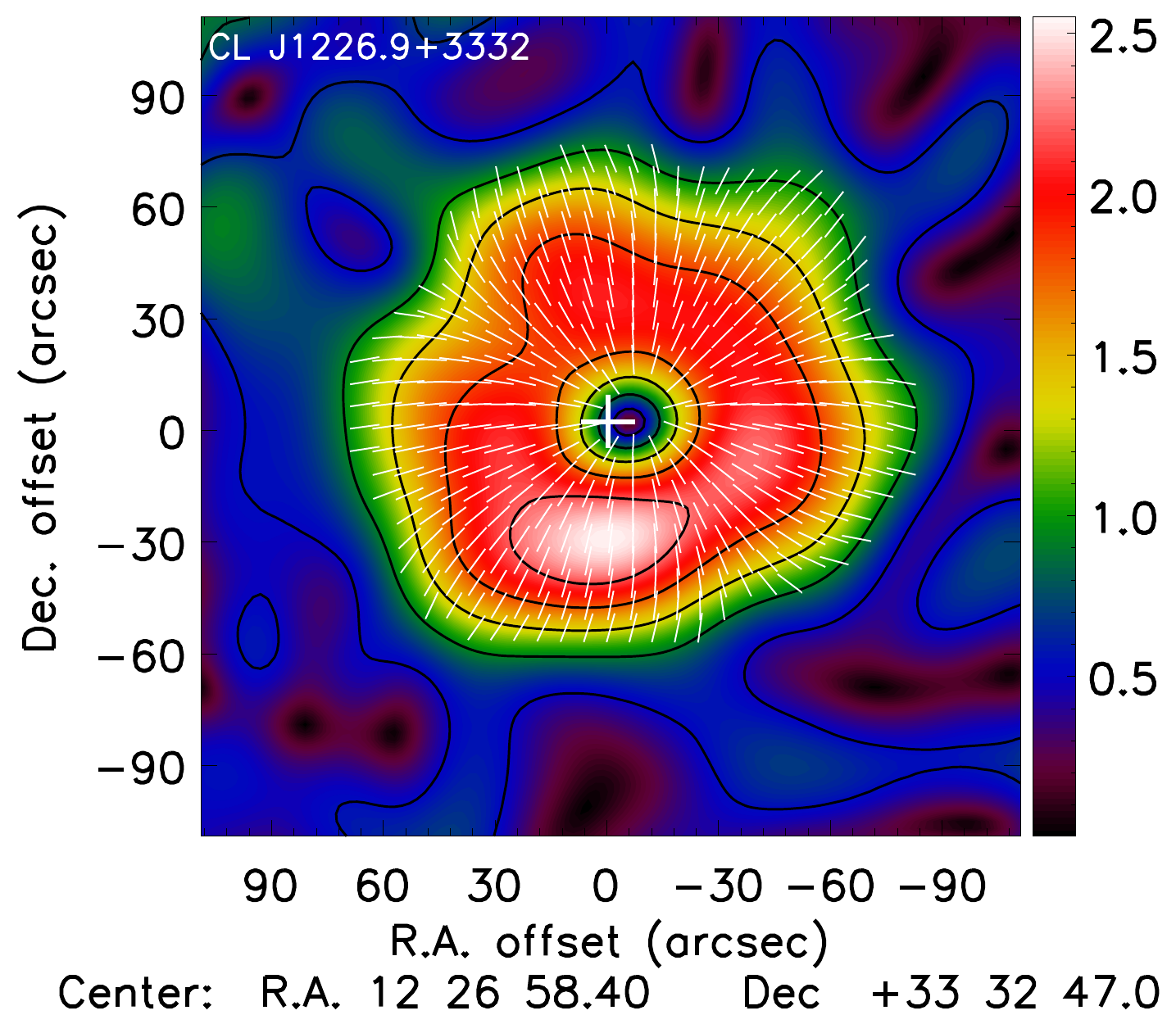} & 
\includegraphics[trim=2.3cm 0.7cm 0cm 0cm, clip=true, scale=1]{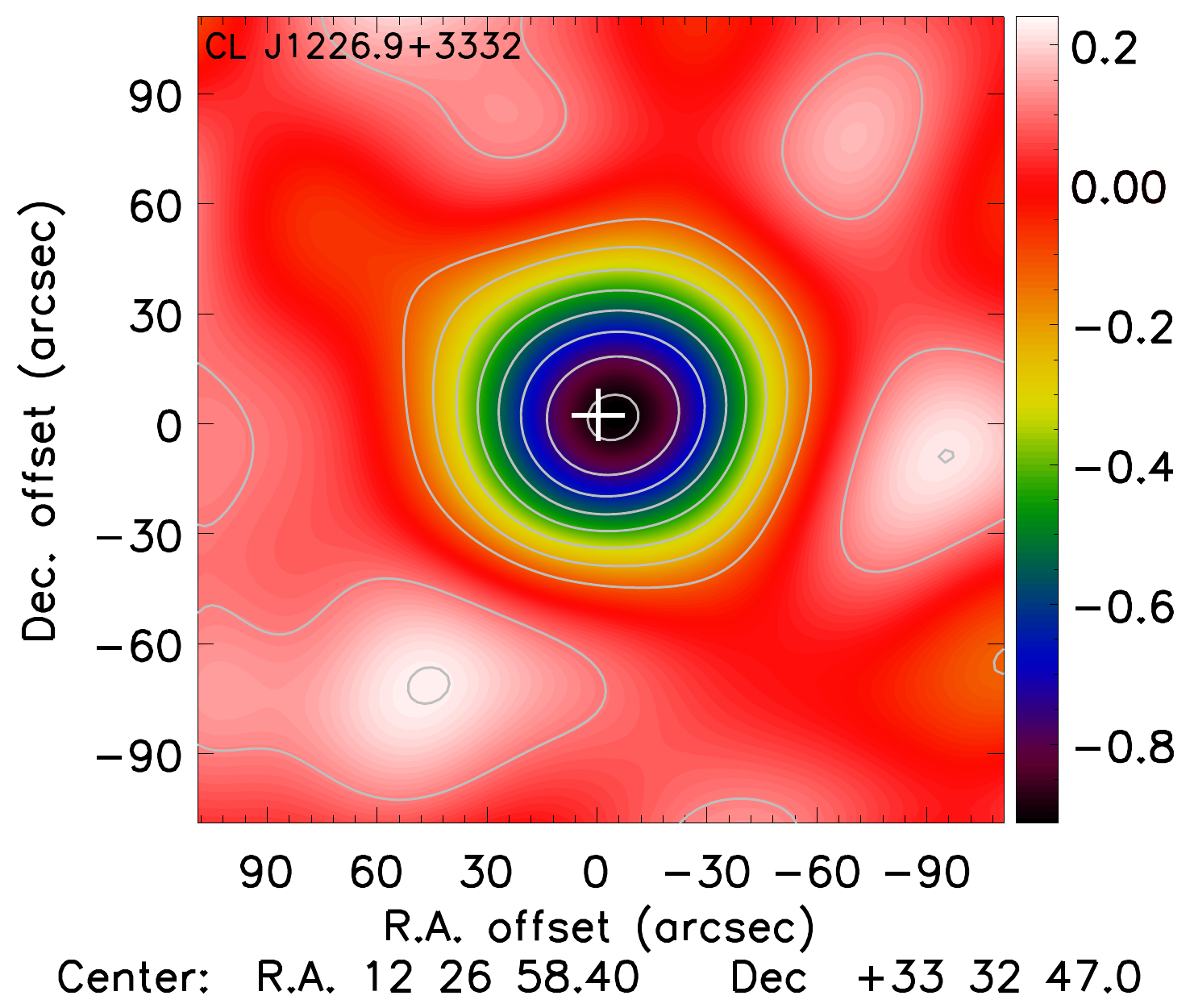}
\end{tabular}}
\caption{\footnotesize{Evolution of the GGM (left) and DoG (right) filtered maps of \mbox{CL~J1226.9+3332} as a function of the filter parameters. From left to right, the parameters are $\left(\theta_0, \theta_1, \theta_2\right) = \left(12, 10, 30\right)$ arcsec, $\left(\theta_0, \theta_1, \theta_2\right) = \left(15, 15, 45\right)$ arcsec, $\left(\theta_0, \theta_1, \theta_2\right) = \left(20, 30, 60\right)$ arcsec and $\left(\theta_0, \theta_1, \theta_2\right) = \left(40, 50, 100\right)$ arcsec.}}
\label{fig:filtered_NIKA_maps_evolution}
\end{figure}

In Figure \ref{fig:filtered_NIKA_maps_evolution}, we provide the extracted filtered signal as a function of the filtering parameters in the case of \mbox{CL~J1226.9+3332}. As the size of the filters increases, the detection strength increases, but the small scale signal we aim at extracting get washed out. The baseline parameters chosen in this paper correspond to the second row.

\section{Transfer function deconvolution}\label{sec:Transfer_function_deconvolution}
\begin{figure*}[h]
\centering
\resizebox{\textwidth}{!} {
\begin{tabular}{llllll}
\includegraphics[trim=0cm 2.2cm 0cm 0cm, clip=true, scale=1]{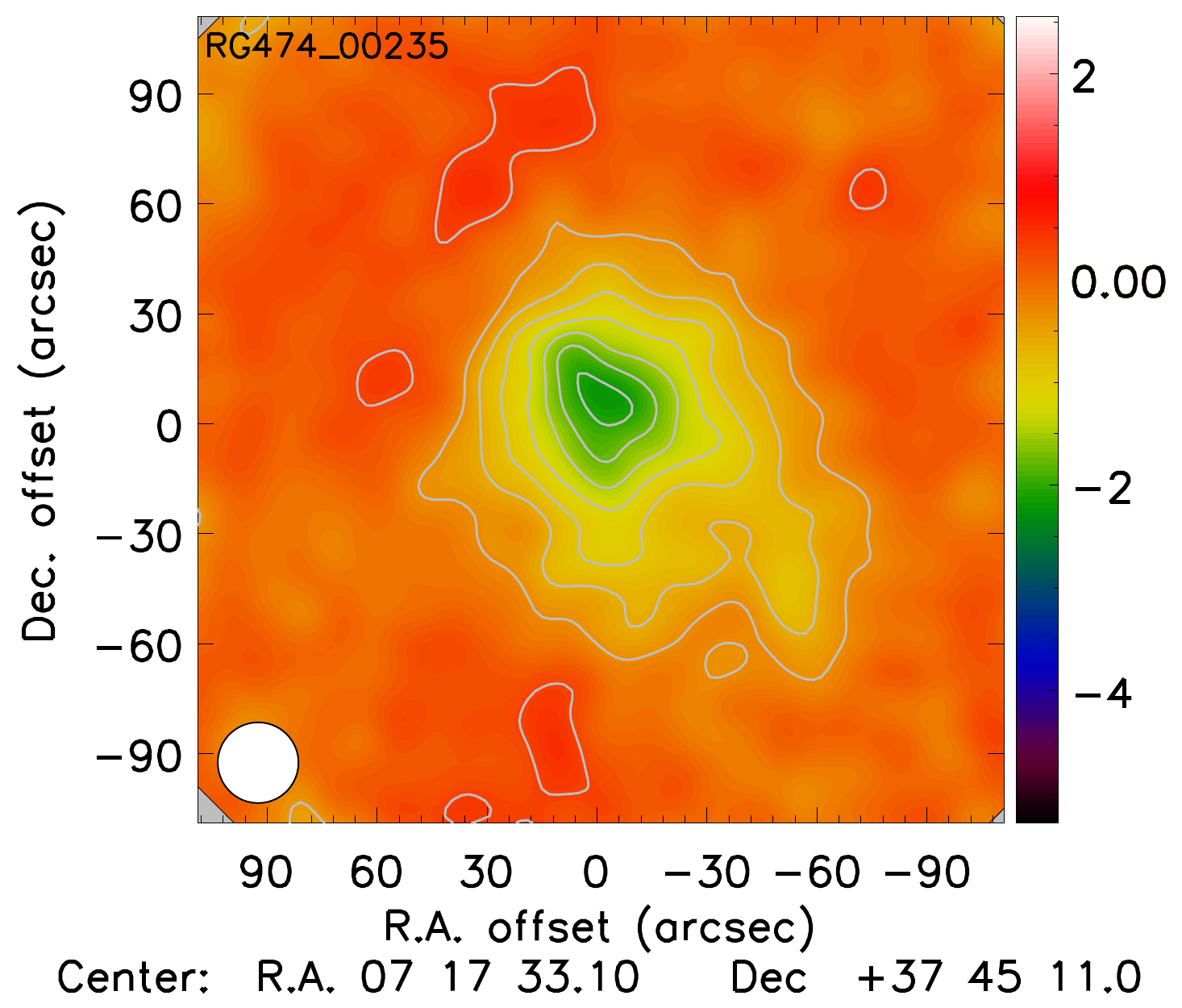} 
\put(-60,310){\makebox(0,0){\rotatebox{0}{\LARGE mJy/beam}}} &  
\includegraphics[trim=2.3cm 2.2cm 0cm 0cm, clip=true, scale=1]{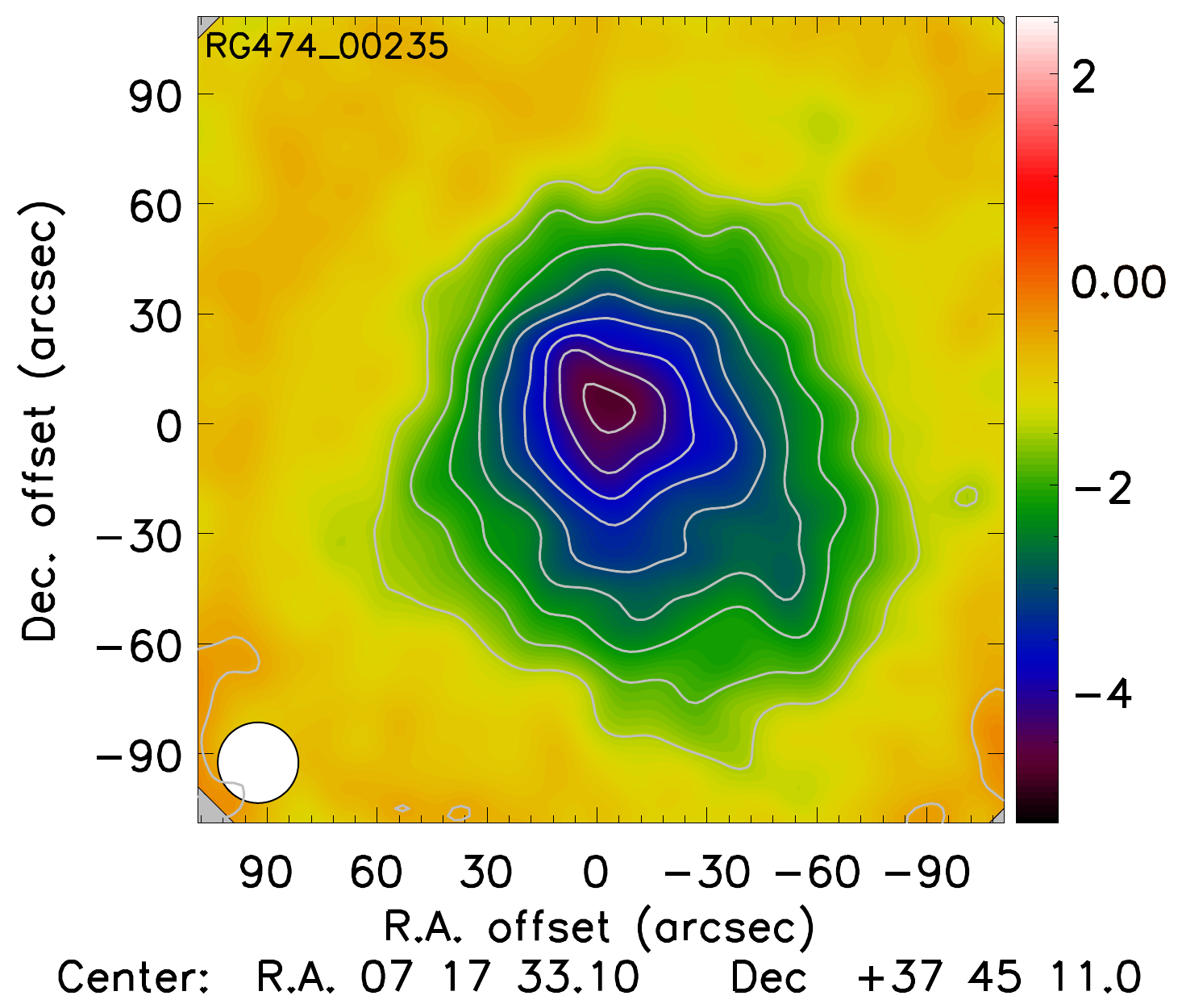} 
\put(-60,310){\makebox(0,0){\rotatebox{0}{\LARGE mJy/beam}}} & 
\includegraphics[trim=2.3cm 2.2cm 0cm 0cm, clip=true, scale=1]{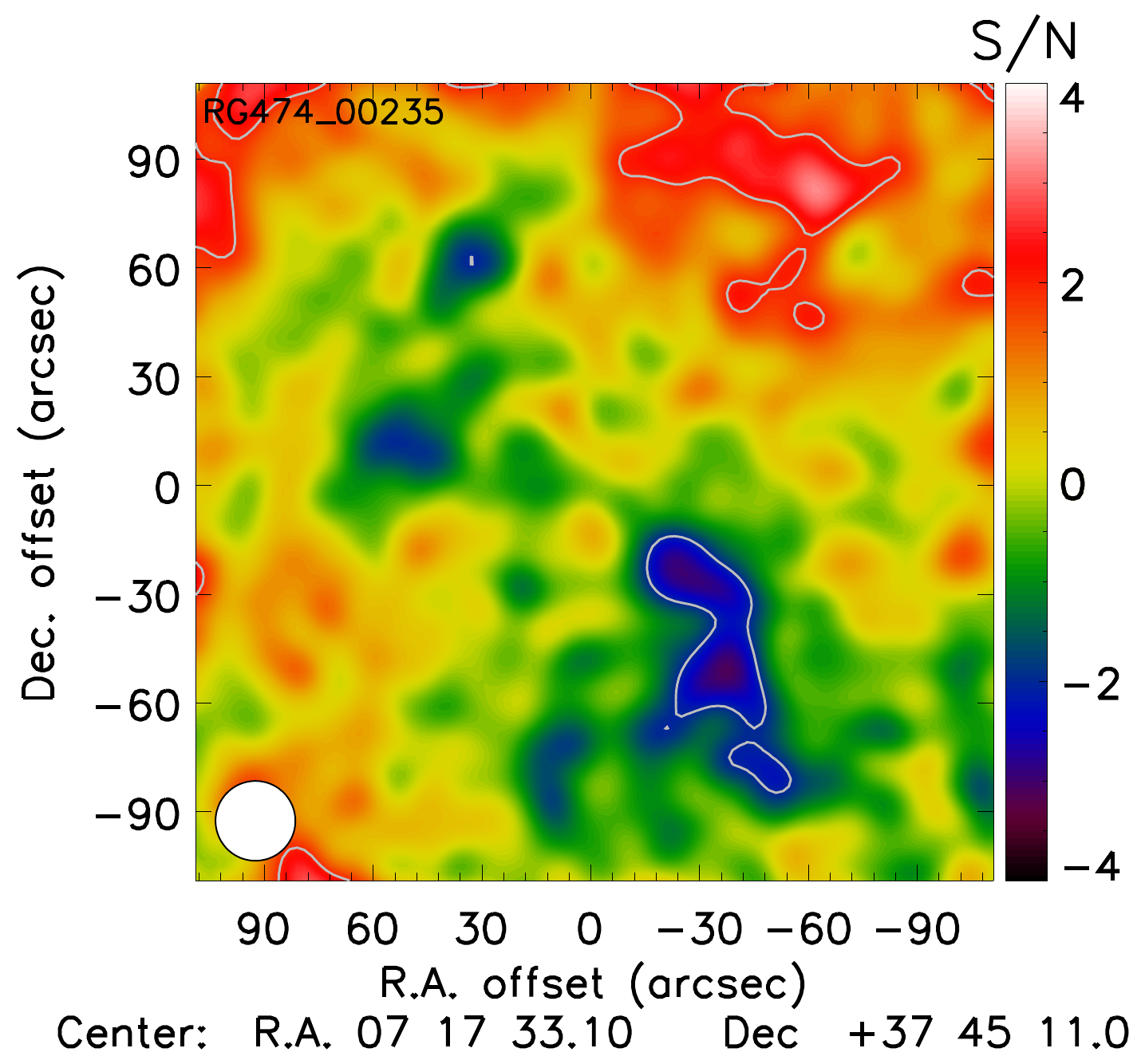}  & 
\includegraphics[trim=2.3cm 2.2cm 0cm 0cm, clip=true, scale=1]{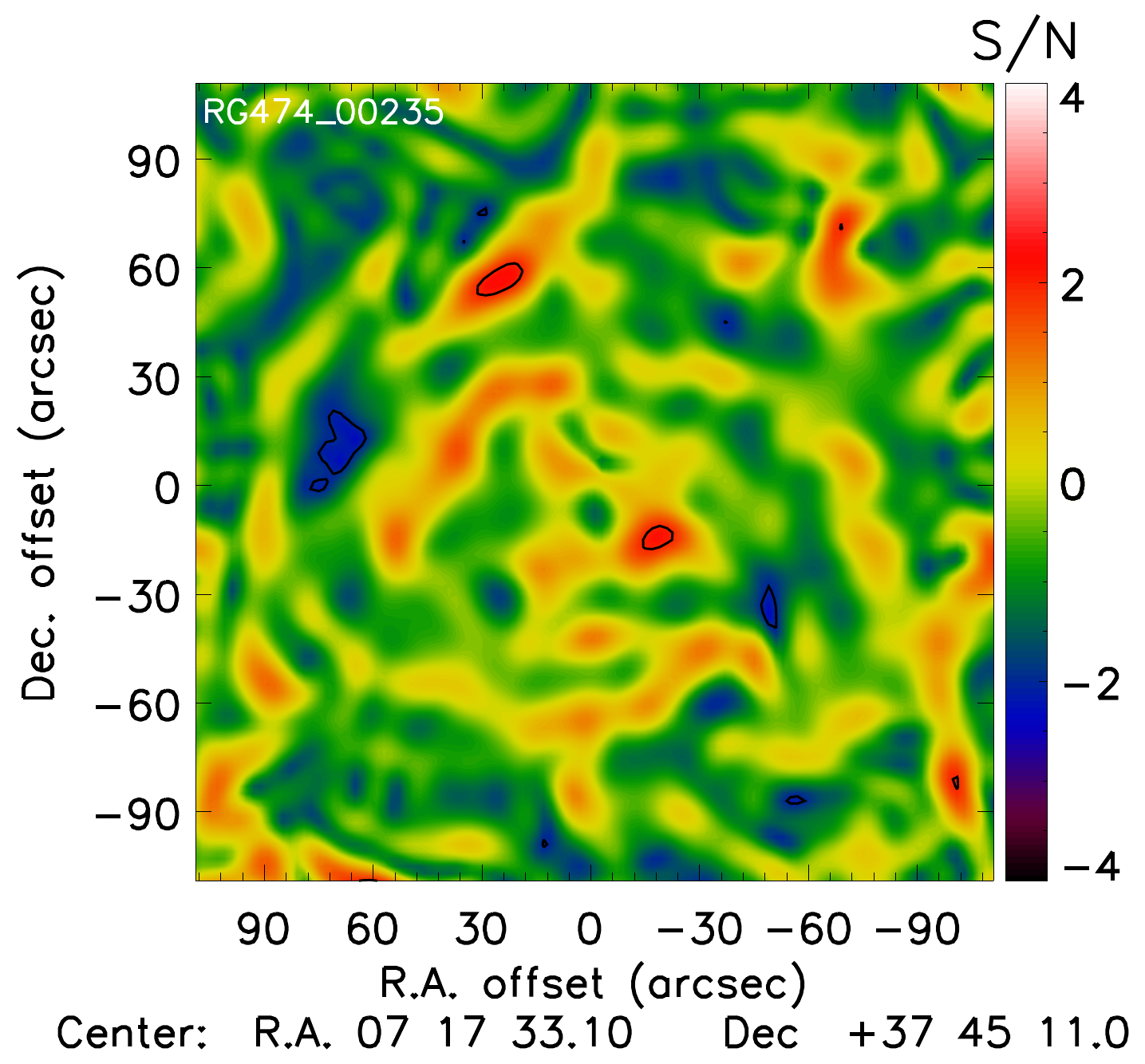}  & 
\includegraphics[trim=2.3cm 2.2cm 0cm 0cm, clip=true, scale=1]{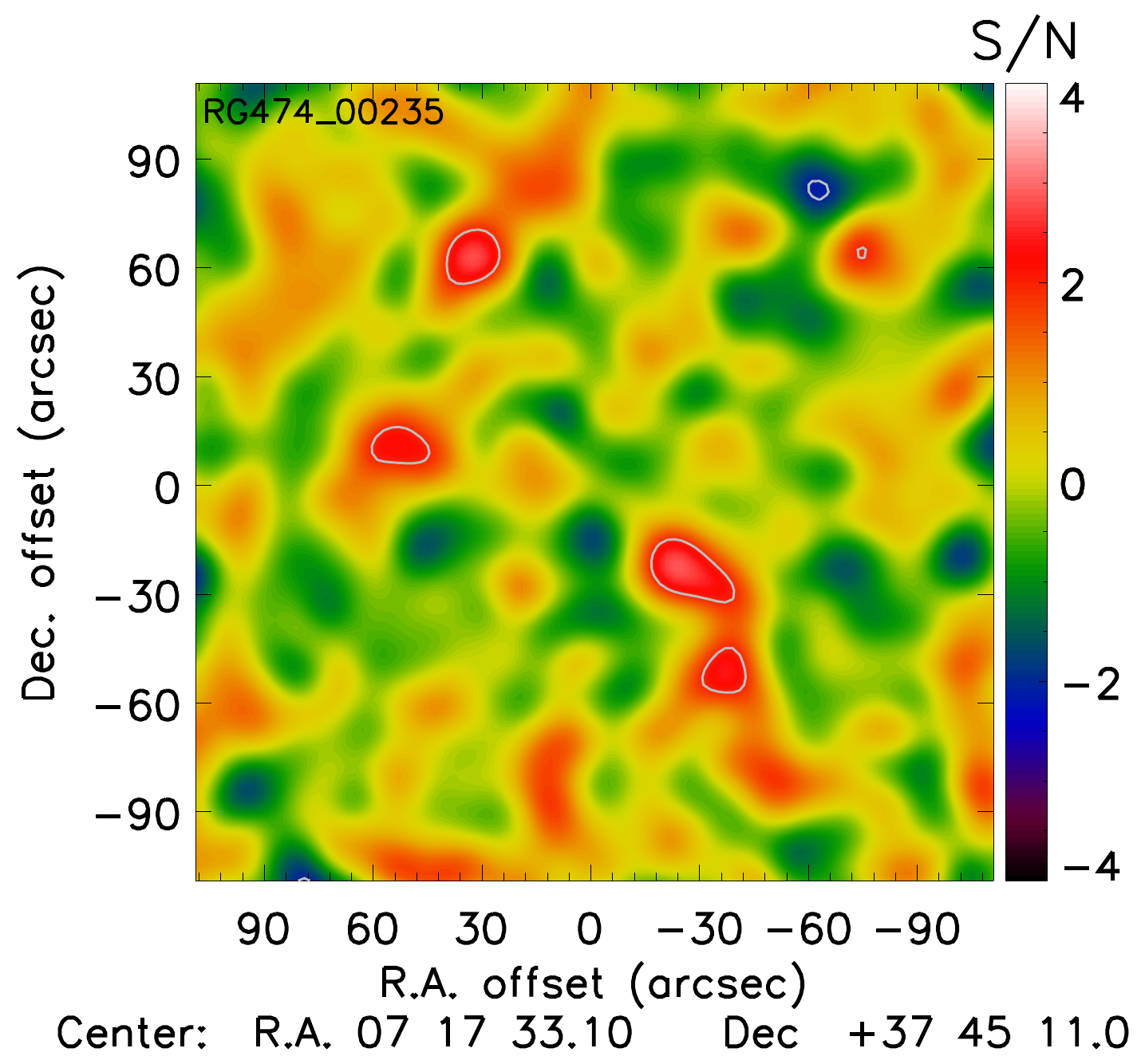} \\
\includegraphics[trim=0cm 0.7cm 0cm 0cm, clip=true, scale=1]{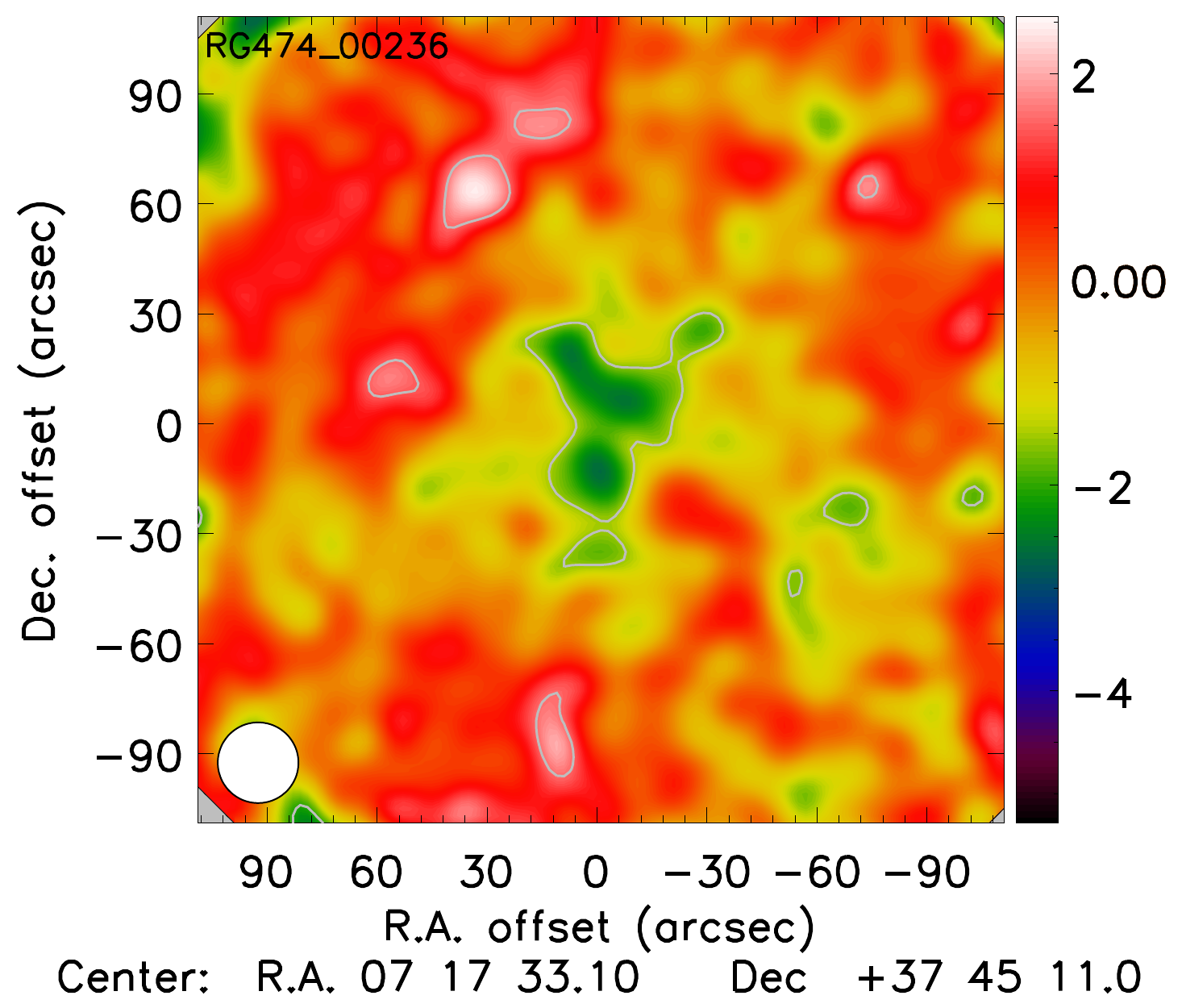} & 
\includegraphics[trim=2.3cm 0.7cm 0cm 0cm, clip=true, scale=1]{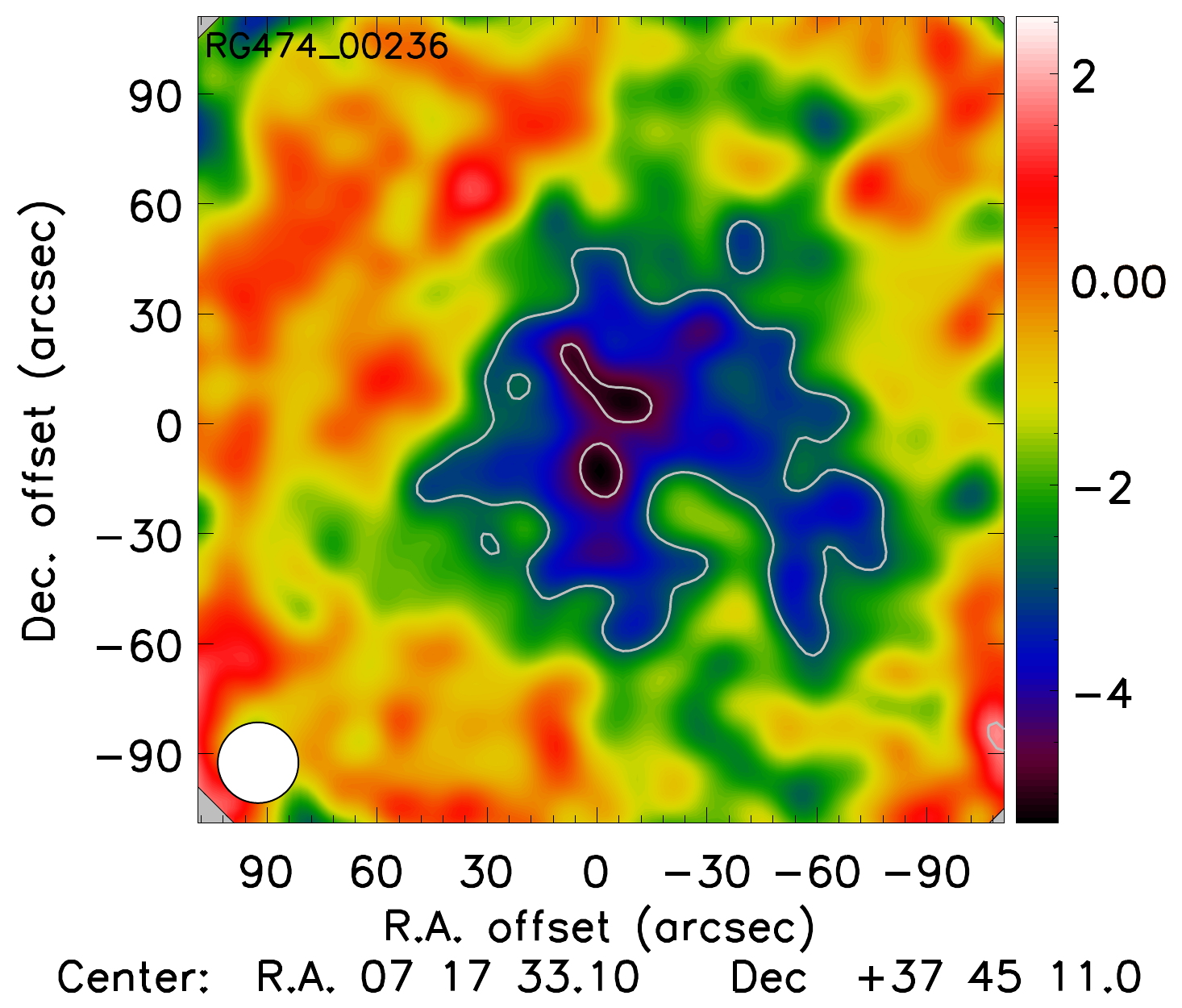} & 
\includegraphics[trim=2.3cm 0.7cm 0cm 0cm, clip=true, scale=1]{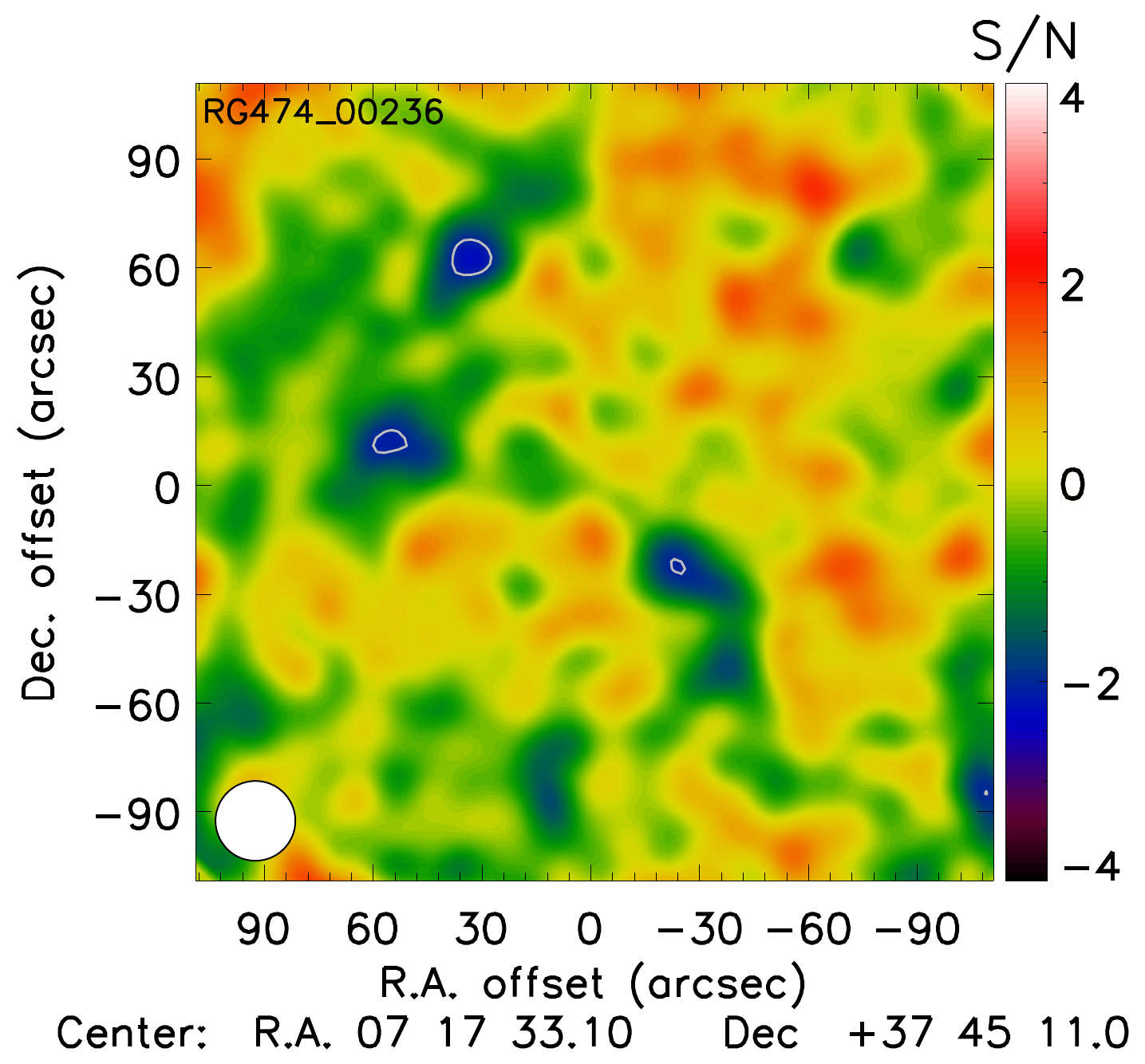} & 
\includegraphics[trim=2.3cm 0.7cm 0cm 0cm, clip=true, scale=1]{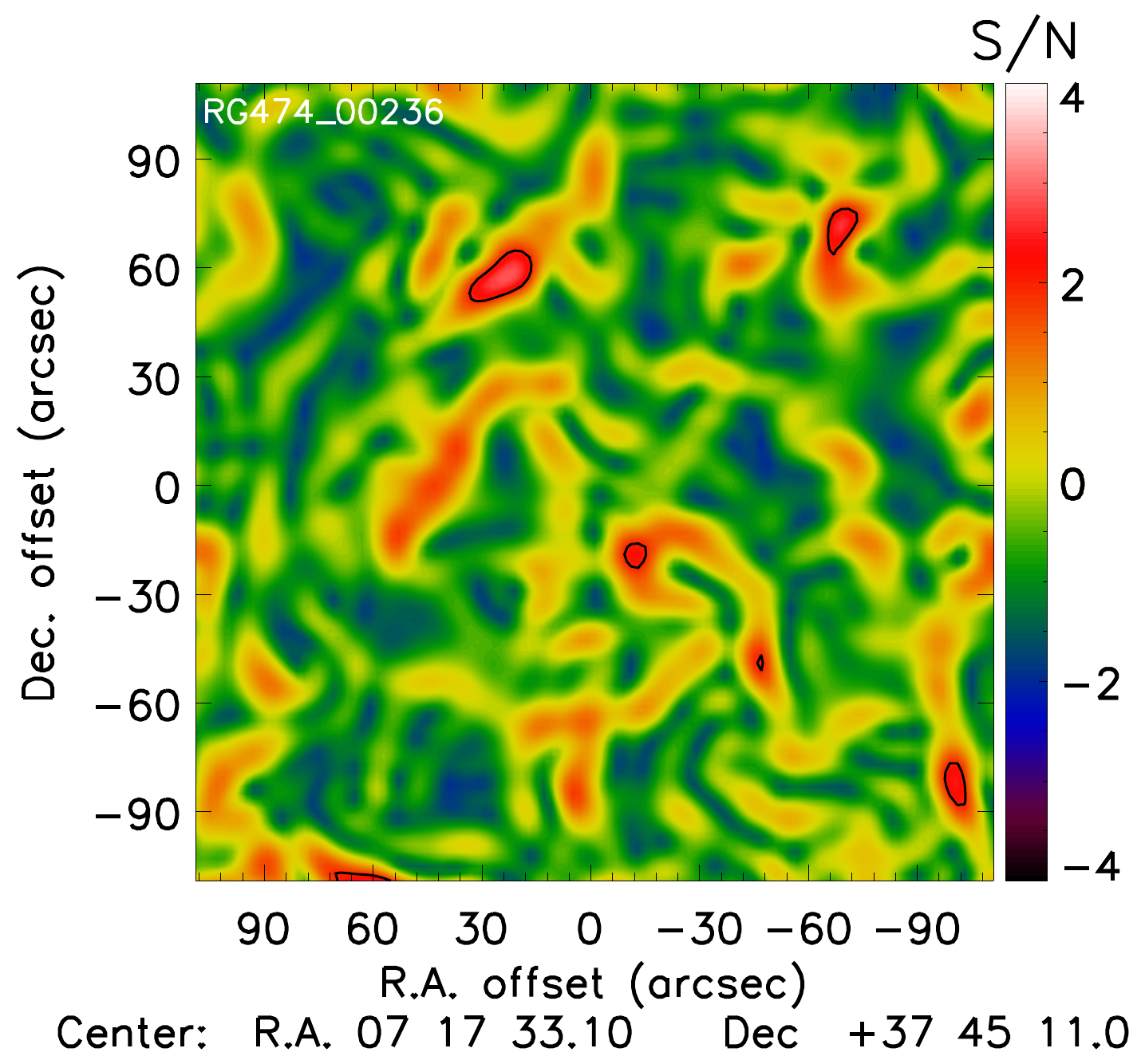} & 
\includegraphics[trim=2.3cm 0.7cm 0cm 0cm, clip=true, scale=1]{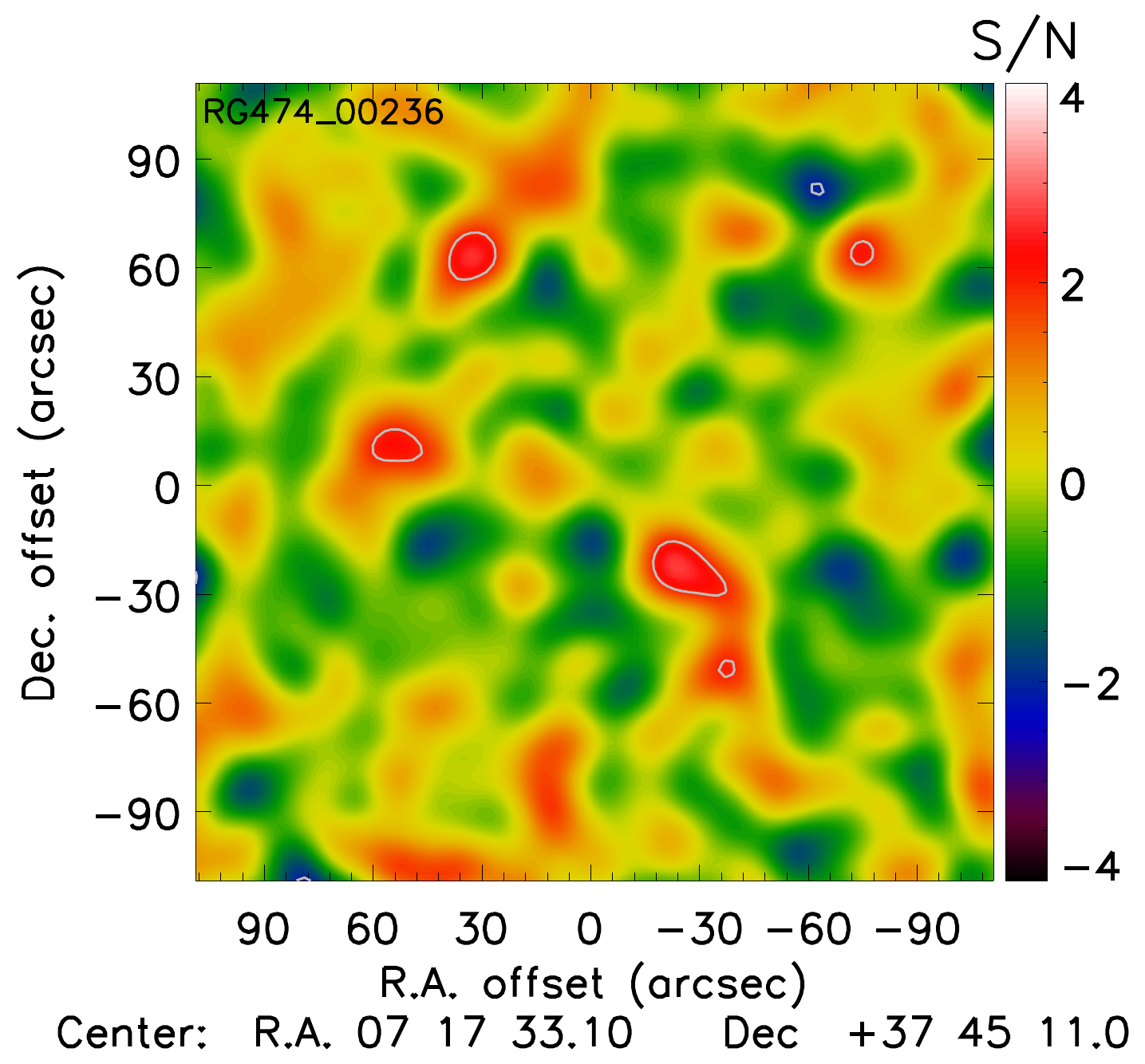} 
\end{tabular}}
\caption{\footnotesize{Effect of the deconvolution on the processed map (first two collumns) and difference between the expected signal and the recovered signal (3 last columns), in the case of the very massive and nearby test cluster RG474\_00235. The top panel correspond to the high signal-to-noise case, and the bottom panel to the low signal-to-noise case, with a peak signal-to-noise of about 20 and 3, respectively, at the 22 arcsec resolution. From left to right, we provide: the raw processed surface brightness, the deconvolved surface brightness, the deconvolved surface brightness difference map, the GGM difference map, and the DoG difference map. In all cases, contours provide the signal-to-noise ratio, starting at $\pm 2 \sigma$ and increasing by $2 \sigma$ steps.}}
\label{fig:transfer_function_effect}
\end{figure*}

In order to account for the NIKA processing transfer function, we compute the fraction of filtering, as a function of wave number $k$, by comparing input simulations with known signal (typically cluster toy models that contain signal mostly on large scales, to which we add white noise that contains signal at all scales). We refer to \cite{Adam2015,Adam2016a}, for more details on the procedure. The surface brightness images are deconvolved from the filtering by dividing the data by the transfer function in Fourier space. We also refer to the NIKA sample data release web page\footnote{\url{http://lpsc.in2p3.fr/NIKA2LPSZ/nika2sz.release.php}}, where we provide an {\tt IDL} script to account for the transfer function. The noise is boosted on large scales, and this is included in our analysis by considering the deconvolved noise Monte Carlo realization when using the deconvolved images. We note that the data processing might generally lead to bouncing artifacts around the brightest signal regions, which could propagate into the filtered maps. However, these effects are strongly mitigated by the masking of the source that is done by iteratively flagging the highest signal-to-noise regions \citep[see][]{Adam2015}. Such artifacts remain insignificant with respect to the noise, as shown in Figure \ref{fig:transfer_function_effect} and discussed below. Further details and more illustrations about the data reduction can be found in the Ph.D. thesis by \cite{Adam2015Thesis}, in particular in chapter 7.

The two first columns of Figure \ref{fig:transfer_function_effect} show the processed surface brightness images before and after deconvolution, in the case of the relatively nearby and very massive test case RHAPSODY-G cluster RG474\_00235. This cluster is chosen because of its large angular size on the sky, such that it is the most affected by transfer function effects, and corresponds to an upper limit case with respect to the NIKA clusters. The top row corresponds to a case with signal-to-noise ratio peaking at about 20 on the surface brightness, at 22~arcsec resolution (scaled to the best NIKA detection), and the bottom row to a peak signal-to-noise ration of 4 (the lowest NIKA detection). The three last columns of Figure \ref{fig:transfer_function_effect} provide the difference maps between the processed data and the true expected maps (see Figure \ref{fig:RG_cluster_sample}), for the deconvolved surface brightness, the GGM filter and the DoG filter. The difference maps are provided in terms of signal-to-noise ratio.

As shown in Figure \ref{fig:transfer_function_effect}, the differences between the maps are consistent with noise. In the surface brightness map, most of the residual power is on large scales. This is because the noise is boosted by the deconvolution, mainly on large scales, but also because of the small inaccuracies in the transfer function that lead to differences between the true initial map and the one we recover. The latter is particularly true for this test cluster, in the high signal-to-noise case, because the signal is very large on large scales, where the filtering of the transfer function is important. For the low signal-to-noise case, the effect of errors in the deconvolution has the same amplitude, but is subdominant compared to the noise that is larger. In the case of the GGM and the DoG filters, the scales that we consider are smaller than those where the deconvolution is critical and thus, the maps are not affected significantly by the deconvolution. Both residual maps do not show any structure above the noise, even in the high signal-to-noise regime. In the low signal-to-noise regime, the GGM filter map presents more structure (still well below detection limits) that arise because of inaccuracies in the input map, which is used in the filter ($\hat{S}$ in equation \ref{eq:GGM_filter_noise}), as discussed in Section \ref{sec:Noise_spatial_correlations_and_propagation_through_the_filters} and shown in Figure \ref{fig:noise_statistics2}.

In summary, the systematic effects introduced by the deconvolution are negligible with respect to the noise, even in the highest signal-to-noise cases available from the NIKA sample. In the low signal-to-noise regime, the deconvolution does not introduce any uncontrolled artifacts. Therefore, systematics associated to the transfer function can be safely neglected in the main paper analysis.

\section{Impact of kSZ signal in \mbox{MACS~J0717.5+3745}}\label{sec:Impact_of_kSZ}
\begin{figure*}[h]
\centering
\resizebox{0.75\textwidth}{!} {
\begin{tabular}{lll}
\includegraphics[trim=0cm 0.7cm 0cm 0cm, clip=true, scale=1]{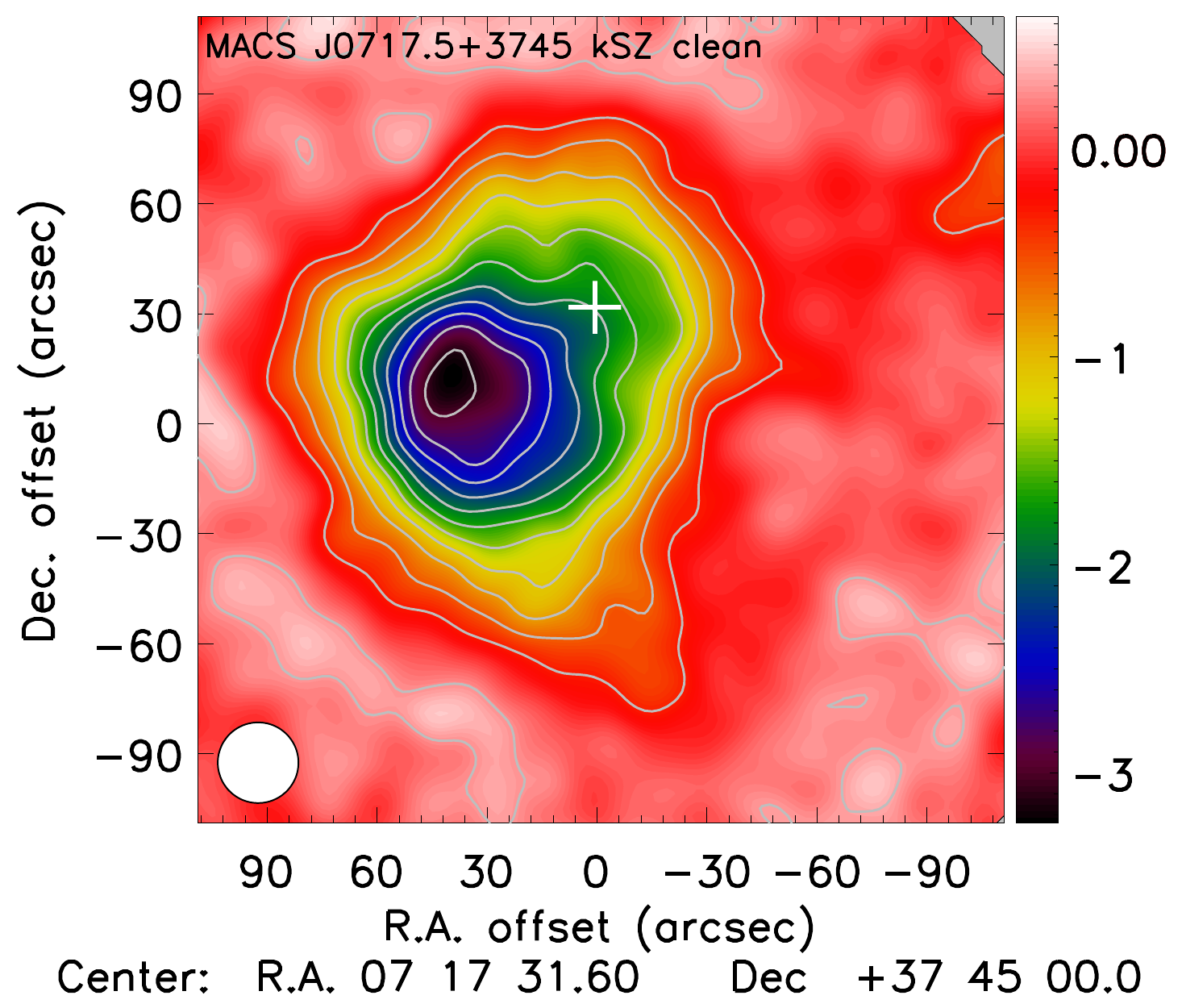}  
\put(-60,350){\makebox(0,0){\rotatebox{0}{\LARGE mJy/beam}}} & 
\includegraphics[trim=2.3cm 0.7cm 0cm 0cm, clip=true, scale=1]{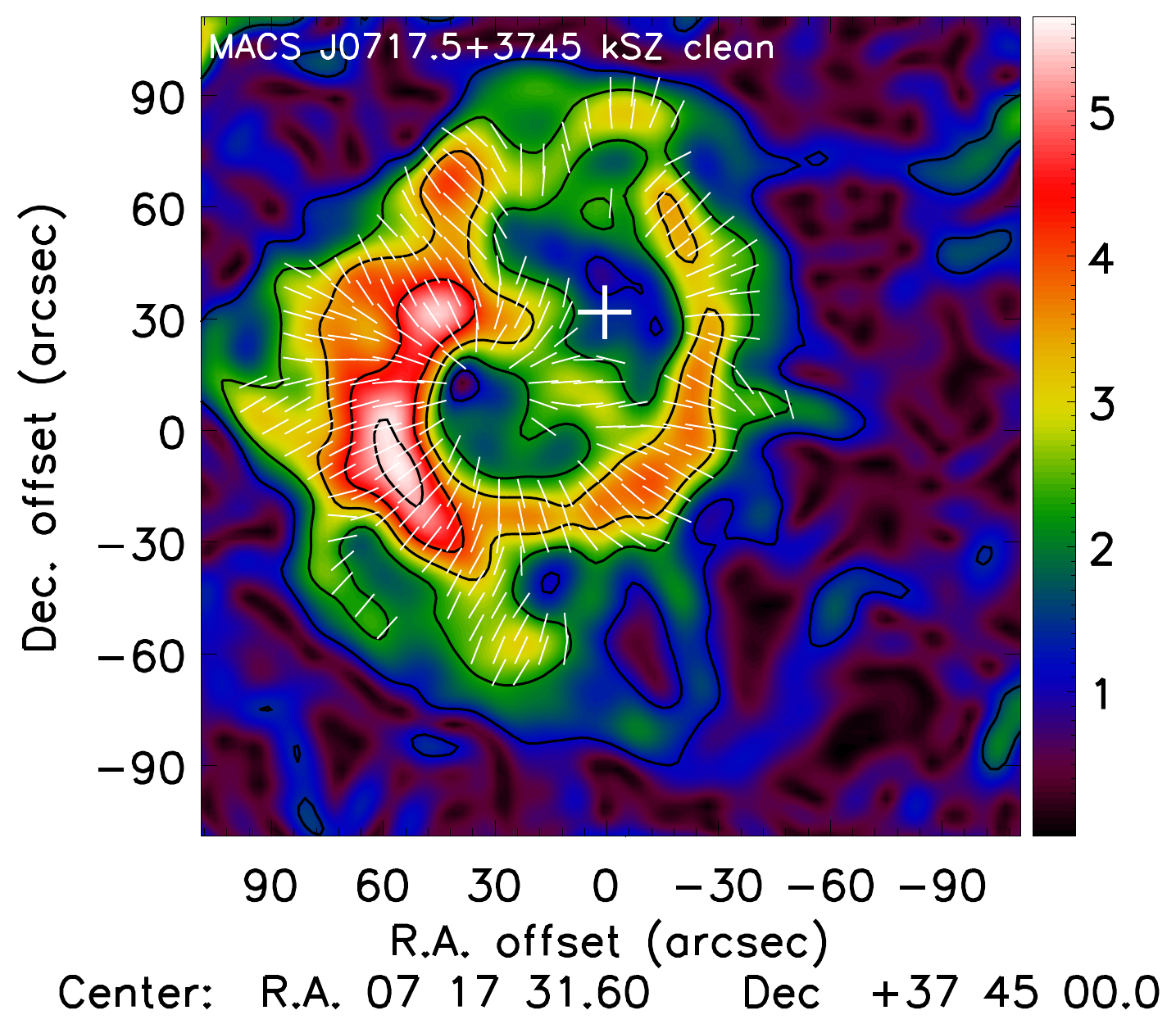}  
\put(-60,350){\makebox(0,0){\rotatebox{0}{\LARGE mJy/beam/arcmin}}} & 
\includegraphics[trim=2.3cm 0.7cm 0cm 0cm, clip=true, scale=1]{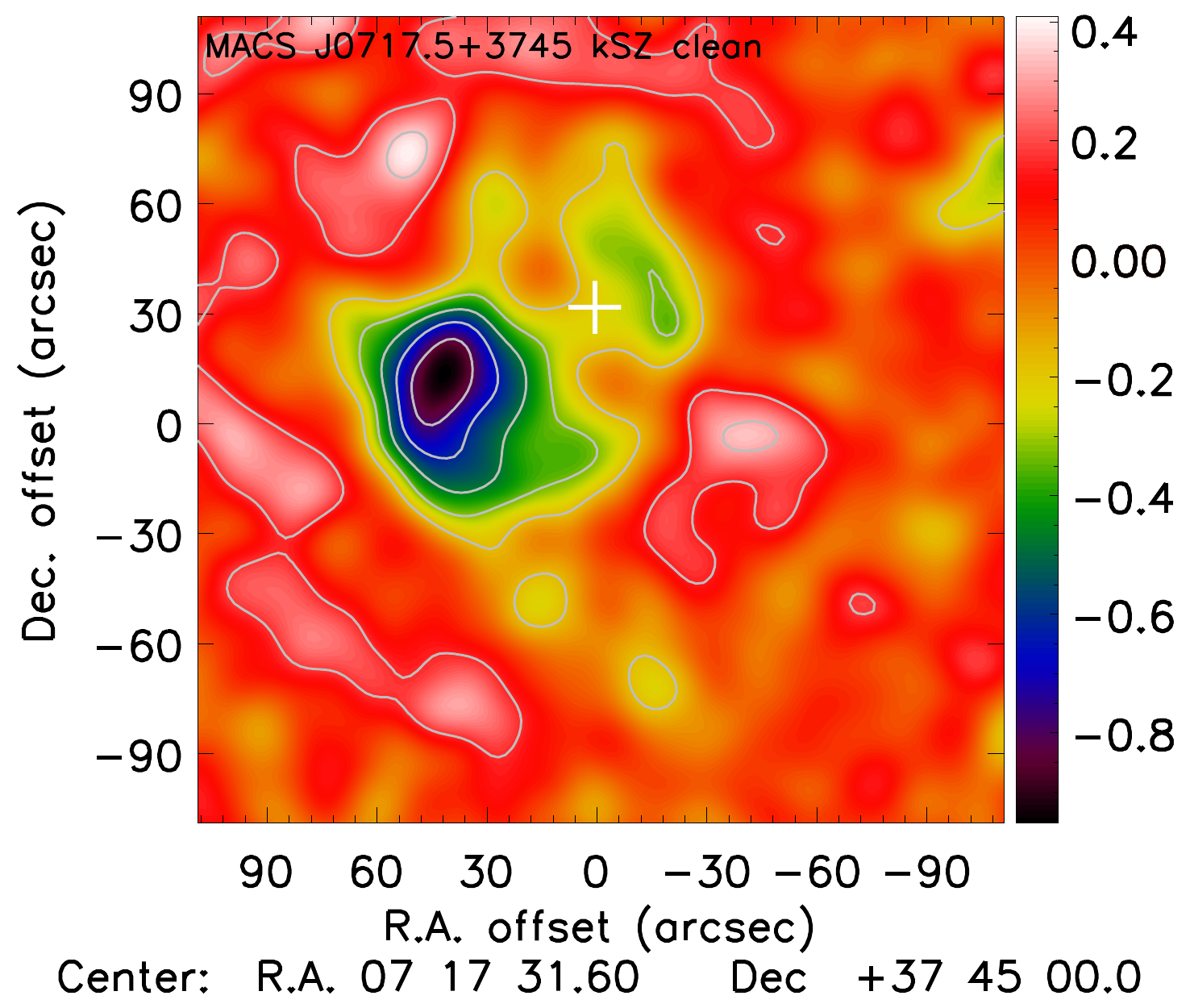}  
\put(-60,350){\makebox(0,0){\rotatebox{0}{\LARGE mJy/beam}}}
\end{tabular}}
\caption{\footnotesize{Surface brightness (left), GGM (middle) and DoG (right) maps of \mbox{MACS~J0717.5+3745} in the case of kSZ correction applied. See figure \ref{fig:NIKA_cluster_sample} for further details.}}
\label{fig:MACSJ0717_kSZ}
\end{figure*}

The surface brightness of \mbox{MACS~J0717.5+3745} is affected by kSZ signal, and it is the only cluster for which an individual kSZ detection has been obtained to date \citep[][]{Mroczkowski2012,Sayers2013,Adam2016b}. The kSZ signal is related to the gas density and line-of-sight velocity and has to be considered in order to study the pressure structure via the tSZ signal. \cite{Mroczkowski2012}, \cite{Sayers2013} and \cite{Adam2016b} have obtained several kSZ signal models, but they are all limited by the large uncertainties in their best-fit parameters and are subject to strong assumptions in the gas modeling. Therefore, it is only possible to test the impact of the kSZ signal on our result by comparing the recovered sub-structures in the cases with and without kSZ correction. We thus produce a kSZ clean map of \mbox{MACS~J0717.5+3745} using the best-fit kSZ model from \cite{Adam2016b} to do so. Figure \ref{fig:MACSJ0717_kSZ} provides the surface brightness and filtered maps of \mbox{MACS~J0717.5+3745} in the case where the kSZ signal model has been subtracted to the data. The overall structure is similar to that observed in Figure \ref{fig:NIKA_cluster_sample}, where no correction is applied, but the magnitude of the different structures is significantly affected. See Section \ref{sec:MACSJ0717} for discussions.

\end{document}